\documentclass[11pt,a4paper]{article}

\usepackage{jheppub}
\usepackage[usenames,dvipsnames]{xcolor}
\usepackage[mathscr]{eucal}
\usepackage{amssymb} 
\usepackage{MnSymbol}
\usepackage{pgfplots}
\pgfplotsset{compat=1.17}
\usetikzlibrary{pgfplots.fillbetween}
\usepackage{amsmath}
\usepackage{mathtools}
\usepackage{amsfonts}  
\usepackage{dsfont}
\usepackage{pdfpages}
\usepackage{verbatim}
\usepackage{stmaryrd}
\usepackage{graphicx}
\usepackage{tabularx}        
\usetikzlibrary{positioning}
\usepackage{hhline}
\usepackage{multirow}
\usepackage[most]{tcolorbox}
\usepackage{tikzlings}
\usetikzlibrary{decorations.markings,arrows.meta}
\usetikzlibrary{matrix}

\usepackage{tensor}
\usepackage{textgreek} 
\usepackage[smalltableaux]{ytableau}
\ytableausetup{centertableaux}
\usepackage{tikz}
\usetikzlibrary{decorations.pathreplacing, decorations.markings,calc,shapes.misc,decorations.pathmorphing,patterns.meta, math}
\usetikzlibrary{decorations.pathmorphing}
\usetikzlibrary{decorations.markings}
\usepackage{slashed}
\usepackage{datetime}
\usepackage{hyperref}
\usepackage{braket}
\hypersetup{
    pdfencoding=unicode,
	colorlinks=true,
	urlcolor=Maroon,
	linkcolor=RoyalBlue,
	citecolor=Maroon,
	pdftitle={Notes on de Sitter space},
	pdfauthor={Beatrix M\"uhlmann},
	pdfdisplaydoctitle=true,
	pdfstartview=FitH,
	linktocpage=true
}
\usepackage{pgfplots}
\pgfplotsset{compat=1.17}

\usetikzlibrary{shapes}
\usetikzlibrary{arrows}
\usetikzlibrary{decorations.pathmorphing}
\usetikzlibrary{decorations.markings}
\usetikzlibrary{shapes.misc}
\tikzset{snake it/.style={decorate, decoration=snake}}
\tikzset{cross/.style={cross out, draw=black, minimum size=2*(#1-\pgflinewidth), inner sep=0pt, outer sep=0pt},
cross/.default={1pt}}
\usetikzlibrary{shapes.geometric}
\tikzset{
    partial ellipse/.style args={#1:#2:#3}{
        insert path={+ (#1:#3) arc (#1:#2:#3)}
    }
}

\usepackage{subcaption}

\newcommand{\ba}{\begin{align}}

\newcommand{\be}{\begin{equation}}
\newcommand{\ee}{\end{equation}}
\def\bd{\begin{tikzpicture}}
\def\ed{\end{tikzpicture}}

\DeclareMathOperator\tr{tr}

\renewcommand\Im{\mathop{\text{Im}}}

\renewcommand\Re{\mathop{\text{Re}}}

\allowdisplaybreaks

\def\XXint#1#2#3{{\setbox0=\hbox{$#1{#2#3}{\int}$}
     \vcenter{\hbox{$#2#3$}}\kern-.5\wd0}}

\definecolor{light-gray}{gray}{0.75}

\newcommand\arccosh{\text{arccosh}}

\renewcommand\d{\text{d}}

\newcommand{\e}{\mathrm{e}}

\newcommand{\R}{\mathcal{R}}

\renewcommand{\ge}{\geqslant}
\renewcommand{\leq}{\leqslant}
\renewcommand{\geq}{\geqslant}

\definecolor{bleudefrance}{rgb}{0.19, 0.55, 0.91}

\definecolor{vert}{rgb}{0.1367 0.543 0.1367}

\definecolor{pink}{rgb}{1.0, 0.13, 0.32}

\begin{document}

\vspace*{2.5cm}
\begin{center}
{\LARGE {\textsc{Notes On De Sitter Space}}}

\vspace*{1.7cm}

{\bf
\mbox{
Beatrix M\"uhlmann}
}

\vspace*{0.6cm}

{\footnotesize
School of Natural Sciences, Institute for Advanced Study, Princeton, NJ 08540, USA
}

\vspace*{0.4cm}

{\footnotesize\textsf{
beatrix@ias.edu
}}

\vspace*{0.6cm}
\end{center}

\vspace*{1.5cm}
\begin{abstract}
\noindent
We discuss classical and quantum features of spacetimes with a positive cosmological
constant. After introducing basic aspects of de Sitter cosmology and the
classical geometry of de Sitter space, we consider de Sitter space both
as a rigid background, on which quantum fields propagate, and as a
fluctuating spacetime in its own right. We review the wavefunction of
an expanding universe, black holes in de Sitter space, and the de Sitter entropy,
and compare the cosmological horizon with the black hole horizon.
The final three sections are dedicated to concrete models of de Sitter
space in two, three and four spacetime dimensions.

\end{abstract}

\newpage

\tableofcontents

\section{Introduction}
Establishing a theory that captures the quantum aspects of spacetimes undergoing an accelerated phase of expansion is one of the central challenges in theoretical physics. At the center of this challenge sits de Sitter space, the maximally symmetric solution of Einstein's equations 
\begin{equation}\label{eq: Einstein equations}
   \mathcal{G}_{\mu\nu}\equiv R_{\mu\nu} - \frac{1}{2}Rg_{\mu\nu}  = -\Lambda g_{\mu\nu}~,
\end{equation}
with positive cosmological constant $\Lambda>0$. Eighty years after its theoretical discovery \cite{Sitter1917OnTR}, observations of Type Ia supernovae revealed that the expansion of our Universe is accelerating \cite{SupernovaCosmologyProject:1998vns, SupernovaSearchTeam:1998fmf}, driven by a tiny, yet positive cosmological constant $\Lambda$. If $\Lambda>0$ and this acceleration persists we will ultimately live in a pure de Sitter Universe. 

As space expands toward the future, it also contracts toward the past. In the 1960s, Penzias and Wilson discovered the cosmic microwave background (CMB)—a nearly uniform radiation field with a temperature, currently measured at 2.725\,K, providing strong evidence for a hot Big Bang origin of our Universe. The CMB is interpreted as the afterglow of a hot, dense plasma that cooled as the Universe expanded, with recombination occurring at approximately \( T \approx 3000\,\text{K} \). Detailed maps of the CMB anisotropies with tiny temperature inhomogeneities at the level of $\tfrac{\delta T}{T} \sim 10^{-5}$
were first reported by the COBE satellite in 1992. These measurements were later refined with increasing precision by WMAP and the Planck satellite. The primordial spectrum of fluctuations inferred from these anisotropies is nearly scale invariant.
A leading hypothesis that explains this scale invariance is inflation \cite{Guth:1980zm, Linde:1981mu, Albrecht:1982wi}: in its earliest phase, the Universe underwent a period of accelerated expansion governed by an approximately de Sitter geometry. 

In the early Universe at very high energies, quantum fluctuations affected both matter fields and the metric itself and left imprints that later seeded the formation of large-scale structures. These fluctuations, amplified during inflation, are encoded in the CMB and the matter power spectrum we observe today. 
At the opposite extreme, in the deep infrared, the cosmological constant $\Lambda \sim 2.9\times 10^{-122} \ell_{\mathrm{{Planck}}}^{-2}$ should
 receive contributions from the zero-point energies of all quantum fields. This connection between vacuum fluctuations and gravity, first emphasized by Zeldovich in the 1960s \cite{Zeldovich:1968ehl, Weinberg:1988cp} (see also \cite{Bousso:2007gp}), gives rise to the cosmological constant problem: why do these enormous quantum contributions nearly cancel to produce such a tiny—but nonzero—value for $\Lambda$? 
A positive cosmological constant drives an exponential expansion of space to the extent that not even light can travel all around. This leads to the formation of a cosmological event horizon at a distance of roughly $16$--$17 \mathrm{Gly}$ and bounds what we can ever see or influence.

Both inflation and the current accelerated expansion of our Universe are well approximated by a pure de Sitter phase with a constant $\Lambda$. A natural and interesting question is how far one can deviate from this idealized picture. On the observational side, ongoing and upcoming experiments are probing whether the cosmological constant is truly constant, e.g.\ through the dark energy equation of state $w(z)$ \cite{DESI:2024mwx, Laureijs:2011gra, Euclid:2024yrr, DES:2024jxu}.

Although many questions about de Sitter space still await an answer, recent years have seen significant progress on quantum de Sitter
space: low-dimensional quantum gravity;
Euclidean quantum gravity and the quantum-corrected de Sitter entropy; the
representation theory of the de Sitter isometry group; the algebraic description
of the static patch and its observers; the wavefunction of the universe; and the
non-perturbative de Sitter bootstrap, to name just a few directions.

These notes are an extended version of lectures presented at the 31st W.E.~Heraeus
``Saalburg'' Summer School (Bayrischzell, September 2025), and aim to present the
basic features of de Sitter space and to combine them with a selection of these
developments. The selection inevitably reflects my own familiarity;
for directions treated in less detail, references are provided throughout. We do
not cover inflation or string compactifications, but instead refer the reader to \cite{Baumann:2009ds} and \cite{Denef:2008wq}, respectively, and references therein.
Excellent reviews on de Sitter quantum gravity include \cite{Anninos:2012qw, Galante:2023uyf, Spradlin:2001pw, Witten:2001kn, Bousso:2002fq} and in particular \cite{Anninos:2025lectures} which served as a foundation of these notes.

These notes are organized as follows. We start in section \ref{sec: cosmology} with some basics of cosmology; section \ref{sec: geometry} discusses the geometry of de Sitter space, its coordinate patches, and their Penrose diagrams. Section \ref{sec:QFT} reviews quantum field theory on a fixed de Sitter background: irreducible representations, Harish-Chandra characters, and bulk and boundary two-point functions. Section \ref{sec:dS wavefunction} discusses the wavefunction of the universe — both the Bunch–Davies and the Hartle–Hawking wavefunction — through a series of examples. Section \ref{sec:dSentropy} revisits the Gibbons–Hawking proposal that the sphere partition function computes the cosmological horizon entropy: we go through the one-loop evaluation in detail, including the heat kernel analysis and the conformal mode problem, and discuss the algebraic approach to the de Sitter entropy. Section \ref{sec:dSBH} summarizes the black hole solutions of Einstein gravity with $\Lambda>0$ and their thermodynamics, while in section \ref{sec: dS vs BH} we compare the de Sitter and black hole horizons. Finally, sections \ref{sec:2D}, \ref{sec:3D} and \ref{sec:4D} turn to explicit models: timelike Liouville theory and dilaton gravity models in two dimensions, the complex Liouville string and its dual matrix model in three, and higher-spin de Sitter holography in four. Appendix \ref{app:compendium} collects various useful equations.

\section{de Sitter Cosmology}\label{sec: cosmology}

The aim of this short section is to remind the reader of the basic properties of cosmological spacetimes. 

\subsection{FLRW metric}

A very broad class of cosmological spacetimes is described by the
Friedmann--Lemaître--Robertson--Walker (FLRW) metrics. These metrics arise naturally once
we impose the two key principles of {homogeneity} and {isotropy}. Homogeneity
requires that the universe looks the same at every spatial point. Isotropy requires that the universe looks the same in every
direction, corresponding to rotational invariance around any given point. Together, these
two symmetries uniquely determine the metric up to the choice of a scale factor $a(t)$ and
a constant $k$ that characterizes the spatial curvature. For the spatial geometry we have three options, corresponding to the three
values of the curvature parameter $k$:
\begin{equation}
   k = \begin{cases}
     +1 & \quad S^3 \quad \text{(positively curved space)} \\
     \;\;0 & \quad \mathbb{R}^3 \quad \text{(flat space)} \\
     -1 & \quad \mathbb{H}^3 \quad \text{(negatively curved space)} 
    \end{cases}
\end{equation}
The resulting line element can be
written as
\begin{equation}\label{eq:FLRW metric 1}
    ds^2 = - \d t^2 + a(t)^2 \d \Sigma_k^2~,
    \qquad 
    \d \Sigma_k^2 = \frac{\d r^2}{f(r)} + r^2 \d\Omega_2^2~,
    \qquad 
    f(r) = 1-k r^2~,
\end{equation}
where $a(t) >0$ is the scale factor, which encodes the expansion or contraction of the universe and is determined dynamically by the matter content.

The guiding assumption of modern cosmology, often called the {Cosmological Principle},
states that there exists at each spacetime point a hypothetical observer for whom the
universe appears isotropic and homogeneous. Such observers are referred to as
{comoving observers}. In a comoving coordinate system these
observers remain at rest, meaning that their spatial coordinates do not change in time.
Physically, this means that comoving coordinates expand together with the universe, following the evolution of the scale factor $a(t)$.
The scale factor is related to the {Hubble parameter} as follows
\begin{equation}
   H(t) \equiv \frac{\dot{a}(t)}{a(t)}~.
\end{equation}
The {cosmological redshift} is defined by
\begin{equation}
    z = \frac{a(t_0)}{a(t_\mathrm{em})}-1 
    \approx \left(1+ \frac{\dot{a}(t_0)}{a(t_0)}(t_0-t_\mathrm{em}) + \ldots \right)-1 
    \approx  H_0 \delta t~,\quad \delta t\equiv t_0 - t_{\mathrm{em}}
\end{equation}
where $t_{\mathrm{em}}$ denotes the time at which the light was emitted; $H_0 \approx 67.4 \,\mathrm{km}\,\mathrm{s}^{-1}\,\mathrm{Mpc}^{-1}$ (with
$1\,\mathrm{Mpc}\approx 3.09 \times 10^{19}\,\mathrm{km}$).  If we normalize the
scale factor today as $a(t_0)=1$, then
\begin{equation}
    a(t_\mathrm{em}) = \frac{1}{1+z}~.
\end{equation}
For example, nearby galaxies with $z\sim 0.01$ correspond to 
\begin{equation}
    a(t_{\mathrm{em}}) \sim 0.99~,
\end{equation}
meaning that the universe was roughly $99\%$ of its present size at the time of emission.
By contrast, at the epoch of recombination, corresponding to $z\approx 1100$, we find
\begin{equation}
    a(t_{\mathrm{em}}) \sim 9\times 10^{-4}~,
\end{equation}
so the universe was only about $0.1\%$ of its current size.

\paragraph{FLRW equations.}
The Einstein equations with matter are
\begin{equation}
    \mathcal G_{\mu\nu} + \Lambda g_{\mu\nu}= 8\pi G_N T_{\mu\nu}~.
\end{equation}
For the stress–energy tensor $T_{\mu\nu}$ to be compatible with homogeneity and isotropy it needs to take the form
\begin{equation}
   T_{\mu\nu} = \begin{pmatrix}
    \rho(t) & 0 & 0 & 0 \\
    0 & & & \\
    0 & \multicolumn{3}{c}{{p(t)}{a(t)^2} \tilde{g}_{ij}} \\    
    0 & & &
\end{pmatrix}~,
\end{equation}
where $\tilde{g}_{ij}$ denotes the time independent metric on the spatial three-manifold $\Sigma_k$ (\ref{eq:FLRW metric 1}).
The Einstein equations then reduce to two independent relations, known as the FLRW equations which are given by:
\begin{subequations}\label{eq:FLRW}
    \begin{align}\label{eq:FLRW1}
        \dot{\rho} +3\frac{\dot{a}}{a}(p+\rho) &=0~,\\ \label{eq:FLRW2}
        -\left(\frac{\dot{a}}{a}\right)^2 + \left(\frac{8\pi G_N}{3}\rho - \frac{k}{a^2} + \frac{\Lambda}{3}\right)&=0~.
    \end{align}
\end{subequations}
Equation~\eqref{eq:FLRW1} expresses the conservation law $\nabla^\mu T_{\mu\nu}=0$. 
Equation~\eqref{eq:FLRW2}
is the Hamiltonian constraint. Its structure reflects the fact that in general relativity,
time translations are part of diffeomorphism invariance and thus appear as a constraint
rather than as an independent equation of motion. Differentiating (\ref{eq:FLRW2}) with respect to $t$ and using (\ref{eq:FLRW1}) we obtain
\begin{equation}\label{eq: acceleration equation}
    \frac{\ddot a}{a} = -\frac{4\pi G_N}{3}(\rho+3p) + \frac{\Lambda}{3}~.
\end{equation}
Note in particular that the curvature $k$ drops out.

\paragraph{Matter matters.}
Different types of matter lead to different behaviors of the scale factor $a(t)$. For example,
\begin{subequations}
    \begin{align}
        \text{dust (pressureless matter):} \qquad &p =0 \quad \Rightarrow \rho \sim a^{-3}~,\\
        \text{radiation:} \qquad &p =\tfrac{\rho}{3} \quad \Rightarrow \rho \sim a^{-4}~,\\
        \text{vacuum energy:} \qquad &p =-\rho \quad \Rightarrow \rho = \text{constant}~.       
    \end{align}
\end{subequations}
Neglecting the $ka^{-2}$ curvature term in \eqref{eq:FLRW2}, we obtain the approximate solutions
\begin{equation}
    a_{\mathrm{rad}}(t)= \left(\frac{t}{t_0}\right)^{1/2}~,\qquad a_{\mathrm{dust}}(t) = \left(\frac{t}{t_0}\right)^{2/3}~
\end{equation}
for the scale factor. The dominant matter contribution of the Universe thus influences its expansion rate. Another important solution is the Milne solution $a_{\mathrm{Milne}}(t) = t$. It corresponds to a purely curvature driven solution with $k=-1$ and $\rho=0$.

A more general parametrization is obtained by introducing an equation of state $p=w\rho$. Assuming a constant $w$ and plugging this into the FLRW equation (\ref{eq:FLRW1}) we obtain
\begin{equation}
    \rho \propto a^{-3(1+w)}~.
\end{equation}
Examples are (see figure \ref{fig: FLRW data})
\begin{align}
    \mathrm{radiation}~\rightarrow~w = \frac{1}{3}~, \quad \quad &\mathrm{dust}~\rightarrow~w = 0~,\cr
    \mathrm{cosmological~constant}~\rightarrow ~w =-1~,\quad \quad &\mathrm{inflation}~\rightarrow~w \approx -1~.
\end{align}
In the $\Lambda\mathrm{CDM}$ model of modern cosmology $w=-1$. In general we have $\rho = \sum_i \rho_i$ each with its own equation of state parameter $w_i$:
\begin{equation}
    \rho_i (a)= \rho_i(t_0)\left(\frac{a_0}{a}\right)^{3(1+w_i)}~,\quad \Omega_i = \frac{\rho_i(t_0)}{\rho_{\mathrm{crit}}}~,\quad \rho_{\mathrm{crit}} =\frac{3H_0^2}{8\pi G_N} \approx 8.5 \times 10^{-27}\mathrm{kg\, m}^{-3}~,
\end{equation}
where $i \in \{\mathrm{rad}, \mathrm{dust}, \Lambda\}$ and $\rho_\Lambda \equiv \Lambda/8\pi G_N$. 
\begin{figure}[ht]
\centering
\begin{tikzpicture}
\matrix (m) [
  matrix of nodes,
  nodes={minimum height=8.4mm, anchor=center, inner xsep=3mm},
  column sep=2mm, row sep=0mm,
  column 1/.style={nodes={minimum width=47mm}},
]{
  component & $w_i$ & $3(1+w_i)$ & dilution & today \\
  vacuum energy ($\Lambda$) & $-1$ & $0$ & constant & $\Omega_\Lambda\approx 0.685$ \\
  dust (baryons $+$ CDM) & $0$ & $3$ & $(a_0/a)^3$ & $\Omega_{\mathrm{dust}}\approx 0.315$ \\
  radiation & $1/3$ & $4$ & $(a_0/a)^4$ & $\Omega_{\mathrm{rad}}\approx 9\times 10^{-5}$ \\
};
\draw[line width=0.5pt] (m.west |- m-1-1.south) -- (m.east |- m-1-1.south);
\end{tikzpicture}
\caption{Summary of the equation of state parameter $w_i$, dilution and current value for various components of our Universe.}
\label{fig: FLRW data}
\end{figure}
Moreover curvature enters the FLRW equations as the term $-k/a^2$, not as matter. We use the definition
\begin{equation}
    \Omega_k \equiv -\frac{k}{(a_0H_0)^2}~
\end{equation}
for bookkeeping. From the FLRW equation (\ref{eq:FLRW2}) it follows that
\begin{equation}\label{eq:density}
    H(a)^2 = H_0^2\left(\Omega_\Lambda + \Omega_{\mathrm{rad}}\left(\frac{a_0}{a}\right)^4 + \Omega_{\mathrm{dust}}\left(\frac{a_0}{a}\right)^3 + \Omega_k \left(\frac{a_0}{a}\right)^2\right)~,\quad a_0 \equiv a(t_0)~.
\end{equation}
Evaluating this at $a=a_0$, where $H=H_0$ by definition we obtain
\begin{equation}
    \Omega_\Lambda +\Omega_{\mathrm{rad}} + \Omega_{\mathrm{dust}} + \Omega_k =1~.
\end{equation}
The crucial feature of \eqref{eq:density} is that the $\Omega_\Lambda$ term comes with no power of $a_0/a$: the vacuum energy density $\rho_\Lambda$ does not dilute with expansion, whereas all other contributions do. As $a(t)$ increases, the vacuum component therefore eventually dominates.

\paragraph{Positive $\Lambda$.}
In four spacetime dimensions, the empty ($\rho=p=0$) FLRW solutions with $\Lambda = 3/\ell^2>0$ are
\begin{subequations}
\begin{align}
   k=+1:\quad  \frac{ds^2}{\ell^2} &= - \d \tau^2 + \cosh^2(\tau) \d\Omega_3^2~,\\ \label{eq: conformal coordinates}
   k=0:\quad   \frac{ds^2}{\ell^2} &= - \d \mathsf{t}^2 + \e^{2\mathsf{t}} \d \vec{x}^2~,\\
   k=-1:\quad  \frac{ds^2}{\ell^2} &= - \d T^2 + \sinh^2(T) \d H_3^2~,
 \end{align}   
\end{subequations}
where $\d\Omega_3^2$, $\d\vec{x}^2$ and $\d H_3^2$ are the line elements on $S^3, \mathbb{R}^3$ and $\mathbb{H}^3$ respectively. All three describe the same spacetime, de Sitter space, in different coordinates; only the $k=+1$ slicing covers it globally, as we will see in the next section.

\section{de Sitter Geometry}\label{sec: geometry}
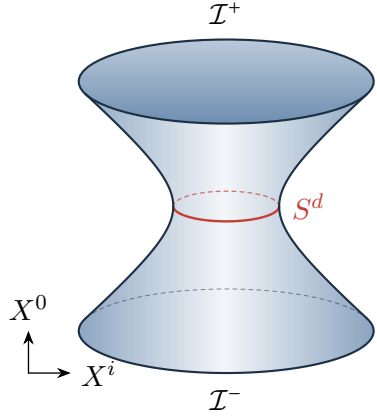
\begin{figure}[ht]
\centering
\begin{tikzpicture}[scale=.7, line cap=round, line join=round, >=Stealth]
  \definecolor{dsblue}{RGB}{62,104,150}
  \definecolor{dsred}{RGB}{196,64,52}
  \pgfmathsetmacro{\e}{0.30}                 
  \pgfmathsetmacro{\c}{sqrt(1-\e*\e)}         
  \pgfmathsetmacro{\tn}{\e/\c}                
  \pgfmathsetmacro{\ctwo}{\c*\c-\e*\e}        
  \pgfmathsetmacro{\kv}{0.95}                 
  \pgfmathsetmacro{\Z}{2.6}                  
  \pgfmathsetmacro{\R}{sqrt(1+\Z*\Z)}         
  \pgfmathsetmacro{\V}{\Z*\ctwo/\c}           
  \pgfmathsetmacro{\aZ}{\Z*\tn/\R}            
  \pgfmathsetmacro{\sT}{asin(\aZ)}            
  \pgfmathsetmacro{\vtop}{\kv*\Z*\c}          

  \shade[top color=dsblue!35, bottom color=dsblue!75]
    (0,\vtop) ellipse [x radius=\R, y radius=\kv*\R*\e];

  \shade[left color=dsblue!60, right color=dsblue!60, middle color=dsblue!6]
      plot[domain=180+\sT:360-\sT, samples=80, variable=\s]
        ({\R*cos(\s)}, {\kv*(\R*\e*sin(\s)+\Z*\c)})
   -- plot[domain=\V:-\V, samples=80, variable=\t]
        ({sqrt(1+\t*\t/\ctwo)}, {\kv*\t})
   -- plot[domain=360+\sT:180-\sT, samples=80, variable=\s]
        ({\R*cos(\s)}, {\kv*(\R*\e*sin(\s)-\Z*\c)})
   -- plot[domain=-\V:\V, samples=80, variable=\t]
        ({-sqrt(1+\t*\t/\ctwo)}, {\kv*\t})
   -- cycle;

  \draw[dsred, line width=0.5pt, dash pattern=on 1.5pt off 1.5pt, opacity=0.7]
    plot[domain=0:180, samples=70, variable=\s] ({cos(\s)}, {\kv*\e*sin(\s)});
  \draw[dsred, line width=0.9pt]
    plot[domain=180:360, samples=70, variable=\s] ({cos(\s)}, {\kv*\e*sin(\s)});

  \draw[dsblue!45!black, line width=0.8pt]
    (0,\vtop) ellipse [x radius=\R, y radius=\kv*\R*\e];
  \draw[dsblue!45!black, line width=0.35pt, dash pattern=on 1.5pt off 1.5pt, opacity=0.55]
    plot[domain=\sT:180-\sT, samples=70, variable=\s]
      ({\R*cos(\s)}, {\kv*(\R*\e*sin(\s)-\Z*\c)});
  \draw[dsblue!45!black, line width=0.8pt]
    plot[domain=180-\sT:360+\sT, samples=80, variable=\s]
      ({\R*cos(\s)}, {\kv*(\R*\e*sin(\s)-\Z*\c)});
  \draw[dsblue!45!black, line width=0.8pt]
    plot[domain=-\V:\V, samples=80, variable=\t] ({sqrt(1+\t*\t/\ctwo)}, {\kv*\t});
  \draw[dsblue!45!black, line width=0.8pt]
    plot[domain=-\V:\V, samples=80, variable=\t] ({-sqrt(1+\t*\t/\ctwo)}, {\kv*\t});

  \node[above=2pt] at (0,{\vtop+\kv*\R*\e}) {$\mathcal{I}^{+}$};
  \node[below=2pt] at (0,{-\vtop-\kv*\R*\e}) {$\mathcal{I}^{-}$};
  \node[dsred, right=1pt] at (1,0) {$S^{d}$};
  \begin{scope}[shift={({-\R-0.95},{-\vtop-\kv*\R*\e})}]
    \draw[->, line width=0.5pt] (0,0) -- (0,0.8) node[above] {$X^{0}$};
    \draw[->, line width=0.5pt] (0,0) -- (0.8,0) node[right] {$X^{i}$};
  \end{scope}
\end{tikzpicture}
\caption{de Sitter space as the hyperboloid $-(X^0)^2+\sum_i (X^i)^2=\ell^2$ in
$\mathbb{R}^{1,d+1}$. Slices of constant $X^0$ are spheres $S^d$, which shrink to the minimal
sphere (red) at $X^0=0$ and grow toward $\mathcal{I}^\pm$.}
\label{fig:dS hyperboloid}
\end{figure}

In this section we review the geometry of $(d+1)$-dimensional de Sitter space, various useful coordinate systems, and the de Sitter Penrose diagram. We also discuss asymptotically de Sitter spacetimes and boundary conditions.
\\ \\
 \((d{+}1)\)-dimensional de Sitter space, denoted as dS$_{d+1}$ is a maximally symmetric solution to Einstein’s equations,
\begin{equation}
    \mathcal{G}_{\mu\nu}  + \Lambda g_{\mu\nu} = 0\,,\quad  \Lambda = \frac{d(d{-}1)}{2\ell^2}\,,
\end{equation}
where $\ell$ is the de Sitter radius. 
de Sitter space can be realized as the Lorentzian hyperboloid (see figure \ref{fig:dS hyperboloid})
\begin{equation}\label{eq: de SItter hyperboloid general d}
    -(X^0)^2 + (X^1)^2 + \cdots + (X^{d+1})^2 = \ell^2
\end{equation}
embedded in a flat \((d{+}2)\)-dimensional Minkowski space $\mathbb{R}^{1,d+1}$ with metric
\begin{equation}\label{eq:induced metric}
    ds^2 = \eta_{AB}\d X^A \d X^B=-\,(\d X^0)^2 + \sum_{i=1}^{d+1}(\d X^i)^2~.
\end{equation}
For each constant $X^0$ the spatial slicing is a $d$-sphere. In other words, the Cauchy slices in de Sitter are compact surfaces.  

From the embedding picture, it follows that the isometry group of de Sitter space is \(\mathrm{SO}(1,d{+}1)\). This group is isomorphic both to the Euclidean conformal group in \(\mathbb{R}^d\) and to the Lorentz group in \((d{+}2)\) dimensions. Its dimension is
\begin{equation}\label{eq:dimSO1dp1}
    \dim\big(\mathrm{SO}(1,d{+}1)\big) = \frac{(d+1)(d+2)}{2}\,,
\end{equation}
and its maximal compact subgroup is \(\mathrm{SO}(d{+}1)\). The Killing vectors of the de Sitter isometry group are 
\begin{equation}\label{eq: dS KV}
    L_{AB} =-i( X_A\partial_{X^B} - X_B \partial_{X^A})~,
\end{equation}
which satisfy the commutation relations
\begin{equation}\label{eq:commutations}
    [L_{AB},L_{CD}] = i\left(\eta_{AC}L_{BD} +\eta_{BD}L_{AC} -\eta_{AD}L_{BC} -\eta_{BC}L_{AD}\right)~.
\end{equation}
We stress that indices in $L_{AB}$ are raised and lowered with the metric $\eta_{AB}= \mathrm{diag}(-1,1,1,\ldots)$ where there are $(d+2)$ diagonal entries.

\paragraph{Four-dimensional de Sitter space.}
The hypersurface
\begin{equation}\label{eq:embedding picture}
    -(X^0)^2 + (X^1)^2 + \ldots + (X^4)^2 = \ell^2
\end{equation}
embedded in five-dimensional Minkowski space
\begin{equation}
    ds^2 = -(\d X^0)^2 + (\d X^1)^2+ (\d X^2)^2+ (\d X^3)^2+ (\d X^4)^2~
\end{equation}
represents dS$_4$.
This geometry preserves the group \(\mathrm{SO}(1,4)\) of dimension
\begin{equation}
    \dim(\mathrm{SO}(1,4)) = \frac{4 \cdot 5}{2} = 10~,
\end{equation}
which is also the five-dimensional Lorentz group. If we include also fermionic fields we need to consider its double cover $\mathrm{Spin}(1,4)$. 

\paragraph{Two-dimensional de Sitter space.}
Two-dimensional de Sitter space is defined by the equation of a hyperboloid in three-dimensional Minkowski space:
\begin{equation}
    -(X^0)^2 + (X^1)^2 + (X^2)^2 = \ell^2~.
\end{equation}
 Its isometry group is
\begin{equation}\label{eq:dS2AdS2}
    \mathrm{SO}(1,2) \cong \mathrm{SO}(2,1)\,,
\end{equation}
with corresponding Lie algebra $\mathfrak{so}(1,2) \cong \mathfrak{sl}(2,\mathbb{R})$. 
At the level of groups this equivalence holds up to a \(\mathbb{Z}_2\) quotient:
\begin{equation}
    \mathrm{SO}(1,2) \cong \mathrm{PSL}(2,\mathbb{R}) \equiv \mathrm{SL}(2,\mathbb{R})/\mathbb{Z}_2~.
\end{equation}
(\ref{eq:dS2AdS2}) implies that the isometry groups of Lorentzian \(\mathrm{dS}_2\) and Euclidean \(\mathrm{AdS}_2\) coincide.

\subsection*{Coordinate Systems on de Sitter Space}

\paragraph{Global coordinates.}
A convenient parametrization of (\ref{eq: de SItter hyperboloid general d}) is
\begin{equation}
    X^0 = \ell \sinh(\tau)\,,\qquad X^j= \ell \cosh(\tau)\,y^j\,,\qquad y_jy^j =1~,
\end{equation}
so that the induced metric (\ref{eq:induced metric}) becomes
\begin{equation}\label{eq: global patch}
     \frac{ds_{\mathrm{gl}}^2}{\ell^2} = -\d\tau^2 +  \cosh^2 \tau \d\Omega_{d}^2\,,\qquad \d\Omega_{d}^2= \d\theta^2 + \sin^2\theta \d\Omega_{d-1}^2~.
\end{equation}
Here \(\tau\in \mathbb{R}\) and $\theta \in [0,\pi]$. This is a closed Universe. Constant \(\tau\) slices are compact \(d\)-spheres that shrink from infinite past denoted as \(\mathcal{I}^-\) at \(\tau=-\infty\) to a minimal radius at \(\tau=0\), and re-expand again toward \(\mathcal{I}^+\) as \(\tau\to\infty\).

As a simple example for global dS$_2$ we have 
\begin{equation}
    X^0= \ell \sinh(\tau)~,\quad X^1= \ell \cosh(\tau)\cos(\varphi)~,\quad X^2 = \ell \cosh(\tau)\sin(\varphi)~,
\end{equation}
and the line element is
\begin{equation}
    \frac{ds_{\mathrm{gl}}^2}{\ell^2} = -\d \tau^2 + \cosh^2(\tau)\d\varphi^2~.
\end{equation}
Its Killing vectors follow from (\ref{eq: dS KV}) and are given by
\begin{align}\label{eq:dS2 KV}
    L_{12} &= -i\partial_\varphi~,\cr
    L_{01} &= i(\cos(\varphi)\partial_\tau -\sin(\varphi) \tanh \tau\partial_\varphi)~,\cr
    L_{02} &= i(\sin(\varphi)\partial_\tau +\cos(\varphi) \tanh \tau\partial_\varphi)~
\end{align}
whose norms are
\begin{align}
    \|L_{12}\|^2 &\equiv g_{\mu\nu}(L^\mu)^\dagger L^\nu = \ell^2 \cosh^2(\tau)~,\cr
    \|L_{01}\|^2&= \ell^2 (-\cos^2(\varphi) + \sinh^2(\tau) \sin^2
    (\varphi))~,\cr
    \|L_{02}\|^2 &= \ell^2(-\sin^2 (\varphi) + \sinh^2 (\tau) \cos^2(\varphi))~.
\end{align}
Moreover if we define the combinations 
\begin{equation}
    L_0 \equiv L_{12} = -i \partial_\varphi~,\quad L_{\pm}\equiv -(L_{02}\pm i L_{01})=  \e^{\mp i\varphi}(-i\tanh(\tau)\partial_\varphi \pm \partial_\tau)~
\end{equation}
we can map the $L_{AB}$ to the generators $\{L_0,L_{\pm}\}$ of $\mathfrak{sl}(2,\mathbb{R})$
\begin{equation}
  [L_\pm,L_0] = \pm L_{\pm}~,\quad [L_{+},L_{-}] =2L_0~.
\end{equation}

\begin{figure}[ht]
\begin{center}
\begin{tikzpicture}[scale=0.95, line cap=round, line join=round, >=Stealth,
  worldline/.style={line width=1.5pt, postaction={decorate},
     decoration={markings, mark=at position 0.7 with {\arrow{Stealth[length=5pt,width=4pt]}}}}]
  \definecolor{dsblue}{RGB}{62,104,150}
  \definecolor{dsred}{RGB}{196,64,52}
  \definecolor{dsorange}{RGB}{228,140,38}
  \definecolor{dspatch}{RGB}{198,184,208}

  \begin{scope}
    \fill[dsorange!22] (0,0) -- (0,4) -- (4,4) -- cycle;   
    \fill[dsblue!16]   (0,0) -- (4,0) -- (0,4) -- cycle;   
    \fill[dspatch]     (0,0) -- (2,2) -- (0,4) -- cycle;   
    \foreach \k in {1,2,3} {
      \draw[dsorange!85!black, line width=0.45pt] (0,\k) -- (4-\k,4);
      \draw[dsblue!75!black,   line width=0.45pt] (0,\k) -- (\k,0);
    }
    \draw[dsorange!85!black, line width=0.8pt] (0,0) -- (4,4);   
    \draw[dsblue!75!black,   line width=0.8pt] (0,4) -- (4,0);   
    \draw[dsred, line width=0.9pt] (0,2) -- (4,2);
    \fill[black] (3,2) circle (1.5pt);
    \node[above, inner sep=2pt] at (3,2) {$S^{d-1}$};
    \draw[line width=1pt] (0,0) rectangle (4,4);
    \draw[line width=1pt] (0,0) -- (0,4);
    \node[above=2pt] at (2,4) {$\mathcal{I}^{+}$};
    \node[below=2pt] at (2,0) {$\mathcal{I}^{-}$};
    \node[left=3pt]  at (0,2.9) {SP};
    \node[right=3pt] at (4,2.9) {NP};
    \node[dsred, right=2pt, scale=0.85] at (4,2) {$S^{d}$};
    \draw[->, line width=0.5pt] (-0.9,-0.9) -- (-0.15,-0.9) node[right] {$\theta$};
    \draw[->, line width=0.5pt] (-0.9,-0.9) -- (-0.9,-0.15) node[above] {$\sigma$};
  \end{scope}

  \begin{scope}[shift={(6.6,0)}]
    \fill[dsblue!24] (0,0) -- (2,2) -- (0,4) -- cycle;
    \fill[dsred!18]  (4,0) -- (2,2) -- (4,4) -- cycle;
    \draw[black!65, line width=0.6pt] (0,0) -- (4,4);
    \draw[black!65, line width=0.6pt] (0,4) -- (4,0);
    \draw[line width=1pt] (0,0) rectangle (4,4);
    \draw[line width=1pt] (0,0) -- (0,4);
    \draw[line width=1pt] (4,0) -- (4,4);
    \fill[black] (2,2) circle (1.6pt);
    \node[rotate=90,  scale=0.8, dsblue!40!black] at (0.62,2) {static patch};
    \node[rotate=-90, scale=0.8, dsred!60!black]  at (3.38,2) {static patch};
    \node[rotate=90,  scale=0.8] at (-0.32,2) {SP observer};
    \node[rotate=-90, scale=0.8] at (4.32,2)  {NP observer};
    \node[above=2pt] at (2,4) {$\mathcal{I}^{+}$};
    \node[below=2pt] at (2,0) {$\mathcal{I}^{-}$};
  \end{scope}
\end{tikzpicture}

\end{center}
\caption{The whole square is the Penrose diagram of global de Sitter. The overlap of past (blue) and future (orange) light rays is the de Sitter static patch. Each Cauchy slice of de Sitter, indicated by the red line is an $S^{d}$. Each point on the red line is a $S^{d-1}$ worth of points. Exceptions to this are only the north and south pole where $\sin\theta=0$. These are single points. Moving along the red slice each $S^{d-1}$ grows from a point at the SP to a maximal equator at $\theta = \pi/2$ and shrinks back to a point at the NP. }
\label{fig:global Penrose}
\end{figure}
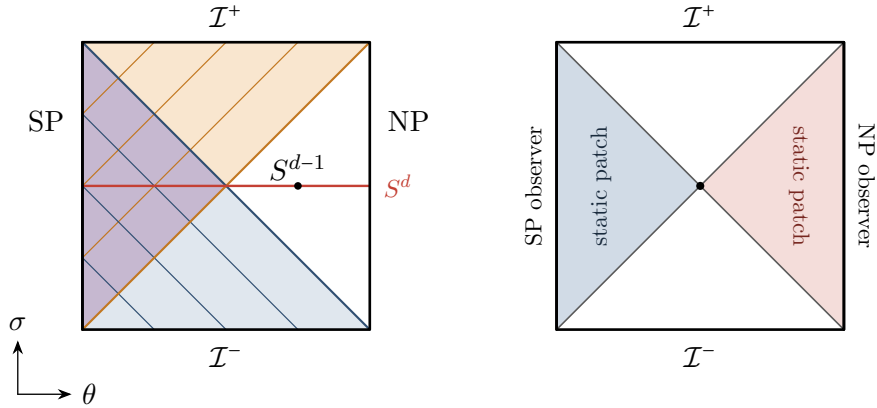

\paragraph{Conformal coordinates.} To understand the Penrose diagram of global de Sitter space we perform a conformal transformation leading to the conformal coordinate system.
Conformal coordinates $(\sigma,\theta)$ are related to the global coordinates via
\begin{equation}
    \sigma = \arctan (\sinh\tau)~,\quad \sigma \in (-\frac{\pi}{2},\frac{\pi}{2})~,
\end{equation}
in which the metric takes the form
\begin{equation}\label{eq:conformal coordinates 2}
    \frac{ds_{\mathrm{con}}^2}{\ell^2} = \frac{1}{\cos^2(\sigma)}\big(-\d \sigma^2 + \d\theta^2 + \sin^2\theta \d\Omega_{d-1}^2\big)~.
\end{equation}
We can now read off the properties of the Penrose diagram of de Sitter space. Since both $-\tfrac{\pi}{2}<\sigma <\tfrac{\pi}{2}$ and  $\theta \in [0,\pi]$ the Penrose diagram of global de Sitter space is a square (see figure \ref{fig:global Penrose}).

Each point on a horizontal line in the Penrose diagram represents an \(S^{d-1}\), except for the left and right boundaries where $\theta\in \{0,\pi\}$. These points, corresponding to $\sin\theta =0$ are single points in the Penrose diagram and are denoted as the north and south poles (NP and SP). Light rays propagate at \(45^\circ\), while timelike and spacelike surfaces appear more vertical or horizontal, respectively. The spacelike surfaces \(\mathcal{I}^\pm\) denote future and past infinity.
Future infinity $\mathcal{I}^+$ in de Sitter is a Euclidean surface, but arises at the end point of a Lorentzian spacetime evolution. 

We can now realize the Killing vectors of dS$_2$ in a conformal coordinate system, see figure \ref{fig:KV}.
As can be seen none of them is everywhere timelike. This holds true for general dimensions. As a consequence, there is no good notion of energy in global de Sitter. 
\begin{figure}[ht]
\centering
\begin{tikzpicture}[scale=1.55,
  flow/.style={postaction={decorate},decoration={markings,
      mark=at position #1 with {\arrow{Stealth[length=4.6pt]}}}},
  flow/.default=0.56,
  wolf/.style={postaction={decorate},decoration={markings,
      mark=at position #1 with {\arrowreversed{Stealth[length=4.6pt]}}}},
  wolf/.default=0.56,
  orbit/.style={line width=0.5pt,streamcol},
  hor/.style={dash pattern=on 2.8pt off 2.2pt,line width=0.85pt,horizoncol},
  scri/.style={line width=1.3pt}]
\definecolor{timelikecol}{HTML}{C9DFF2}
\definecolor{streamcol}{HTML}{25455F}
\definecolor{horizoncol}{HTML}{8A1C1C}

\newcommand\Frame[1]{%
  \draw[scri] (-2,1)--(2,1) node[right=2pt]{$\mathcal{I}^{+}$};
  \draw[scri] (-2,-1)--(2,-1) node[right=2pt]{$\mathcal{I}^{-}$};
  \draw[line width=0.5pt] (-2,-1)--(-2,1) (2,-1)--(2,1);
  \node[above=8pt,scale=0.92] at (0,1) {#1};
  \foreach \x/\l in {-2/{-\pi},-1/{-\frac{\pi}{2}},0/{0},1/{\frac{\pi}{2}},2/{\pi}}{%
    \draw[line width=0.4pt] (\x,-1)--(\x,-1.05);
    \node[below=3pt,scale=0.66] at (\x,-1) {$\l$};}
  \foreach \y/\l in {-1/{-\frac{\pi}{2}},0/{0},1/{\frac{\pi}{2}}}{%
    \draw[line width=0.4pt] (-2,\y)--(-2.05,\y);
    \node[left=2pt,scale=0.66] at (-2.05,\y) {$\l$};}
  \node[below=14pt,scale=0.78] at (0,-1) {$\varphi\ \ (\varphi\sim\varphi+2\pi)$};
    \node[scale=0.85] at (-2.55,0) {$\sigma$};}

\newcommand\BoostPattern{%
  \fill[timelikecol] (-1,0)--(0,1)--(1,0)--(0,-1)--cycle;
  \fill[timelikecol] (1,0)--(2,1)--(2,-1)--cycle;
  \fill[timelikecol] (-1,0)--(-2,1)--(-2,-1)--cycle;
  \draw[hor] (0,-1)--(2,1);\draw[hor] (0,1)--(2,-1);
  \draw[hor] (0,-1)--(-2,1);\draw[hor] (0,1)--(-2,-1);
  \draw[orbit,flow] (0,-1)--(0,1);
  \foreach \C in {0.45,0.8,0.965}{
    \draw[orbit,flow] plot[domain=-90:90,samples=51,variable=\t]
      ({asin(\C*cos(\t))/90},{\t/90});
    \draw[orbit,flow] plot[domain=-90:90,samples=51,variable=\t]
      ({-asin(\C*cos(\t))/90},{\t/90});}
  \draw[orbit,wolf] (2,-1)--(2,1);\draw[orbit,wolf] (-2,-1)--(-2,1);
  \foreach \C in {0.45,0.8,0.965}{
    \draw[orbit,wolf] plot[domain=-90:90,samples=51,variable=\t]
      ({2-asin(\C*cos(\t))/90},{\t/90});
    \draw[orbit,wolf] plot[domain=-90:90,samples=51,variable=\t]
      ({-2+asin(\C*cos(\t))/90},{\t/90});}
  \foreach \C in {1.35,2.3,4.5}{
    \draw[orbit,wolf] plot[domain=4:176,samples=51,variable=\t]
      ({\t/90},{acos(sin(\t)/\C)/90});
    \draw[orbit,flow] plot[domain=4:176,samples=51,variable=\t]
      ({\t/90-2},{acos(sin(\t)/\C)/90});
    \draw[orbit,flow] plot[domain=4:176,samples=51,variable=\t]
      ({\t/90},{-acos(sin(\t)/\C)/90});
    \draw[orbit,wolf] plot[domain=4:176,samples=51,variable=\t]
      ({\t/90-2},{-acos(sin(\t)/\C)/90});}
  \foreach \p in {(1,0),(-1,0)}{
    \fill[black] \p circle (1.7pt);\draw[white,line width=0.5pt] \p circle (1.7pt);}}

\begin{scope}
  \foreach \y in {-0.78,-0.52,-0.26,0,0.26,0.52,0.78}
    \draw[orbit,flow] (-2,\y)--(2,\y);
  \Frame{$L_{12}=-\mathrm{i}\,\partial_\varphi$\quad(spacelike everywhere)}
\end{scope}

\begin{scope}[shift={(0,-3.45)}]
  \BoostPattern
  \Frame{$L_{01}=\mathrm{i}\left(\cos\varphi\,\partial_\tau-\sin\varphi\tanh\tau\,\partial_\varphi\right)$}
\end{scope}

\begin{scope}[shift={(0,-6.9)}]
  \begin{scope}
    \clip (-2,-1) rectangle (2,1);
    \begin{scope}[shift={(1,0)}]\BoostPattern\end{scope}
    \begin{scope}[shift={(-3,0)}]\BoostPattern\end{scope}
  \end{scope}
  \Frame{$L_{02}=\mathrm{i}\left(\sin\varphi\,\partial_\tau+\cos\varphi\tanh\tau\,\partial_\varphi\right)$}
\end{scope}
\end{tikzpicture}
\caption{Penrose diagram of dS$_2$ in conformal coordinates with its Killing vectors (\ref{eq:dS2 KV}) (using $\sigma = \arctan (\sinh\tau)$). Shaded regions are timelike ($\|L\|^2<0$),
     white regions spacelike, dashed null (horizons) and dots $=$ fixed points.}
 \label{fig:KV}    
\end{figure}
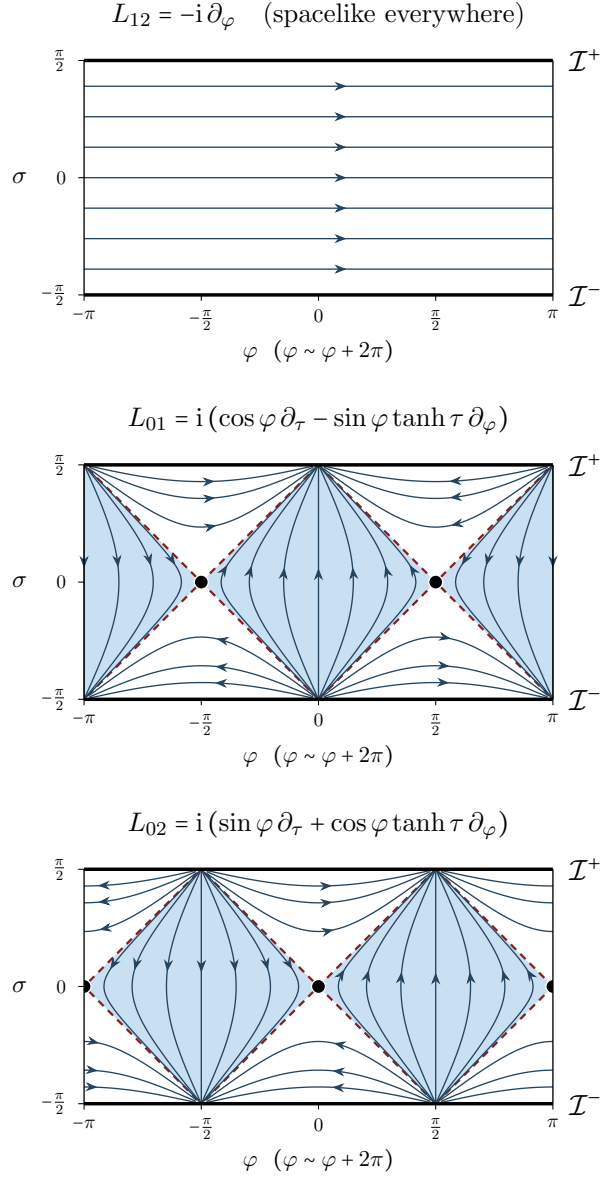

\begin{figure}[ht]
    \centering
\begin{tikzpicture}
  \definecolor{dsred}{RGB}{196,64,52}
    \draw[thick] (-3,0)--(3,0);
    \draw[thick] (-3,3)--(3,3);
    \draw[thick] (-3,0) -- (-3,3);
    \draw[thick] (3,0)-- (3,3);
    \draw[thick] (-3,3) -- (0,0);
    \draw[thick] (-3,0) -- (0,3);
    \draw[thick] (0,3)-- (3,0);
    \draw[thick] (0,0) -- (3,3);
    \draw[thick] (-3.1,1.8) -- (-2.9,2);
    \draw[thick] (-3.1,1.9) -- (-2.9,2.1);
        \draw[thick] (2.9,1.8) -- (3.1,2);
    \draw[thick] (2.9,1.9) -- (3.1,2.1);
    \draw[thick, dsred] (-3,1.5)-- (3,1.5);
    \fill[black] (.7,1.5) circle (0.5mm);
    \node[scale=.9] at (.5,1.7) {1 point};
    \node[scale=.9,dsred] at (-.3,1.3) {$S^1$};
\end{tikzpicture}
\caption{For dS$_2$ it is often convenient to ``open-up" the Penrose diagram. Instead of a square the Penrose diagram becomes a rectangle. Since $S^0$ is two points we can double the square, leading to the above rectangle where each point on the red $S^1$ is now indeed just a single point. Since the red line is a circle $S^1$ where we identify $\varphi = 0$ and $\varphi =2\pi$ we need to identify the two edges. }
\label{fig:penrose dS2}
\end{figure}
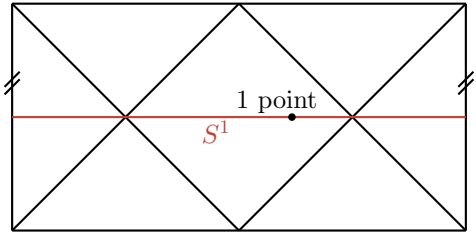

\paragraph{Static patch.}
The exponential expansion in de Sitter restricts the region an observer can see or influence. An observer indeed has causal access only to a portion of global de Sitter space, known as the static patch.  This corresponds to the intersection of the causal future and past of an observer sitting at the SP (or NP). On the left in figure \ref{fig:global Penrose} the orange lines are future, the blue past light rays. In the figure on the right we show the static patches associated to observers living at the SP and NP, which are spacelike separated. 
Two possible metrics of the de Sitter static patch are\footnote{In terms of embedding coordinates we obtain (\ref{eq: static patch 1}) from $X^0 = \ell \cos(\rho)\sinh(t)$, $X^{d+1}= \ell \cos(\rho)\cosh(t)$, and $X^i = \ell \sin(\rho)n^i$ where $n_i n^i =1$ are unit vectors on $S^{d-1}$}
\begin{subequations}\label{eq:static patches}
\begin{align}\label{eq: static patch 1}
    \frac{ds_{\mathrm{st}}^2}{\ell^2} &= -\d t^2\cos^2\rho + \d\rho^2+ \sin^2\rho \d\Omega_{d-1}^2~,\\ \label{eq: static patch}
    \frac{ds_{\mathrm{st}}^2}{\ell^2} &= - \left(1-{r^2}\right)\d t^2 + \frac{\d r^2}{\left(1-{r^2}\right)}  + r^2 \d\Omega_{d-1}^2~,
\end{align}    
\end{subequations}

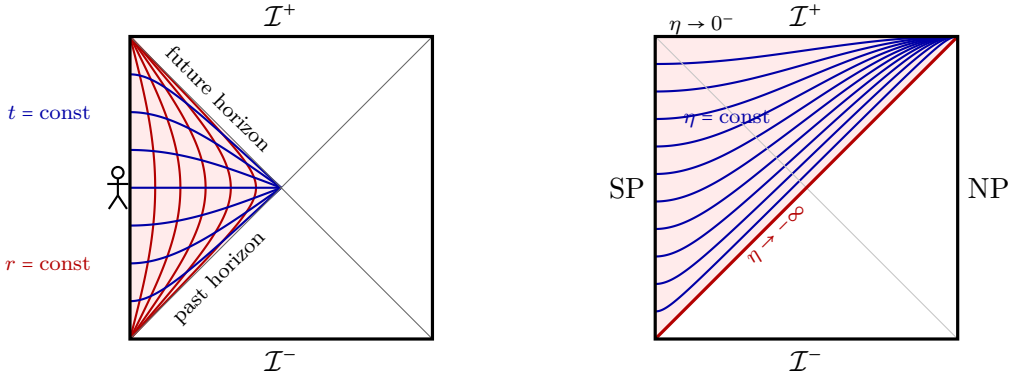
\begin{figure}[ht]
\centering
\begin{minipage}[b]{0.48\textwidth}
\centering
\begin{tikzpicture}[scale=1]
  \fill[red!8] (0,0) -- (2,-2) -- (0,-4) -- cycle;
  \foreach \k in {1,...,5}{
    \pgfmathsetmacro{\R}{sin(\k*15)}
    \draw[red!70!black, thick, domain=-180:0, samples=121, smooth]
      plot ({asin(-\R*sin(\x))/45}, {\x/45});
  }
  \foreach \k in {1,...,7}{
    \pgfmathsetmacro{\tautt}{cos(\k*22.5)}
    \draw[blue!65!black, thick, domain=0:90, samples=61, smooth]
      plot ({\x/45}, {-acos(\tautt*cos(\x))/45});
  }
  \draw[black!60] (0,0) -- (4,-4);
  \draw[black!60] (0,-4) -- (4,0);
  \draw[very thick] (0,-4) rectangle (4,0);
  \node[above] at (2,0)  {$\mathcal{I}^{+}$};
  \node[below] at (2,-4) {$\mathcal{I}^{-}$};
  \node[rotate=-45] at (1.18,-0.82) {\scriptsize future horizon};
  \node[rotate=45]  at (1.18,-3.18) {\scriptsize past horizon};
  \node[left, blue!65!black] at (-0.38,-1.0) {\scriptsize $t=\mathrm{const}$};
  \node[left, red!70!black]  at (-0.38,-3.0) {\scriptsize $r=\mathrm{const}$};
  \draw[thick] (-0.16,-1.80) circle (0.075);
  \draw[thick] (-0.16,-1.875) -- (-0.16,-2.10);
  \draw[thick] (-0.30,-1.95) -- (-0.02,-1.95);
  \draw[thick] (-0.16,-2.10) -- (-0.28,-2.28);
  \draw[thick] (-0.16,-2.10) -- (-0.04,-2.28);
\end{tikzpicture}
\end{minipage}\hfill
\begin{minipage}[b]{0.48\textwidth}
\centering
\begin{tikzpicture}[scale=1]
  \fill[red!8] (0,-4) -- (0,0) -- (4,0) -- cycle;
  \foreach \k in {1,...,10}{
    \pgfmathsetmacro{\etaa}{-tan(\k*90/11)}
    \draw[blue!65!black, thick, domain=0:180, samples=91, smooth]
      plot ({\x/45}, {(-\k*90/11 + asin(\etaa*cos(\x)/sqrt(1+\etaa*\etaa)))/45});
  }
  \draw[gray!45] (0,0) -- (4,-4);
  \draw[red!70!black, very thick] (0,-4) -- (4,0);
  \draw[very thick] (0,-4) rectangle (4,0);
  \node[above] at (2,0)  {$\mathcal{I}^{+}$};
  \node[below] at (2,-4) {$\mathcal{I}^{-}$};
  \node[left]  at (0,-2) {SP};
  \node[right] at (4,-2) {NP};
  \node[] at (0.62,0.16) {\scriptsize $\eta\to 0^-$};
  \node[rotate=45, red!70!black] at (1.62,-2.62) {\scriptsize $\eta\to-\infty$};
  \node[blue!65!black] at (0.95,-1.08) {\scriptsize $\eta=\mathrm{const}$};
\end{tikzpicture}
\end{minipage}
\caption{Left: the static patch of the south-pole observer, foliated by constant-$t$ (blue) and
constant-$r$ (red) slices; the diagonals are the past and future horizons. Right: the planar patch,
foliated by constant-$\eta$ slices; the red diagonal is the horizon $\eta\rightarrow-\infty$ of the
flat slicing, and the slices approach $\mathcal{I}^+$ as $\eta\rightarrow0^-$.}
\label{fig: patches}
\end{figure}

The name static patch is chosen because the metrics such as (\ref{eq:static patches}) are time independent, with Killing vector \(\partial_t\).
Here \(t\in\mathbb{R}\) and \(\rho\in[0,\pi/2)\) (for \(d\ge2\)), while for \(d=1\), \(\rho\in(-\pi/2,\pi/2)\).  On the other hand $r\in [0,1)$. For $\rho=\pi/2$ or $r=1$, $g_{tt}=0$ indicating an event horizon of area
\begin{equation}\label{eq: horizon area}
    A_{\mathrm{hor}}=\ell^{d-1}\int_{S^{d-1}} \d\Omega_{d-1} =
    \begin{cases}
    2\pi \ell~, & d=2~,\\
    4\pi \ell^2~, & d=3~.
    \end{cases}
\end{equation}
The observer is located at \(\rho=0\) or $r=0$. 

\paragraph{Flat slicing.}
Flat slicing coordinates cover half of de Sitter space and are particularly natural for cosmological (inflationary) observers. The line element is given by
\begin{equation}\label{eq:flatslicing}
    \frac{ds_{\mathrm{flat}}^2}{\ell^2} = \frac{-\d\eta^2+ \d\vec{x}^2}{\eta^2}\,,\qquad \eta\in(-\infty,0)~,\quad \vec{x}\in \mathbb{R}^d~.
\end{equation}
For the four-dimensional case the $\mathrm{SO}(1,4)$ isometries for the flat slicing are
\begin{equation}\label{eq: So14 isometries}
\begin{aligned}
        \vec{x} &\rightarrow \lambda \vec{x}~,
        &\quad \eta &\rightarrow \lambda \eta~,\\
        \vec{x} &\rightarrow {M} \vec{x} + \vec{a}~,&\quad   \eta &\rightarrow \eta~,\\
        \vec{x} &\rightarrow \frac{\vec{x} + \vec{b}(-\eta^2 + \vec{x}^2)}{1+2\vec{b}\cdot\vec{x} + b^2 (-\eta^2 + \vec{x}^2)}~,&\quad {\eta} &\rightarrow \frac{\eta}{1+2\vec{b}\cdot \vec{x} + b^2 (-\eta^2 + \vec{x}^2)}~,
\end{aligned}
\end{equation}
where $b^2\equiv \vec b \cdot \vec b$.
In the above we have three translations $\vec{a}$, three degrees of freedom in the rotation matrix ${M} \in \mathrm{SO}(3)$, three special conformal transformations $\vec{b}$ and one dilation $\lambda>0$.
These transformations leave the metric (\ref{eq:flatslicing}) invariant. 
de Sitter space has a conformal boundary, where the de Sitter isometries act as conformal symmetries. Explicitly, for the four-dimensional case in the limit $\eta \rightarrow 0^-$ it is straightforward to check that (\ref{eq: So14 isometries}) become the symmetry transformations of a conformal field theory on $\mathbb{R}^3$:
\begin{subequations}\label{eq:conformal transformations}
    \begin{align}
        \vec{x} &\rightarrow \lambda \vec{x}~,\\
        \vec{x} &\rightarrow {M} \vec{x} + \vec{a}~,\\
        \vec{x} &\rightarrow \frac{\vec{x} + \vec{b}\,\vec{x}^{\,2}}{1+2\vec{b}\vec{x} + b^2 \vec{x}^2}~.
    \end{align}
\end{subequations}

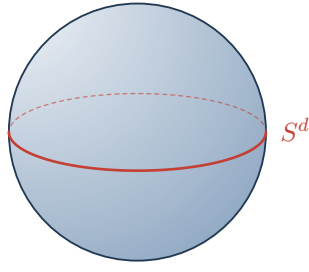
\begin{figure}[ht]
\centering
\begin{tikzpicture}[scale=1.7, line cap=round, line join=round, >=Stealth]
  \definecolor{dsblue}{RGB}{62,104,150}
  \definecolor{dsred}{RGB}{196,64,52}
  \pgfdeclareradialshading{dssphere}{\pgfpoint{-1.0cm}{1.1cm}}{%
    color(0cm)=(white); color(0.45cm)=(dsblue!18); color(1.4cm)=(dsblue!48);
    color(2.3cm)=(dsblue!78); color(3.0cm)=(dsblue!95!black); color(3.4cm)=(dsblue!95!black)}
  \shade[shading=dssphere] (0,0) circle (1);
  \draw[dsblue!55!black, line width=0.7pt] (0,0) circle (1);
  \pgfmathsetmacro{\e}{0.30}
  \draw[dsred, line width=0.5pt, dash pattern=on 1.5pt off 1.5pt, opacity=0.6] plot[domain=0:180, samples=70, variable=\s] ({cos(\s)}, {\e*sin(\s)});
  \draw[dsred, line width=1pt] plot[domain=180:360, samples=90, variable=\s] ({cos(\s)}, {\e*sin(\s)});
  \node[dsred, right=2pt, scale=0.9] at (1,0.02) {$S^{d}$};
\end{tikzpicture}
\caption{Both the static as well as the global patch of dS$_{d+1}$ rotate in Euclidean signature to
the $(d+1)$-dimensional sphere $S^{d+1}$. The equator (red) is the $S^d$ at $\tau_E=0$, the throat of
the hyperboloid in figure \ref{fig:dS hyperboloid}.}
\label{fig:Euclidean sphere}
\end{figure}

\paragraph{Euclidean continuation.}
Wick-rotating global (\ref{eq: global patch}) or static (\ref{eq:static patches}) de Sitter to Euclidean signature amounts to 
\begin{equation}
    t\rightarrow -i t_E\,,\qquad \tau \rightarrow -i\tau_E~,
\end{equation}
This implies that the line elements of the Euclidean global (Egl) and the Euclidean static (Est) patch become 
\begin{subequations}\label{eq: Euclidean dS}
\begin{align}
     \frac{ds_{\mathrm{Egl}}^2}{\ell^2} &= \d \tau_E^2 +  \cos^2 (\tau_E) \d\Omega_{d}^2~,\\
     \frac{ds_{\mathrm{Est}}^2}{\ell^2} &= \d t_E^2\cos^2\rho + \d\rho^2+ \sin^2\rho \d\Omega_{d-1}^2~.
\end{align}     
\end{subequations}
For regularity at $\rho = \pi/2$ we identify $t_E \sim t_E+ 2\pi$. This is the hallmark of a thermal behaviour, and therefore we can read off the de Sitter temperature to be $\beta_{\mathrm{dS}}= 2\pi \ell$ \cite{Figari:1975km}.
Euclidean de Sitter (\ref{eq: Euclidean dS}) thus has the line element of the round sphere (figure \ref{fig:Euclidean sphere}): Explicitly we have
\begin{equation}
    \mathrm{dS}_{d+1} \;\xrightarrow{\;\mathrm{Euclidean}\;}\; S^{d+1}\,.
\end{equation}

\subsection{Asymptotically de Sitter}
Pure de Sitter is a highly symmetric instance of a vast solution space of Einstein's equations. For convenience we restrict to $d=3$. At late times the metric of a general asymptotically de Sitter spacetime can be organized in the Fefferman--Graham form as follows 
\begin{equation}\label{eq:late time expansion}
    \frac{ds^2}{\ell^2} = - \frac{\d\eta^2}{\eta^2} + \frac{1}{\eta^2} \left(g_{ij}^{(0)} + \eta^2 g_{ij}^{(2)} + \eta^3 g_{ij}^{(3)} + \ldots \right)\d x^i \d x^j~.
\end{equation}
Pure dS$_4$ corresponds to $g_{ij}^{(0)}$ the round metric on $S^3$ and $g_{ij}^{(3)}=0$.
The boundary data at future infinity $\mathcal{I}^+$ is encoded in $g_{ij}^{(0)}$.
The physical data is encoded in the conformal class of the $g_{ij}^{(0)}$ and $g_{ij}^{(3)}$:
\begin{equation}
    (g_{ij}^{(0)}, g_{ij}^{(3)})
    \sim
    (\e^{2\omega(x)} g_{ij}^{(0)}, \e^{-\omega(x)} g_{ij}^{(3)})~.
\end{equation}
Einstein's equations lead to the constraints
\begin{equation}
    \nabla^{i}_{g^{(0)}} g_{ij}^{(3)} = 0, \qquad g^{(0)\,ij} g_{ij}^{(3)} = 0~.
\end{equation}
Boundary diffeomorphisms tangent to $\mathcal{I}^+$ remove three components of
$g^{(0)}_{ij}$. In addition, residual bulk diffeomorphisms that preserve
Fefferman--Graham gauge act at $\mathcal{I}^+$ as local Weyl transformations,
\begin{equation}
    g^{(0)}_{ij} \;\to\; \e^{2\omega(x)}\, g^{(0)}_{ij}~.
\end{equation}
This redundancy allows one to fix one further local degree of freedom of
$g^{(0)}_{ij}$, for instance by fixing its determinant locally.
In four dimensions, after quotienting by boundary diffeomorphisms and Weyl transformations, $g^{(0)}_{ij}$ 
  contains two degrees of freedom. Similarly, the tensor $g^{(3)}_{ij}$, subject to transverse and traceless constraints, also contains two degrees of freedom:
\begin{equation}
    (6+6) - (3+1 + 3+1) = 2+2~.
\end{equation}
The $2$ represents the two polarizations of the graviton (the other 2 is its conjugate momentum). Other terms in the expansion (\ref{eq:late time expansion}) such as $g_{ij}^{(2)}$ are fully fixed in terms of $g_{ij}^{(0)}$; $g^{(4)}_{ij}$, $g^{(5)}_{ij}$ and so on are fully determined in terms of $(g_{ij}^{(0)},g_{ij}^{(3)})$.  
That $(g_{ij}^{(0)},g_{ij}^{(3)})$ is the complete set of data is the content of Friedrich's theorem \cite{Friedrich:1986qfi}: for any conformal class of metrics on a compact three-manifold
and any transverse-traceless tensor on it, the vacuum Einstein equations with $\Lambda>0$ have a
unique solution in a neighbourhood of $\mathcal{I}^+$ with these as boundary data, so the
asymptotic initial value problem at $\mathcal{I}^+$ is well posed. Moreover Friedrich established the classical stability of four-dimensional de Sitter. 

There exist choices of $(g_{ij}^{(0)},g_{ij}^{(3)})$ at $\mathcal{I}^+$ that correspond to geometries with past singularities. Some other choices lead to a bounce in the past, such as the $\Lambda-$Taub--NUT spacetimes. 
The Taub--NUT metric is a non-linear example of a bounce solution of the Einstein equations with positive cosmological constant given by \cite{Beyer_2008, Osuga:2016fts}
\begin{equation}
    ds^2 = \frac{3D_0}{\Lambda} \left( -\frac{\d\tau^2}{f(\tau)} + \frac{f(\tau)}{4}\omega_3^2 + \frac{\tau^2 +1}{4} (\omega_1^2 + \omega_2^2)\right)~,
\end{equation}
where 
\begin{equation}\label{eq: f in TaubNUT}
    f(\tau) = \frac{D_0 \tau^4 + 2(3D_0 -2)\tau^2 + C_0 \tau +4 -3 D_0}{(1+\tau^2)}~.
\end{equation}
$\omega_1, \omega_2, \omega_3$ are SU(2)-invariant forms, explicitly 
\begin{subequations}\label{eq:omegai}
    \begin{align}
        \omega_1 &= \cos\psi \d\theta + \sin\psi \sin\theta \d\phi~,\\
        \omega_2 &= \sin\psi \d\theta - \cos\psi \sin\theta \d\phi~,\\
        \omega_3&= \d\psi + \cos\theta \d\phi~.
    \end{align}
\end{subequations}
Within the space of Taub--NUT metrics, pure de Sitter corresponds to $D_0=1$ and $C_0=0$ and there exists a finite neighbourhood around $(D_0, C_0) = (1,0)$ for which $f(\tau)$ is positive $\forall \tau \in\mathbb{R}$ and at $\mathcal{I}^{\pm}$ we have 
\begin{equation}\label{eq:taubnut asymptotic}
    ds^2 \approx \frac{3D_0}{\Lambda} \left( - \frac{\d\tau^2}{D_0 \tau^2} + \frac{\tau^2}{4}\Big[D_0 \omega_3^2 + (\omega_1^2 +\omega_2^2)\Big]\right)~.
\end{equation}
The term in square brackets is a homogeneous anisotropic space and corresponds to a squashed $S^3$. 
Beyond leading order we have with $\tau =-1/\eta$
\begin{equation}
    F(\eta)\equiv \eta^2 f = \frac{D_0+2(3D_0-2)\eta^2 - C_0\eta^3+(4-3D_0)\eta^4}{1+\eta^2}
 = D_0+(5D_0-4)\,\eta^2 - C_0\,\eta^3+\mathcal{O}(\eta^4)~.
\end{equation}
The full metric is then
\begin{equation}\label{eq: dS taub NUT}
    ds^2 =  -\frac{3D_0}{\Lambda F(\eta)}\frac{\d\eta^2}{\eta^2} + \frac{3D_0}{\Lambda} \frac{1}{\eta^2}\left(\frac{F(\eta)}{4} \omega_3^2 +\frac{1+\eta^2}{4}(\omega_1^2 +\omega_2^2)\right)~.
\end{equation}
Comparing this with (\ref{eq:late time expansion}) we see that the term at order $\eta^2$ (i.e. $g_{ij}^{(2)}$) is fully determined in terms of $D_0$, the boundary data of $g_{ij}^{(0)}$. The first coefficient not fixed by $D_0$ is $C_0$ the term multiplying $\eta^3$, corresponding to $g_{ij}^{(3)}$. In this regime the full metric (\ref{eq: dS taub NUT}) describes a geometry that contracts from an asymptotically de Sitter phase at
$\mathcal{I}^-$, reaches a minimal squashed $S^3$ near $\tau=0$, and re-expands to the
asymptotically de Sitter phase \eqref{eq:taubnut asymptotic} at $\mathcal{I}^+$. 
Away from the regime discussed above $f(\tau)$ (\ref{eq: f in TaubNUT}) can also develop zeros \cite{Beyer_2008}.

\section{de Sitter Quantum Field Theory}\label{sec:QFT}

We now consider quantum field theory in a fixed de Sitter background with perturbative matter. In de Sitter space time-translation invariance is broken (as we have seen in the last section there is no everywhere timelike generator). Consequently, the notions of energy conservation and classifying states according to Hamiltonian eigenvalues do not apply in the usual sense.  
In the static patch, the situation is further complicated because of the presence of a cosmological horizon. The relevant excitations are quasinormal modes, which themselves pose interpretational challenges.  
A further difference from flat space arises in the failure of cluster decomposition (at least for light fields): in Minkowski space, spatially separated events factorize, enabling the definition of an $S$-matrix. In de Sitter space, this property no longer holds, and accordingly there is no mathematically complete $S$-matrix formulation. Moreover, the S-matrix in flat space is built from a complete basis of asymptotically free states in the far past and far future, whereas in cosmology we work with a particular state.

In asymptotically Minkowski spacetime, in the presence of gravity the energy, momentum and angular momentum are all defined as surface integrals at spatial infinity. Since de Sitter space is a closed universe, there is no spatial infinity and no way to define conserved charges associated to spacetime symmetries. The dS symmetries become constraints that annihilate physical states.

In light of a quantum field theoretic Hilbert space,
the goal of this section is to review the representation theory of the dS$_{d+1}$ isometry group $\mathrm{SO}(1,d+1)$. We explicitly derive the irreducible representations of $\mathrm{SO}(1,2)$ and present them for $d\geq 2$. We also discuss the group characters, which are known as the Harish-Chandra characters and relate them to counting states -- the quasinormal modes -- in the de Sitter static patch. Finally we discuss bulk and boundary two-point functions.

\subsection{Group theory in de Sitter}

Global de Sitter space has no spatial asymptotia or time-translation invariance.  
Its isometry group is $\mathrm{SO}(1,d{+}1)$, rather than the Poincaré group. Although the appropriate organizing principle is less clear than in Minkowski space, physical quantities should nevertheless transform properly under $\mathrm{SO}(1,d{+}1)$. Of particular importance are its unitary irreducible representations (UIRs).

The de Sitter isometry group $\mathrm{SO}(1,d{+}1)$ is generated by $L_{AB} = -L_{BA}$ which satisfy the commutation relations (\ref{eq:commutations}). 
We furthermore realize the generators as Hermitian operators
\begin{equation}\label{eq:antihermitianL}
    L_{AB}^\dagger = L_{AB}~.
\end{equation}  

In terms of the conformal algebra of $\mathbb{R}^d$, one may write
\begin{equation}
    L_{0,d+1}=H~,\quad L_{d+1,i} = \frac{1}{2}(P_i+K_i)~,\quad L_{0,i} = \frac{1}{2}(P_i-K_i)~,\quad 
    L_{ij} = M_{ij}~,\quad i=1,\ldots, d~.
\end{equation}
We identify $P_i$ as translations, $K_i$ as special conformal transformations, and $M_{ij}=-M_{ji}$ as spatial rotations.
Unlike AdS, the generator $H$ in de Sitter is not positive definite: it generates a boost in $\mathrm{SO}(1,1)$ and is not bounded below.
In AdS, by contrast, $H$ plays the role of the Hamiltonian and generates the compact $\mathrm{SO}(2)$ with discrete spectrum.  
The condition (\ref{eq:antihermitianL}) further implies, for example, $P_i^\dagger = P_i$ (while in Euclidean AdS one has $P_i^\dagger = K_i$).  
The quadratic Casimir, which commutes with all the generators $L_{AB}$ takes the form
\begin{equation}\label{eq:quadratic Casimir}
    \mathcal{C}_2 \equiv \tfrac{1}{2}L_{AB}L^{AB}= -H^2+\frac{1}{2}(P_i K_i +K_i P_i) + \frac{1}{2}M_{ij}M^{ij}~,
\end{equation}
where $\frac{1}{2}M_{ij}M^{ij}$ is the quadratic Casimir of $\mathrm{SO}(d)$. For a spin $s$ representation it is given by $s(d+s-2)$.

The relevant group-theoretic correspondence is
\begin{align}
    \mathrm{dS}_{d+1}&:~ \text{isometry group } \mathrm{SO}(1,d{+}1)~,\\
    \mathrm{Euclidean~CFT}_d&:~ \text{conformal group } \mathrm{SO}(1,d{+}1)~,
\end{align}
where the de Sitter future boundary is a spacelike slice.  
In contrast,
\begin{align}
    \mathrm{Lorentzian~AdS}_{d+1}&:~ \text{isometry group } \widetilde{\mathrm{SO}}(2,d)~,\\
    \mathrm{Lorentzian~CFT}_d&:~ \text{conformal group } \widetilde{\mathrm{SO}}(2,d)~,
\end{align}
where $\widetilde{\mathrm{SO}}(2,d)$ is the universal cover of ${\mathrm{SO}}(2,d)$. For AdS the boundary is timelike and its isometry group is the conformal group on $\mathbb{R}^{1,d-1}$.

\begin{table}[ht]
    \centering
    \small
    \setlength{\arrayrulewidth}{.7pt}
    \renewcommand{\arraystretch}{1.2}

    \begin{tabular}{|c!{\vrule width .8pt}c!{\vrule width .8pt}c!{\vrule width .8pt}c!{\vrule width .8pt}c|}
\hline
        \multirow{2}{*}{$\mathrm{Lorentzian}\vspace{6mm}$}
        & $\mathrm{Isometry}$
        & $\mathrm{at~conformal~bdy}~$
        & $\mathrm{Max.~compact}$
        & $\mathrm{Hamiltonian}$ \\
        \hhline{|~|~|~|~|~|}
        $\mathrm{Spacetime}$
        & $\mathrm{Group}$
        & 
        & $\mathrm{subgroup}$
        &  \\
        \hhline{|=|=|=|=|=|}

        \multirow{2}{*}{AdS$_{d+1}$}
        & $\widetilde{\mathrm{SO}}(2,d)$
        & $\mathrm{Lorentzian~CFT}$
        & $\mathrm{SO}(2)\times\mathrm{SO}(d)$
        & $H\in\mathrm{SO}(2)$ \\
        \hhline{|~|~|~|~|~|}
        &
        &
        $\mathrm{in}~\mathbb{R}^{1,d-1}$
        &
        &
        $\mathrm{compact}$ \\
        \hhline{|=|=|=|=|=|}

        \multirow{2}{*}{dS$_{d+1}$}
        & $\mathrm{SO}(1,d+1)$
        & $\mathrm{Euclidean~CFT}$
        & $\mathrm{SO}(d+1)$
        & $H\in\mathrm{SO}(1,1)$ \\
        \hhline{|~|~|~|~|~|}
        &
        &
        $\mathrm{in}~\mathbb{R}^{d}$
        &
        &
        $\mathrm{non\text{-}compact}$ \\
        \hline
    \end{tabular}

    \caption{Comparison of AdS$_{d+1}$ and dS$_{d+1}$.}
    \label{tab:AdS_vs_dS}
\end{table}

\paragraph{dS$_2$ Representations.}\label{subsec:dS2 QFT}

In two dimensions ($d=1$), the de Sitter isometry group is $\mathrm{SO}(1,2) \cong \mathrm{SL}(2,\mathbb{R})/\mathbb{Z}_2$ where
\begin{equation}
    \mathrm{SL}(2,\mathbb{R}) = \bigg\{
    \begin{pmatrix} a & b \\ c & d\end{pmatrix} ;~ a,b,c,d\in \mathbb{R},~ ad-bc=1
    \bigg\}~.
\end{equation}
The group is non-compact and non-abelian, implying that any nontrivial unitary irreducible representation is infinite-dimensional.\footnote{Non-unitary, finite-dimensional representations of $\mathrm{SL}(2,\mathbb{R})$ are isomorphic to those of $\mathrm{SU}(2)$. } The maximal compact subgroup of $\mathrm{SO}(1,2)$ is $\mathrm{SO}(2)$. At the level of Lie algebras, $\mathfrak{so}(1,2) \cong \mathfrak{sl}(2,\mathbb{R})$.
The Lie algebra $\mathfrak{sl}(2,\mathbb{R})$ is generated by $\{P,H,K\}$ satisfying
\begin{equation}
    [H,P] =-iP~,\quad [H,K] = iK~,\quad [K,P] = -2iH~.
\end{equation}
We define the combinations
\begin{equation}
    L_0 = -\tfrac{P+K}{2} ~,\quad 
    L_{\pm} = \mp \tfrac{i}{2}(P-K) - H~,
\end{equation}
which satisfy
\[
[L_+,L_-] = 2L_0~, \qquad [L_\pm, L_0]= \pm L_\pm~, \qquad L_+^\dagger = L_-~,\quad L_0^\dagger = L_0~.
\]
$L_0$ generates the compact $\mathrm{SO}(2)$ subgroup and takes integer values for a representation of $\mathrm{SO}(1,2)$.  
The quadratic Casimir is (\ref{eq:quadratic Casimir}) \footnote{Using $[AB,C]=A[B,C]+[A,C]B$, one verifies that $\mathcal{C}_2$ commutes with all generators.}
\begin{align}
    \mathcal{C}_2 &= \frac{1}{2}L_{AB}L^{AB} = L_0^2 -\frac{1}{2}(L_+ L_- +L_- L_+) = L_0(L_0-1) -L_- L_+ =L_0(L_0+1) -L_+ L_- ~,
\end{align}
where we used $\eta_{AB}$ to lower the indices and $A,B\in \{ 0,1,2\}$.
We label a state by an integer $n\in \mathbb{Z}$ with $L_0$ eigenvalue $-n$  and the eigenvalue of the quadratic Casimir
\begin{equation}\label{eq: Casimir SO12}
    L_0|\Delta,n\rangle = -n |\Delta,n\rangle~,\quad\quad  \mathcal{C}_2|\Delta,n\rangle = \Delta(\Delta-1)|\Delta,n\rangle~.
\end{equation}
For a unitary irreducible representation the Casimir should have real eigenvalues, however for non-compact groups (such as the dS isometry groups) it does not need to be positive definite. 

From this we conclude the action of the lowering and raising operators 
\begin{equation}\label{eq:raising and lowering}
    L_{\pm}|\Delta,n\rangle = -(n\pm \Delta)|\Delta,n\pm 1\rangle~.
\end{equation}
A highest-weight representation $D_\Delta^-$ is a representation where $L_-$ truncates. It satisfies
\begin{align}
  D_\Delta^-:\quad   L_-|\Delta,\Delta\rangle &= 0~, \quad
    L_0|\Delta,\Delta\rangle = -\Delta|\Delta,\Delta\rangle~~\Rightarrow \Delta \in \mathbb{Z}_+~.
\end{align}
The states are generated by repeatedly applying $L_+$: $L_+^k|\Delta,\Delta\rangle$, $k= 1,\ldots$. As an example we have $L_+|\Delta,\Delta\rangle= -2\Delta |\Delta,\Delta+1\rangle$ and acting with the quadratic Casimir 
\begin{equation}
  \mathcal{C}_2  |\Delta,\Delta+1\rangle = (L_0(L_0-1) -L_- L_+)|\Delta,\Delta+1\rangle = \Delta(\Delta-1)|\Delta,\Delta+1\rangle~.
\end{equation}
We can repeat this for $k\geq 1$ and confirm that within a given irreducible representation the quadratic Casimir acts like the identity up to a multiplicative constant which is its eigenvalue. The $L_0$ eigenvalues of the highest weight representation 
$D_\Delta^-$  are the tower $\{-\Delta,-\Delta-1,-\Delta-2,\ldots\}$ which are bounded from above and unbounded from below. 
On the other hand, a lowest weight representation $D_\Delta^+$ satisfies
\begin{align}
     D_\Delta^+:\quad   L_+|\Delta,-\Delta\rangle &= 0~, \quad
    L_0|\Delta,-\Delta\rangle = \Delta|\Delta,-\Delta\rangle~~\Rightarrow \Delta \in \mathbb{Z}_+~,
\end{align}
with states generated by acting with $L_-$: $L_-^k|\Delta,-\Delta\rangle$, $k= 1,\ldots$. The $L_0$ spectrum is the tower $\{\Delta, \Delta+1,\ldots\}$ which is bounded from below and unbounded from above.
Both $D_\Delta^\pm$ are unitary and infinite-dimensional.

On top of the discrete series irreducible representations $D_\Delta^\pm$, $\mathrm{SO}(1,2)$ has two more irreducible infinite dimensionsal representation. To obtain these we look at the norm of the states $L_\pm |\Delta, n\rangle$. Using $L_-= L_+^\dagger$ we obtain
\begin{align}\label{eq: LpLm}
   |L_-|\Delta,n\rangle|^2&=  \langle n,\Delta|L_+L_-|\Delta,n\rangle = (n(n-1) -\Delta(\Delta-1))\langle n, \Delta|\Delta, n\rangle \geq 0~,\\
   |L_+|\Delta,n\rangle|^2&=  \langle n,\Delta|L_-L_+|\Delta,n\rangle = (n(n +1) -\Delta(\Delta-1))\langle  n, \Delta|\Delta, n\rangle\geq 0~,
\end{align}  
Reality and positivity require
\begin{subequations}
\begin{align}
   \pi_\nu:\quad  \Delta &= \tfrac{1}{2}+i\nu~, \quad \nu \in \mathbb{R}_+ ~,\\
   \gamma_\Delta:\quad \Delta &\in (0,1) ~.
\end{align}
\end{subequations}
These are the principal $\pi_\nu$ and complementary $\gamma_\Delta$ irreducible unitary representations of $\mathrm{SO}(1,2)$. From (\ref{eq:raising and lowering}) and the reality condition $L_+^\dagger = L_-$ we infer
\begin{align}
-(n+1-\Delta) \langle n,\Delta|\Delta,n\rangle&  =\langle n, \Delta|L_-|\Delta, n+1\rangle = \langle n+1,\Delta|L_+|\Delta, n\rangle^*\cr
    &=  -(n+\Delta^*) \langle n+1,\Delta|\Delta,n+1\rangle~.
\end{align}
For the principal series $\overline{\Delta}\equiv 1- \Delta= \Delta^*$ and hence we can take the normalization $\langle n,\Delta|\Delta,n'\rangle = \delta_{nn'}$. 
 For the complementary series $\Delta \in \mathbb{R}_{(0,1)}$ with $\Delta^*\neq 1-\Delta$, and hence we must modify the inner product to 
\begin{equation}
    \langle n,\Delta|\Delta,n'\rangle = N_n \delta_{nn'} ~,\quad \frac{N_{n+1}}{N_n} = {\frac{n+\overline{\Delta}}{n+\Delta}}\Rightarrow N_n = {\frac{\Gamma(n+\overline{\Delta})}{\Gamma(n+\Delta)}}~.
\end{equation}

\begin{table}[ht]
    \centering
    \small
         \setlength{\arrayrulewidth}{1.2pt}
    \renewcommand{\arraystretch}{1.5}
    \begin{tabular}{|c!{\vrule width .8pt}c!{\vrule width .8pt}c|}
        \hline
        Irrep & Range of $\Delta$ & Range of $n$  \\
        \hhline{|=|=|=|}
        $\pi_\nu$ 
        & $\Delta = \tfrac{1}{2} + i\nu,\ \nu \in \mathbb{R}$ 
        & $n\in \mathbb{Z}$  \\
        \hhline{|=|=|=|}
       $\gamma_\Delta$ 
        & $0<\Delta<1$ 
        & $n\in \mathbb{Z}$  \\
        \hhline{|=|=|=|}
        $D_\Delta^{\pm}$ 
        & $\Delta = 1+t\,,~t \in \mathbb{N}_0$ 
        & $n= \mp \Delta, \mp (\Delta+1), \ldots $   \\
        \hline
    \end{tabular}
    \caption{Summary of bosonic UIRs of $\mathrm{SO}(1,2)$.}
    \label{tab:scalar-reps-dS2}
\end{table}
\noindent

In dS$_2$ there is no intrinsic spin, since the rotation group is trivial.

\paragraph{dS$_{d+1}$ representations.} For $d\geq 2$ an infinite-dimensional representation of $\mathrm{dS}_{d+1}$ is labeled by $(\Delta, s)$, where $\Delta$ is the weight and $s$ the spin.  
The quadratic Casimir is
\begin{equation}
    \mathcal{C}_2 = \Delta(\Delta-d) + s(d+s-2)~,\quad s \in \mathrm{SO}(d)~.
\end{equation}
The unitary representations are classified as follows (for reviews see e.g. \cite{Basile:2016aen,Sun:2021thf, Hinterbichler:2026xqf}):
\begin{align}\label{eq:higher D rep}
    &\mathcal{F}_{\Delta,s} \equiv
    \begin{cases}
        \text{Principal series } \pi_{\nu,s}: &
        \Delta \in \tfrac{d}{2} + i\mathbb{R}_+~, \quad s \ge 0~, \\[4pt]
        \text{Complementary series } \gamma_{\Delta,s}: &
        \begin{cases}
            \tfrac{d}{2} < \Delta < d~, & s=0~,\\
            \tfrac{d}{2} < \Delta < d-1~, & s \ge 1~,
        \end{cases}
    \end{cases} \\[6pt]
    &\mathcal{E}_{\Delta,s} \equiv
    \begin{cases}
        \text{Exceptional type I}: &
        \Delta = d + p - 1~, \quad s=0~, \quad p \ge 1~,\\[4pt]
        \text{Exceptional type II}: &
        \Delta = d + t - 1~, \quad s \ge 1~, \quad t = 0,1,\ldots,s-1~.
    \end{cases}
\end{align}
We restrict to $\Delta \geq \tfrac{d}{2}$ for $\mathcal{F}_{\Delta,s}$ since it is isomorphic to the representation with shadow weight $\overline{\Delta} \equiv d-\Delta$: $\mathcal{F}_{\Delta,s} \cong \mathcal{F}_{\overline{\Delta},s}$.
\begin{figure}[ht]
  \centering
  \begin{tikzpicture}
  \draw[thin,->] (-2,0) -- (5.2,0) node[right,scale=.7] {$\mathrm{Re}\,\Delta$};
  \draw[thin,->] (-1,-2) -- (-1,3.4) node[right,scale=.7] {$\mathrm{Im}\,\Delta$};
  \draw[thin] (-2,0) -- (5,0);
\draw[very thick, red] (.25,0)-- (.25,3);
\draw[very thick, blue] (.25,0)-- (1.5,0);
  \fill (1.5,0) circle (2.5pt);
  \fill (3,0) circle (2.5pt);  
  \fill (4.5,0) circle (2.5pt);  
  \draw[thin] (-1,-2) -- (-1,3); 
\draw[thick,->] (2,2)-- (1,.2);  
\draw[thick,->] (2,2)-- (.4,1.2);  

\draw[thick,->] (3,-1.5)-- (1.52,-.15);  
\draw[thick,->] (3,-1.5)-- (3,-.15); 
\draw[thick,->] (3,-1.5)-- (4.5,-.15);  

\node[scale=.8] at (.25,-.32) {$\frac{d}{2}$};
\node[scale=.7] at (1.4,-.4) {$d$};
\node[scale=.7] at (3,-1.8) {Exceptional series I};  

\node[scale=.7] at (2.5,.9) {Complementary}; 
\node[scale=.7] at (.9,1.9) {Principal}; 

\draw[thin,->] (7,0) -- (13.6,0) node[right,scale=.7] {$\mathrm{Re}\,\Delta$};
  \draw[thin,->] (8,-2) -- (8,3.4) node[right,scale=.7] {$\mathrm{Im}\,\Delta$};
   \draw[thin] (7,0) -- (13.5,0);
\draw[very thick, red] (9.25,0)-- (9.25,3);
\draw[very thick, blue] (9.25,0)-- (10,0);
  \fill (10,0) circle (2.5pt);
  \fill (11.5,0) circle (2.5pt);  
  \fill (13,0) circle (2.5pt);   
    \draw[thin] (8,-2) -- (8,3); 

\node[scale=.8] at (9.25,-.32) {$\frac{d}{2}$};
\node[scale=.7] at (9.89,-.4) {$d\!-\!1$};
\node[scale=.7] at (11.5,-1.8) {Exceptional series II};    
    
\draw[thick,->] (11.5,-1.5)-- (10,-.15);  
\draw[thick,->] (11.5,-1.5)-- (11.5,-.15); 
\draw[thick,->] (11.5,-1.5)-- (13,-.15);

\draw[thick,->] (11,2)-- (9.6,.2);  
\draw[thick,->] (11,2)-- (9.4,1.2);  

\node[scale=.7] at (11.3,.9) {Complementary}; 
\node[scale=.7] at (9.9,1.9) {Principal};

  \end{tikzpicture}
  \caption{Left: $\Delta$ in scalar UIRs. Right: $\Delta$ in spinning UIRs.}
  \label{fig:myfig4.2}
\end{figure}
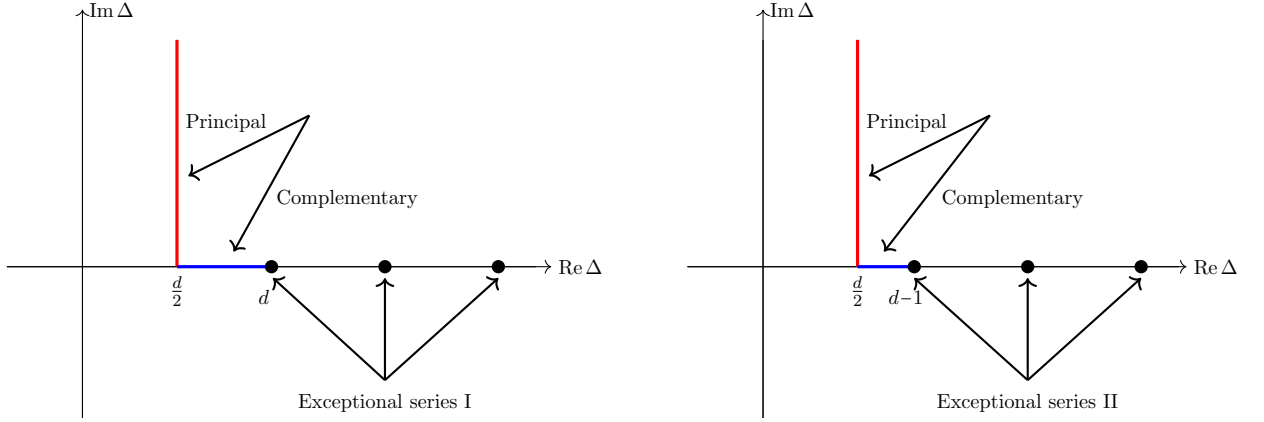
\noindent

\paragraph{dS$_3$ Representations.}

In three-dimensional de Sitter space, only the principal and complementary series irreducible unitary representations occur:
\begin{subequations}
    \begin{align}
        \pi_{\nu,s} &:\quad \Delta = 1+i\nu~,\quad \nu \in  \mathbb{R}_+~,\quad s\geq 0~,\\
        \gamma_\Delta&:\quad  \Delta\in (1,2)~,\quad s=0~.
    \end{align}
\end{subequations}
Spin $s$ labels an $\mathrm{SO}(2)$ representation. The principal and complementary series irreducible representations are isomorphic to their shadow representations $\pi_\nu \cong \pi_{-\nu}$ and $\gamma_{\Delta} \cong \gamma_{\overline{\Delta}}$, where $\overline{\Delta} \equiv 2-\Delta$. Therefore we take $\nu \in \mathbb{R}_+$ and $\Delta\geq 1$. Since $\frac{2}{2}= 2-1$ we infer from (\ref{eq:higher D rep}) that in dS$_3$ there is no complementary series irreducible representation for $s\geq 1$. There are however principal series representations with $s\geq 1$.

\paragraph{dS$_4$ Representations.}

The isometry group of $\mathrm{dS}_4$ is $\mathrm{SO}(1,4)$.  
In addition to the weight $\Delta$, the unitary representations are labeled by a spin $s$ associated with the $\mathrm{SO}(3)$ rotation subgroup of $\mathrm{SO}(1,4)$.  
A spin-$s$ field has Casimir eigenvalue
\begin{equation}
    \mathcal{C}_2 = \Delta(\Delta-3) + s(s+1)~.
\end{equation}
The $\mathrm{SO}(1,4)$ irreducible representations are summarized in table \ref{tab:scalar-reps-dS4}. For completeness we also summarise the fermionic representations of the $\mathrm{SO}(1,4)$ double cover $\mathrm{Spin}(1,4)$ in table \ref{tab:fer-reps-dS4}.

\begin{table}[ht]
    \centering
    \small
         \setlength{\arrayrulewidth}{1.2pt}
    \renewcommand{\arraystretch}{1.5}
    \begin{tabular}{|c!{\vrule width .8pt}c!{\vrule width .8pt}c|}
        \hline
        Irrep & Range of $\Delta$ & Range of $s$ \\
        \hhline{|=|=|=|}
        $\pi_{\nu,s}$ 
        & $\Delta = \tfrac{3}{2} + i\nu,\ \nu \in \mathbb{R}$ 
        & $s = 0,1,2,\ldots$ \\
        \hhline{|=|=|=|}
        \multirow{2}{*}{$\gamma_{\Delta,s}$}
        & $0 < \Delta < 3$
        & $s = 0$ \\
        \hhline{|~|~|~|}
        & $1 < \Delta < 2$
        & $s \geq 1$ \\
        \hhline{|=|=|=|}
        $\mathcal{E}_{\Delta,0}$ 
        & $\Delta = 2 + p,\ p \geq 1$ 
        & $s = 0$ \\
        \hline
        $D^\pm_{t,s}$ 
        & $\Delta = 2 + t,\ t = 0,1,\ldots,s-1$ 
        & $s \geq 1$ \\
        \hline
    \end{tabular}
    \caption{Summary of bosonic UIRs of $\mathrm{SO}(1,4)$. The fermionic counterparts are presented in table \ref{tab:fer-reps-dS4}.}
    \label{tab:scalar-reps-dS4}
\end{table}
\begin{table}[ht]
    \centering
    \small
         \setlength{\arrayrulewidth}{1.2pt}
    \renewcommand{\arraystretch}{1.5}
    \begin{tabular}{|c!{\vrule width .8pt}c!{\vrule width .8pt}c|}
        \hline
        Irrep & Range of $\Delta$ & Range of $s$ \\
        \hhline{|=|=|=|}
        $\pi_{\nu,s}$ 
        & $\Delta = \tfrac{3}{2}+i\nu,~\nu \in \mathbb{R}_+$ 
        & $s=\frac{1}{2},\frac{3}{2},\frac{5}{2},\ldots$ \\
        \hhline{|=|=|=|}
        $\gamma_{\Delta,s}$ 
        & $\mathsf{X}$ 
        &  $\mathsf{X}$  \\
        \hhline{|=|=|=|}
        $D_{t,s}^\pm$ 
        & $\Delta = 2+t~,~t= -\frac{1}{2},\frac{1}{2},\ldots, s-1$
        &  $s=\frac{1}{2},\frac{3}{2},\ldots$   \\
        \hline
    \end{tabular}
    \caption{Summary of fermionic UIRs of $\mathrm{Spin}(1,4)$. There is no complementary series for half-integer spin \cite{Anninos:2025mje}.}
    \label{tab:fer-reps-dS4}
\end{table}

In four-dimensional de Sitter space $s$ labels a representation in $\mathrm{SO}(3)$ and the exceptional type II representation is reducible into $D_{t,s}^\pm$. As we will see in a moment these are the so called partially massless fields, and $t$ is called the depth. In higher than four spacetime dimension the exceptional series is no longer reducible, the spin transforms in a representation of $\mathrm{SO}(d)$, $d\geq 4$ and we have mixed spin fields.

On a final remark we note that the tensor products of the unitary irreducible representations of $\mathrm{SO}(1,2)$ and $\mathrm{SO}(1,3)$ have been obtained in \cite{Penedones:2023uqc}. Understanding the tensor product for fermionic representations is to date an open problem. Interestingly  we have \cite{Anninos:2025mje}
\begin{equation}
\mathcal{H}_{\mathrm{gravitino} }\subset \mathcal{H}_{\mathrm{photon}} \otimes \mathcal{H}_{\mathrm{electron}}~.
\end{equation}

\subsection{Quantum field theory in de Sitter}\label{subsec:QFT in de Sitter}
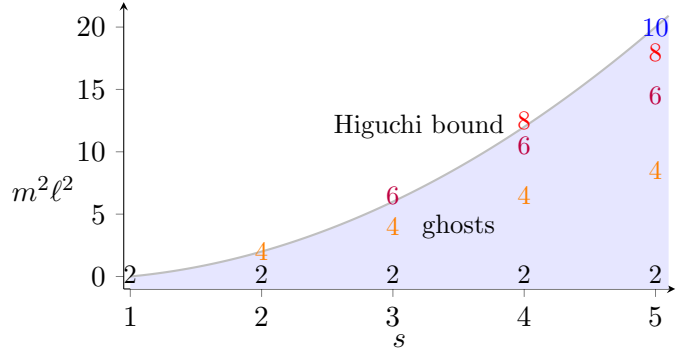
\begin{figure}[h]
\centering
\begin{tikzpicture}[scale=.7]
  \begin{axis}[
    width=12cm,
    height=7cm,
    xmin=.95, xmax=5.15,
    ymin=-1, ymax=22,
    xlabel={$s$},
    ylabel={$m^2\ell^2$},
    ylabel style={rotate=270, at={(axis description cs:-.1,0.5)}},
    axis x line=bottom,
    axis y line=left,
    tick align=outside,
    xtick={1,2,3,4,5},      
    minor x tick num=0,
  ]

  \addplot[
    name path=curve,
    domain=1:5.1,
    samples=200,
    thick,
    gray!50
  ] {x*x - x};

  \addplot[
    name path=baseline,
    domain=1:5.1,
    samples=2,
    draw=none
  ] {-1};

  \addplot[blue!10] fill between[of=curve and baseline];

  \node[font=\small] at (3.2,12) {Higuchi bound};
    \node[font=\small] at (3.5,4) {ghosts};

\node[font= \small] at (1.,.2) {2};
\node[font= \small] at (2.,.2) {2};
\node[font= \small] at (3.,.2) {2};
\node[font= \small] at (4.,.2) {2};
\node[font= \small] at (5.,.2) {2};

\node[font= \small, orange] at (2.,2) {4};

\node[font= \small, orange ] at (3.,4) {4};
\node[font= \small, purple] at (3.,6.5) {6};

\node[font= \small, orange] at (4.,6.5) {4};
\node[font= \small, purple] at (4.,10.5) {6};
\node[font= \small, red] at (4.,12.5) {8};

\node[font= \small, orange] at (5.,8.5) {4};
\node[font= \small, purple] at (5.,14.5) {6};
\node[font= \small, red] at (5.,18) {8};
\node[font= \small, blue] at (5.,20) {10};

  \end{axis}
\end{tikzpicture}
\caption{Higuchi bound for dS$_4$ along with some partially massless fields and their degeneracies.}
\label{fig:HiguchiBound}
\end{figure}

In dimensions $d\geq 2$, the relation between the mass and weight for the irreducible representations $\mathcal{F}_{\Delta,s}$ (\ref{eq:higher D rep}) reads 
\begin{align}
    m^2\ell^2 &= \Delta(d-\Delta)~, \qquad && s=0~, \label{eq:msquared F}\\
    m^2\ell^2 &= (\Delta+s-2)\,(d+s-2-\Delta)~, \qquad && s \ge 1~.
\end{align}
Since the mass is measured in units of the de Sitter length $\ell^2$ we find for example:
\begin{subequations}
    \begin{align}\nonumber
        \pi_{\nu,s}:\quad &\mathrm{neutrino},~\mathrm{quarks}~,\ldots~,\cr
        \gamma_\Delta: \quad &\mathrm{inflaton}~,\cr
        D_{t,s}^{\pm}:\quad &\mathrm{photon},~\mathrm{graviton}~,\ldots~.
    \end{align}
\end{subequations}
For minimally coupled fields in dS$_2$ the relation between mass and weight is given by
\begin{equation}
    \Delta = \frac{1}{2}(1\pm \sqrt{1-4m^2\ell^2})~.
\end{equation}
From this we infer 
\begin{equation}
    \pi_\nu:~~ m^2\ell^2>\frac{1}{4}~,\quad\quad \gamma_\Delta:~~0<m^2\ell^2<\frac{1}{4}~.
\end{equation}
Finally for the discrete series $\Delta = 1+t$ with $t\in \mathbb{N}_0$ and 
\begin{equation}
    D_\Delta^\pm:~~m^2 \ell^2 = - t(t+1)~.
\end{equation}
\paragraph{Partially massless fields.}
Particles in $\mathcal{E}_{\Delta,s}$, $s\geq 1$ (\ref{eq:higher D rep}) are called partially massless fields. Partially massless fields are totally symmetric spin $s$ fields $\phi_{\mu_1 \ldots \mu_s}$ which in $d+1$ spacetime dimensions satisfy
\begin{equation}
    \left(\nabla^2 -m^2 +\frac{(s-1)(d+s-3)-(d+s-1)}{\ell^2}\right)\phi_{\mu_1 \ldots \mu_s}=0~,\quad \nabla^\nu \phi_{\nu\mu_2 \ldots \mu_s}=0~,\quad \phi^\nu_{~\nu\mu_3\ldots \mu_s}=0~.
\end{equation}
For the fields to furnish unitary representations, the mass needs to satisfy what is called the Higuchi bound \cite{Higuchi:1986py}, which is given by:
\begin{equation}
   m^2 \ell^2 \geq  m_{\mathrm{H.b.}}^2\ell^2 =  (s-1)(d+s-3)~.
\end{equation}
Below this value one of the Stueckelberg fields that implements the transverse traceless condition develops a ghost-like kinetic term. At special values of the mass, however, given by (see e.g. \cite{Brust:2016zns})
\begin{equation}\label{eq: mst}
    m^2_{s,t}\ell^2 = (s-t-1)(d+s+t-3)~,\quad t= 0,1,2,\ldots , s-1~
\end{equation}
the particles develop enough gauge redundancy to remove these ghosts and furnish unitary irreducible representations. In (\ref{eq: mst})
$t$ is called the depth. The lowest depth is $t=0$, where $m^2_{s,0}\ell^2 = m_{\mathrm{H.b.}}^2\ell^2$, while the highest depth $t=s-1$ describes massless gauge fields. To give some examples, the photon is a highest depth spin $s=1$ partially massless field, the graviton is a highest depth partially massless spin $s=2$ field. 
A partially massless field has $D_{s}^{(d)}- D_{t}^{(d)}$ (\ref{eq:degeneracy Sdp1}) propagating degrees of freedom. These degrees of freedom are indicated in figure \ref{fig:HiguchiBound}.

\paragraph{Harish-Chandra Character.}
For a representation $\mathcal{R}$ of $\mathfrak{so}(1,d+1)$ classified by weight and spin $(\Delta,s)$ we have \cite{HarishChandra:1955, HarishChandra:1965}
\begin{equation}\label{eq:HC character}
    \chi_{\Delta,s}(\mathfrak{t}) = \tr_{\Delta,s} \e^{-i H \mathfrak{t}} = \int_{\mathbb{R}} \d\omega \rho(\omega)\e^{-i\omega \mathfrak{t}}~.
\end{equation}
We restrict to $H \in \mathfrak{so}(1,1)$ which generates time translations in the static patch.  In de Sitter no everywhere timelike Killing vectors exist but if we restrict to one of the static patches $H$ generates a timelike Killing vector (see figure \ref{fig:KV}).
The Harish-Chandra character generalizes the group character to non-compact semisimple Lie groups. Due to non-compactness, the Harish-Chandra character is in fact a {distribution}, and its rigorous definition involves smearing over the group.
Examples of Harish-Chandra characters are 
\begin{subequations}
\begin{align}
    D_{\Delta}^{\pm}~\text{in dS$_2$:} \quad
    &\chi_{\Delta,0}(\mathfrak{t}) = \frac{\e^{-\Delta\mathfrak{t}}}{(1-\e^{-\mathfrak{t}})}~,\quad \Delta \in \mathbb{Z}_+~,\\   \label{eq:massive scalar}
    \mathcal{F}_{\Delta,0}~\text{in dS$_{d+1}$:} \quad
    &\chi_{\Delta,0}(\mathfrak{t}) = \frac{\e^{-\Delta \mathfrak{t}} + \e^{-(d-\Delta)\mathfrak{t}}}{(1-\e^{-\mathfrak{t}})^d}~,\quad \Delta = \frac{d}{2}+ i\nu~,~ \nu \in \mathbb{R}\quad\mathrm{or}\quad \Delta \in (0,d)~, \\
        D_{1,2}^{\pm}~\text{in dS$_{4}$:}  \quad &\chi_{3,2}(\mathfrak{t}) = \frac{5\e^{-3\mathfrak{t}} -3\e^{-4\mathfrak{t}}}{(1-\e^{-\mathfrak{t}})^3}~,  \\
    D^\pm_{-\frac{1}{2},\frac{1}{2}}~\mathrm{in~dS}_4:\quad &\chi_{-\frac{1}{2},\frac{1}{2}}(\mathfrak{t}) = \frac{\e^{-\frac{3\mathfrak{t}}{2}}}{(1-\e^{-\mathfrak{t}})^3}~.
\end{align}
\end{subequations}
The Harish-Chandra character has the series expansion
\begin{equation}\label{eq:expansion chi}
    \chi(\mathfrak{t}) = \sum_r N_r \e^{-r\mathfrak{t}}~.
\end{equation}
For example a field in the complementary series $\mathcal{F}_{2,0}$ of $\mathrm{SO}(1,4)$ satisfies
\begin{equation}\label{eq: chi20}
    \chi_{2,0}(\mathfrak{t}) = \frac{\e^{-2\mathfrak{t}}+ \e^{-\mathfrak{t}}}{(1-\e^{-\mathfrak{t}})^3} = \e^{-\mathfrak{t}} + 4\e^{-2\mathfrak{t}} + 9 \e^{-3\mathfrak{t}} + 16 \e^{-4\mathfrak{t}} + \ldots~.
\end{equation} 
Note that the operator $H$ in (\ref{eq:HC character}) is Hermitian and hence all its eigenvalues $\omega$ are real. On the other hand, comparing (\ref{eq:HC character}) and (\ref{eq:expansion chi}) we see that for the latter $\omega$ has to be complex. These $\omega$'s correspond not to eigenstates but resonances of the theory and are called quasinormal modes. We will discuss them in more detail in section \ref{sec: dS vs BH}.

\subsection{Late time conformal structure}

We now turn to conformal operators. When working at $\mathcal{I}^+$ in de Sitter, it is particularly convenient to go to momentum space and consider the Fourier modes of local operators, for which correlation functions take a simple diagonal form. Unlike in AdS, local operator insertions at $\mathcal{I}^+$ do not define normalizable states but create non-normalizable states due to coincident point singularities. This reflects the absence of a standard operator--state correspondence in de Sitter and stands in sharp contrast to AdS, where Euclidean boundary operators create normalizable states via radial quantization \cite{Paulos:2016fap}.

We proceed in two steps. We first construct boundary operators using only $\mathfrak{sl}(2,\mathbb{R})$ representation theory and the invariance of the vacuum, which already fixes their two-point function up to two constants. We then solve the bulk wave equation near $\mathcal{I}^+$ and show that the same operators appear as the coefficients of the two late-time falloffs.

\paragraph{Boundary operators from the algebra.}
We can construct a class of operators $\widehat{\mathcal{O}}_n^{(\Delta)}$ that transform as conformal primaries of weight $\Delta$ under $\mathfrak{sl}(2,\mathbb{R})$. From these we build local operators \cite{Anninos:2023lin}
\begin{equation}\label{eq:local boundary op}
\widehat{\mathcal{O}}^{(\Delta)}(\theta) = \sum_{n\in \mathbb{Z}}\e^{-in \theta}\widehat{\mathcal{O}}_n^{(\Delta)}~,\quad \theta \in (0,2\pi]~.
\end{equation}
We will focus on the principal series irrep, $\Delta = \tfrac{1}{2}+i\nu$ with $\nu \in \mathbb{R}$, for which
\begin{equation}\label{eq: Ls on conformal operators}
    [L_0,\widehat{\mathcal{O}}_n^{(\Delta)}] =- n \widehat{\mathcal{O}}_n^{(\Delta)}~,\quad [L_{\pm}, \widehat{\mathcal{O}}_n^{(\Delta)}] =- (n\pm \Delta)\widehat{\mathcal{O}}_{n\pm 1}^{(\Delta)}~.
\end{equation}
For a $\mathrm{SL}(2,\mathbb{R})$ invariant state $|0\rangle$ (i.e. a state that is annihilated by all the de Sitter generators) we have
\begin{equation}\label{eq:first mmp}
    0 = \langle 0| [L_0,\widehat{\mathcal{O}}_n^{(\Delta)}\widehat{\mathcal{O}}_{n''}^{(\Delta')}]|0\rangle = -(n+n'')c_{n\,n''}^{\Delta\,\Delta'}~,
\end{equation}
where we defined
\begin{equation}
    c_{n\,n''}^{\Delta\,\Delta'} \equiv \langle 0| \widehat{\mathcal{O}}_n^{(\Delta)}\widehat{\mathcal{O}}_{n''}^{(\Delta')}|0\rangle~.
\end{equation}
We also have
\begin{equation}\label{eq:conformal operators eq2}
    0= \langle 0| L_\pm \widehat{\mathcal{O}}_n^{(\Delta)}\widehat{\mathcal{O}}_{n'}^{(\Delta')}|0\rangle \Rightarrow c_{n\pm 1,n'}^{\Delta \Delta'}(\pm \Delta+n) + c_{n,n'\pm 1}^{\Delta \Delta'}(\pm\Delta'+n') =0~.
\end{equation}
From (\ref{eq:first mmp}) we inferred that $c_{n\,n''}^{\Delta\,\Delta'}$ is only non-vanishing if $n= -n''$. Applying this to (\ref{eq:conformal operators eq2}) we conclude $n'= - n\mp 1$, leading to
\begin{equation}
    c_{n\pm 1}^{\Delta \Delta'}(\pm \Delta+n) + c_{n}^{\Delta \Delta'}(\pm (\Delta'-1)-n) =0~,
\end{equation}
where we use the notation $c_n^{\Delta\,\Delta'} \equiv c_{n,-n}^{\Delta\,\Delta'}$.
This equation determines the coefficients $c_n^{\Delta \Delta'}$ recursively:
\begin{equation}\label{eq:recursion}
    \frac{c_{n+ 1}^{\Delta \Delta'}}{c_{n}^{\Delta \Delta'}}= \frac{n+1-\Delta'}{n+\Delta}~,\quad \frac{c_{n+ 1}^{\Delta \Delta'}}{c_{n}^{\Delta \Delta'}}= \frac{n+1-\Delta}{n+\Delta'}~.
\end{equation}
The two recursions in \eqref{eq:recursion} are compatible only when $\Delta'(1-\Delta') = \Delta(1-\Delta)$, that is when $\Delta'=\Delta$ or $\Delta' = 1-\Delta$. This is consistent with the quadratic Casimir $\mathcal{C}_2$ of $\mathrm{SO}(1,2)$ (\ref{eq: Casimir SO12}): a two-point function can only be non-vanishing if the two operators sit in the same representation. The solutions in the two cases are
\begin{equation}\label{eq:cn solution}
    c_{n,n'}^{\Delta \Delta'}= \mathcal{N}_\Delta\,\delta_{\Delta \Delta'} \delta_{n',-n}
    \frac{\Gamma(\overline{\Delta}+n)}{\Gamma(\Delta+n)}
    + \mathcal{M}_\Delta\, \delta_{\Delta,1- \Delta'}\delta_{n',-n}~,
\end{equation}
where $\mathcal{N}_\Delta$ and $\mathcal{M}_\Delta$ are constants and the definition $\overline{\Delta} = 1-\Delta$ has been used.
The local operator \eqref{eq:local boundary op} hence satisfies
\begin{equation}\label{eq:OO two point}
    \langle 0|\widehat{\mathcal{O}}^{(\Delta)}(\theta)\widehat{\mathcal{O}}^{(\Delta')}(0)|0 \rangle
    = \xi_\Delta\,\big(\sin^2(\theta/2)\big)^{-\overline{\Delta}}\,\delta_{\Delta \Delta'}
    + \zeta_\Delta\, \delta_{\Delta', 1-\Delta}\,\delta(\theta)~,
\end{equation}
with $\xi_\Delta = \mathcal{N}_\Delta\,\Gamma(2\overline{\Delta})\cos(\pi \overline{\Delta})/2^{2\overline{\Delta}-1}$
and $\zeta_\Delta = 2\pi \mathcal{M}_\Delta$, where we used
\begin{equation}
    \sum_{n\in \mathbb{Z}} \frac{\Gamma(\overline{\Delta}+n)}{\Gamma(\Delta+n)}\, \e^{in\theta}
    = \frac{\Gamma(2\overline{\Delta})\cos(\pi \overline{\Delta})}{2^{2\overline{\Delta}-1}}
    \left(\frac{1}{\sin^2(\frac{\theta}{2})}\right)^{\overline{\Delta}}
    \qquad\text{and}\qquad \sum_{n\in\mathbb{Z}}\e^{in\theta} = 2\pi\,\delta(\theta)~.
\end{equation}
We note the appearance of a contact term for the principal series in (\ref{eq:OO two point}).

\paragraph{Single particle states.}

Given an $\mathrm{SL}(2,\mathbb{R})$ invariant state $|0\rangle$ we can define a single particle state at $\mathcal{I}^+$ as $\widehat{\mathcal{O}}_n^{(\Delta)\dagger}|0\rangle \equiv |\Delta,- n\rangle$, with $n \in \mathbb{Z}$ and $\Delta = \tfrac{1}{2}+i \nu$. We take the commutator
\begin{equation}\label{eq:modeCCR}
    [\widehat{\mathcal{O}}_n^{(\Delta)}, \widehat{\mathcal{O}}_{n'}^{(\Delta)\dagger}] = \delta_{nn'}
\end{equation}
and interpret $\widehat{\mathcal{O}}_n^{(\Delta)\dagger}$ and $\widehat{\mathcal{O}}_n^{(\Delta)}$ as creation and annihilation operators.
Note that the Hermiticity relation $(\widehat{\mathcal{O}}^{(\Delta)}(\theta))^\dagger = \widehat{\mathcal{O}}^{(\overline{\Delta})}(\theta)$ is equivalent to
\begin{equation}\label{eq:adjoint is shadow}
    \big(\widehat{\mathcal{O}}_n^{(\Delta)}\big)^\dagger = \widehat{\mathcal{O}}_{-n}^{(\overline{\Delta})}~,
\end{equation}
so the creation operators are the modes of the shadow. 
We can construct field operators
\begin{equation}\label{eq:Phi mode expansion}
    \widehat{\Phi}(\sigma,\theta) = \sum_{n\in \mathbb{Z}} \e^{-in \theta} \widehat{\mathcal{O}}_n^{(\Delta)} \Phi_n^{(\Delta)}(\sigma) + \mathrm{h.c.}~,
\end{equation}
where $\Phi_n^{(\Delta)}$ are classical field modes that do not need to transform unitarily, and which are determined by the wave equation solved below. At $\mathcal{I}^+$ ($\sigma \rightarrow \pi/2$) we would like $\widehat{\Phi}(\sigma,\theta)$ to transform as a linear combination of creation and annihilation operators.

\paragraph{From bulk modes to boundary operators.}
Having fixed the boundary two-point function by symmetry alone, we now ask what bulk data these operators encode. Solving the wave equation near $\mathcal{I}^+$ identifies them with the coefficients of the two asymptotic falloffs, and their algebra is then inherited from the bulk canonical commutation relations.

The Klein--Gordon equation in dS$_2$ is
\begin{equation}\label{eq:KGin 2D}
    -\nabla^2 \Phi(\sigma,\theta) =\Delta(\Delta-1)\Phi(\sigma,\theta)~,\quad \sigma \in (-\pi/2,\pi/2)~,\quad \theta \in (0,2\pi]~.
\end{equation}
In conformal coordinates (\ref{eq:conformal coordinates 2}) we obtain
\begin{equation}
\Phi''_n(\sigma) + \Big[n^2 -\frac{\Delta(\Delta-1)}{\cos^2\sigma}\Big]\Phi_n(\sigma)=0~,
\end{equation}
whose solutions are associated Legendre functions
\begin{equation}
    \Phi_n(\sigma) = a_1\cos^{1/2}(\sigma)P_{n-\frac{1}{2}}^{\frac{1}{2}-\Delta}(\sin(\sigma)) + a_2\cos^{1/2}(\sigma)P_{n-\frac{1}{2}}^{\frac{1}{2}-\Delta}(-\sin(\sigma))~.
\end{equation}
On the $S^1$ slicing the momentum is quantised, $n \in \mathbb{Z}$, and these are the modes $\Phi_n^{(\Delta)}$ appearing in \eqref{eq:Phi mode expansion}.
Asymptotically, we can have linear combinations of the solutions that satisfy
\begin{equation}\label{eq:falloff dS2}
    \Phi_n(\sigma) \approx \alpha^\pm_n \cos^\Delta(\sigma) + \beta^\pm_n \cos^{\overline{\Delta}}(\sigma)~,\quad \sigma \rightarrow \pm \frac{\pi}{2}~.
\end{equation}
We now append a field theoretic Hilbert space to each Cauchy slice $\Sigma$. It decomposes into the unitary irreps of the dS isometry group: principal $\pi_\nu$ (massive), complementary $\gamma_\Delta$ (light), and exceptional $\mathcal{E}_{\Delta,s}$ (massless/gauge fields).
Promoting $\Phi$ to an operator $\widehat{\Phi}$, the expansion \eqref{eq:falloff dS2} becomes
\begin{equation}\label{eq:dS2 latetime ops}
    \widehat{\Phi}(\sigma, \theta) \approx \widehat{\mathcal{O}}^{(\Delta)}(\theta)\, \cos^\Delta\sigma + \widehat{\mathcal{O}}^{(\overline{\Delta})}(\theta)\, \cos^{\overline{\Delta}}\sigma~,\quad \sigma \rightarrow \frac{\pi}{2}~.
\end{equation}
We use the same symbol as in \eqref{eq:local boundary op} because the falloff coefficients transform as weight-$\Delta$ and weight-$\overline{\Delta}$ primaries under \eqref{eq: Ls on conformal operators}.
We further assume that $\widehat{\Phi} = \widehat{\Phi}^\dagger$ is Hermitian and satisfies the equal-time relation
\begin{equation}\label{eq:dS2 CCR}
    [\widehat \Phi(\sigma,\theta),\partial_\sigma \widehat{\Phi}(\sigma,\theta')] = i\, \delta(\theta-\theta')~.
\end{equation}
Inserting \eqref{eq:dS2 latetime ops} into \eqref{eq:dS2 CCR} and matching the $\cos^{\Delta+\overline{\Delta}-1}\sigma = \cos^0\sigma$ terms gives
\begin{equation}\label{eq:dS2 commutator}
    \big[\widehat{\mathcal{O}}^{(\Delta)}(\theta),\widehat{\mathcal{O}}^{(\overline{\Delta})}(\theta')\big] = \frac{i}{\Delta - \overline{\Delta}}\,\delta(\theta-\theta')~,\qquad
    \big[\widehat{\mathcal{O}}^{(\Delta)},\widehat{\mathcal{O}}^{(\Delta)}\big] = \big[\widehat{\mathcal{O}}^{(\overline{\Delta})},\widehat{\mathcal{O}}^{(\overline{\Delta})}\big]=0~.
\end{equation}
The reality properties of the two operators now depend on the unitary irreducible representation of $\Delta$:
\begin{enumerate}
    \item For $\Delta = \frac{1}{2}+i\nu \in \pi_\nu$ we have $\overline{\Delta} = \Delta^*$, so Hermiticity of $\widehat{\Phi}$ implies
    \begin{equation}
      \big(\widehat{\mathcal{O}}^{(\Delta)}\big)^\dagger = \widehat{\mathcal{O}}^{(\overline{\Delta})}~.
    \end{equation}
  The two operators are complex conjugates of one another rather than individually Hermitian, and since $\Delta - \overline{\Delta} = 2i\nu$ the factor of $i$ in \eqref{eq:dS2 commutator} cancels,
    \begin{equation}\label{eq:principal comm dS2}
        \big[\widehat{\mathcal{O}}^{(\Delta)}(\theta), \big(\widehat{\mathcal{O}}^{(\Delta)}\big)^\dagger(\theta')\big] = \frac{1}{2\nu}\,\delta(\theta- \theta')~.
    \end{equation}
    \item For $\Delta \in (0,1) \in \gamma_\Delta$ we have $\Delta^* = \Delta \neq \overline{\Delta}$, so Hermiticity of $\widehat{\Phi}$ instead implies
       \begin{equation}
      \big(\widehat{\mathcal{O}}^{(\Delta)}\big)^\dagger = \widehat{\mathcal{O}}^{(\Delta)}~,\quad \big(\widehat{\mathcal{O}}^{(\overline{\Delta})}\big)^\dagger = \widehat{\mathcal{O}}^{(\overline{\Delta})}~.
    \end{equation}
    Both operators are separately Hermitian, and the imaginary unit survives:
    \begin{equation}
         \big[\widehat{\mathcal{O}}^{(\Delta)}(\theta), \widehat{\mathcal{O}}^{(\overline{\Delta})}(\theta')\big] = \frac{i}{2\Delta-1}\,\delta(\theta- \theta')~.
    \end{equation}
\end{enumerate}

\paragraph{Comparison to the planar patch.}
We can compare the dS$_2$ discussion above with a massless
scalar field in dS$_4$ in planar coordinates, for which the mode functions are elementary.
In planar coordinates, $\eta\in(-\infty,0)$, the massless wave equation reads
\begin{equation}
    \left(-\partial_\eta \frac{1}{\eta^2}\partial_\eta + \frac{1}{\eta^{2}}\partial_{\vec{x}}^2\right)\Phi(\eta,\vec{x})=0~.
\end{equation}
Since the spatial slices are $\mathbb{R}^3$, the Fourier transform is an integral over
continuous momenta (in contrast with the discrete mode sum of the global $S^1$ slicing),
\begin{equation}
    \Phi(\eta,\vec{x}) = \int_{\mathbb{R}^3} \frac{\d^3 \vec{k}}{(2\pi)^3}\, \e^{i\vec{k}\cdot \vec{x}}\,\Phi_{\vec{k}}(\eta)~,
\end{equation}
and the wave equation reduces to
\begin{equation}
    \left(\partial_\eta \frac{1}{\eta^{2}}\partial_\eta + \frac{\vec{k}^2}{\eta^2} \right)\Phi_{\vec{k}}(\eta)=0~,
\end{equation}
with solutions
\begin{equation}\label{eq: phik dS4}
    \Phi_{\vec{k}}(\eta) =  c_1 \e^{-ik \eta}(1+ik\eta) + c_2 \e^{ik\eta}(1-ik\eta)~,\qquad k\equiv |\vec{k}|~.
\end{equation}
Expanding \eqref{eq: phik dS4} at late times we find
\begin{equation}\label{eq: massless late}
    \Phi_{\vec{k}}(\eta) = \mathcal{O}^{(0)}_{\vec{k}}\left(1+\tfrac{1}{2} k^2\eta^2+\ldots\right)
    + \mathcal{O}^{(3)}_{\vec{k}}\,(-\eta)^3\left(1+\ldots\right)~,\qquad \eta \rightarrow 0^-~,
\end{equation}
with $\mathcal{O}^{(0)}_{\vec{k}} = c_1+c_2$ and $\mathcal{O}^{(3)}_{\vec{k}} = \tfrac{i}{3}k^3(c_1-c_2)$.
This exhibits the two falloffs $\Delta = 0$ and $\overline{\Delta} = 3$ of a massless field in
dS$_4$; the $\Delta=0$ branch is constant --- the mode freezes at late times. More generally,
light fields furnish the complementary series with $\Delta\in(0,d)$ and decay as
$(-\eta)^{\min(\Delta,\overline{\Delta})}$.
Promoting the field to a Hermitian operator, the late-time behaviour in
dS$_{d+1}$ is
\begin{equation}\label{eq: latetime ops}
     \widehat{\Phi} (\eta,\vec{x}) \approx \widehat{\mathcal{O}}^{(\Delta)}(\vec{x})\, (-\eta)^{\Delta}
      + \widehat{\mathcal{O}}^{(\overline{\Delta})}(\vec{x})\,
     (-\eta)^{\overline{\Delta}}~.
\end{equation}
The algebra of the
late-time operators is inherited from the bulk canonical commutation relations: the momentum
conjugate to $\Phi$ is $\Pi = (\ell/(-\eta))^{d-1}\partial_\eta\Phi$, so that
\begin{equation}\label{eq: bulk CCR}
    \big[\widehat\Phi(\eta,\vec{x}),\partial_\eta\widehat\Phi(\eta,\vec{y})\big]
    = \frac{i}{\ell^{d-1}}\,(-\eta)^{d-1}\,\delta^{(d)}(\vec{x}-\vec{y})~.
\end{equation}
Inserting \eqref{eq: latetime ops}, the cross terms scale as
$(-\eta)^{\Delta+\overline{\Delta}-1}=(-\eta)^{d-1}$, matching the right-hand side of
\eqref{eq: bulk CCR}, while the same-branch terms scale as $(-\eta)^{2\Delta-1}$ and
$(-\eta)^{2\overline{\Delta}-1}$ and must cancel separately. Matching coefficients yields
\begin{equation}\label{eq: latetime commutator}
     \big[\widehat{\mathcal{O}}^{(\Delta)}(\vec{x}),\widehat{\mathcal{O}}^{(\overline{\Delta})}(\vec{y})\big]
     = \frac{i}{\ell^{d-1}(\Delta-\overline{\Delta})}\, \delta^{(d)}(\vec{x} - \vec{y})~,\qquad
     \big[\widehat{\mathcal{O}}^{(\Delta)},\widehat{\mathcal{O}}^{(\Delta)}\big]
     =\big[\widehat{\mathcal{O}}^{(\overline{\Delta})},\widehat{\mathcal{O}}^{(\overline{\Delta})}\big]=0~,
\end{equation}
where we also used $[\widehat{\Phi}(\eta,\vec x),\widehat{\Phi}(\eta,\vec y)]=0$.
The reality structure is the $d$-dimensional version of the dS$_2$ discussion around \eqref{eq:dS2 commutator}: for light fields $\Delta-\overline{\Delta}=2\Delta-d$ is real and the imaginary unit survives, consistent with $\widehat{\mathcal{O}}^{(\Delta)}$ and $\widehat{\mathcal{O}}^{(\overline{\Delta})}$ being individually Hermitian, while for the principal series $\Delta-\overline{\Delta}=2i\nu$ is imaginary, the $i$ cancels, and $(\widehat{\mathcal{O}}^{(\Delta)})^\dagger=\widehat{\mathcal{O}}^{(\overline{\Delta})}$.
For the massless example \eqref{eq: massless late} one finds
$[\widehat{\mathcal{O}}^{(0)}(\vec{x}),\widehat{\mathcal{O}}^{(3)}(\vec{y})]
=-\tfrac{i}{3\ell^{2}}\,\delta^{(3)}(\vec{x}-\vec{y})$.

\begin{figure}[t]
\centering
\begin{tikzpicture}[
    scale=.9, line join=round, line cap=round,
    boundary/.style={line width=0.9pt, edgecol},
    hidden/.style={line width=0.6pt, edgecol!50, dash pattern=on 2.4pt off 2pt},
    curve/.style={line width=0.8pt, edgecol},
    op/.style={circle, fill=opcol, inner sep=1.35pt},
    label/.style={scale=0.65}
]
\definecolor{edgecol}{HTML}{25455F}
\definecolor{facetop}{HTML}{C9DFF2}
\definecolor{opcol}{HTML}{8A1C1C}
\coordinate (A) at (0,0);    \coordinate (B) at (5,0);
\coordinate (C) at (1,1.3);  \coordinate (D) at (6,1.3);
\coordinate (E) at (1,-2.4); \coordinate (F) at (4,-2.4);
\coordinate (G) at (5,-1.5); \coordinate (H) at (2,-1.5);
\fill[facetop!30] (A) to[bend left=15] (E) -- (F) to[bend left=15] (B) -- cycle;   
\fill[facetop!45] (B) to[bend right=15] (F) -- (G) to[bend left=10] (D) -- cycle;  
\fill[facetop]    (A) -- (B) -- (D) -- (C) -- cycle;                               
\draw[hidden] (C) to[bend left=10] (H);
\draw[hidden] (H) -- (E);
\draw[hidden] (G) -- (H);
\draw[boundary] (A) -- (B) -- (D) -- (C) -- cycle;
\draw[curve, bend left=15]  (A) to (E);
\draw[curve, bend right=15] (B) to (F);
\draw[curve, bend right=10] (D) to (G);
\draw[boundary] (E) -- (F) -- (G);
\node[op] (x1) at (1.70,0.95) {};
\node[op] (x2) at (4.50,0.62) {};
\node[op] (x3) at (2.60,0.22) {};
\node[label, right=2.5pt of x1] {$\widehat{\mathcal{O}}^{(\Delta)}(\vec{x}_1)$};
\node[label, left=2.5pt of x2]  {$\widehat{\mathcal{O}}^{(\Delta)}(\vec{x}_2)$};
\node[label, right=2.5pt of x3] {$\widehat{\mathcal{O}}^{(\Delta)}(\vec{x}_3)$};
\node[label, anchor=east] at (0.42,0.75) {$\mathcal{I}^{+}$};
\end{tikzpicture}
\caption{At $\mathcal{I}^+$ operators are organized in terms of $\mathrm{SO}(1,d+1)$ irreducible
representations acting as the conformal group on the late time slice. The late time operators are
generally non-commuting.}
\label{fig: late time physics}
\end{figure}
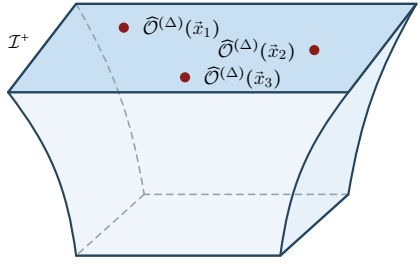

\paragraph{Bulk two-point function.}

We can compare the late-time two-point function with the two-point function from a bulk perspective. For a de Sitter invariant state $|0\rangle$, the Wightman function is defined as
\begin{equation}
   G_W(x,x')\equiv \langle 0|\Phi(x)\Phi(x')|0\rangle~,
\end{equation}
We will solve this in terms of the invariant distance\footnote{A related quantity is the geodesic distance, $\mathsf{D}$, which for spacelike separated points is given  by $P(x_1,x_2) = \cos(\mathsf{D}(x_1,x_2)/\ell)$.}
\begin{equation}
    P(x_1,x_2) =\frac{\eta_{AB}X_1^A(x_1) X_2^B(x_2)}{\ell^2} ~,
\end{equation}
where $X(x)$ are the embedding coordinates in (\ref{eq:embedding picture}). 
Away from coincident points, the Wightman function satisfies the Klein–Gordon equation
\begin{equation}
    (-\nabla^2 + m^2)G_W(P) = 0~.
\end{equation}
This leads to the differential equation
\begin{equation}\label{eq:GW equation}
   (1-P^2)G_W''(P) - (d+1)P G_W'(P) -m^2\ell^2 G_W(P)=0~,
\end{equation}
The change of variables $\xi= \tfrac{1+P}{2}$ maps (\ref{eq:GW equation}) to a hypergeometric differential equation whose solutions are
\begin{equation}\label{eq: Wightman}
    G_W(P) = a_1 \,_2F_1\left(\Delta_+,\Delta_-, \frac{d+1}{2}, \frac{1+P}{2}\right) + a_2\,_2F_1\left(\Delta_+,\Delta_-, \frac{d+1}{2}, \frac{1-P}{2}\right)~,
\end{equation}
with
\begin{equation}
    \Delta_{\pm} = \frac{d}{2}\pm \sqrt{\frac{d^2}{4}-m^2\ell^2}~.
\end{equation}
Linear combinations with non-trivial $a_1$ and $a_2$ yield the Green's function for so called $\alpha$-vacua \cite{Mottola:1984ar, Allen:1985ux}. The solution with $a_2=0$ has only coincident point singularities (but no anti-podal singularities). Among all  the de Sitter invariant vacua $|0\rangle$ we single out the Bunch-Davies vacuum \cite{Bunch:1978yq, Mottola:1984ar, Allen:1985ux, 
Schomblond:1976xc,Chernikov:1968zm}, which we will discuss in the next section, as it is the vacuum which has only the coincident point singularity, and no antipodal singularity.\footnote{One can ask if there is a single de Sitter invariant state that satisfies this condition (known as the Hadamard condition). In \cite{Aguilera-Damia:2026dbk} the authors obtained a model that has multiple distinct de Sitter invariant states satisfying the Hadamard condition.} This requirement is called the Hadamard condition, visible for $\tfrac{1+P}{2}\rightarrow 1$.  On the other hand the two point function is regular for antipodal points where $\tfrac{1+P}{2}\rightarrow 0$. Imposing the same short distance behaviour as in Minkowski space sets the coefficient 
\begin{equation}
    a_1= \frac{\Gamma(\Delta_+)\Gamma(\Delta_-)}{(4\pi)^{\frac{d+1}{2}}\Gamma\left(\frac{d+1}{2}\right)}\ell^{1-d}~.
\end{equation}
The large-separation behaviour of the Wightman function was analyzed in detail in \cite{Anninos:2022ujl}.

\paragraph{KMS condition.}\label{subsec:KMS} We can now study what property correlation functions obey in thermal systems with inverse temperature $\beta$. 
The thermal two point function is an analytic function in the strip
$0<\mathrm{Im}(t)<\beta$ and we can consider 
\begin{equation}
    G_\beta (t) \equiv \langle \widehat{O}_1(0)\widehat{O}_2(t)\rangle_\beta =\frac{1}{Z} \tr \e^{-\beta \widehat{H}}\widehat{O}_1(0)\widehat{O}_2(t)
    = \frac{1}{Z}\tr \e^{-\beta \widehat{H}}  \widehat{O}_1(0) \e^{it \widehat{H}} \widehat{O}_2(0)\e^{-it \widehat{H}}~.
\end{equation}
Using the cyclicity of the trace we obtain
\begin{align}
    G_\beta(t+i\beta) &=\frac{1}{Z} \tr \e^{-\beta \widehat{H} } \widehat{O}_1(0)\e^{it H}\e^{-\beta \widehat{H}}\widehat{O}_2(0)\e^{-it\widehat{H}}\e^{\beta \widehat{H}}
    = \frac{1}{Z}\tr \widehat{O}_1(0)\e^{it \widehat{H}}\e^{-\beta \widehat{H}}\widehat{O}_2(0) \e^{-it \widehat{H}} \cr
    &=\frac{1}{Z} \tr \e^{-\beta \widehat{H}}\widehat{O}_2(0)\widehat{O}_1(-t) =\frac{1}{Z} \tr \e^{-\beta \widehat{H}} \widehat{O}_2(t)\widehat{O}_1(0)~.
\end{align}
In particular we have 
\begin{equation}
    \langle \widehat{O}_1(0)\widehat{O}_2(t+i\beta)\rangle_\beta = \langle \widehat{O}_2(t)\widehat{O}_1(0) \rangle_\beta~. 
\end{equation}
This analyticity and periodicity property of correlation functions is the KMS
condition \cite{Kubo:1957mj, Martin:1959jp};  
The two operator orderings thus arise at $\mathrm{Im}(t)=0$ and
$\mathrm{Im}(t)=\beta$ respectively.

Now let us consider two points at timelike separation on the worldline of the
static observer at $\rho=0$, say at times $t$ and $0$. In embedding coordinates
(see footnote below \ref{eq:static patches})
\begin{equation}
    X^0 = \ell \sinh(t/\ell)~,\quad X^i=0~,\quad X^{d+1}= \ell \cosh(t/\ell)~,
    \qquad Y^0=Y^i=0~,\quad Y^{d+1}=\ell~,\qquad i=1,\ldots,d~.
\end{equation}
The de Sitter invariant distance between the two points is then
\begin{equation}\label{eq: periodicity of P}
    P(t)= \frac{X\cdot Y}{\ell^2} = \cosh(t/\ell)~,\qquad P(t+2\pi i \ell)=P(t)~.
\end{equation}
Since the Bunch--Davies correlator $G_W$ depends on the two points only through
$P$, it is automatically periodic in imaginary static time with period
$2\pi\ell$. Moreover, $G_W(P)$ is analytic away from the cut $P\geq 1$
($\leftrightarrow \frac{1+P}{2}\in (1,\infty)$)
(coincident and timelike separated points), which the strip
$0<\mathrm{Im}(t)<2\pi\ell$ only touches at its edges.
The Bunch--Davies correlators restricted to the static patch therefore satisfy
precisely the KMS condition derived above, with
\begin{equation}
    \beta_{\mathrm{dS}} = 2\pi \ell ~.
\end{equation}
In this sense the static patch of a fixed de Sitter background, in the
Bunch--Davies state, is thermal at the de Sitter temperature
\begin{equation}
    T_{\mathrm{dS}} = \frac{1}{2\pi\ell}~.
\end{equation}
\begin{center}
*************
\end{center}

Ordinarily, in flat-space QFT one computes $\langle\mathrm{out}|\mathrm{in}\rangle$ amplitudes. In de
Sitter space a conventional S-matrix is not available and one instead computes
expectation values in a fixed state: the $\langle\mathrm{in}|\mathrm{in}\rangle$, or
Schwinger--Keldysh, formalism. The basic building block is the Wightman function (\ref{eq: Wightman}).
For the Bunch--Davies choice $a_2=0$ it is analytic in $P$ up to a branch cut at
$\tfrac{1+P}{2}\in(1,\infty)$, i.e.\ at timelike separation. The Schwinger--Keldysh contour ordering dictates the side from which
each two-point function approaches the cut: the Wightman function, the Feynman propagator and the
retarded Green's function are the standard boundary values, and the commutator is the discontinuity
across the cut.

Beyond free fields, considerable progress has been made at the perturbative level through Mellin-space
methods \cite{Sleight:2019mgd}, perturbative bootstrap techniques and geometric approaches \cite{Arkani-Hamed:2018kmz, Arkani-Hamed:2017fdk}; beyond
perturbation theory much less is known, notable exceptions being low-dimensional models such as the
Schwinger model \cite{Anninos:2024fty, Galati:2026btt, Aguilera-Damia:2026dbk} and stochastic and resummation techniques~\cite{Starobinsky:1986fx, Starobinsky:1994bd, Marolf:2010zp, Gorbenko:2019rza, Anninos:2014lwa}. A central non-perturbative tool \cite{Bros:1995js, Bros:1998ik, Anninos:2023lin, Hogervorst:2021uvp, DiPietro:2021sjt, Sleight:2020obc} is the
spectral representation of the two-point function: in dS$_4$,
\begin{equation}\label{interacting2pt}
\langle \Omega | \Phi(x') \Phi(x) | \Omega \rangle
=
\int_{\mathcal{C}} \frac{\d \Delta}{2\pi i}\,
\varrho(\Delta)\, G_\Delta(P_{x',x})\,,\qquad
G_\Delta(P)=\frac{\Gamma(\Delta)\Gamma(3-\Delta)}{16\pi^2}\,
 _2F_1\left(\Delta,3-\Delta;2;\frac{1+P}{2}\right),
\end{equation}
with $G_\Delta$ the free Bunch--Davies propagator of weight $\Delta$ and $\mathcal{C}$ running along
the principal series $\Delta\in\tfrac32+i\mathbb{R}$ (a priori for more general fields and dimensions it could also include the complementary and discrete series). This is the
K\"all\'en--Lehmann representation: decomposing the Hilbert space into unitary irreducible
representations, unitarity implies $\varrho(\Delta)\ge0$ along $\mathcal{C}$, while sufficiently light (complementary)
states contribute isolated terms at real $\Delta$. For a free field $\varrho$ collapses to a delta function at the weight of a free field $\Delta_0$.

\section{de Sitter Wavefunction}\label{sec:dS wavefunction}
A major open problem in cosmology is the need for an initial state. Even if we require de Sitter invariance for QFT in fixed de Sitter, there could still be many choices. 

In this section we review a particular de Sitter invariant wavefunction for both de Sitter quantum field theory (we will refer to it as the Bunch-Davies wavefunction \cite{Bunch:1978yq, Mottola:1984ar, Allen:1985ux, 
Schomblond:1976xc,Chernikov:1968zm}) and de Sitter quantum gravity (we will refer to it as the Hartle-Hawking wavefunction \cite{Hartle:1983ai, Hartle:2010vi}). We provide some examples in both cases.

\subsection{Wavefunction for rigid de Sitter}
We start by discussing a rigid de Sitter background and studying the wavefunction of a de Sitter
quantum field theory \cite{Anninos:2014lwa}. Consider a scalar field $\phi$ propagating in the flat slicing
(\ref{eq:flatslicing}) of dS$_{d+1}$,
\begin{equation}
\d s^2 = \frac{\ell^2}{\eta^2}\left(-\d\eta^2 + \d\vec{x}^{\,2}\right)~,\qquad
\vec{x} \in \mathbb{R}^d~,\quad \eta \in (-\infty, 0)~.
\end{equation}
The Bunch--Davies wavefunction is defined by the path integral
\begin{equation}\label{eq:PsiBD}
    \Psi_{\rm BD}[\varphi(\vec{x}),\eta_c]
\equiv \int [\mathcal{D}\phi]\,
\e^{i S_L[\phi]}~,
\end{equation}
where $S_L$ is the Lorentzian action and we integrate over all field configurations that satisfy the
boundary conditions
\begin{equation}\label{eq:BD boundary condition}
    \phi( \eta = \eta_c,\vec{x}) = \varphi(\vec{x})~,\qquad
    \phi( \eta \rightarrow -\infty(1-i\epsilon),\vec{x}) = 0~,
\end{equation}
for some $\eta_c$ that is often pushed to future infinity, $\eta_c=0$. The deformation
$\eta \to -\infty(1-i\epsilon)$ is the analogue of the $i\epsilon$ prescription that projects onto the
Minkowski ground state. The Bunch--Davies wavefunction satisfies the Schr\"odinger equation
\begin{equation}
 i\,\partial_{\eta_c}\Psi_{\mathrm{BD}}[\varphi(\vec{x}),\eta_c]=\widehat{H}\,\Psi_{\mathrm{BD}}[\varphi,\eta_c]~,
\end{equation}
where $\widehat{H}$ is the Hamiltonian constructed from the action $S_L$ in (\ref{eq:PsiBD}). Decomposing the field as
\begin{equation}
\phi(\eta,\vec{x}) = \int_{\mathbb{R}^d}\frac{\d^d\vec{k}}{(2\pi)^d}\,
\e^{i\vec{k}\cdot\vec{x}}\phi_{\vec{k}}(\eta)~,
\end{equation}
we promote the field to an operator and expand
\begin{equation}\label{eq: mode expansion}
    \hat{\phi}(\eta,\vec{x}) = \int \frac{\d^d \vec{k}}{(2\pi)^d} \left(\hat{a}_{\vec{k}}
    u_{\vec{k}}(\eta)\,\e^{i \vec{k}\cdot\vec{x}} + \hat{a}_{\vec{k}}^\dagger
    u_{\vec{k}}^*(\eta)\,\e^{-i \vec{k}\cdot\vec{x}}\right)~,
\end{equation}
where the mode function $u_{\vec{k}}(\eta) \sim \e^{-ik\eta}$ is chosen to behave as a positive-frequency
plane wave in the infinite past, $k\eta \rightarrow -\infty$, with $k\equiv |\vec{k}|$, and the operators
$\hat{a}_{\vec{k}}$ satisfy
\begin{equation}
    [\hat{a}_{\vec{k}}, \hat{a}_{\vec{k}'}^\dagger] = (2\pi)^d\delta^{(d)}(\vec{k}- \vec{k}')~.
\end{equation}
This choice of mode functions defines the Bunch--Davies vacuum $|\Omega\rangle$ through
\begin{equation}
    \hat{a}_{\vec{k}} |\Omega\rangle=0~,\qquad \forall\, {\vec{k}}~.
\end{equation}
In the Schr\"odinger picture we introduce field eigenstates on the slice $\eta=\eta_c$,
\begin{equation}
    \hat{\phi}(\eta_c,\vec{x})|\varphi\rangle =  {\varphi}(\vec{x})|\varphi\rangle~,
\end{equation}
in terms of which the Bunch--Davies wavefunctional is the overlap
\begin{equation}
  \Psi_{\mathrm{BD}}[\varphi(\vec{x}),\eta_c]= \langle \varphi|\Omega\rangle ~.
\end{equation}
We now study the
Bunch--Davies wavefunction $\Psi_{\mathrm{BD}}$ in three examples:
\begin{enumerate}
    \item a massless scalar in dS$_4$~,
    \item a conformally coupled scalar in dS$_4$~,
    \item a gauge field in dS$_2$~.
\end{enumerate}

\paragraph{1. Massless scalar in dS$_4$.}
Consider the Lorentzian action for a scalar field in a dS$_4$ background:
\begin{equation}
S_L= -\frac{1}{2}\int \d^4 x \sqrt{-g}\left(g^{\mu \nu}\partial_\mu\phi \partial_\nu \phi +V(\phi)\right)~.
\end{equation}
In the flat slicing
\begin{equation}
d s^2 = \frac{\ell^2}{\eta^2} \left(-\d \eta^2 + \d \vec{x}^{\,2}\right)~,\qquad \vec x \in \mathbb{R}^3~,\quad \eta \in (-\infty, 0)~,
\end{equation}
this action is given by
\begin{equation}\label{eq: scalar flat slicing dS4}
S_L = \frac{\ell^2}{2} \int_{\mathbb{R}^3} \d^3 \vec x\int \frac{\d\eta}{\eta^2} \left((\partial_\eta \phi(\eta,\vec x))^2 - (\partial_{\vec x}\phi(\eta,\vec{x}))^2 - \frac{\ell^2}{\eta^2} V(\phi(\eta,\vec{x}))\right)~.
\end{equation}
Decomposing the field as
\begin{equation}\label{eq:dS4 Fourier}
\phi(\eta,\vec{x}) = \int_{\mathbb{R}^3}\frac{\d^3\vec{k}}{(2\pi)^3}\,
\e^{i\vec{k}\cdot\vec{x}}\phi_{\vec{k}}(\eta)~,
\end{equation}
the equations of motion follow from varying \eqref{eq: scalar flat slicing dS4} with $V=0$:
\begin{equation}
-\,\partial_\eta\!\left(\frac{1}{\eta^2}\partial_\eta\phi_{\vec{k}}\right)
- \frac{k^2}{\eta^2}\phi_{\vec{k}}=0~,
\end{equation}
where $k^2 \equiv \vec{k}\cdot \vec{k}$. The general solution is given by
\begin{equation}
\phi_{\vec{k}} = c_1\,|\eta|^{3/2}H^{(1)}_{3/2}(-k\eta)
+ c_2\,|\eta|^{3/2}H^{(2)}_{3/2}(-k\eta)~,
\end{equation}
where $H^{(1,2)}$ are Hankel functions that satisfy
\begin{equation}
    |\eta|^{3/2} H^{(1,2)}_{\nu}(-k\eta)
\;\sim\; \sqrt{\tfrac{2}{\pi k}}\;|\eta|\;
\e^{\mp ik\eta}\,\e^{\mp \frac{i\pi}{4}(2\nu+1)}~,\qquad k\eta\rightarrow-\infty~.
\end{equation}
The first solution is the Bunch--Davies mode function $u_{\vec k}$ of \eqref{eq: mode expansion},
which defines the state $|\Omega\rangle$ through
$\hat a_{\vec k}|\Omega\rangle = 0$. The wavefunction
$\Psi_{\rm BD} = \langle\varphi|\Omega\rangle$ is instead controlled by the
conjugate solution: the boundary condition \eqref{eq:BD boundary condition} sets $c_1 = 0$, selecting
$u^*_{\vec k}$ as the saddle. That the saddle contains no
$u_{\vec k}$-component is the path-integral statement of
$\hat a_{\vec k}|\Omega\rangle = 0$. We hence choose
\begin{equation}
\phi_{\vec{k}} = \varphi_{\vec{k}}
\left(\frac{\eta}{\eta_c}\right)^{3/2}\frac{H_{3/2}^{(2)}(-k\eta)}{H_{3/2}^{(2)}(-k\eta_c)}~.
\end{equation}
After integrating by parts, the on-shell action becomes
\begin{align}
S_{\rm on\text{-}shell}
&= \frac{\ell^2}{2\eta_c^2}\int_{\mathbb{R}^3}\d^3\vec{x}
\,\phi\,\partial_\eta\phi\,\big|_{\eta=\eta_c} = \frac{\ell^2}{2}\int_{\mathbb{R}^3}\frac{\d^3\vec{k}}{(2\pi)^3}
\frac{k^2}{\eta_c(1-ik\eta_c)}\,
\varphi_{\vec{k}}\varphi_{-\vec{k}}~.
\end{align}
Substituting this into \eqref{eq:PsiBD} gives
\begin{equation}
\Psi_{\rm BD}[\varphi_{\vec{k}},\eta_c]
=
\prod_{\vec{k}\in \mathbb{R}^3}\mathcal{N}_{\vec{k}}\,\exp\!\left[\frac{i\ell^2}{2}\int_{\mathbb{R}^3}
\frac{\d^3\vec{k}}{(2\pi)^3}
\frac{k^2}{\eta_c(1-ik\eta_c)}\varphi_{\vec{k}}\varphi_{-\vec{k}}\right]~,
\end{equation}
where $\mathcal{N}_{\vec{k}}$ is a normalization constant that we fix such that
\begin{align}\label{eq:BD norm}
1
&= \int\!\prod_{\vec{k}\in \mathbb{R}^3}\d\varphi_{\vec{k}}\,
|\Psi_{\rm BD}[\varphi_{\vec{k}},\eta_c]|^2 = \int\!\prod_{\vec{k}\in \mathbb{R}^3}\d\varphi_{\vec{k}}\,
|{\mathcal{N}}_{\vec{k}}|^2
\exp\!\left[-\ell^2\int_{\mathbb{R}^3}\frac{\d^3\vec{k}}{(2\pi)^3} \frac{k^3}{1+k^2 \eta_c^2}|\varphi_{\vec{k}}|^2\right]~.
\end{align}
Performing the Gaussian integrals gives
\begin{equation}
|\mathcal{N}_{\vec{k}}|
= \left(\frac{\ell^2 k^3}{\pi}\right)^{1/4}
\frac{1}{(1+k^2\eta_c^2)^{1/4}}~.
\end{equation}
The condition \eqref{eq:BD norm} fixes $\mathcal{N}_{\vec{k}}$ only up to a phase; demanding that
$\Psi_{\rm BD}$ solve the Schr\"odinger equation \eqref{eq:BD Schrodinger} below fixes the residual
$\eta_c$-dependent phase,\footnote{The $\varphi$-independent part of \eqref{eq:BD Schrodinger} gives
$\partial_{\eta_c}\ln\mathcal{N}_{\vec{k}} = -k^2\eta_c/2(1-ik\eta_c)$, and
$\int k^2\eta\, \d\eta/(1-ik\eta)=\ln(1-ik\eta)-(1-ik\eta)$, so that
$\mathcal{N}_{\vec{k}}\propto \e^{-ik\eta_c/2}/\sqrt{1-ik\eta_c}$.}
so that the normalized Bunch--Davies wavefunction reads
\begin{equation}
\Psi_{\rm BD}[\varphi_{\vec{k}},\eta_c]
= \prod_{\vec{k}\in \mathbb{R}^3}
\left(\frac{\ell^2 k^3}{\pi}\right)^{1/4}
\frac{\e^{-\frac{i k \eta_c}{2}}}{\sqrt{1-ik\eta_c}}\,
\exp\!\left[\frac{i\ell^2}{2}\int_{\mathbb{R}^3}
\frac{\d^3\vec{k}}{(2\pi)^3}
\frac{k^2}{\eta_c(1-ik\eta_c)}\varphi_{\vec{k}}\varphi_{-\vec{k}}\right]~.
\end{equation}
The wavefunction satisfies the Schr\"odinger equation
\begin{equation}\label{eq:BD Schrodinger}
    \sum_{\vec{k} \in \mathbb{R}^3} \bigg(-\frac{\eta_c^2}{2\ell^2}\frac{\delta}{\delta \varphi_{\vec{k}}}\frac{\delta}{\delta \varphi_{-\vec{k}}} +\frac{\ell^2 k^2}{2\eta_c^2}\varphi_{\vec{k}}\varphi_{-\vec{k}}\bigg) \Psi_{\rm BD}[\varphi_{\vec{k}},\eta_c] = i\partial_{\eta_c}\Psi_{\rm BD}[\varphi_{\vec{k}},\eta_c]~.
\end{equation}
Using the BD wavefunction we can in particular obtain expectation values
\begin{equation}
    \langle \varphi_{\vec{k}_1}\ldots \varphi_{\vec{k}_n}\rangle  = \frac{\int \prod_{\vec{k}} \d\varphi_{\vec{k}} |\Psi_{\mathrm{BD}}[\varphi_{\vec{k}},\eta_c]|^2 \varphi_{\vec{k}_1}\cdots \varphi_{\vec{k}_n}}{\int\prod_{\vec{k}} \d\varphi_{\vec{k}}|\Psi_{\mathrm{BD}}[\varphi_{\vec{k}}, \eta_c]|^2}~,
\end{equation}
where the denominator equals one with the normalization above. In particular, the two-point function
reads
\begin{equation}
\langle\varphi_{\vec{k}}\,\varphi_{\vec{k}'}\rangle
= (2\pi)^3\delta^{(3)}(\vec{k}+\vec{k}')\,\frac{1+k^2\eta_c^2}{2k^3\ell^2}~,
\end{equation}
which in the late-time limit $\eta_c\rightarrow 0^-$ reduces to the scale-invariant spectrum
\begin{equation}\label{eq: k3}
\langle\varphi_{\vec{k}}\,\varphi_{\vec{k}'}\rangle
\approx \frac{(2\pi)^3\delta^{(3)}(\vec{k}+\vec{k}')}{2k^3\ell^2}~.
\end{equation}
Although this BD vacuum is derived for $V=0$, it serves as the standard initial state for fluctuations
in inflationary cosmology, where its two-point function matches the observed nearly scale-invariant
CMB spectrum. The two-point function diverges as $k \to 0$, reflecting an infrared divergence of the
massless scalar in de Sitter: the $\vec{k}=0$ zero mode has vanishing Gaussian weight, and the
wavefunctional is correspondingly not normalizable in the strict infrared.

\paragraph{2. A conformally coupled scalar in dS$_4$.}
A conformally coupled scalar field in dS$_4$ has $m^2 \ell^2 = 2$. In the flat slicing it has the action
\begin{equation}\label{eq: flat dS4 x}
S_L = \frac{\ell^2}{2} \int_{\mathbb{R}^3} \d^3 \vec{x}\int_{-\infty}^{\eta_c} \frac{\d \eta}{\eta^2} \left((\partial_\eta \phi)^2 - (\partial_{\vec{x}} \phi)^2 - \frac{{m^2\ell^2}}{\eta^2}\phi^2\right)~.
\end{equation}
The equations of motion are
\begin{equation}\label{eq: eom cc x}
 -\partial_\eta\left(\frac{1}{\eta^2}\partial_\eta \phi (\eta,\vec{x})\right) + \frac{1}{\eta^2}\partial_{\vec{x}}^2\phi(\eta,\vec{x}) - \frac{m^2\ell^2}{\eta^4}\phi(\eta,\vec{x}) = 0~.
\end{equation}
We expand the field in Fourier modes (\ref{eq:dS4 Fourier}) and obtain the solutions
\begin{equation}
\phi_{\vec k}(\eta) = c_1\,|\eta|^{3/2} H^{(1)}_{\nu}(-k\eta)
                    + c_2\,|\eta|^{3/2} H^{(2)}_{\nu}(-k\eta)~,
\qquad \nu \equiv \tfrac12\sqrt{9-4m^2\ell^2} = \frac{1}{2}~.
\end{equation}
The first solution is purely positive frequency in the far past and is hence identified with the
Bunch--Davies mode function $u_{\vec k}$ of (\ref{eq: mode expansion}); Klein--Gordon normalization
fixes
\begin{equation}
  u_{\vec k}(\eta)=-i\frac{\sqrt{\pi}}{2\ell}\,(-\eta)^{3/2} H^{(1)}_{\frac{1}{2}}(-k\eta)~,
\end{equation}
where we choose the phase for later convenience.
The second solution is its complex conjugate: it is the unique solution that vanishes along the
deformed contour $\eta \to -\infty(1-i\epsilon)$, and therefore the one selected as the saddle of the
Bunch--Davies path integral (\ref{eq:PsiBD})--(\ref{eq:BD boundary condition}). Performing a calculation analogous to the massless scalar we obtain 
\begin{equation}
    \Psi_{\mathrm{BD}}[\varphi_{\vec k},\eta_c] = \prod_{\vec k \in \mathbb{R}^3} \left(\frac{k\ell^2}{\pi \eta_c^2}\right)^{\frac{1}{4}} \e^{-\frac{ik\eta_c}{2}} \mathrm{exp}\bigg[{\frac{i\ell^2}{2}\int \frac{\d^3 \vec k}{(2\pi)^3} \frac{1+i k\eta_c}{\eta_c^3}\varphi_{\vec k}\varphi_{-\vec k}}\bigg]~.
\end{equation}
We hence obtain
\begin{equation}\label{eq: variance conformally coupled}
\langle\varphi_{\vec{k}}\,\varphi_{\vec{k}'}\rangle
= (2\pi)^3\delta^{(3)}(\vec{k}+\vec{k}')\,\frac{\eta_c^2}{2k\ell^2}~.
\end{equation}

\paragraph{Conformal structure at late times.}
At late times $\eta \rightarrow 0^-$ we observed that the de Sitter isometries act as conformal
transformations on the late-time slice (\ref{eq:conformal transformations}). In this limit the spatial
derivative in \eqref{eq: eom cc x} can be ignored and the solutions take the form
\begin{equation}
   \phi(\eta, \vec{x}) \approx (-\eta)^{\Delta_+} \mathcal{O}^{(\Delta_+)}(\vec{x}) + (-\eta)^{\Delta_-} \mathcal{O}^{(\Delta_-)}(\vec{x})~,\qquad \Delta_\pm = \frac{3}{2} \pm \sqrt{\frac{9}{4} - m^2 \ell^2}~.
\end{equation}
The boundary data $\mathcal{O}^{(\Delta_\pm)}$ furnish the complementary series representation of
$\mathrm{SO}(1,4)$ with $\Delta(3-\Delta)=m^2\ell^2$ --- the shadow pair $\Delta_\pm$ labels the same
UIR (see table \ref{tab:scalar-reps-dS4}). We can now derive the relation between
$\mathcal{O}^{(\Delta_\pm)}$ and the operators $\hat{a}_{\vec{k}}$, $\hat{a}^\dagger_{\vec{k}}$ (\ref{eq: mode expansion}). For the
conformally coupled scalar, $\nu = \tfrac12$, the Hankel function truncates to an elementary function,
\begin{equation}
   H^{(1)}_{\frac{1}{2}}(-k\eta) = -i\sqrt{\frac{2}{\pi k}}\,|\eta|^{-1/2}\,\e^{-ik\eta}~.
\end{equation}
The modes are plane waves at all times, not just in the far past, and the canonically normalized mode
function reduces to
\begin{equation}
   u_{\vec k}(\eta) = \frac{\e^{-ik\eta}}{\sqrt{2k}}\,\frac{\eta}{\ell}~.
\end{equation}
Writing the exponentials in terms of trigonometric functions, the field operator at finite time is
\begin{equation}
    \hat{\phi}(\eta,\vec{x}) = \int \frac{\d^3\vec{k}}{(2\pi)^3}
    \frac{\eta}{\ell\sqrt{2k}}
    \left((\hat{a}_{\vec{k}} + \hat{a}_{-\vec{k}}^\dagger) \cos(k\eta)
    - i(\hat{a}_{\vec{k}} - \hat{a}_{-\vec{k}}^\dagger)\sin(k\eta) \right)
    \e^{i\vec{k}\cdot\vec{x}}~,
\end{equation}
which at late times behaves as
\begin{align}
    \hat{\phi}(\eta,\vec{x})
    &\approx \int \frac{\d^3 \vec{k}}{(2\pi)^3} \left(
    \eta\,\frac{\hat{a}_{\vec{k}} + \hat{a}_{-\vec{k}}^\dagger}{\ell\sqrt{2k}}
    - i\,\frac{\eta^2}{\ell} \sqrt{\frac{k}{2}}\,
    (\hat{a}_{\vec{k}} - \hat{a}_{-\vec{k}}^\dagger)\right) \e^{i \vec{k}\cdot\vec{x}} \cr
    &= \int \frac{\d^3 \vec{k}}{(2\pi)^3} \left(
    (-\eta)^{\Delta_-} \widehat{\mathcal{O}}^{(\Delta_-)}_{\vec{k}}
    + (-\eta)^{\Delta_+} \widehat{\mathcal{O}}^{(\Delta_+)}_{\vec{k}}\right)
    \e^{i \vec{k}\cdot\vec{x}}~,\qquad \eta\rightarrow0^-~,
\end{align}
where $\Delta_\mp = \tfrac32 \mp \nu = 1,\,2$ and
\begin{equation}\label{eq: boundary relations x}
    \widehat{\mathcal{O}}^{(\Delta_-)}_{\vec{k}}
    \equiv -\frac{1}{\ell\sqrt{2k}}\,(\hat{a}_{\vec{k}} + \hat{a}_{-\vec{k}}^\dagger)~,
    \qquad
    \widehat{\mathcal{O}}^{(\Delta_+)}_{\vec{k}}
    \equiv -\frac{i}{\ell}\sqrt{\frac{k}{2}}\,(\hat{a}_{\vec{k}} - \hat{a}_{-\vec{k}}^\dagger)~.
\end{equation}
The operators $\widehat{\mathcal{O}}^{(\Delta_\pm)}_{\vec{k}}$ are the Fourier modes of boundary fields
$\mathcal{O}^{(\Delta_\pm)}(\vec{x})$ that transform as conformal primaries with dimension
$\Delta_{\pm}$; they are Hermitian in position space,
$\big(\widehat{\mathcal{O}}^{(\Delta_\pm)}_{\vec{k}}\big)^\dagger=\widehat{\mathcal{O}}^{(\Delta_\pm)}_{-\vec{k}}$.
From \eqref{eq: boundary relations x} and $[\hat a_{\vec k},\hat a^\dagger_{\vec k'}]
=(2\pi)^3\delta^{(3)}(\vec k-\vec k')$ one finds
\begin{equation}
  [\widehat{\mathcal O}^{(\Delta_\mp)}_{\vec k},\widehat{\mathcal O}^{(\Delta_\mp)}_{\vec k'}]=0~,\qquad
  [\widehat{\mathcal O}^{(\Delta_-)}_{\vec k},\widehat{\mathcal O}^{(\Delta_+)}_{\vec k'}]
  =-\frac{i}{\ell^{2}}\,(2\pi)^3\delta^{(3)}(\vec k+\vec k')~.
\end{equation}
This is the Fourier transform of (\ref{eq: latetime commutator}) for $d=3$ where the $(2\pi)^3$ comes from the transformation and we use $\Delta_\pm$ instead of $\Delta$ and $\bar \Delta$.

\paragraph{3. Gauge field in dS$_2$.}
We now specialize to two-dimensional de Sitter space. The Lorentzian action for a scalar field with
mass squared $m^2$ is
\begin{equation}\label{eq:2dscalar general x}
S_L= -\frac{1}{2}\int \d^2 x \sqrt{-g}\left(g^{\mu \nu}\partial_\mu\phi \partial_\nu \phi +m^2 \phi^2\right)~.
\end{equation}
In flat slicing coordinates this becomes
\begin{equation}\label{eq: flat dS2 x}
S_L = \frac{1}{2} \int_{\mathbb{R}} \d x\int_{-\infty}^{\eta_c} \d \eta \left((\partial_\eta \phi)^2 - (\partial_x \phi)^2 - \frac{m^2\ell^2}{\eta^2}\phi^2\right)~,
\end{equation}
with unit kinetic prefactor owing to the conformal flatness of two dimensions. The equations of motion
are
\begin{equation}
 -\partial_\eta^2 \phi + \partial_x^2 \phi - \frac{m^2\ell^2}{\eta^2}\phi = 0~.
\end{equation}
Decomposing $\phi(\eta,x)=\int_{\mathbb{R}}\frac{\d k}{2\pi}\,\e^{ikx}\phi_k(\eta)$, the solutions are
Hankel functions,
\begin{equation}
\phi_k(\eta) = c_1\,|\eta|^{1/2} H^{(1)}_{\nu}(-k\eta)
             + c_2\,|\eta|^{1/2} H^{(2)}_{\nu}(-k\eta)~,
\qquad \nu = \tfrac12\sqrt{1-4m^2\ell^2}~,\quad k\equiv |k|~.
\end{equation}
As before, the deformed contour of \eqref{eq:BD boundary condition} sets $c_1=0$, and the condition at
$\eta_c$ fixes the saddle
\begin{equation}\label{eq: phik BD 2d x}
\phi_{k}(\eta) = \varphi_{k}
\left(\frac{\eta}{\eta_c}\right)^{1/2}\frac{H_{\nu}^{(2)}(-k\eta)}{H_{\nu}^{(2)}(-k\eta_c)}~.
\end{equation}
Discrete series (gauge) fields in dS$_2$ have $m^2 \ell^2 = -t(t+1)$, $t\geq 1$ (see table
\ref{tab:scalar-reps-dS2}). As an example we take $t=1$, for which $\nu=\tfrac32$ and the Hankel
functions in \eqref{eq: phik BD 2d x} reduce to elementary functions,
\begin{equation}
\phi_k(\eta) = \varphi_k \, \e^{ik(\eta-\eta_c)}\frac{1 + \frac{i}{k\eta}}{1 + \frac{i}{k\eta_c}}~.
\end{equation}
The on-shell action \eqref{eq: flat dS2 x} becomes
\begin{align}
S_{\rm on\text{-}shell} &= \frac{1}{2} \int_{\mathbb{R}} \d x \int_{\mathbb{R}} \frac{\d k}{2\pi} \int_{\mathbb{R}} \frac{\d k'}{2\pi} \e^{i(k+k')x}\,\phi_{k}(\eta_c)\, \partial_\eta \phi_{k'}(\eta)\Big|_{\eta = \eta_c} \cr
&= \frac{1}{2} \int_{\mathbb{R}} \frac{\d k}{2\pi}\, \phi_{-k}(\eta_c)\, \partial_\eta \phi_{k}(\eta)\Big|_{\eta = \eta_c}
= \frac{1}{2} \int_{\mathbb{R}} \frac{\d k}{2\pi}\, \frac{1}{\eta_c}\left(i k\eta_c-\frac{1}{1-i k\eta_c} \right)\varphi_k \varphi_{-k}~.
\end{align}
Fixing the normalization as in example 1 --- the modulus from $\int\prod_k\d\varphi_k\,|\Psi_{\rm BD}|^2=1$,
the phase from the Schr\"odinger equation below --- the normalized Bunch--Davies wavefunction reads
\begin{equation}
  \Psi_{\rm BD} [\varphi_{ k}, \eta_c] = \prod_{k\in \mathbb{R}} \left(\frac{k^3 \eta_c^2}{\pi}\right)^{1/4} \frac{\e^{-\frac{ik \eta_c}{2}}}{\sqrt{1- ik \eta_c}}\, \exp \left(-\frac{i}{2}\int_{\mathbb{R}}\frac{\d k}{2\pi}\,\frac{1}{\eta_c}\left(\frac{1}{1-ik\eta_c} -ik\eta_c\right)\varphi_k \varphi_{-k} \right)~,
\end{equation}
and it solves the Schr\"odinger equation
\begin{equation}
    \sum_{k\in \mathbb{R}}\left(- \frac{1}{2}\frac{\delta^2}{\delta \varphi_{{k}}\delta \varphi_{-{k}}} +\frac{k^2}{2} \varphi_{{k}}\varphi_{-{k}} -\frac{1}{\eta_c^2}\varphi_{{k}}\varphi_{-{k}}\right)\Psi_{\mathrm{BD}}[\varphi_{{k}},\eta_c] = i \partial_{\eta_c} \Psi_{\mathrm{BD}}[\varphi_{{k}},\eta_c]~.
\end{equation}
At late times the $\varphi$-dependent part of the logarithm of the wavefunction behaves as
\begin{equation}
   \log  \Psi_{\rm BD} [\varphi_{ k}, \eta_c] \approx \frac{i}{2} \int_{\mathbb{R}} \frac{\d k}{2\pi}\left(-\frac{1}{\eta_c} + k^2 \eta_c + i k^3 \eta_c^2 + \ldots \right)\varphi_k \varphi_{-k}~,\qquad \eta_c\rightarrow0^-~,
\end{equation}
so that the two-point function scales as
\begin{equation}
    \langle \varphi_{{k}}\varphi_{k'} \rangle = 2\pi\delta(k+k')\,\frac{1}{2k^3\eta_c^2}~.
\end{equation}
In general, for a discrete series field with $m^2\ell^2 = -t(t+1)$ the late-time two-point function
scales as
\begin{equation}\label{eq:k2t+1}
    \langle \varphi_{{k}}\varphi_{k'} \rangle = 2\pi\delta(k+k')\,\frac{((2t-1)!!)^2}{2k^{2t+1}\eta_c^{2t}}~.
\end{equation}
We can compare the $\eta_c$ and $k$ dependence of the late-time two point function (\ref{eq:k2t+1}) and (\ref{eq: k3}), (\ref{eq: variance conformally coupled}).

\subsection{Wavefunction for de Sitter gravity}

Starting from the Einstein–Hilbert action,
\begin{equation}
    S_{\mathrm{EH}} = \frac{1}{16\pi G_N}\int \d^4 x\sqrt{-g}\,(R^{(\rm 4D)}-2\Lambda)~,
\end{equation}
we perform the Arnowitt-Deser-Misner (ADM) decomposition,
\begin{equation}
    ds^2 = - N^2 \d t^2+ h_{ij}(N^i \d t + \d x^i)(N^j \d t+ \d x^j)~,
\end{equation}
where $N>0$ is the lapse and $N^i$ the shift vector. The action becomes
\begin{equation}
    S_{\mathrm{EH}} = \kappa \int \d^4 x\,\sqrt{h}\, N\left(K_{ij}K^{ij} -K^2  +R -2\Lambda\right)~,
    \quad K_{ij} = \frac{1}{2N}(\dot{h}_{ij} - \nabla_i N_j - \nabla_j N_i)~,
\end{equation}
where $\kappa \equiv \tfrac{1}{16\pi G_N}$ and $R$ is the Ricci scalar of $h_{ij}$.
The conjugate momenta are
\begin{subequations}
\begin{align}
  \pi^{ij} &= \frac{\delta \mathcal{L}}{\delta \dot{h}_{ij}} 
  = \frac{\sqrt{h}}{16\pi G_N}(K^{ij} - h^{ij} K)~,\\
  \pi^i &= \frac{\delta \mathcal{L}}{\delta \dot{N}_i} = 0~, 
  \qquad 
  \pi = \frac{\delta \mathcal{L}}{\delta \dot{N}} = 0~.
\end{align}
\end{subequations}
Thus $N$ and $N^i$ are Lagrange multipliers enforcing constraints.
The Hamiltonian reads
\begin{align}
  H &= \int \d^3 x \left(
  N\left[\frac{1}{2\kappa \sqrt{h}} G_{ij,kl}\pi^{ij}\pi^{kl} 
  +\kappa \sqrt{h} (2\Lambda- R)\right] 
  + 2\nabla_i N_j\,\pi^{ij}\right)~,\\
  G_{ij,kl} &\equiv h_{ik}h_{jl} +h_{il}h_{jk} - h_{ij}h_{kl}~.
\end{align}
The tensor $G_{ij,kl}$ defines the DeWitt metric on superspace, the space of spatial metrics.
Variation with respect to $N$ and $N^i$ yields the Hamiltonian and momentum constraints,
\begin{subequations}\label{eq: classical Hamiltonian constraint}
\begin{align}
    \frac{1}{2\kappa \sqrt{h}} G_{ij,kl}\pi^{ij}\pi^{kl} 
    +\kappa \sqrt{h} (2\Lambda- R)&=0~,\\
    \nabla_i \pi^{ij}&=0~.
\end{align}
\end{subequations}
The momentum constraint generates spatial diffeomorphisms, while the Hamiltonian constraint enforces invariance under deformations of the spatial slice. Unlike in Yang–Mills theory, the Hamiltonian constraint is quadratic in the momenta.
Canonical quantization promotes (we restore an $\hbar$ dependence to order the semiclassical expansion below)
\begin{equation}
    \pi^{ij} \;\to\; - i \hbar \frac{\delta}{\delta h_{ij}}~,
\end{equation}
leading to the Wheeler–DeWitt equations \cite{DeWitt:1967yk}
\begin{subequations}
\begin{align}\label{eq:Hamilton constraint}
    \left(-\frac{\hbar^2}{2\kappa \sqrt{h}} 
    G_{ij,kl}\frac{\delta}{\delta h_{ij}}\frac{\delta}{\delta h_{kl}} 
    + \kappa \sqrt{h} (2\Lambda-R)\right)\Psi[h_{ij}] &=0~,\\
    -2 i \hbar\, \nabla_i \frac{\delta \Psi[h_{ij}]}{\delta h_{ij}} &=0~.
\end{align}
\end{subequations}
The second equation implies that the wavefunctional is invariant under spatial diffeomorphisms,
\begin{equation}
    \Psi[h_{ij} + 2\nabla_{(i}\upsilon_{j)}] = \Psi[h_{ij}]~,
\end{equation}
so that
\begin{equation}
    \Psi = \Psi\big[\frac{h_{ij}}{\mathrm{Diff}}\big]~.
\end{equation}
We now consider a semiclassical ansatz
\begin{equation}
    \Psi = \exp\!\left(\frac{i}{\hbar} W[h_{ij}]\right)~,
\end{equation}
under which the Hamiltonian constraint reduces to the Hamilton–Jacobi equation
\begin{subequations}
\begin{align}
    -\frac{1}{2\kappa \sqrt{h}} 
    G_{ij,kl} \frac{\delta W}{\delta h_{ij}}\frac{\delta W}{\delta h_{kl}} 
    + \kappa \sqrt{h}(R- 2\Lambda) &=0~,\\
    2\nabla_i \frac{\delta W}{\delta h_{ij}} &=0~.
\end{align}
\end{subequations}
In the absence of an intrinsic notion of time, the role of late times is played by large volume \cite{Pimentel:2013gza}. We write
\begin{equation}
    h_{ij} = a^2 \tilde{h}_{ij}~, \qquad a \to \infty~.
\end{equation}
In terms of $\tilde h_{ij}$, the Hamilton–Jacobi equation becomes
\begin{equation}
     -\frac{1}{2\kappa a^3\sqrt{\tilde{h}}} 
     \widetilde{G}_{ij,kl} 
     \frac{\delta W}{\delta \tilde{h}_{ij}}\frac{\delta W}{\delta \tilde{h}_{kl}} 
     + \kappa \sqrt{\tilde{h}}(a\widetilde{R}- 2a^3\Lambda) =0~.
\end{equation}
We solve this equation using the ansatz
\begin{equation}
    W = a^3 \alpha \int \d^3 x \sqrt{\tilde{h}} 
    + a \beta \int \d^3 x \sqrt{\tilde{h}} \widetilde{R} 
    + W_0[\tilde h] + \mathcal{O}(a^{-1})~.
\end{equation}
Matching powers of $a$ determines
\begin{equation}\label{eq: alpha beta wavefunction}
    \alpha = 4\kappa \sqrt{\frac{\Lambda}{3}}~, 
    \qquad 
    \beta = - \kappa \sqrt{\frac{3}{\Lambda}}~, 
    \qquad 
    \tilde{h}_{ij}\frac{\delta W_0}{\delta \tilde{h}_{ij}}=0~.
\end{equation}
The last condition shows that $W_0$ is Weyl invariant and independent of $a$. In general, $W_0$ is a non-local functional of the metric.
In the large-$a$ limit, the wavefunction takes the form
\begin{equation}
    \Psi[h_{ij}] \sim 
    \exp\!\left(\frac{i}{\hbar} a^3 \alpha \int \sqrt{\tilde{h}} 
    + \frac{i}{\hbar} a \beta \int \sqrt{\tilde{h}} \widetilde{R}\right)
    \times \exp\!\left(\frac{i}{\hbar}W_0\right)~.
\end{equation}
Since $\alpha$ and $\beta$ are real, the first factor is a local, purely phase contribution, and therefore drops out of $|\Psi|^2$, leaving only the non-local piece $W_0$, in particular
\begin{equation}
    |\Psi[h_{ij}]|^2= \e^{-\frac{2}{\hbar} \mathrm{Im}(W_0[\tilde{h}])}~.
\end{equation}

Finally, we consider the norm of the wavefunction. Since the DeWitt metric $G_{ij,kl}$ is not positive definite, the natural inner product is subtle, similar to the Klein–Gordon case. 
Moreover, the late-time condition $\tilde{h}_{ij}\delta W_0/\delta \tilde{h}_{ij}=0$ enforces that $W_0$ is invariant under Weyl rescalings. This motivates treating Weyl transformations as a redundancy at $\mathcal{I}^+$, where the de Sitter isometry group $\mathrm{SO}(1,d+1)$ acts as the conformal group. From this perspective, observables at $\mathcal{I}^+$ are expected to be invariant under $\mathrm{Diff} \times \mathrm{Weyl}$, although these symmetry considerations fix the measure, they do not uniquely fix the wavefunction $\Psi$.
This motivates that at $\mathcal{I}^+$ the norm is given by \cite{Higuchi:1991tm} (see also \cite{Witten:2001kn,Maldacena:2002vr,Anninos:2012qw, Anninos:2020hfj,Chakraborty:2023yed, Collier:2025lux, Cotler:2025gui})
\begin{equation}\label{eq: norm at Ip}
    \langle \Psi | \Psi \rangle 
    \equiv \int \frac{[\mathcal{D}h_{ij}]}{\mathrm{vol}_{\mathrm{Diff}\times \mathrm{Weyl}}} 
    \, |\Psi[h_{ij}]|^2~.
\end{equation}

\paragraph{Connection to FLRW.}
Before moving on, we relate the classical Hamiltonian constraint to the Friedmann constraint
(\ref{eq:FLRW2}) discussed in Section~\ref{sec: cosmology}. The FLRW metric
\begin{equation}
    \d s^2 = - \d t^2 + a(t)^2 \gamma_{ij}\d x^i \d x^j~,
\end{equation}
with $\gamma_{ij}$ the unit-curvature metric on $\Sigma_k$ (so that $R[\gamma]=6k$), is of ADM form
with
\begin{equation}
    N= 1~, \qquad N^i=0~, \qquad h_{ij} = a(t)^2 \gamma_{ij}~.
\end{equation}
In this case, the extrinsic curvature is
\begin{equation}
    K_{ij}= \frac{\dot{a}}{a}h_{ij}~,
    \qquad
    K= 3\frac{\dot{a}}{a}~,
\end{equation}
and the conjugate momentum becomes
\begin{equation}
    \pi^{ij} = \kappa \sqrt{h}(K^{ij} - h^{ij}K)
    = -2 \kappa \sqrt{h}\,\frac{\dot{a}}{a}\, h^{ij}~.
\end{equation}
Using $G_{ij,kl}h^{ij}h^{kl}=-3$, one finds
\begin{equation}
    \frac{1}{2\kappa\sqrt{h}}G_{ij,kl} \pi^{ij}\pi^{kl}
    = -6 \kappa \sqrt{h}\left(\frac{\dot{a}}{a}\right)^2~.
\end{equation}
Note the sign: the scale factor carries negative kinetic energy --- the minisuperspace
manifestation of the indefiniteness of the DeWitt metric, and the simplest incarnation of the
conformal mode problem we encounter in Euclidean signature in subsection \ref{subsec:conformal mode problem}. The classical Hamiltonian
constraint (\ref{eq: classical Hamiltonian constraint}) then reduces to
\begin{equation}
   -\left(\frac{\dot{a}}{a}\right)^2 +\frac{\Lambda}{3} - \frac{k}{a^2}=0~,
\end{equation}
where we used $R[h] = \tfrac{6k}{a^2}$. This is the Friedmann constraint (\ref{eq:FLRW2}) with
$\rho=0$.

\subsection{Path integrals and the Hartle--Hawking wavefunction}

\begin{figure}[t]
\centering
\begin{tikzpicture}
[
    scale=1,
    boundary/.style={thick},
    bulk/.style={thick, dashed},
    curve/.style={thick},
    dcurve/.style={thick, dashed},
    label/.style={scale=0.7}
]
      \draw[dashed,  thick, opacity=0.6] (1.8,1) arc [start angle=0, end angle=180, x radius=.8cm, y radius=0.2cm];
  \draw[opacity=0.8,  thick] (0.2,1) arc [start angle=180, end angle=360, x radius=.8cm, y radius=0.2cm];
  \draw[opacity=0.8, thick, blue] (0.2,1) arc [start angle=180, end angle=360, x radius=.8cm, y radius=.8cm];
\coordinate (A) at (1.8,1);
\coordinate (B) at (0.2,1);
\coordinate (C) at (-1.5,2.5);
\coordinate (D) at (3.5,2.5);
\coordinate (E) at (3.5,-.5);
\draw[curve, bend right=25, red] (B) to (C);
\draw[curve, bend left=25, red] (A) to (D);
      \draw[dashed, very thick, opacity=0.9, gray] (3.5,2.5) arc [start angle=0, end angle=180, x radius=2.5cm, y radius=0.3cm];
  \draw[opacity=0.9, very thick, gray] (-1.5,2.5) arc [start angle=180, end angle=360, x radius=2.5cm, y radius=0.3cm];
 \fill[gray!30, opacity=0.8]
  (-1.5,2.5)
  arc[start angle=180, end angle=360, x radius=2.5cm, y radius=0.3cm]
  arc[start angle=0, end angle=180, x radius=2.5cm, y radius=0.3cm]
  -- cycle;
\draw[<-,thick] (1.75,.5)--(2,.5);
\node[label] at (2.15,.5) {$\Bigg\{$};
\node[label] at (3.2,.85) {Preparation of };
 \node[label] at (3.45,.52) {Euclidean vacuum};
 \node[label] at (2.95,.15) {state $|\Omega_{\mathrm{E}}\rangle$};
\node[label] at (-1.4,1.85) {Lorentzian};
 \node[label] at (-1.45,1.52) {evolution};
\node[label] at (-.6,.65) {Euclidean};
 \node[label] at (-.6,.32) {evolution};
\node[label] at (1,3.1) {$\mathcal{I}^+$};
\draw[thick,->] (6,.5)--(10,.5);
\draw[thick,->] (6.5,0) -- (6.5, 2.5);
\draw[very thick, ->,blue] (6.5, 1.5) -- (6.5,.5);
\draw[very thick, red, decoration={markings, mark=at position 0.5 with {\arrow{>}}},
      postaction={decorate}] (6.5,.5) -- (9.9,.5);
\draw[thick] (6.42, 1.5)  -- (6.58, 1.5);
\node at (6.2, 1.45) {$\frac{i\pi}{2}$};
\node[label] at (9.7,2.5) {$\tau$};
\draw[] (9.5,2.3) -- (9.8,2.3);
\draw[] (9.5,2.3) -- (9.5,2.6);
\end{tikzpicture}
\caption{The Hartle--Hawking construction. Left: the Euclidean vacuum $|\Omega_{\mathrm{E}}\rangle$
is prepared by gluing half of the Euclidean sphere (blue) onto Lorentzian de Sitter evolution (red).
Right: the corresponding contour in the complex time plane, descending from $\tau=i\pi/2$ along the
imaginary axis (Euclidean evolution, blue) and continuing along real time (Lorentzian evolution,
red).}
\label{fig:HH contour}
\end{figure}
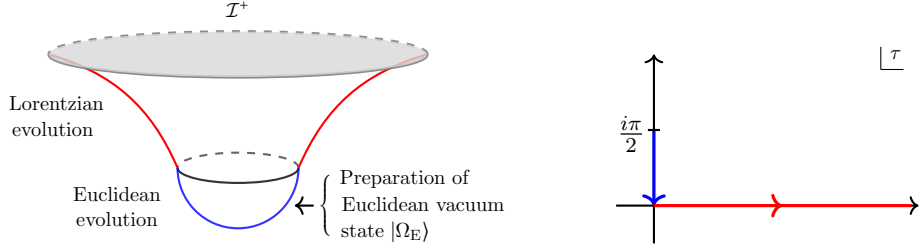

In ordinary quantum mechanics, the transition amplitude for a time-independent Hamiltonian
$\widehat{H}$ is
\begin{subequations}
\begin{align}
    K(x,x';t) &= \langle x| \e^{-it\widehat{H}} |x'\rangle = \langle x,t|x',0\rangle~,\\
    K(x,x';0) &= \delta(x-x')~,
\end{align}
\end{subequations}
which propagates the wavefunction according to
\begin{equation}
    \psi(x,t) = \int \d x' \, K(x,x';t)\,\psi(x',0)~.
\end{equation}
The transition amplitude admits a path integral representation
\begin{equation}
    K(x,x';t) = \int_{\substack{x(0)=x'\\ x(t)=x}} [\mathcal{D}x]\, \e^{iS[x]}~,
\end{equation}
which satisfies the Schr\"odinger equation
\begin{equation}
    i\partial_t K(x,x';t) = \widehat{H} K(x,x';t)~.
\end{equation}
Assuming for simplicity a discrete spectrum, inserting a complete set of energy eigenstates,
\begin{equation}
    \mathbb{I} = \sum_E |E\rangle \langle E|~,
\end{equation}
gives
\begin{equation}
    K(x,x';t) = \sum_E \psi_E(x)\psi_E^*(x')\, \e^{-iEt}~,
\end{equation}
where $\psi_E(x)=\langle x|E\rangle$. Taking $t\to\infty(1-i\epsilon)$ suppresses the excited states
and, up to normalization, projects onto the ground state,
\begin{equation}
    K(x,x';t) \sim \psi_{E_0}(x)\psi_{E_0}^*(x')\, \e^{-iE_0 t}~.
\end{equation}
From the path integral perspective, this corresponds to a contour deformation into the complex time
plane, illustrated in figure~\ref{fig:HH contour}. 

\begin{figure}[t]
    \centering
\begin{tikzpicture}
[
    scale=.7,
    boundary/.style={thick},
    bulk/.style={thick, dashed},
    curve/.style={thick},
    dcurve/.style={thick, dashed},
    label/.style={scale=0.7}
]
\fill[gray!20, opacity=0.5]
  (0,0)
  arc [start angle=180, end angle=360, x radius=2cm, y radius=0.5cm]
  -- (4,0)
  arc [start angle=0, end angle=180, x radius=2cm, y radius=0.5cm]
  -- cycle;
\shade[top color=gray!10, bottom color=gray!80, shading angle=90, opacity=0.6]
  (0,0)
  arc[start angle=180, end angle=360, x radius=2cm, y radius=2cm]
  -- (4,0.)
  arc[start angle=360, end angle=180, x radius=2cm, y radius=0.5cm]
  -- cycle;
 \draw[opacity=0.8, thick, black] (0,0) arc [start angle=180, end angle=360, x radius=2cm, y radius=2cm];
  \draw[opacity=0.8, very thick, red] (0,0) arc [start angle=180, end angle=360, x radius=2cm, y radius=0.5cm];
  \draw[very thick, opacity=0.8, red] (4,0) arc [start angle=0, end angle=180, x radius=2cm, y radius=0.5cm];
\fill[gray!20, opacity=0.5]
  (6,0)
  arc [start angle=180, end angle=360, x radius=2cm, y radius=0.5cm]
  -- (10,0)
  arc [start angle=0, end angle=180, x radius=2cm, y radius=0.5cm]
  -- cycle;
\shade[top color=gray!10, bottom color=gray!80, shading angle=90, opacity=0.6]
  (6,0)
  arc[start angle=180, end angle=360, x radius=2cm, y radius=2cm]
  -- (10,0.)
  arc[start angle=360, end angle=180, x radius=2cm, y radius=0.5cm]
  -- cycle;
 \draw[opacity=0.8, thick, black] (6,0) arc [start angle=180, end angle=360, x radius=2cm, y radius=2cm];
 \draw[opacity=0.8, very thick, red] (6,0) arc [start angle=180, end angle=360, x radius=2cm, y radius=0.5cm];
 \draw[very thick, opacity=0.8, red] (10,0) arc [start angle=0, end angle=180, x radius=2cm, y radius=0.5cm];
\coordinate (A) at (7,-1.2);
\coordinate (B) at (8,-1.2);
\coordinate (C) at (7.15,-1.28);
\coordinate (D) at (7.85,-1.28);
\draw[curve, bend right=45] (A) to (B);
\draw[curve, bend left=45] (C) to (D);
\fill[gray!20, opacity=0.5]
  (12,0)
  arc [start angle=180, end angle=360, x radius=2cm, y radius=0.5cm]
  -- (16,0)
  arc [start angle=0, end angle=180, x radius=2cm, y radius=0.5cm]
  -- cycle;
\shade[top color=gray!10, bottom color=gray!80, shading angle=90, opacity=0.6]
  (12,0)
  arc[start angle=180, end angle=360, x radius=2cm, y radius=2cm]
  -- (16,0.)
  arc[start angle=360, end angle=180, x radius=2cm, y radius=0.5cm]
  -- cycle;
 \draw[opacity=0.8, thick, black] (12,0) arc [start angle=180, end angle=360, x radius=2cm, y radius=2cm];
 \draw[opacity=0.8, very thick, red] (12,0) arc [start angle=180, end angle=360, x radius=2cm, y radius=0.5cm];
 \draw[very thick, opacity=0.8, red] (16,0) arc [start angle=0, end angle=180, x radius=2cm, y radius=0.5cm];
\coordinate (A) at (13,-1.2);
\coordinate (B) at (14,-1.2);
\coordinate (C) at (13.15,-1.28);
\coordinate (D) at (13.85,-1.28);
\draw[curve, bend right=45] (A) to (B);
\draw[curve, bend left=45] (C) to (D);
\coordinate (E) at (14.8,-1.5);
\coordinate (F) at (15.4,-1);
\coordinate (G) at (14.9,-1.5);
\coordinate (H) at (15.35,-1.1);
\draw[curve, bend right=45] (E) to (F);
\draw[curve, bend left=45] (G) to (H);
\end{tikzpicture}
\caption{The Hartle--Hawking wavefunction fixes a single boundary (red); the manifold that is filled
in can have arbitrary topology \cite{Chen:2020tes, Betzios:2020nry,Anninos:2020geh,Anninos:2026eqv}.}
\label{fig:HH topologies}
\end{figure}
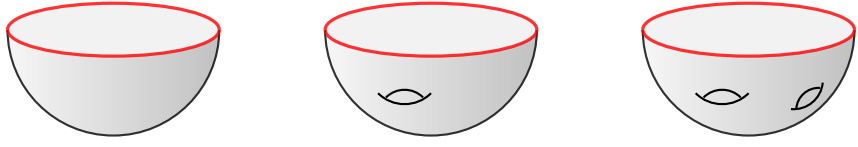

Hartle and Hawking proposed an analogous construction in gravity \cite{Hartle:1983ai}. Formally, one may define a
transition amplitude between spatial metrics,
\begin{equation}
    \langle h'_{ij}|h_{ij}\rangle
    = \int_{\substack{g|_{\Sigma_i}=h\\ g|_{\Sigma_f}=h'}}
    [\mathcal{D}g_{\mu\nu}]\, \e^{iS_L[g_{\mu\nu}]}~,
\end{equation}
which can be viewed as a functional of the boundary data and is expected to satisfy the
Wheeler--DeWitt equation as a constraint on its dependence on the boundary metrics. In de Sitter
space, it is natural to prepare a state via a Euclidean path integral. In global coordinates,
\begin{equation}
    \d s^2 = \ell^2\left(-\d\tau^2 + \cosh^2\tau \, \d\Omega_d^2\right)~,
\end{equation}
the Wick rotation $\tau \to i\theta$ gives the Euclidean metric on the sphere,
\begin{equation}
    \d s^2 = \ell^2\left(\d\theta^2 + \cos^2\theta \, \d\Omega_d^2\right)~.
\end{equation}
The analogue of the $i\epsilon$ prescription is to require regularity of the geometry at the ``south
pole'' of the sphere. For free fields on a rigid de Sitter background this is exactly the regularity
condition that selects the Bunch--Davies mode functions, so the Euclidean preparation reproduces the
Bunch--Davies wavefunction of the previous subsections --- the Hartle--Hawking proposal is its
gravitational completion. The Hartle--Hawking (HH) wavefunction is defined by a Euclidean path
integral over geometries that smoothly cap off,
\begin{equation}
    \Psi_{\mathrm{HH}}[h_{ij}]
    = \sum_{\mathcal{M}} \int_{\mathcal{M}} [\mathcal{D}g_{\mu\nu}] \, \e^{-S_E[g_{\mu\nu}]}~,
\end{equation}
where $S_E$ is the Euclidean action and $h_{ij}$ the induced metric on the boundary
$\partial\mathcal{M}$. The path integral is taken over manifolds that are regular in the interior and
match $h_{ij}$ at the boundary. In principle, this includes a sum over different topologies (see
figure~\ref{fig:HH topologies}). Moreover the Euclidean gravitational action is Gaussian unsuppressed, this is the conformal mode problem, and we return to this in subsection \ref{subsec:conformal mode problem}.

\noindent
\paragraph{A simple example.}
As a simple example, we consider the four-dimensional Euclidean Einstein Hilbert action on a compact four-dimensional manifold $\mathcal{M}$ with an $S^3$ boundary metric of the form
\begin{equation}\label{eq: HH bdy conditions}
    \tilde h_{ij}\d x^i \d x^j = a^2 \d\Omega_3^2~.
\end{equation}
The Einstein Hilbert action with a GHY boundary term is given by
\begin{equation}
    S_{\mathrm{EH}}[g_{\mu\nu}] = -\frac{1}{16\pi G_N} \int_{\mathcal{M}} \d^4 x \sqrt{g}(R-2\Lambda) -\frac{1}{8\pi G_N} \int_{\partial \mathcal{M}} \d^3 x\sqrt{h}K~,
\end{equation}
where $\Lambda = \frac{3}{\ell^2}$ in four-dimensions. On a four sphere with radius $\ell$ we have the embedding coordinates
\begin{equation}
    X_1^2 + X_2^2 + X_3^2 + X_4^2 + X_5^2 = \ell^2~,
\end{equation}
where
\begin{align}
    X_5 &=\ell \cos\chi~,\cr
    X_4 &=\ell \sin\chi \cos\psi~,\cr
    X_3 &= \ell \sin\chi \sin\psi \cos\theta~,\cr
    X_2 &= \ell \sin\chi \sin\psi \sin\theta \sin\varphi~,\cr
    X_1 &= \ell \sin\chi \sin\psi \sin\theta \cos\varphi~,
\end{align}
where $\varphi \in [0,2\pi)$ and $\chi, \psi,\theta \in [0,\pi]$. This leads to the metric and measure
\begin{equation}\label{eq: HH S4 measure}
    \frac{ds^2}{\ell^2} = \d\chi^2 + \sin^2\chi \d\Omega_3^2~,\quad \quad \sqrt{g} = \ell^4 \sin^3\chi \sin^2 \psi \sin\theta~.
\end{equation}
As $\chi$ goes from 0 to $\pi$ the sphere goes from one pole to the other. If we fix $X_5$ we have a sphere of radius $\sqrt{\ell^2 -X_5^2}= \ell \sin\chi$ and the boundary conditions (\ref{eq: HH bdy conditions}) fix
\begin{equation}\label{eq:boundary sinchi}
    \ell \sin \chi_b = a~,
\end{equation}
where $\chi_b$ is the boundary angle. Depending on the ratio $\frac{a}{\ell}$ the equation (\ref{eq:boundary sinchi}) has two real, one real, or only complex solutions. 
We first consider the case $a<\ell$, in which case we find two solutions: either we cap the sphere off below or above the equator.  The bulk part of the Einstein Hilbert action evaluates to 
\begin{equation}
    S_{\mathrm{EH}}[g_{\mu\nu}] = -\frac{3}{8\pi G_N \ell^2} \int_{0}^{2\pi} \d\varphi\int_0^\pi\d\theta\int_0^\pi\d\psi \int_0^{\chi_b}\d\chi \sqrt{g} = -\frac{\pi\ell^2}{4G_N} (2-3c+c^3)~,
\end{equation}
where $\sqrt{g}$ is given in (\ref{eq: HH S4 measure}) and $\chi_b$ and $c$ follow from 
\begin{equation}\label{eq: pm c}
    \cos\chi_b = \cos\arcsin\left(\frac{a}{\ell}\right) = c\quad \mathrm{or} \quad \cos\chi_b = \cos\left(\pi -\arcsin\left(\frac{a}{\ell}\right) \right) =-c~,\quad c\equiv   \sqrt{1-\frac{a^2}{\ell^2}}~.
\end{equation}
The extrinsic curvature is $K= \frac{3}{\ell}\cot \chi_b$ which leads to the boundary contribution 
\begin{equation}
    S_{\mathrm{GHY}}[h_{ij}] = -\frac{\pi\ell^2}{4G_N} (3c-3c^3)~.
\end{equation}
Adding the EH and the GHY term we obtain the on-shell action
\begin{equation}\label{eq: a<l solution}
   S^{\mp, a<\ell}_{\mathrm{EH}} = -\frac{\pi\ell^2}{2G_N} (1-c^3) = -\frac{\pi\ell^2}{2G_N} \Big[1\mp \Big(1-\frac{a^2}{\ell^2}\Big)^{\frac{3}{2}}\Big]~,\quad a<\ell~.
\end{equation}
The saddle that contributes in the Hartle-Hawking contour is $S^{-, a<\ell}_{\mathrm{EH}}$ \cite{Hartle:1983ai}. 
For $a>\ell$ the continuation $c\rightarrow \pm i (\frac{a^2}{\ell^2}-1)^{1/2}$
turns the solutions into a complex-conjugate pair leading to
\begin{equation}\label{eq: a>l solution}
    \e^{-S_{\mathrm{EH}}^{+,a>\ell}} + \e^{-S_{\mathrm{EH}}^{-,a>\ell}} = 2\e^{\frac{\pi\ell^2}{2G_N}} \cos \bigg[\frac{\pi\ell^2}{2G_N}\left(\frac{a^2}{\ell^2}-1\right)^{\frac{3}{2}} \bigg]~,\quad a>\ell~.
\end{equation}
Moreover if we expand the phase in (\ref{eq: a>l solution}) for large $a$ we obtain
\begin{equation}
   \frac{\pi\ell^2}{2G_N}\left(\frac{a^2}{\ell^2}-1\right)^{\frac{3}{2}} \simeq a^3\frac{1}{\ell}\frac{4}{16\pi G_N}2\pi^2 -a\ell \frac{1}{16\pi G_N}12\pi^2 +\mathcal{O}(a^{-1})~.
\end{equation}
which using $\int \d^3x \sqrt{\tilde{h}} = 2\pi^2$ and $\widetilde R=6$ matches the coefficients (\ref{eq: alpha beta wavefunction}).

Instead of Dirichlet boundary conditions we could also consider conformal boundary conditions \cite{Witten:2018lgb, Anninos:2024wpy}. Instead of fixing the value of the metric at the boundary this fixes the trace of the extrinsic curvature and the conformal class of the metric. The coefficient of the boundary term in the Euclidean action also gets adjusted leading to
\begin{equation}
    S_{\mathrm{EH}}[g_{\mu\nu}] = -\frac{1}{16\pi G_N} \int_{\mathcal{M}} \d^4 x \sqrt{g}(R-2\Lambda) -\frac{1}{24\pi G_N} \int_{\partial \mathcal{M}} \d^3 x\sqrt{h}K~,
\end{equation}
Evaluating the action with $c = \cos \chi_b$ we obtain
\begin{equation}
S_{\mathrm{EH}} = -\frac{\pi\ell^2}{2G_N}(1- \cos \chi_b) =-\frac{\pi\ell^2}{2G_N} \left(1-\frac{k}{\sqrt{1+k^2}}\right)~, \quad k\equiv \frac{K\ell}{3}~.
\end{equation}
where we used that $K= \frac{3}{\ell}\cot \chi_b$. This leads to 
\begin{equation}
    \Psi[K,\mathrm{round}~S^3] \sim \e^{\frac{\pi\ell^2}{2G_N} \left(1-\frac{k}{\sqrt{1+k^2}}\right)}
\end{equation}
which in particular satisfies (see figure \ref{fig: York norm})
\begin{equation}\label{eq: York norm}
    \Psi[K,\mathrm{round}~S^3]\Psi[-K,\mathrm{round}~S^3] = \e^{\frac{\pi\ell^2}{G_N}}~.
\end{equation}
This is a saddle point result\footnote{In a two-dimensional model with de Sitter saddles such a pairing was obtained to one-loop order \cite{Anninos:2025fer}.} but it is indicative that the inner product should pair the wavefunctions with opposite $K$. We can interpret $K$ as the York time \cite{York:1972sj} and perhaps we should see this as an indication that in the inner product we should only integrate over the conformal class of metric across the interface \cite{Anninos:2026hia}. This might be a sharper version of the proposal in (\ref{eq: norm at Ip}).

\begin{figure}
\centering
\begin{tikzpicture}
  \def\R{2}          
  \def\zc{0.6}       
  \pgfmathsetmacro\rc{sqrt(\R*\R-\zc*\zc)}   
  \pgfmathsetmacro\thc{atan2(\zc,\rc)}       
  \pgfmathsetmacro\bc{0.22*\rc}              
  \pgfmathsetmacro\be{0.22*\R}               
  \fill[blue!10] (0,0) circle (\R);                          
  \fill[red!12] ({180-\thc}:\R) arc({180-\thc}:\thc:\R)      
                arc(0:-180:{\rc} and \bc) -- cycle;
  \draw[densely dotted, gray!90, line width=0.5pt] (\R,0) arc(0:-180:{\R} and \be);
  \draw[dashed, gray!60, very thin] (0,0) -- (0,\R);
  \draw[dashed, gray!60, very thin] (0,0) -- (\thc:\R);
  \draw[gray!60, thin] (0,0.55) arc(90:\thc:0.55);
  \node[black!60] at (52:0.85) {\scriptsize $\chi_b$};
  \fill[gray!60] (0,0) circle (0.035);
  \draw[black!80, line width=0.8pt] (0,0) circle (\R);
  \draw[purple!70!black, dashed, line width=0.6pt]
        (-\rc,\zc) arc(180:0:{\rc} and \bc);                 
  \draw[purple!70!black, line width=1pt]
        (\rc,\zc) arc(0:-180:{\rc} and \bc);                 
  \node[black, below left] at (2.72,0.28) {$S^4$};        
  \node[purple!70!black, below left, font=\scriptsize] at (-1.22,0.88) {$S^3$};
  \node[red!55!black]  at (-0.85,1.32) {$\Psi[K]$};
  \node[blue!50!black] at (0,-1.15)    {$\Psi[-K]$};
  \node at (0,-2.75) {$\Psi[K]\,\Psi[-K]\;=\;\e^{S_{\rm dS}}\;=\;\mathcal Z_{S^4}$};
\end{tikzpicture}
  \caption{For conformal boundary conditions there is one real saddle, and the two caps are labelled by $\pm K$.}
  \label{fig: York norm}
\end{figure}

\section{de Sitter Entropy}\label{sec:dSentropy}

The accelerated expansion of our Universe implies the existence of a cosmological event horizon. Observers in de Sitter space have access only to a finite region of spacetime, bounded by a horizon beyond which signals cannot reach them. For present-day cosmological parameters, this horizon lies at a distance of order $10^{10}$ light years. In this section we discuss the Gibbons-Hawking entropy $S_{\mathrm{dS}}$ \cite{Gibbons:1977mu}, conjecturally capturing the entropy of the de Sitter cosmological horizon. 
In particular we explain the conjecture $S_{\mathrm{dS}_{d+1}}= \log \mathcal{Z}_{\mathrm{grav}}^{S^{d+1}}$. $\mathcal{Z}_{\mathrm{grav}}^{S^{d+1}}$ is the Euclidean sphere partition function. 
That a partition function computes an entropy rather than the free energy has a simple explanation. For a black hole in asymptotically flat space the Euclidean path integral
at inverse temperature $\beta$ computes
\begin{eqnarray}\label{eq: thermo Law}
    S -\beta E= \log \mathcal{Z} ~.
\end{eqnarray}
Euclidean de Sitter space is a sphere, and the spatial slices are closed manifolds and hence 
there is no location to anchor the ADM energy.
In de Sitter (\ref{eq: thermo Law}) reduces to \cite{Gibbons:1976ue}
\begin{eqnarray}\label{eq: thermo Law dS}
    S= \log \mathcal{Z} ~.
\end{eqnarray}

\paragraph{The cosmological horizon}
The presence of a cosmological horizon is manifest in the static patch metrics (\ref{eq: static patch 1}) and (\ref{eq: static patch}), where the timelike Killing vector becomes null at the horizon. In particular, $g_{tt}=0$ at $\rho=\pi/2$ in (\ref{eq: static patch 1}), and at $r=1$ in (\ref{eq: static patch}). The area of the cosmological horizon is (\ref{eq: horizon area})
\begin{equation}
    A_{\mathrm{hor}} = \ell^{d-1} \int_{S^{d-1}} \d\Omega_{d-1}
    =
    \begin{cases}
    2\pi \ell~, & d=2~,\\
    4\pi \ell^2~, & d=3~.
    \end{cases}
\end{equation}

Inspired by the Bekenstein--Hawking area law for black holes, one assigns an entropy to the de Sitter horizon,
\begin{equation}
    S_{\mathrm{dS}} = \frac{A_{\mathrm{hor}}}{4G_N}~.
\end{equation}
In dS$_4$, restoring dimensions,
\begin{equation}\label{eq: SdS 4D}
   S_{\mathrm{dS}_4}= \frac{\pi \ell^2 c^3}{\hbar G_N}~. 
\end{equation}
Using $\Lambda = 3/\ell^2$, this gives $S_{\mathrm{dS}_4} \sim 10^{122}$ for the observed Universe. This enormous value suggests a Hilbert space of dimension $\sim 
\e^{10^{122}}$. As a comparison, the entropy of all stars in the observable Universe is $\approx 10^{81}$, the entropy of the CMB is $\approx 10^{89}$. The entropy of super massive black holes in the center of galaxies is $10^{104}$.

The microscopic interpretation of the de Sitter entropy remains unclear. In particular it is not known what degrees of freedom are being counted or whether it has a completely different interpretation. Understanding the origin of the de Sitter entropy therefore remains a central open problem. 
This motivates the question: What are the macroscopic constraints on a microscopic realization of the de Sitter entropy \cite{Gibbons:1977mu,Bousso:2000nf, Banks:2000fe,Fischler:2000,Parikh:2004wh, Dong:2018cuv, Anninos:2020hfj, Shaghoulian:2021cef,Chandrasekaran:2022cip, Maldacena:2024spf}? 

\begin{figure}[ht]
\centering
    \begin{tikzpicture}[scale=1.2]
        \begin{scope}[shift = {(5,0)}]

\coordinate (A) at (0.02, 0.02);
    \coordinate (B) at (0.02,2.98);
    \coordinate (C) at (1.493, 1.493);

\coordinate (D) at (2.98, 0.02);
    \coordinate (E) at (2.98,2.98);
    \coordinate (F) at (1.493, 1.493);    

\shade[right color=gray!30, left color=gray!30] (A) -- (B) -- (C) -- cycle;

\draw[very thick](0,0) -- (0,3.02);   
\draw[very thick](0,3) -- (3.02,3.);       
\draw[very thick](3,0) -- (3,3);   
\draw[very thick](3.02,0) -- (-.02,0);   
\draw[very thick](3,3) -- (0,0);   
\draw[very thick](0,3) -- (3,0);

\node[scale=.7, rotate = 315] at (.9,2.3) {event horizon}; 
\node[scale=.7, rotate = 45] at (.9,.7) {$\rho = \pi/2$};
\node[scale=.7, rotate = 90] at (-.15,1.5) {$\rho = 0$};

\node[scale=.8] at (1.5,3.2) {$\mathcal{I}^+$}; 

\node[scale=.8] at (1.5,-0.2) {$\mathcal{I}^-$}; 

        \end{scope}
    \end{tikzpicture}
\caption{An observer at $\rho=0$ is surrounded by an event horizon, created by the exponential accelerated expansion of spacetime.}
\label{fig:penrose}
\end{figure}
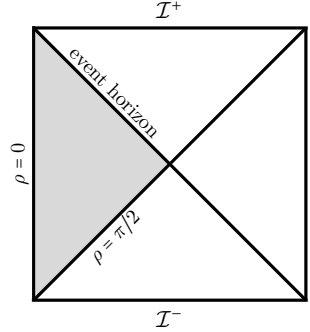

\subsection{Static Patch Density Matrix and Thermal Structure}
If one assumes that the Hilbert space factorizes as
\begin{equation}
    \mathcal{H} = \mathcal{H}_L \otimes \mathcal{H}_R \,,
\end{equation}
one may define a reduced density matrix associated with an observer following a static patch worldline, 
\begin{equation}
    \hat{\rho}_L = \tr_{\mathcal{H}_R} \big( |0\rangle \langle 0| \big) \,,
\end{equation}
where $|0\rangle$ denotes a de Sitter invariant state, such as the Bunch--Davies (Hartle--Hawking) vacuum.
A natural question is whether this reduced density matrix takes a thermal form,
\begin{equation}
    \hat{\rho}_L = \frac{\e^{-\beta_{\mathrm{dS}} \mathsf{H}_L}}{Z_{\mathrm{thermal}}} \,,
\end{equation}
with inverse temperature $\beta_{\mathrm{dS}} = 2\pi \ell$ and $\mathsf{H}_L$ the static patch Hamiltonian. For quantum fields in the Bunch–Davies state the answer
is affirmative \cite{Unruh:1976db, Gibbons:1977mu, Figari:1975km}. In a quantum field theory however the identification  $S_{\mathrm{dS}}= - \tr\hat{\rho}_L \log\hat{\rho}_L$ cannot hold literally as the entanglement entropy across the horizon
is dominated by ultraviolet modes and diverges. The conjecture is that in the gravitational theory
this divergence is absorbed into the renormalization of Newton's constant. 

It is important to emphasize, however, that the factorization 
$\mathcal{H}=\mathcal{H}_L\otimes\mathcal{H}_R$ 
is subtle for any local quantum field theory. The division into left and right static patches breaks down at the cosmological horizon, in part due to the infinite redshift experienced by a static observer and the presence of gauge constraints \cite{tHooft:1984kcu}. 
One could try to define a Hilbert space for global de Sitter space, but for the static patch this is more difficult because particles can enter and leave the horizon (and we encounter ultraviolet divergences near the horizon).
In algebraic language the obstruction has a precise name: the observables of the static patch
generate a von Neumann algebra of type III, which admits no trace, hence no density matrices
and no von Neumann entropy. Recent work \cite{Chandrasekaran:2022cip} has shown that including an observer’s degrees of freedom,
and imposing the gravitational constraints, improves the algebra to type II$_1$:
density matrices and entropies then exist, defined up to an additive
constant, and the maximal-entropy state is the de Sitter-invariant one \cite{Bousso:2000nf,Bousso:2000md, Maeda:1997fh}.
In this section
we discuss two methods that implement these ideas concretely: algebraic quantum field theory
and Euclidean techniques.

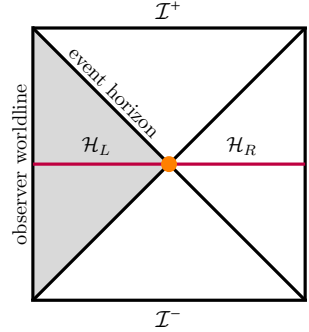
\begin{figure}[ht]
\centering
    \begin{tikzpicture}[scale=1.2]
        \begin{scope}[shift = {(5,0)}]

\coordinate (A) at (0.02, 0.02);
    \coordinate (B) at (0.02,2.98);
    \coordinate (C) at (1.493, 1.493);

\coordinate (D) at (2.98, 0.02);
    \coordinate (E) at (2.98,2.98);
    \coordinate (F) at (1.493, 1.493);    

\shade[right color=gray!30, left color=gray!30] (A) -- (B) -- (C) -- cycle;

\draw[very thick](0,0) -- (0,3.02);   
\draw[very thick](0,3) -- (3.02,3.);       
\draw[very thick](3,0) -- (3,3);   
\draw[very thick](3.02,0) -- (-.02,0);   
\draw[very thick](3,3) -- (0,0);   
\draw[very thick](0,3) -- (3,0);   

\draw[very thick, purple](0,1.5) -- (3,1.5);  
\fill[orange] (1.5,1.5) circle (2.5pt);

\node[scale=.7] at (.7,1.7) {$\mathcal{H}_L$}; 
\node[scale=.7] at (2.3,1.7) {$\mathcal{H}_R$}; 
\node[scale=.7, rotate = 315] at (.9,2.3) {event horizon}; 
\node[scale=.7, rotate = 90] at (-.15,1.5) {observer worldline};

\node[scale=.8] at (1.5,3.2) {$\mathcal{I}^+$}; 

\node[scale=.8] at (1.5,-0.2) {$\mathcal{I}^-$}; 

        \end{scope}
    \end{tikzpicture}
\caption{Static patch decomposition of de Sitter space. An observer has access only to the left patch $\mathcal{H}_L$, while degrees of freedom beyond the cosmological horizon are traced over.}
\label{fig:penrose entropy section}
\end{figure}

To motivate a thermal structure of the partition function, it is useful to recall the Euclidean path integral representation of the thermal partition function in quantum mechanics. Wick rotating $t \to -i\beta$ gives
\begin{equation}
    \mathcal{Z}(\beta) 
    = \tr \, \e^{-\beta \mathsf{H}}
    =
    \int_{\mathrm{PBC}} [\mathcal{D}x(\tau)] \,
    \e^{-S_{\mathrm{E}}[x(\tau)]} \,,
\end{equation}
where PBC stands for periodic boundary conditions  $x(\tau+\beta)=x(\tau)$.
For a bosonic and fermionic harmonic oscillator of frequency $\omega$, one finds upon regularization
\begin{subequations}
\begin{align}
    \mathcal{Z}_{\mathrm{bos}}(\beta)
    &= \det\nolimits^{-1/2}
    \left(-\frac{\d^2}{\d\tau^2} + \omega^2 \right)
    = \frac{1}{2\sinh\!\left(\frac{\beta \omega}{2}\right)}~, \\
    \mathcal{Z}_{\mathrm{fer}}(\beta)
    &= \det
    \left(\frac{\d}{\d\tau} + \omega\right)
    = 2\cosh\!\left(\frac{\beta \omega}{2}\right)~,
\end{align}
\end{subequations}
with boundary conditions
\begin{equation}
   x(\tau+\beta)=x(\tau), 
   \qquad 
   \psi(\tau+\beta)=-\psi(\tau)~.
\end{equation}

A free quantum field can be viewed as a collection of harmonic oscillators labeled by their frequency $\omega$. Summing over modes with spectral density $\rho(\omega)$ leads to
\begin{equation}\label{eq: Zcan}
    \log Z_{\mathrm{can}}(\beta)
    =
    \int_0^\infty \d\omega
    \left(
    -\rho_{\mathrm{bos}}(\omega)
    \log \!\left(
    2\sinh\!\frac{\beta\omega}{2}
    \right)
    +
    \rho_{\mathrm{fer}}(\omega)
    \log \!\left(
    2\cosh\!\frac{\beta\omega}{2}
    \right)
    \right).
\end{equation}

In de Sitter space, the naive spectral density is ill-defined due to the infinite redshift near the cosmological horizon. A more appropriate definition is obtained from the Harish--Chandra character (see subsection \ref{subsec:QFT in de Sitter}) of the de Sitter isometry group,\footnote{Of course this integral still has UV divergences which however can be treated locally. The IR divergences however are removed.}
\begin{equation}\label{eq: rhow}
    \rho(\omega)
    =
    \frac{1}{2\pi}
    \int_{-\infty}^{\infty}
    \d \mathfrak{t} \,
    \chi_{\Delta, s}(\mathfrak{t})
    \e^{i\omega \mathfrak{t}} \,,
\end{equation}
which effectively replaces the divergent density of states by a well-defined group-theoretic quantity. As a reminder for scalars the Harish-Chandra character is given by
\begin{equation}
    \chi_{\Delta,0}(\mathfrak{t}) = \frac{\e^{-\Delta \mathfrak{t}} + \e^{-\bar \Delta \mathfrak{t}}}{(1-\e^{-\mathfrak{t}})^{d}}~,\quad \bar \Delta \equiv d- \Delta~,
\end{equation}
a more general list is given e.g. in \cite{Anninos:2020hfj}.
In (\ref{eq: rhow}) $\omega$ is defined along the whole real axis, whereas in (\ref{eq: Zcan}) we only keep $\omega \in\mathbb{R}_{\geq 0}$.
Substituting this representation into the canonical partition function yields for de Sitter
\begin{equation}
    \log Z_{\mathrm{bulk}}
    =
    \int_0^\infty
    \frac{\d \mathfrak{t}}{2\mathfrak{t}}
    \left(
    \frac{1+\e^{-\frac{2\pi \mathfrak{t}}{\beta_{\mathrm{dS}}}}}
         {1- \e^{-\frac{2\pi \mathfrak{t}}{\beta_{\mathrm{dS}}}}}
    \chi_{\Delta,s}^{\mathrm{bos}}(\mathfrak{t})
    -
    \frac{2\e^{-\frac{\pi \mathfrak{t}}{\beta_{\mathrm{dS}}}}}
         {1-\e^{-\frac{2\pi \mathfrak{t}}{\beta_{\mathrm{dS}}}}}
    \chi_{\Delta,s}^{\mathrm{fer}}(\mathfrak{t})
    \right).
\end{equation}
In the next section we will see that the one-loop sphere partition function takes a form closely resembling $Z_{\mathrm{bulk}}$ at the equilibrium temperature of the static patch, $\beta_{\mathrm{dS}} = 2\pi \ell$. Interestingly, however, the Harish-Chandra character $\chi_{\Delta,s}$ in $Z_{\mathrm{bulk}}$ needs to be replaced by $\chi_{\mathrm{tot}} \equiv \chi_{\Delta,s} - \chi_{\mathrm{edge}}$, where $\chi_{\mathrm{edge}}$ has a codimension two UV divergence compared to $\chi_{\Delta,s}$.

\subsection{Gibbons--Hawking conjecture}

In a theory of gravity, diffeomorphism invariance implies that local observables are not gauge-invariant. Instead, one may consider gauge- and diff-invariant, as well as local field redefinition invariant quantities such as the Euclidean partition function,
\begin{equation}\label{eq:logZ}
    \mathcal{Z} = \int [\mathcal{D}g]\, \e^{-S_{\mathrm{EH}}[g,\Lambda]} \, \mathcal{Z}_{\mathrm{matter}}~,
\end{equation}
evaluated semiclassically around the Euclidean de Sitter saddle. This corresponds to a path integral on the sphere $S^{d+1}$, obtained by Wick rotation from Lorentzian de Sitter space.

Gibbons and Hawking proposed that the de Sitter entropy is related to this partition function. 
At leading order, the path integral is dominated by the classical saddle,
\begin{equation}
    \log \mathcal{Z} \approx -S_{\mathrm{EH}}[g_*,\Lambda]~,
\end{equation}
where $g_*$ is the round sphere metric. Using
\begin{equation}
    \Lambda = \frac{d(d-1)}{2\ell^2}~,
\end{equation}
the on-shell Ricci scalar is
\begin{equation}
    R_* = \frac{2(d+1)}{d-1}\Lambda~,
\end{equation}
which evaluates to the on-shell action
\begin{equation}\label{eq: tree level EH}
    -S_{\mathrm{EH}}[g_{\mu\nu,*},\Lambda] = \frac{1}{16\pi G_N}\frac{2d}{\ell^2}\ell^{d+1} \times \mathrm{vol}(S^{d+1})~.
\end{equation}
For $d=3$, using $\mathrm{vol}(S^4)=\frac{8\pi^2}{3}$, one finds
\begin{equation}
    -S_{\mathrm{EH}}[g_*,\Lambda] = \frac{\pi \ell^2}{G_N}~.
\end{equation}
 This provides a semiclassical derivation of the area law (\ref{eq: SdS 4D}).
\paragraph{One-loop.}
More generally, the partition function (\ref{eq:logZ}) admits a loop expansion. 
The one-loop contribution of the graviton and matter fields can be written as \cite{Anninos:2020hfj}
\begin{equation}\label{eq: thermal structure}
    \log\mathcal{Z}^{(1)}=  \log \Big(2\pi \sqrt{\frac{8\pi G_N}{A_{\mathrm{hor}}}}\Big)^{\mathrm{dim}(G_0)} \!\!- \log \mathrm{vol}(G_0)+\int_0^\infty \frac{\d \mathfrak{t}}{2\mathfrak{t}} 
    \left(\frac{1+\e^{-\frac{\mathfrak{t}}{\ell}}}{1-\e^{-\frac{\mathfrak{t}}{\ell}}} \chi_{\mathrm{tot}}^{\mathrm{bos}}(\mathfrak{t}) 
    - \frac{2\e^{-\frac{\mathfrak{t}}{2\ell}}}{1-\e^{-\frac{\mathfrak{t}}{\ell}}} \chi_{\mathrm{tot}}^{\mathrm{fer}}(\mathfrak{t}) \right)~,
\end{equation}
where $G_0$ are the diffeomorphisms that act trivially on the saddle metric $g_*$, i.e. the isometry group of the round sphere $G_0= \mathrm{SO}(d+2)$.
As an example for $S^3$ we have
\begin{equation}
    \log \left(2\pi \sqrt{\frac{8\pi G_N}{A_{\mathrm{hor}}}}\right)^6  - 4\log (2\pi) = 5 \log(2\pi) -3\log \frac{\pi \ell}{2G_N}~,
\end{equation}
where we used that $G_0 = \mathrm{SO}(4)$ for the three-sphere with $\mathrm{dim}(G_0)=6$ and $\mathrm{vol}(\mathrm{SO}(4))= (2\pi)^4$. In odd spacetime dimension the character integral is UV-finite and
the log-terms come entirely from the group-volume factor. 
$\chi_{\mathrm{tot}}$ is determined by the particle spectrum and decomposes into bulk and edge contributions,
\begin{equation}
    \chi_{\mathrm{tot}} = \chi_{\mathrm{bulk}} - \chi_{\mathrm{edge}}~.
\end{equation}
The bulk character $\chi_{\mathrm{bulk}}$ 
\begin{equation}
    \chi_{\mathrm{bulk}}(\mathfrak{t}) \equiv \chi_{\Delta,s}(\mathfrak{t})= \mathrm{tr}_{\Delta,s}\, \e^{-i\mathfrak{t}H}
\end{equation}
is the Harish--Chandra character of the de Sitter isometry group $\mathrm{SO}(1,d+1)$, where $H \in \mathfrak{so}(1,1)$. If it were just this contribution that we find in the one-loop contribution, the sphere partition function could be written as a trace with a natural counting interpretation of quasinormal modes (see subsec. \ref{subsection:Fields in static patch}).

However, $\chi_{\mathrm{tot}}$ includes a second contribution $\chi_{\mathrm{edge}}$. 
The edge contribution $\chi_{\mathrm{edge}}$ is in general not a Harish-Chandra character and arises from the structure of the gravitational path integral.
The two contributions bulk/edge can be distinguished by their short-time behavior. For dS$_{d+1}$, the bulk character diverges as $\mathfrak{t}^{-d}$, while the edge contribution exhibits a codimension-two divergence scaling as $\mathfrak{t}^{-(d-2)}$.
Above we have seen that the one-loop sphere partition function can ``almost" be realized in terms of a trace. The almost is obstructed by the edge contributions. One way around this would be to introduce something that exactly removes this contribution. In subsection \ref{subsec:algebras} we discuss such a new feature.

\subsection{Technical details}\label{subsec:Technical details}
In this somewhat technical subsection, we explain, following \cite{Anninos:2020hfj}, how to arrive at $\log\mathcal{Z}^{(1)}$. 

\begin{figure}[ht]
    \centering
\begin{tikzpicture}[
    scale=1.2,
    boundary/.style={thick},
    bulk/.style={thick, dashed},
    curve/.style={thick},
    dcurve/.style={thick, dashed},
    label/.style={scale=0.7}
]
    \draw[] (-2,0) -- (2,0);
    \draw[densely dashed] (0,.4) -- (0,2);
    \draw[densely dashed] (0,-.4) -- (0,-2);
    \draw[thick, blue,->] (-2,.15) -- (2,.15);
  \draw[thick, red!90] (-.15,.4) -- (-.15,2);
    \draw[thick, red!90,->] (.15,.4) -- (.15,2);
    \draw[fill=yellow] (0,-.4) circle (0.05);
        \draw[fill=yellow] (0,.4) circle (0.05);
\coordinate (A) at (-.15,.41);
\coordinate (B) at (.15,.41);
\draw[red!90, thick] (B) arc[start angle=180, end angle=0, radius=-0.15];
 \draw[fill=orange] (-1.5,0) circle (0.05);
  \draw[fill=orange] (-1,0) circle (0.05);
   \draw[fill=orange] (-.5,0) circle (0.05);
    \draw[fill=orange] (0,0) circle (0.05);
     \draw[fill=orange] (1.5,0) circle (0.05);
  \draw[fill=orange] (1,0) circle (0.05);
   \draw[fill=orange] (.5,0) circle (0.05);

\node[scale=.7,red] at (.4,1.45) {$B$};    
\node[scale=.7,blue] at (1.3,.45) {$A$}; 
\node[scale=.7] at (-.4,.45) {$i\epsilon$}; 
\node[scale=.7] at (-.43,-.35) {$-i\epsilon$};
\node[scale=.7] at (1.8,1.8) {$\underline{u}$};

    \draw[thick,->, magenta] (6.5,0) -- (8,0);
    \node[scale=.7,magenta] at (7.4,.2) {$C$}; 
    \draw[fill=yellow] (6.5,0) circle (0.05);
    \node[scale=.7] at (6.5,.2) {$\epsilon$}; 
        \draw[fill=yellow] (5.5,0) circle (0.05);
           \node[scale=.7] at (5.4,.2) {$-\epsilon$}; 
    \draw[fill=orange] (6,1.5) circle (0.05);
\draw[fill=orange] (6,1) circle (0.05);
 \draw[fill=orange] (6,.5) circle (0.05);
  \draw[fill=orange] (6,0) circle (0.05);
   \draw[fill=orange] (6,-.5) circle (0.05);
   \draw[fill=orange] (6,-1) circle (0.05);
 \draw[fill=orange] (6,-1.5) circle (0.05);
 \node[scale=.7] at (7.8,1.8) {$\underline{t}$};
    
\end{tikzpicture}
\caption{Integration contours for one-loop sphere partition functions \cite{Anninos:2020hfj}.}
\label{fig:contour A}
\end{figure}
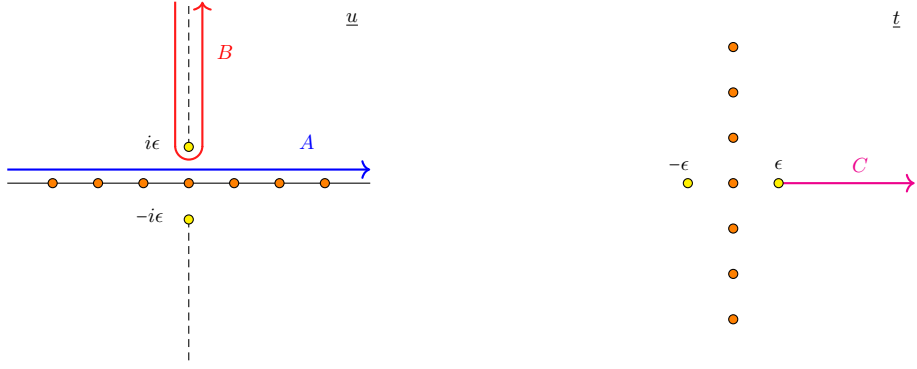

When evaluating the sphere partition function of a scalar field, a gauge field or a fermion beyond tree level we one way or another have to deal with infinite products arising from the one loop determinant. 
Going back to the conformally coupled scalar in four dimensions we have
\begin{equation}\label{eq:Z01 sphere}
   \mathcal{Z}^{(1)}_{(2,0)}=  {\det}^{-\frac{1}{2}}\Big(\frac{-\nabla^2 +2 }{\Lambda_{\text{u.v.}}}\Big) = \prod_l\left(\frac{\Lambda_{\text{u.v.}}}{l(l+3)+ 2}\right)^{\frac{1}{2}D_l^{(5)}}~,
\end{equation}
where $-\nabla^2$ is the spherical Laplacian on the four-sphere. Its eigenvalues and their degeneracies are known 
\begin{equation}
    -\nabla^2 Y_{l,m}= {l(l+3)}Y_{l,m}~,\quad D_l^{(5)} = \frac{1}{6}(1+l)(2+l)(3+2l)~,
\end{equation}
where $D_l^{(5)}$ is the degeneracy of the $l^{\mathrm{th}}$ eigenvalue. We have set the radius of the four-sphere to one and also set the de Sitter length $\ell=1$. The purpose of this section is to acquire a machinery to evaluate this functional determinant. We will make extended use of the identity $\tr \log\mathcal{O} = \log\det\mathcal{O}$ for some operator $\mathcal{O}$ as well as
\begin{equation}\label{eq: log trick}
    -\frac{1}{2}\log \frac{x\epsilon^2\e^{2\gamma_E}}{4}= \int_{0}^\infty\frac{\d{t}}{2{t}}\,\e^{-\frac{\epsilon^2}{4{t}}}\e^{-x {t}} ~,
\end{equation}
where $\epsilon>0$ is a cutoff for small $t$.
Before delving into the extended machinery let's see how we could proceed in (\ref{eq:Z01 sphere}) using (\ref{eq: log trick}). We have
\begin{align}\label{eq: Z20 1 loop}
   \log\mathcal{Z}^{(1)}_{(2,0)}= -\frac{1}{2}\sum_{l\geq 0} D_l^{(5)} \log\left(\frac{l(l+3)+ 2}{\Lambda_{\text{u.v.}}}\right) &= \int_0^\infty \frac{\d {t} }{2{t} }\,\e^{-\frac{\epsilon^2}{4t}} \sum_{l\geq 0}D_l^{(5)}\e^{-(l(l+3)+2){t}}~\cr
   &= \int_0^\infty \frac{\d {t}}{2{t}}\e^{\frac{-\epsilon^2}{4{t}}} \sum_{l\geq 0}D_l^{(5)} \e^{-(-\frac{1}{4}){t}}\e^{-(l+\frac{3}{2})^2 {t}}~.
\end{align}
where $\epsilon \equiv 2 \e^{-\gamma_E}/\sqrt{\Lambda_{\mathrm{u.v.}}}$.
This generalizes easily for a scalar on a $(d+1)$-dimensional sphere $S^{d+1}$:
\begin{align}\label{eq: PI epsilon}
    \hspace{-4mm}\log \mathcal{Z}_{(\Delta,0)} =\int_0^\infty \frac{\d{t}}{2{t}}\e^{-\frac{\epsilon^2}{4{t}}} F_\nu({t})~,\quad F_\nu({t})  &= \sum_{l\geq 0} D_{l}^{(d+2)} \e^{-\left(l+\frac{d}{2}+i\nu\right)\left(l+\frac{d}{2} -i\nu\right){t}}~\cr
    &=\sum_{l \geq 0} D_{l}^{(d+2)} \e^{-\nu^2 {t}}\e^{-\left(l+\frac{d}{2}\right)^2{t}}~,\quad \Delta = \frac{d}{2}+i\nu~,
\end{align}
where $D_l^{(d+2)}$ (\ref{eq:degeneracy Sdp1}) is the degeneracy of the $l^{\mathrm{th}}$ eigenvalue on $S^{d+1}$.
For the conformally coupled scalar on $S^4$ (\ref{eq: Z20 1 loop}), $d=3$ and $\nu =\sqrt{2-9/4}=i/2$.
To perform the sum over $l$ we perform a Hubbard-Stratonovich trick 
\begin{equation}
    \sum_{l=0}^\infty D_{l}^{(d+2)} \e^{-\left(l+\frac{d}{2}\right)^2{t}}= \int_A\! \frac{\d u}{\sqrt{4\pi {t}}}\e^{-\frac{u^2}{4{t}}}f(u)~,\quad f(u) = \sum_{l=0}^\infty D_l^{(d+2)} \e^{iu(l+\frac{d}{2})}~,
\end{equation}
where $A= \mathbb{R}+i\delta$, $0<\delta <\epsilon$ is shown in figure \ref{fig:contour A}. We can now perform the ${t}$ integral first
\begin{equation}
    \int_A \frac{\d u}{2\sqrt{u^2+\epsilon^2}}\e^{-\nu\sqrt{u^2+\epsilon^2}}f(u)
\end{equation}
then fold the contour into contour $B$. Changing variables $u= i\mathfrak{t}$ we transform into contour $C$ and obtain
\begin{align}
    \log \mathcal{Z}_{(\Delta,0)} &= \int_\epsilon^\infty \frac{\d \mathfrak{t}}{2\sqrt{\mathfrak{t}^2-\epsilon^2}} \sum_n D_n^{(d+2)} \left(\e^{-(n+\frac{d}{2})\mathfrak{t} -i\nu \sqrt{\mathfrak{t}^2-\epsilon^2}} + \e^{-(n+\frac{d}{2})\mathfrak{t} +i\nu \sqrt{\mathfrak{t}^2-\epsilon^2}}\right)~\\
    &=\int_\epsilon^\infty \frac{\d \mathfrak{t}}{2\sqrt{\mathfrak{t}^2-\epsilon^2}} \frac{1+\e^{-\mathfrak{t}}}{1-\e^{-\mathfrak{t}}} \frac{\e^{-\frac{d}{2}\mathfrak{t} -i\nu \sqrt{\mathfrak{t}^2-\epsilon^2}} +\e^{-\frac{d}{2}\mathfrak{t} +i\nu \sqrt{\mathfrak{t}^2-\epsilon^2}} }{(1-\e^{-\mathfrak{t}})^d}~.
\end{align}
For $\epsilon=0$ we find
\begin{equation}\label{eq:Zepsilon0}
    \log \mathcal{Z}_{(\Delta,0)}  = \int_0^\infty \frac{\d \mathfrak{t}}{2\mathfrak{t}} \frac{1+\e^{-\mathfrak{t}}}{1-\e^{-\mathfrak{t}}} \frac{\e^{-(\frac{d}{2}+ i\nu)\mathfrak{t}} + \e^{-(\frac{d}{2}-i\nu)\mathfrak{t}}}{(1-\e^{-\mathfrak{t}})^d}~.
\end{equation}
For the conformally coupled scalar on $S^4$ this leads to
\begin{equation}
   \log\mathcal{Z}^{(1)}_{(2,0)} = \int_0^\infty \frac{\d \mathfrak{t}}{2\mathfrak{t}} \frac{1+\e^{-\mathfrak{t}}}{1-\e^{-\mathfrak{t}}} \frac{\e^{-(\frac{3}{2}+i\nu)\mathfrak{t}} + \e^{-(\frac{3}{2}-i\nu)\mathfrak{t}}}{(1-\e^{-\mathfrak{t}})^3} =\int_0^\infty \frac{\d \mathfrak{t}}{2\mathfrak{t}} \frac{1+\e^{-\mathfrak{t}}}{1-\e^{-\mathfrak{t}}}\frac{\e^{-2\mathfrak{t}}+ \e^{-\mathfrak{t}}} {(1-\e^{-\mathfrak{t}})^3}  ~.
\end{equation}
We recognize $\chi_{\Delta,s}(\mathfrak{t})= \chi_{2,0}(\mathfrak{t})$, which is the character of a field transforming in the complementary series irreducible representation of $\mathrm{SO}(1,4)$. This is consistent with the mass-weight relation for a conformally coupled scalar in dS$_4$: $\Delta = \frac{3}{2} + \sqrt{\frac{9}{4}-2}=2$, $\overline{\Delta}\equiv 3-\Delta=1$.
\paragraph{Heat-kernel analysis.} 
Finally, let us discuss how to obtain the UV finite part of the path integral $\mathcal{Z}_{(\Delta,0)}$. 
We have (\ref{eq: PI epsilon})
\begin{equation}
    F_\nu(\mathfrak{t}) =\sum_{\Delta_\pm}\sum_{l=0}^\infty P_{\Delta_\pm}(l) \e^{-(l+\Delta_\pm)\mathfrak{t}}~,\quad \frac{1}{2t}F_\nu(\mathfrak{t}) =\frac{1}{\mathfrak{t}}\sum_{k=0}^{d+1} b_{k} \mathfrak{t}^{-(d+1-k)} + \mathcal{O}(\mathfrak{t}^0)~,
\end{equation}
and $b_k(\nu) = \sum_{\ell} b_{k\ell} \nu^\ell$. We obtain the exact heat kernel partition function with regulator $\e^{-\tfrac{\epsilon^2}{4\tau}}$ from
\begin{align}\label{eq: regularization scheme}
    \log \mathcal{Z}_{(\Delta,0)}^{\mathrm{u.v.}} &= \frac{1}{2}\sum_{\Delta_\pm} P_{\Delta_\pm}(\hat{\delta}-\Delta_\pm) \zeta'(0,\Delta_\pm) - \sum_{\ell=0}^{d+1} b_{d+1,\ell} (H_\ell - \frac{1}{2}H_{\ell/2})\nu^\ell \cr
    &\quad + b_{d+1}(\nu) \log \Big(\frac{2\e^{-\gamma_{\mathrm{E}}}}{\epsilon}\Big) + \frac{1}{2} \sum_{k=0}^d \sum_{\ell=0}^k b_{k\ell} B\left(\frac{d+1-k}{2}, \frac{\ell+1}{2}\right)\nu^\ell \epsilon^{-(d+1-k)}~,
\end{align}
where $B(x,y) = \frac{\Gamma(x)\Gamma(y)}{\Gamma(x+y)}$ and $H_x$ is the $x^{\mathrm{th}}$ harmonic number; $\hat{\delta}$ is the unit shift operator acting on the first argument of the Hurwitz zeta function.

Again we can go back to the conformally coupled scalar on $S^4$. We have
\begin{equation}
    \frac{1+\e^{-\mathfrak{t}}}{1-\e^{-\mathfrak{t}}} \frac{\e^{-(\frac{3}{2}+i\nu)\mathfrak{t}} + \e^{-(\frac{3}{2}-i\nu)\mathfrak{t}}}{(1-\e^{-\mathfrak{t}})^3} = \sum_{l=0}^\infty D_l^{(5)}\e^{-(l+\Delta_+)\mathfrak{t}}+\sum_{l=0}^\infty D_l^{(5)}\e^{-(l+\Delta_-)\mathfrak{t}}~,\quad \Delta_\pm=\frac{3}{2}\pm i\nu~,
\end{equation}
and hence $P_{\Delta_\pm}(l)= D_l^{(5)}$. Furthermore we have the expansion
\begin{equation}
    \frac{1}{2\mathfrak{t}}\frac{1+\e^{-\mathfrak{t}}}{1-\e^{-\mathfrak{t}}} \frac{\e^{-(\frac{3}{2}+i\nu)\mathfrak{t}} + \e^{-(\frac{3}{2}-i\nu)\mathfrak{t}}}{(1-\e^{-\mathfrak{t}})^3} = \frac{1}{\mathfrak{t}}\left(\frac{2}{\mathfrak{t}^4} - \frac{1+12\nu^2}{12\mathfrak{t}^2} +\frac{240\nu^4 + 120\nu^2 -17}{2880} \right) + \mathcal{O}(\mathfrak{t})~.
\end{equation}
We can hence read off
\begin{equation}
    b_0 =2~,\quad b_2=-\frac{1}{12}-\nu^2~,\quad b_4=-\frac{17}{2880} +\frac{\nu^2}{24} + \frac{\nu^4}{12} ~.
\end{equation}
The $b_{k\ell}$ then follow immediately from the coefficients of $\nu$. We obtain
\begin{align}
    &\frac{1}{2}\sum_{\Delta_\pm} D^{(5)}_{\hat{\delta} -\Delta_\pm}\zeta'(0,\Delta_\pm) = \frac{1}{2}\sum_{\Delta_\pm}\bigg[\frac{1}{3}\zeta'(-3,\Delta_\pm) +\left(\frac{3}{2}-\Delta_\pm\right)\zeta'(-2,\Delta_\pm) \cr
    &\quad + \left(\frac{13}{6} +\Delta_\pm(\Delta_\pm-3)\right)\zeta'(-1,\Delta_\pm)  -\frac{1}{6}(\Delta_\pm -2)(\Delta_\pm -1)(2\Delta_\pm -3)\zeta'(0,\Delta_\pm)\bigg]\cr
    &\quad = \frac{1}{3}\zeta'(-3) + \frac{1}{6}\zeta'(-1) = \frac{1}{3}\zeta'(-3) +\frac{1}{72} -\frac{1}{6}\log A~,
\end{align}
where $A$ is Glaisher's constant. Next using $H_1=1$, $H_2=3/2$ and $H_4= 25/12$ we obtain 
\begin{equation}
    -\frac{1}{24}\nu^2 - \frac{1}{9}\nu^4 +  \frac{240\nu^4 + 120\nu^2 -17}{2880}\log \frac{2\e^{-\gamma_E}}{\epsilon} +\frac{4}{3\epsilon^4} -\frac{1}{12\epsilon^2} -\frac{\nu^2}{3\epsilon^2}~.
\end{equation}
Using that $\nu = i/2$ for the conformally coupled scalar the $\epsilon^{-2}$ divergence cancels and we obtain 
\begin{align}
     \log\mathcal{Z}^{(\mathrm{u.v.})}_{(2,0)}  = \frac{5}{288} + \frac{1}{3}\zeta'(-3) - \frac{1}{6}\log A  -\frac{1}{90}\log \frac{2\e^{-\gamma_E}}{\epsilon} + \frac{4}{3\epsilon^4}~.
\end{align}
As expected in an even spacetime dimensional theory we obtain a logarithmic divergence. The logarithmic coefficient is fixed by the Weyl anomaly as $-4a$ with $a= \frac{1}{360}$.

\paragraph{Examples}
We give two more examples
\begin{enumerate}
 \item Gauge field in 2D
 \item Graviton in 4D
\end{enumerate}

\paragraph{1. Gauge field in 2D.}
We discuss a gauge field in dS$_2$ with $m^2\ell^2=-2$. Examples of this are the fluctuations of the Weyl factor in Liouville theory, or the fluctuation of the Dilaton in JT gravity around an $S^2$ saddle. It is important that we view this particle not as a tachyonic scalar field but as a gauge field (it is part of the metric field) that transforms in the discrete series unitary irreducible representation of $\mathrm{SO}(1,2)$. We consider the path integral
\begin{equation}
    S= \frac{1}{2} \int_{S^2} \d^2x \sqrt{g} \,\phi\left(-\nabla^2 -2\right)\phi~.
\end{equation}
Expanding $\phi$ into a basis of spherical harmonics
\begin{equation}
  \phi= \sum_{l,m}\phi_{l,m}Y_{l,m}(\theta,\varphi)~,\quad   -\nabla^2 Y_{l,m}(\theta,\varphi) = l(l+1)Y_{l,m}(\theta,\varphi)~,\quad D_l =2l+1~,
\end{equation}
we see that $l=1$ is a three-fold degenerate zero mode of the action. In Liouville theory or JT gravity this three-fold degenerate zero mode is related to the $\mathrm{PSL}(2,\mathbb{C})$ invariance of the theory on the sphere. The $l=1$ modes need to be treated using a proper FP gauge fixing (see e.g. \cite{Anninos:2021ene}). For now we write the determinant as a character integral for the $l\geq 2$ modes.  The $l=0$ mode we include in the end. We have
 \begin{equation}
 \log{\det}_{l\geq 2}^{-\frac{1}{2}}\Big(\frac{-\nabla^2 -2}{\Lambda_{\mathrm{u.v.}}}\Big)=\int_0^\infty\frac{\d\tau}{2\tau}\,\e^{-\frac{\epsilon^2}{4\tau}}\e^{-\tau\nu^2}\sum_{l=2}^\infty(2l+1)\e^{-\tau\left(l+\frac{1}{2}\right)^2}~,\quad \nu\equiv \sqrt{-\frac{1}{4}-2}~,
 \end{equation}
 where $\epsilon \equiv 2 \e^{-\gamma_E}/\sqrt{\Lambda_{\mathrm{u.v.}}}$
 To perform the sum over $l$ we use Hubbard-Stratonovich and complete the square. We have 
 \begin{equation}
 \sum_{l=2}^\infty(2l+1)\e^{-\tau\left(l+\frac{1}{2}\right)^2}= \int_A\d u\frac{\e^{-\frac{u^2}{4\tau}}}{\sqrt{4\pi\tau}}f(u)~,\quad \quad f(u)\equiv \sum_{l=2}^\infty(2l+1)\e^{iu\left(l+\frac{1}{2}\right)}~,
 \end{equation}
 where $A=\mathbb{R}+i\delta$, $\delta>0$. Evaluating the sum we obtain
 \begin{equation}\label{eq: fu integral 2D}
 \int_A\d u\frac{1}{2\sqrt{u^2+\epsilon^2}}\e^{-\nu\sqrt{u^2+\epsilon^2}}f(u)~,\quad f(u)= \frac{5-3\e^{iu}}{(1-\e^{iu})^2}\e^{\frac{5}{2}iu}~.
 \end{equation}
Changing the contour $A\rightarrow B \rightarrow C$ (see fig. \ref{fig:contour A}) and performing the change of variables $u\rightarrow i\mathfrak{t}$, we obtain for $\epsilon \rightarrow 0$
 \begin{align}
 &\int_{0}^\infty\frac{\d \mathfrak{t}}{2\mathfrak{t}}\frac{5-3\e^{-\mathfrak{t}}}{(1-\e^{-\mathfrak{t}})^2}\left(\e^{-\frac{5}{2}\mathfrak{t}-i\nu \mathfrak{t}}+\e^{-\frac{5}{2}\mathfrak{t}+i\nu \mathfrak{t}}\right) = \int_0^\infty  \frac{\d \mathfrak{t}}{2\mathfrak{t}}\frac{1+\e^{-\mathfrak{t}}}{(1-\e^{-\mathfrak{t}})}\left(\frac{2\e^{-2\mathfrak{t}}}{(1-\e^{-\mathfrak{t}})} + 3\e^{-3\mathfrak{t}} - 5 \e^{-2\mathfrak{t}} + 5 \e^{-\mathfrak{t}}\right)~.
 \end{align}
We can now add the $l=0$ mode. Using (\ref{eq: fu integral 2D}) with $f_{l=0}(u) = \e^{i u/2}$ we obtain
\begin{equation}
    \int_0^\infty \frac{\d \mathfrak{t}}{2\mathfrak{t}}(\e^{-\frac{\mathfrak{t}}{2}-i\nu \mathfrak{t}} + \e^{-\frac{\mathfrak{t}}{2}+i\nu \mathfrak{t}}) = \int_0^\infty \frac{\d \mathfrak{t}}{2\mathfrak{t}} \frac{1+\e^{-\mathfrak{t}}}{(1-\e^{-\mathfrak{t}})}(\e^{\mathfrak{t}} + 2\e^{-\mathfrak{t}} - \e^{-2\mathfrak{t}} -2)~.
\end{equation}
The negative powers $k\leq 0$ of $q^k \equiv \e^{-k\mathfrak{t}}$ make this integral exponentially divergent. We will first discuss the technique -- the flipping formula -- to regulate this and then discuss its physical meaning and origin.
The flipping formula instructs us to perform the following summations
\begin{equation}\label{eq:chi_flipping}
   \chi(q)= \sum_{k\leq 0 } c_k q^k + \sum_{k>  0} c_k q^k \rightarrow [\chi(q)]_{+}\equiv \chi(q) -c_0 - \sum_{k<0} c_k(q^k + q^{-k})~.
\end{equation}
For the case at hand this implies
\begin{equation}
    \int_0^\infty \frac{\d \mathfrak{t}}{2\mathfrak{t}} \frac{1+\e^{-\mathfrak{t}}}{(1-\e^{-\mathfrak{t}})}(\e^{\mathfrak{t}} + 2\e^{-\mathfrak{t}} - \e^{-2\mathfrak{t}} -2)~\rightarrow 
    \int_0^\infty \frac{\d \mathfrak{t}}{2\mathfrak{t}} \frac{1+\e^{-\mathfrak{t}}}{(1-\e^{-\mathfrak{t}})}(- \e^{-2\mathfrak{t}}+\e^{-\mathfrak{t}})~.
\end{equation}
Hence we find
\begin{equation}
   \log {{\det}'}^{-\frac{1}{2}}\left(\frac{-\nabla^2 -2}{\Lambda_{\mathrm{u.v.}}}\right)  = \int_0^\infty  \frac{\d \mathfrak{t}}{2\mathfrak{t}}\frac{1+\e^{-\mathfrak{t}}}{(1-\e^{-\mathfrak{t}})}\left(\frac{2\e^{-2\mathfrak{t}}}{(1-\e^{-\mathfrak{t}})} + 3\e^{-3\mathfrak{t}} - 6 \e^{-2\mathfrak{t}} + 6 \e^{-\mathfrak{t}}\right)~,
\end{equation}
where the prime indicates the omission of the $l=1$ mode.

\paragraph{2. Graviton in 4D.}
The graviton in dS$_4$ is a highest-depth partially massless field (PMF) $D^{\pm}_{1,2}$ (see
table \ref{tab:scalar-reps-dS4}), and as such massless. It has spin $s=2$, depth $t=s-1=1$ and weight
$\Delta =1+s= 3$. A massless spin-$s$ field is a totally symmetric gauge field with linearized
gauge transformations
\begin{equation}\label{eq:gauge transformation}
    \delta\phi_{\mu_1\ldots \mu_s}= \nabla_{(\mu_1} \zeta_{\mu_2\ldots \mu_s)}~,\quad
    \zeta^{\nu}_{~\nu\mu_3\ldots \mu_{s-1}}=0~.
\end{equation}
The weights of the fields are respectively
\begin{equation}
    \Delta_\phi \equiv \frac{d}{2}+i\nu_\phi = d+s-2~,\quad
    \Delta_\zeta \equiv \frac{d}{2}+i\nu_\zeta = d+s-1~.
\end{equation}
For a PMF of depth $0\leq t\leq s-1$, with linearized gauge transformations
$\delta\phi_{\mu_1 \ldots \mu_s} = \nabla_{(\mu_{t+1}}\!\cdots\nabla_{\mu_s}
\zeta_{\mu_1 \ldots \mu_t)}$ involving $s-t$ derivatives this generalizes
to
\begin{equation}
    \Delta_\phi \equiv \frac{d}{2}+i\nu_\phi = d+t-1~,\quad
    \Delta_\zeta \equiv \frac{d}{2}+i\nu_\zeta = d+s-1~.
\end{equation}
The gauge transformations \eqref{eq:gauge transformation} have zero modes: rank-$s'$ Killing
tensors $\bar{\zeta}_{\mu_1 \ldots \mu_{s'}}$, $s'\equiv s-1$, satisfying
$\nabla_{(\mu_1} \bar{\zeta}_{\mu_2\ldots \mu_s)}=0$ --- for the graviton, the Killing
vector of $S^{d+1}$. Due to the gauge redundancy the one-loop contribution of a depth-$t$
PMF is governed by
\begin{equation}
    \chi \equiv \chi_\phi - \chi_\zeta~.
\end{equation}
For a massive spin-$s$ field the bulk and edge characters are \cite{Anninos:2020hfj}
\begin{equation}
    \chi_{\mathrm{bulk}} = D_s^{(d)} \frac{q^{\frac{d}{2}+i\nu} + q^{\frac{d}{2}-i\nu}}{(1-q)^d}~,\quad
    \chi_{\mathrm{edge}} = D_{s-1}^{(d+2)} \frac{q^{\frac{d-2}{2}+i\nu}
    +q^{\frac{d-2}{2}-i\nu}}{(1-q)^{d-2}}~.
\end{equation}
Consequently, with $s'=s-1$, we would predict
\begin{subequations}
\begin{align}
    \hat{\chi}_{\mathrm{bulk}} &= D_s^{(d)} \frac{q^{s'+d-1}+q^{1-s'}}{(1-q)^d}
    - D_{s'}^{(d)}  \frac{q^{s+d-1}+q^{1-s}}{(1-q)^d}~,\\
    \hat{\chi}_{\mathrm{edge}} &= D_{s-1}^{(d+2)}\frac{q^{s'+d-2}+q^{-s'}}{(1-q)^{d-2}}
    - D_{s'-1}^{(d+2)}\frac{q^{s+d-2}+q^{-s}}{(1-q)^{d-2}}~.
\end{align}
\end{subequations}
For the graviton with $s=2$ and $s'=1$ this implies
\begin{equation}
    \hat{\chi}_{\mathrm{bulk}} = 5\frac{q^3+1}{(1-q)^3} - 3 \frac{q^4+q^{-1}}{(1-q)^3}~,\quad
    \hat{\chi}_{\mathrm{edge}} =5 \frac{q^2+ q^{-1}}{(1-q)} -  \frac{q^3+q^{-2}}{(1-q)}~.
\end{equation}
The powers $q^k \equiv \e^{-k \mathfrak{t}}$ with $k\leq 0$ render the character integral
exponentially divergent. The origin of this is the conformal mode problem, which we discuss in
the next paragraph. To deal with the divergence we introduce the full integrand
\begin{equation}
    F(q) \equiv \frac{1+q}{1-q}\,\hat{\chi}_{\mathrm{tot}}(\mathfrak{t})~,\qquad
    \hat{\chi}_{\mathrm{tot}}=\hat{\chi}_{\mathrm{bulk}}-\hat{\chi}_{\mathrm{edge}}~,
\end{equation}
and apply the flipping procedure
\begin{equation}
    \log \mathcal{Z}_{\mathrm{graviton}}^{(1)} \rightarrow
    \int_0^\infty \frac{\d \mathfrak{t}}{2\mathfrak{t}}\,\{F(q)\}_+~,\qquad
    \{F(q)\}_+ \equiv \sum_{k<0} c_k q^{-k} + \sum_{k\geq 0}c_k q^k~,
\end{equation}
where $c_k$ are the Laurent coefficients of $F$: acting on the full integrand, one keeps $c_0$
and reflects the negative powers keeping their signs. Note that this differs from the
flipping $[\,\cdot\,]_+$ of \eqref{eq:chi_flipping}, which acts on characters by dropping $c_0$
and reflecting with a sign flip. The two prescriptions are related by
\begin{equation}\label{eq:flip bridge}
    \{F\}_+\;=\;\frac{1+q}{1-q}\Big([\hat{\chi}_\mathrm{bulk}]_+
    -[\hat{\chi}_\mathrm{edge}]_+ - 2N_1^{\mathrm{KT}_d}\Big)~,
\end{equation}
where the constant is nothing but the constant term of the full integrand,
$c_0 = -2N_1^{\mathrm{KT}_3}=-20$. For the graviton one has
$\hat{\chi}_{\mathrm{bulk}} = -3q^{-1}-4+\ldots$ and
$\hat{\chi}_{\mathrm{edge}} = -q^{-2}+4 q^{-1}+4+ \ldots$, so that \eqref{eq:chi_flipping} gives
\begin{subequations}
\begin{align}
    [\hat{\chi}_{\mathrm{bulk}}]_+&=  \hat{\chi}_{\mathrm{bulk}} -(-4)-(-3)(q^{-1}+q)
    = 2\,\frac{5q^3-3q^4}{(1-q)^3}~,\\
    [\hat{\chi}_{\mathrm{edge}}]_+ &= \hat{\chi}_{\mathrm{edge}} -4 -4(q^{-1}+q) -(-1)(q^{-2}+q^2)
    = 2\,\frac{5q^2 -q^3}{(1-q)}~.
 \end{align}
\end{subequations}
We hence obtain
\begin{equation}\label{eq:logZ graviton}
    \log \mathcal{Z}_{\mathrm{graviton}}^{(1)} = \int_0^\infty \frac{\d \mathfrak{t}}{2\mathfrak{t}}
    \frac{1+q}{1-q}\left(2\,\frac{5q^3-3q^4}{(1-q)^3} - 2\,\frac{5q^2 -q^3}{(1-q)}
    - 2 N_1^{\mathrm{KT}_3}\right)~,
\end{equation}
where $N_1^{\mathrm{KT}_d} = \frac{1}{2}(d+1)(d+2) = \mathrm{dim}(\mathrm{SO}(d+2))$. The
resulting integral still has logarithmic divergences, whose origin are the Killing-vector zero
modes $\bar\zeta$ introduced below \eqref{eq:gauge transformation} --- the same ten modes that
produce the group-volume factor
$\big(2\pi\sqrt{8\pi G_N/A_{\mathrm{hor}}}\big)^{10}/\mathrm{vol}\,\mathrm{SO}(5)$ in (\ref{eq: thermal structure}).

Expanding $F(q)$ at small $q$, $F(q)=\sum_\alpha d_\alpha q^\alpha$, the graviton yields
\begin{equation}
    F_{\mathrm{grav}}(q) = \frac{1}{q^2} - \frac{5}{q} - 20 + \mathcal{O}(q)~,
\end{equation}
where the negative part receives $-3q^{-1}$ from the bulk factor
$\frac{1+q}{1-q}\hat\chi_{\mathrm{bulk}}$ and $-(-q^{-2}+2q^{-1})$ from the edge factor. The
negative coefficients count the phase of the gravitational path integral,
\begin{equation}
    \mathcal{Z}\supset \e^{\pm \frac{i\pi}{2} P_s}~,\qquad P_s = \sum_{\alpha <0}|d_\alpha|~,
\end{equation}
the Polchinski phase. For the graviton $P_2 = 1+5=6$: the two terms match the $l=0$ and $l=1$
conformal modes whose Gaussian rotation produces the phase. (The same counting on $S^3$ gives
$1+4=5$, i.e.\ precisely the $\pm\frac{5\pi i}{2}$ of the three-dimensional entropy formula.)

We have now encountered this flipping procedure twice: for the graviton in 4D and for the gauge
field in 2D. We now explain the origin of the exponentially divergent contributions to the
character integral.

\subsection{Conformal mode problem}\label{subsec:conformal mode problem}
We are interested in the leading contribution to the sphere path integral 
\begin{equation}
    Z_{S^{d+1}}= \int \frac{[Dg]}{\mathrm{vol}_{\mathrm{Diff}}} \e^{-S_{\mathrm{EH}}[g,\Lambda]}~,
\end{equation}
where $S_{\mathrm{EH}}[g,\Lambda]$ is the Euclidean Einstein Hilbert action on a $S^{d+1}$ sphere
\begin{eqnarray}
 - S_{\mathrm{EH}}[g,\Lambda] = \frac{1}{16\pi G_N}\int_{S^{d+1}}\d^{d+1}x \sqrt{g} (R- 2\Lambda)~.
\end{eqnarray}
The tree level EH action evaluates to (\ref{eq: tree level EH}). To go beyond tree level we need to consider fluctuations around the sphere saddle. Explicitly we have $g_{\mu\nu} = g_{\mu\nu}^{S^{d+1}} + h_{\mu\nu}$. We decompose the fluctuation metric $h_{\mu\nu}$ into its traceless and trace parts
\begin{equation}\label{eq: hmunu decomposition}
    h_{\mu\nu} = \phi_{\mu\nu} + \frac{1}{d+1} g_{\mu\nu} h~,
\end{equation}
and fix the gauge covariantly with \cite{Polchinski:1988ua}
\begin{equation}\label{eq: Donder gauge}
    f_{\nu} = \nabla^\mu h_{\mu\nu} -\frac{1}{2}\nabla_\nu h~,\quad h\equiv h_{\lambda}^{~\lambda}~.
\end{equation}
Adding $\frac{1}{32\pi G_N}\int f_\nu f^\nu$ to the action we obtain the quadratic action
\begin{equation}
    S^{(2)}= \frac{1}{64 \pi G_N} \int \d^{d+1}x \sqrt{g}\left[\phi_{\mu\nu} (-\nabla^2 +2)\phi^{\mu\nu} -\frac{d-1}{2(d+1)}h(-\nabla^2 -2d)h\right] + \mathrm{ghosts}~.
\end{equation}
In the de Donder gauge (\ref{eq: Donder gauge}) there are no $\phi-h$ cross terms.
The trace part enters with a wrong sign kinetic term -- the conformal mode problem. In particular if we decompose $h$ into spherical harmonics
\begin{equation}\label{eq: mode expansion}
    h(\Omega) = \sum_{l,m,n}\mathsf{h}_{l,m,n} Y_{l,m,n}(\Omega)~
\end{equation}
where
\begin{equation}
    -\nabla^2Y_{l,m,n}= l(l+d) Y_{l,m,n}~,\quad D^{(d+2)}_l = \frac{(d+2l)\Gamma(d+ l)}{\Gamma(1+d)\Gamma(1+l)}~,
\end{equation}
and $ D^{(d+2)}_l$ is the degeneracy of the $l^{\mathrm{th}}$ eigenvalue on $S^{d+1}$ we obtain
\begin{equation}\label{eq:cases}
  (l(l+d)-2d) = \begin{cases}
      \leq 0 ~,\quad l\leq 1~,\\
      \geq 0~,\quad l\geq 2~.
  \end{cases}  
\end{equation}
The operator $-\nabla^2 -2d$ has $d+3$ negative modes: the constant $l=0$ and the $d+2$ modes at $l=1$. Rotating the contour, $h\rightarrow \pm i h$, renders the trace integral Gaussian suppressed
at the cost of a phase \cite{Polchinski:1988ua, Gibbons:1978ac}
\begin{equation}\label{eq: Zgrav S3}
    \mathcal{Z}_{\mathrm{grav}}^{S^{d+1}} \approx (\pm i)^{d+3}\,\mathrm{e}^{\frac{A_{\mathrm{hor}}}{4G_N}}~.
\end{equation}
The Jacobian of the rotation $h\rightarrow \pm i h$ is an ultralocal term and absorbed in the measure. The $(\pm i)^{d+3}$ comes from the $1+(d+2)$ modes that are Gaussian suppressed before the Wick rotation. 
The ghost term $\int \mathfrak{b}_\mu(-\nabla^2 - d)\mathfrak{c}^\mu$ has zero modes precisely at the Killing vectors of the sphere. These are excluded from the ghost integral and accounted for by the division of $\mathrm{vol}(\mathrm{SO}(d+2))$. A careful treatment of the diffeomorphisms and the functional determinants contributing to the one-loop Einstein Hilbert action can be found in \cite{Volkov:2000ih,Law:2020cpj}. Here we just report the result which is given by 
\begin{equation}\label{eq: Zgrav S3}
    \mathcal{Z}_{\mathrm{grav}}^{S^{d+1}} \approx (\pm i)^{d+3}\,\mathrm{e}^{\frac{A_{\mathrm{hor}}}{4G_N}}\left(\frac{4 G_N}{A_{\mathrm{hor}}}\right)^{\frac{\mathrm{dim}(\mathrm{SO}(d+2))}{2}}\frac{1}{\mathrm{vol}(\mathrm{SO}(d+2))}\frac{{\det}{'}(-\nabla_{(1)}^2-d)^{1/2}}{\det(-\nabla_{(2)}^2+2)^{1/2}}~,
\end{equation}
where $-\nabla_{(s)}^2$ is the generalization of the scalar Laplacian to the spin $s$ transverse traceless Laplacian
\begin{equation}
    -\nabla_{(s)}^2 f_{n,(s)} = \lambda_{n,s}f_{n,(s)}~,\quad \nabla \cdot f_{n,(s)}=0~,\quad \tr f_{n,(s)}=0~.
\end{equation}
The eigenvalues and degeneracies are given in (\ref{eq:eigenvalues spin s Laplacian}).
Combining the one-loop and group volume contributions we obtain for example\footnote{Here $L$ is a scale \cite{Anninos:2020hfj}.}
\begin{subequations}
    \begin{align}\label{eq: SdS3 GH}
\hspace{-9mm}S_{\mathrm{dS}_3}:~&\frac{\pi\ell}{2G_N} -3 \log \frac{\pi\ell}{2G_N} +5\log (2\pi) \pm \frac{5\pi i}{2} ~,\\
        \hspace{-9mm}S_{\mathrm{dS}_4}:~ &\frac{\pi \ell^2}{G_N}- 5 \log \frac{\pi \ell^2}{G_N} -\frac{571}{45}\log\frac{\ell}{L} -\log\frac{8\pi}{3}+ \frac{715}{48} -\frac{47}{3}\zeta'(-1) +\frac{2}{3}\zeta'(-3) \pm \frac{6\pi i}{2}~,
    \end{align}
\end{subequations}
where 
$S_{\mathrm{dS}_{d+1}}= \log \mathcal{Z}_{\mathrm{grav}}^{S^{d+1}}$.
The count of negative modes leading to the phase in $\mathcal{Z}_{\mathrm{grav}}^{S^{d+1}}$ (\ref{eq: Zgrav S3}) depends on the specific topology. As an example, in four dimensions, beyond the leading sphere saddle, the gravitational path integral includes contributions from 
other Einstein metrics with positive curvature, such as $\mathbb{CP}^2$ or $S^2 \times S^2$. Each contributes a topology specific phase \cite{Volkov:2000ih,Ivo:2025yek,Anninos:2025ltd, Shi:2025amq, Law:2025yec, Mukherjee:2025xlt}.

\subsection{Algebras and observers}\label{subsec:algebras}
Entropy, in a certain sense, measures the number of quantum states compatible with a given macroscopic description. For a black hole of mass $M$ there is evidence that \cite{Bekenstein:1973ur, Hawking:1975vcx,Strominger:1996sh, Gibbons:1976ue}
\begin{equation}
    S_{\mathrm{BH}} =  \log N(M)
\end{equation} 
where $N(M)$ denotes the macroscopically indistinguishable microstates with energy $M$.
In de Sitter there is no satisfying analog of this.  

One could conjecture, that the total entropy is the entropy of matter and radiation inside the observer's horizon $S_{\mathrm{vis}}$, plus the entropy of the de Sitter horizon
\begin{equation}
    S_{\mathrm{tot}}= \frac{A_{\mathrm{hor}}}{4G_N}+ S_{\mathrm{vis}}~.
\end{equation}
When matter crosses the horizon, the entropy accessible to the observer decreases, while the horizon contribution compensates this decrease such that $S_{\mathrm{tot}}$ increases. In the far future $S_{\mathrm{vis}} \rightarrow 0$ and the horizon area approaches its maximal value. Conjecturally empty de Sitter space must then be a state of maximum entropy. If empty de Sitter space is the maximum entropy state, what density matrix realizes this entropy? 

In ordinary quantum mechanics we have systems with a finite dimensional Hilbert space $\mathrm{dim}(\mathcal{H})=N$. States are described by density matrices $\rho$ with $\tr \rho=1$ and the von Neumann entropy is defined as
\begin{equation}
    S_{\mathrm{vN}}(\rho) \equiv - \tr (\rho \log \rho)~.
\end{equation}
A pure state is described by $\rho = |\Psi\rangle \langle \Psi|$ which satisfies $\rho^2 =\rho$. For example $\begin{pmatrix}
    1& 0 \\ 0 & 0
\end{pmatrix}$ is pure. A mixed state is a state $\rho = \sum_i p_i |\Psi_i \rangle \langle \Psi_i|$ where $\sum_i p_i=1$. For instance 
\begin{equation}
\rho = \frac{1}{2}
\begin{pmatrix}
1 & 0 \\
0 & 1
\end{pmatrix}
= \frac{1}{2}\big(|\Psi_1\rangle \langle \Psi_1| + |\Psi_2\rangle \langle \Psi_2|\big)~,
\end{equation}
is a mixed state. It cannot be written as $|\Psi\rangle \langle \Psi|$. For example, the superposition
\begin{equation}
|\Psi\rangle = \frac{1}{\sqrt{2}}\big(|\Psi_1\rangle + |\Psi_2\rangle\big)
\end{equation}
gives
\begin{equation}
|\Psi\rangle \langle \Psi| =
\frac{1}{2}
\begin{pmatrix}
1 & 1 \\
1 & 1
\end{pmatrix},
\end{equation}
which contains off-diagonal terms and is therefore different from the mixed state above.
Among all density matrices for a finite dimensional Hilbert space  the entropy is maximized by $\rho_{\mathrm{max}} = \tfrac{\mathbb{I}_N}{N}$, for which $S_{\mathrm{vN}}(\rho_{\mathrm{max}}) = \log N$ and $S(\rho) < \log N = \log \mathrm{dim}(\mathcal{H})$ for any other density matrix. 

Now consider the thermofield double state on a doubled Hilbert space $\mathcal{H} \otimes \widetilde{\mathcal{H}}$. For a system with temperature $\beta$ it is given by
\begin{equation}\label{eq: TFD}
    |\Psi_{\mathrm{TFD}}(\beta)\rangle =\frac{1}{\sqrt{Z(\beta)}} \sum_{n} \e^{-\beta E_n/2} |n\rangle \otimes |\tilde{n}\rangle~.
\end{equation}
We define the density matrix $\rho_{\mathcal{H}\otimes \widetilde{\mathcal{H}}}
= |\Psi_{\mathrm{TFD}}\rangle \langle \Psi_{\mathrm{TFD}}|$. 
Tracing out $\widetilde{\mathcal{H}}$ gives
\begin{align}
    \rho_{\mathcal{H}} = \tr_{\widetilde{\mathcal{H}}} \rho_{\mathcal{H}\otimes \widetilde{\mathcal{H}}} &= \frac{1}{Z(\beta)}\sum_p  \sum_{m,n} \e^{-\beta (E_m+E_n)/2}\delta_{pm}\delta_{pn}|m\rangle \langle n|\cr
    &= \frac{1}{Z(\beta)} \sum_m \e^{-\beta E_m}|m\rangle \langle m|~.
\end{align}
In matrix form,
\begin{equation}
\rho_{\mathcal{H}} = \frac{1}{Z(\beta)}
\begin{pmatrix}
\mathrm{e}^{-\beta E_1} & 0 & \cdots & 0 \\
0 & \mathrm{e}^{-\beta E_2} & \cdots & 0 \\
\vdots & \vdots & \ddots & \vdots \\
0 & 0 & \cdots & \mathrm{e}^{-\beta E_N}
\end{pmatrix}.
\end{equation}
In the limit $\beta \to 0$, this becomes a maximally mixed state $\rho_{\mathcal{H}} \to \frac{\mathbb{I}_N}{N}$.
For operators $\hat a,\hat b \in B(\mathcal{H})$ ($B(\mathcal{H})$ denotes the algebra of bounded linear operators on the Hilbert space) the TFD state satisfies the KMS condition:
\begin{equation}
    \langle \Psi_{\mathrm{TFD}}(\beta)| \hat a(t)\hat b(0)| \Psi_{\mathrm{TFD}}(\beta)\rangle = \langle \Psi_{\mathrm{TFD}}(\beta)| \hat b(0)\hat a(t+i\beta)| \Psi_{\mathrm{TFD}}(\beta)\rangle~.
\end{equation}
In other words in the $\beta =0$ limit the TFD state defines a trace state 
    \begin{equation}
    \tr(\hat a\hat b) = \langle \Psi_{\mathrm{TFD}}(0)| \hat a\hat b |\Psi_{\mathrm{TFD}}(0)\rangle = \langle \Psi_{\mathrm{TFD}}(0)| \hat b\hat a |\Psi_{\mathrm{TFD}}(0)\rangle = \tr(\hat b \hat a)~.
\end{equation}

The relevance for de Sitter is the following: For rigid de Sitter writing formally the global Hilbert space as the
product of the two static patches, $\mathcal{H}_{\mathrm{global}} = \mathcal{H}_L \otimes
\mathcal{H}_R$, the Bunch--Davies vacuum is precisely the thermofield double of the two patches
at inverse temperature $\beta_{\mathrm{dS}}=2\pi\ell$. The Bunch–Davies state is KMS at the fixed temperature $\beta_{\mathrm{dS}}$, as we have seen in subsection \ref{subsec:KMS}. However, the factorization fails in continuum quantum field theory. Local algebras are
von Neumann algebras of type III, no density matrices exist, and the entropy diverges. Since the
de Sitter static patch is bounded by a horizon, this obstruction prevents us from interpreting
empty de Sitter space as a maximum entropy state.

The situation changes once we include gravity. In particular we consider the system consisting
of a quantum field theory, an observer, and weakly coupled dynamical gravity ($G_N\to0$) \cite{Witten:2021unn, Chandrasekaran:2022cip}. 
 In a gravitational theory, time translations become gauge transformations generated by the Hamiltonian constraint. Physical observables must commute with the total Hamiltonian which is now given by
\begin{equation}
\widehat H_{\mathrm{tot}} = \widehat H + \widehat H_{\mathrm{obs}} = \widehat H + \hat{q}~,\quad \hat q\geq 0~,
\end{equation}
where $\hat{q}$ is the observer's energy conjugate to its clock variable: $[\hat{q},\hat{p}]
=i$ with $\hat{p}$ the observer's time. Physical states satisfy the Wheeler-DeWitt constraint $\widehat H_{\mathrm{tot}}|\Psi\rangle=0$, where $H$ is the modular Hamiltonian of the static patch,
$\widehat H = \beta_{\mathrm{dS}} \widehat H_{\mathrm{static}}$. The boost Killing vectors are timelike within the static patch. Solving the constraint dresses the operators to the observer's clock: $\e^{-i \hat{p} \widehat H}$ generates the modular flow, and operators are translated as $\hat a(t) \rightarrow \hat a(t-\hat{p})$. Introducing the observer constructs a bigger algebra, called the crossed product and the type III von Neumann algebra becomes a type II$_1$ algebra. While originally $\tr(\hat a\hat b)$ did not exist, by enlarging the algebra, there exists a state $\Psi_{\mathrm{trace}}$ that satisfies
\begin{equation}
    \widetilde{\mathrm{tr}}(\hat a\hat b) = \langle \Psi_{\mathrm{trace}}| \hat a\hat b |\Psi_{\mathrm{trace}}\rangle = \langle \Psi_{\mathrm{trace}}| \hat b\hat a |\Psi_{\mathrm{trace}}\rangle = \widetilde{\mathrm{tr}}(\hat b\hat a) 
\end{equation}
Explicitly $\Psi_{\mathrm{trace}}$ is given by
\begin{equation}
    |\Psi_{\mathrm{trace}}\rangle = |\Psi_{\mathrm{TFD}}\rangle\otimes \sqrt{\beta}\e^{-\frac{\beta \hat{q}}{2}}|p=0\rangle~,\quad \beta \equiv \beta_{\mathrm{dS}} = 2\pi \ell = \frac{1}{T_{\mathrm{dS}}}~,
\end{equation}
i.e. with normalized observer wavefunction $\psi(q) = \sqrt{\beta}\e^{-\beta q/2}$. The observer factor $\e^{-\beta \hat q/2}$ precisely compensates the thermal weight of the Bunch–Davies state: with respect to the crossed-product algebra, the combined state behaves as the $\beta \rightarrow 0$ TFD did for the finite-dimensional algebra.

Although the crossed product algebra admits a trace, the von Neumann entropy is only defined up to an additive constant. The ultraviolet divergences of type III are cured, what remains is a choice of zero. The physically meaningful quantity is given by the variation of the generalized entropy: 
\begin{equation}
    \delta S_{\mathrm{gen}} = \frac{\delta A}{4G_N} + \delta S_{\mathrm{bulk}}~.
\end{equation}

\paragraph{Observer}
The Lorentzian observer of the previous subsection sits at $\rho=0$ (\ref{eq: static patch 1}). Continuation to Euclidean signature winds the thermal circle, which is a great circle of length $\beta_{\mathrm{dS}}$.\footnote{In embedding coordinates we have $X^0= \ell \cos\rho\sinh t$, $X^{d+1}= \ell \cos \rho \cosh t$ $X^i = \ell\sin\rho n^i$. So the observer at $\rho=0$ satisfies $(X^{d+1})^2 -(X^0)^2=\ell^2$. Wick rotating to Euclidean $t= -i \tau_E$, so that $\sinh t = -i \sin\tau_E~,\cosh t = \cos\tau_E$ and $X^0= -i X_E^0$ leads to $(X_E^0)^2 +(X^{d+1})^2=\ell^2.$} Its Euclidean avatar is a massive particle carrying a clock, and we
consider \cite{Anninos:2017hhn, Maldacena:2024spf, Ivo:2025yek} 
\begin{equation}
   Z_{\mathrm{observer}} = \int \frac{\d\beta}{\beta}Z_{S^{d+1}}(\beta)Z_{\mathrm{part}}(\beta)Z_{\mathrm{clock}}(\beta)~,
\end{equation}
where $\beta$ is the worldline length, $Z_{S^{d+1}}(\beta)$ is the path integral with the observer's circle held at length $\beta$, $Z_{\mathrm{part}} = \e^{-m\beta} \times (\mathrm{transverse~fluctuations})$, $Z_{\mathrm{clock}} = \tr \e^{-\beta H_{\mathrm{clock}}}$ is the thermal trace of the clock, and as we will see below the $\beta$ integral implements the constraint
\begin{equation}\label{eq: constraint euclideal observer}
    H_{\mathrm{dS}} +H_{\mathrm{clock}} + H_{\mathrm{part}} =0~.
\end{equation}
This is the Euclidean avatar of the Wheeler-DeWitt constraint of the previous subsection. 
The great circle is a geodesic of $S^{d+1}$ but not a minimum of the worldline action. Transverse deformations $y^a$ have a second variation $\frac{1}{2}\int\d\tau (\dot y^2 - y^2)$ with eigenvalues $n^2-1$. Hence there is one (per transverse direction) negative mode $(n=0)$, and two zero modes ($n=\pm 1$). The $d$ negative modes contribute $(\pm i)^d$. The $2d$ zero modes are the moduli of great circles and produce $\mathrm{vol}(\mathrm{SO}(d+2)/(\mathrm{SO}(2)\times \mathrm{SO}(d)))$. The volume of $\mathrm{SO}(d+2)$ cancels the volume factor of the pure gravity answer -- the observer breaks the isometries. 
Finally, implementing the constraint (\ref{eq: constraint euclideal observer}) requires rotating $\beta \rightarrow \beta_0 + i s$ which contributes another $\pm i$. 
In total \cite{Maldacena:2024spf} (see also \cite{Ivo:2025yek, Chen:2025jqm, Shi:2025amq})
\begin{equation}
    \underbrace{(\pm i)^{d+3}}_{\mathrm{conformal~modes}}\times \underbrace{(\pm i)^d}_{\mathrm{transverse}} \times \underbrace{(\pm i)}_{\mathrm{lapse}}= (\pm i)^{2d+4} ~.
\end{equation}
The power is even and hence the phase cancels. 
The same observer that rendered the de Sitter entropy well defined algebraically renders the Euclidean path integral real.

\paragraph{Stretched horizon.}
So far the observer sat at the pole. Alternatively one may imagine holographic degrees of freedom residing at the horizon itself \cite{tHooft:1984kcu, Susskind:1993if, Banks:2003cg, Fischler:2024cgm, Anninos:2025zgr}. The horizon is a null surface and carries no clock; displacing it slightly inward — the stretched horizon — makes it timelike.
\begin{figure}[ht]
\centering
\begin{tikzpicture}[scale=.8]
  \draw[very thick] (0,0) circle (2.55);
  \draw[thick, dash pattern=on 3pt off 2.5pt, red!70!black] (0,0) circle (2.2);
  \foreach \a in {0,18,...,342}{
    \pgfmathsetmacro{\tilt}{\a + 55*sin(3.7*\a)}  
    \draw[thick, blue!60!black, -{Stealth[length=3.2pt]}]
      ({2.2*cos(\a)-0.13*cos(\tilt)},{2.2*sin(\a)-0.13*sin(\tilt)}) --
      ({2.2*cos(\a)+0.13*cos(\tilt)},{2.2*sin(\a)+0.13*sin(\tilt)});
  }
  \draw[thick] (0,.32) circle (0.1);
  \draw[thick] (0,.22) -- (0,-.08);
  \draw[thick] (-.16,.1) -- (.16,.1);
  \draw[thick] (0,-.08) -- (-.13,-.3);
  \draw[thick] (0,-.08) -- (.13,-.3);
  \node[scale=.7] at (0,-.55) {$\rho=0$};
  \node[scale=.75] at (0,2.83) {cosmological horizon};
  \draw[thin, red!70!black] (1.35,-1.35) -- (2.05,-2.05);
  \node[scale=.75, red!70!black, below right] at (1.95,-1.95) {stretched horizon};
\end{tikzpicture}
\caption{The stretched horizon. An observer at the pole $\rho=0$ is surrounded
by the cosmological horizon. Displacing the horizon slightly inward yields a
timelike surface --- the stretched horizon --- on which one may imagine
holographic qubit degrees of freedom residing. }
\label{fig: stretched horizon}
\end{figure}
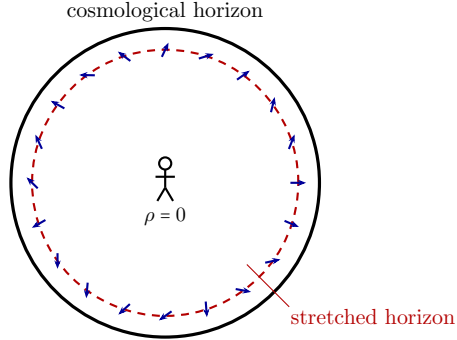

\paragraph{Thickened worldlines and timelike boundaries.} Moreover we can imagine a family of timelike boundaries, that interpolate from a worldline to a thickened worldline to the stretched horizon \cite{Coleman:2021nor, Silverstein:2024xnr, Svesko:2022txo, Anninos:2011zn}. These may serve as anchors for observables in the static patch.

\section{de Sitter Black holes}\label{sec:dSBH}
We will now turn to the study of black holes in de Sitter space. In contrast to anti--de Sitter space, there are no black hole solutions in three-dimensional de Sitter gravity (see however \cite{Emparan:2022ijy}), so we must work in higher dimensions to find such geometries. In four dimensions several important black hole solutions exist.

In this section we will discuss the main examples: the Schwarzschild--de Sitter black hole, which describes a neutral, non-rotating black hole in a de Sitter background; the Reissner--Nordström--de Sitter black hole, which includes electric charge; and finally the Kerr--de Sitter black hole, which generalizes the solution to include rotation. These geometries illustrate the interplay between black hole horizons and the cosmological horizon. We study their near-extremal, near-horizon geometry, some thermodynamic properties and the quasinormal modes of these black holes.

\subsection{Schwarzschild--de Sitter black hole}
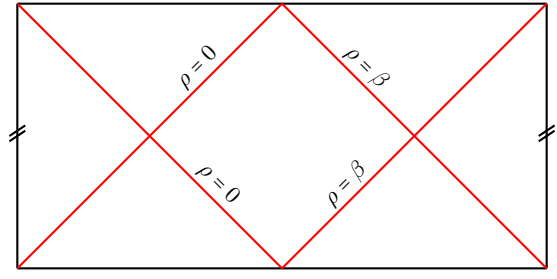
\begin{figure}
\begin{center}
\begin{tikzpicture}[scale=.7]
    \draw[thick] (0,0)-- (10,0);
    \draw[thick] (0,5)-- (10,5);
    \draw[thick] (0,0) -- (0,5);
    \draw[thick] (10,0)-- (10,5);
    \draw[thick, red] (0,0) -- (5,5);
    \draw[thick, red] (0,5) -- (5,0);
    \draw[thick, red] (5,5) -- (10,0);
    \draw[thick, red] (5,0) -- (10,5);
    \draw[thick] (-.15,2.4) -- (.15,2.6);
    \draw[thick] (-.15,2.5) -- (.15,2.7);
    \draw[thick] (9.85,2.4) -- (10.15,2.6);
    \draw[thick] (9.85,2.5) -- (10.15,2.7);
     \node[scale=.8, rotate = 315] at (3.8,1.6) {$\rho =0$}; 
   \node[scale=.8, rotate = 45] at (3.4,3.8) {$\rho =0$}; 
      \node[scale=.8, rotate = 315] at (6.6,3.8) {$\rho =\beta$}; 
   \node[scale=.8, rotate = 45] at (6.2,1.6) {$\rho =\beta$};       
\end{tikzpicture}
\caption{Penrose diagram of dS$_2$. The Nariai limit of the Schwarzschild--de Sitter black hole is dS$_2\times S^2$. In the Penrose diagram we suppress the $S^2$. The red lines are the cosmological $(\rho =\beta)$ and black hole ($\rho=0$) horizons. As discussed in section \ref{sec: geometry} the left and right edges are identified.}
\label{fig:Nariai limit}
\end{center}
\end{figure}

The Schwarzschild--de Sitter black hole is a solution of the four-dimensional Einstein--Hilbert action with positive cosmological constant $\Lambda = 3/\ell^2>0$. Its line element is given by
\begin{equation}
    ds^2 = -f(r)\, \d t^2 + f(r)^{-1} \d r^2 + r^2 \d\Omega_2^2~,
    \qquad 
    f(r)\equiv 1-\frac{2G_NM}{r}- \frac{r^2}{\ell^2}~.
\end{equation}
The parameter $M$ characterizes the mass of the black hole while $\ell$ sets the curvature scale of de Sitter space. In contrast to asymptotically flat or anti--de Sitter case, the presence of a positive cosmological constant introduces a cosmological horizon in addition to the black hole horizon. The horizons are obtained by solving $f(r)=0$ which, assuming $0<3\sqrt{3}M G_N/\ell<1$, leads to the two positive solutions
\begin{subequations}
\begin{align}
        r_b &= \frac{2\ell}{\sqrt{3}}\cos\left(\frac{1}{3}\arccos\left(\frac{3\sqrt{3}M G_N}{\ell}\right) + \frac{\pi}{3}\right)~,\\
        r_c &= \frac{2\ell}{\sqrt{3}}\cos\left(\frac{1}{3}\arccos\left(\frac{3\sqrt{3}M G_N}{\ell}\right) - \frac{\pi}{3}\right)~,
\end{align}
\end{subequations}
where $r_b<r_c$. The radius $r_b$ corresponds to the black hole horizon while $r_c$ corresponds to the cosmological horizon. The third root $r_3\equiv -r_b-r_c$ is negative and therefore unphysical. It is convenient to express $f(r)$ in factorized form
\begin{equation}
    f(r) = -\frac{1}{\ell^2 r}(r-r_b)(r-r_c)(r-r_3)~,
\end{equation}
which makes the structure of the horizons manifest. Both the black hole and cosmological horizons carry thermodynamic properties such as temperature and entropy. In particular the Bekenstein--Hawking entropy is proportional to the area of the horizon and the temperature is given by $T=\frac{1}{4\pi}|f'(r_h)|$. In general the two horizons have different temperatures, so Schwarzschild--de Sitter does not admit a global thermal equilibrium. A special situation arises when the two horizons coincide, $r_b=r_c$, which corresponds to the extremal limit
\begin{equation}\label{eq: Nariai limits SdS}
    M_{\mathrm{ext}}= \frac{\ell}{3\sqrt{3} G_N}~,\quad 
    S_{\mathrm{Nariai}}= \frac{\pi r_c^2}{G_N} = \frac{\pi\ell^2}{3 G_N}~,\quad 
    T_{\mathrm{ext}}= \frac{1}{4\pi}|f'(r_c)|=0~.
\end{equation}
In this limit the surface gravity vanishes and the horizons become degenerate. To study the geometry in this regime it is useful to zoom into the near-horizon region by introducing the rescaled coordinates
\begin{equation}
    \tau = \frac{\lambda t}{r_c}~,\quad 
    \rho = \frac{r-r_b}{\lambda r_c}~,\quad 
    \beta = \frac{r_c-r_b}{r_c\lambda}~,
\end{equation}
and taking the limit $\lambda\rightarrow 0$ while keeping $\beta$ fixed. This scaling focuses on the region between the two nearly coincident horizons and leads to the metric
\begin{equation}
    ds^2 =\frac{\ell^2}{3} 
    \left(
    -\rho(\beta-\rho)\d\tau^2 
    + \frac{\d\rho^2}{\rho(\beta-\rho)} 
    + \d\Omega_2^2
    \right)~,
\end{equation}
which is known as the Nariai geometry \cite{Nariai:1950}. This spacetime factorizes as $\mathrm{dS}_2\times S^2$ and can be interpreted as the near-extremal limit of the Schwarzschild--de Sitter solution, where the two horizons remain at finite separation in the rescaled coordinates (see figure \ref{fig:Nariai limit}). The Nariai geometry provides a particularly useful setting in which various dynamical questions become analytically tractable. For example, while the scalar wave equation in the full Schwarzschild--de Sitter background does not admit a simple analytic solution, in the Nariai limit it reduces to a solvable problem that allows us to determine the spectrum of quasinormal modes. We therefore consider the massless scalar wave equation \cite{Cardoso:2003sw}
\begin{equation}
    \nabla^2 \Phi(\tau,\rho,\Omega)=0~,
\end{equation}
and separate variables using the ansatz
\begin{equation}
\Phi(\tau,\rho,\Omega)=\e^{-i\omega\tau}R(\rho)Y_{l m}(\Omega)~.
\end{equation}
Substituting this ansatz into the wave equation leads to the radial equation
\begin{equation}
  -\partial_\rho\left({\rho(\beta-\rho)}R'(\rho)\right) 
  + {l(l+1)}R(\rho)
  = \frac{\omega^2}{\rho(\beta-\rho)}R(\rho)~,
\end{equation}
whose solutions can be expressed in terms of associated Legendre functions
\begin{equation}
    R(\rho)=c_1 P^{\frac{2i\omega}{\beta}}_{-\frac{1}{2}+i\sqrt{l(l+1)-\frac{1}{4}}}
    \left(-1+\frac{2\rho}{\beta}\right)
    +
    c_2 Q^{\frac{2i\omega}{\beta}}_{-\frac{1}{2}+i\sqrt{l(l+1)-\frac{1}{4}}}
    \left(-1+\frac{2\rho}{\beta}\right)~,
\end{equation}
where $c_{1,2}$ are constants.
A solution that is purely outgoing at the cosmological horizon $\rho=\beta$ is obtained by keeping only the associated Legendre function $P_n^m(z)$ and therefore setting $c_2=0$. Using the relation (\ref{eq:Legendre to hypergeometric}) we express the associated Legendre function in terms of the hypergeometric function $_2F_1(a,b,c;z)$ with
\begin{equation}
    a= \frac{1}{2}-i\sqrt{l(l+1)-\frac{1}{4}}~,\quad 
    b= \frac{1}{2}+i\sqrt{l(l+1)-\frac{1}{4}}~,\quad 
    c=1-\frac{2i\omega}{\beta}~,\quad 
    z=1-\frac{\rho}{\beta}~.
\end{equation}
Expanding this hypergeometric function around $z=1$ using (\ref{eq:compendium hypergeometric}) and imposing the boundary condition that the solution is purely ingoing at the black hole horizon $\rho=0$ leads to the quantization conditions
\begin{equation}
    \frac{1}{2}\pm i\sqrt{l(l+1)-\frac{1}{4}}-\frac{2i\omega}{\beta}=-k~,
    \qquad k\in\mathbb{N}_0~,
\end{equation}
which determine the quasinormal mode frequencies\footnote{As we will see in the next section (see equation \ref{eq: dS2 QNM}) these are also QNM of massive particles in dS$_2$.}
\begin{equation}\label{eq: Nariai QNM}
    \omega_k^{\pm} =
    \frac{\beta}{2}
    \left(
    -\left(k+\frac{1}{2}\right)i
    \pm \sqrt{l(l+1)-\frac{1}{4}}
    \right),
    \qquad k\in\mathbb{N}_0~.
\end{equation}
These frequencies describe the characteristic damped oscillations of scalar perturbations in the near-horizon Nariai geometry and encode how perturbations decay between the black hole and cosmological horizons. 

\paragraph{Dimensional reduction of 4D Schwarzschild--de Sitter.} 
Before moving on we reduce the four-dimensional Einstein--Hilbert action with positive
cosmological constant on the two-sphere, obtaining a two-dimensional dilaton-gravity theory. We start from (we add the superscript 4D here to avoid confusion with the 2D quantities later in this paragraph)
\begin{equation}
S_{\rm E}^{(\rm 4D)}=-\frac{1}{16\pi G_N}\int \d^4x \sqrt{g}\,(R^{(\rm 4D)}- 2\Lambda)
-\frac{1}{8\pi G_N}\int \d^3x\sqrt{{h}}\,K~,\quad \Lambda>0~,
\end{equation}
and make the ansatz (see e.g. \cite{Castro:2022cuo})
\begin{equation}
    ds^2 = \frac{\phi_0}{\phi}\,g_{\alpha\beta}\d x^\alpha \d x^\beta + \phi^2\d\Omega_2^2~,
    \quad \alpha,\beta = 0,1~,
\end{equation}
with $\phi_0$ a fixed constant with dimension of length. The relative Weyl factor $\phi_0/\phi$ is
chosen precisely so that the reduction produces no kinetic term for $\phi$: integrating over the
two-sphere leads to the 2D action
\begin{equation}
    S_{\rm E}^{(\rm 2D)} = -\frac{1}{4G_N}\int  \d^2 x \sqrt{g}\,\phi^2\left(R
    + 2\frac{\phi_0}{\phi^3} - 2\Lambda \frac{\phi_0}{\phi}\right)~,
\end{equation}
where $R$ is the Ricci scalar associated to the two-dimensional metric
$g_{\alpha\beta}$. In
two dimensions any metric on $S^2$ is Weyl-equivalent to the round one, so without loss of
generality we write $g_{\alpha\beta}= \e^{2\rho}\tilde{g}_{\alpha \beta}$ with $\tilde g$ the round
metric of fiducial radius $r$, $\widetilde{R}=2/r^2$. Using $\sqrt{g}\,R
=\sqrt{\tilde g}\,(\widetilde{R}-2\widetilde\nabla^2\rho)$ we obtain
\begin{equation}
    -S_{\rm E}^{(\mathrm{2D})} = \frac{1}{4G_N}\int  \d^2 x \sqrt{\widetilde{g}}\,\phi^2
    \left((\widetilde{R}- 2\widetilde{\nabla}^2\rho) + 2\e^{2\rho}\frac{\phi_0}{\phi^3}
    - 2\e^{2\rho}\Lambda \frac{\phi_0}{\phi}\right)~.
\end{equation}
The equations of motion for $\phi$ and $\rho$ are
\begin{subequations}
    \begin{align}
2\phi(\widetilde{R}- 2\widetilde{\nabla}^2\rho) -2\e^{2\rho}\frac{\phi_0}{\phi^2}
-2 \e^{2\rho}\Lambda\phi_0&=0~,\\
-2 \tilde{\nabla}^2\phi^2 + 4\e^{2\rho}\frac{\phi_0}{\phi} -4 \e^{2\rho}\Lambda \phi_0\phi&=0~.
    \end{align}
\end{subequations}
On constant-dilaton configurations the $\rho$ equation forces $\phi_*^2=\Lambda^{-1}$, and the
$\phi$ equation then fixes $\e^{2\rho_*}=\phi_*/(\phi_0\Lambda r^2)$. Choosing the ansatz constant
as $\phi_0=\phi_*=\Lambda^{-1/2}$ we arrive at
\begin{equation}\label{eq: solutions eom 2d SdS}
    \e^{2\rho_*}= \frac{1}{\Lambda r^2}
    ~\Rightarrow ~R_* = \e^{-2\rho_*}\widetilde{R}= 2\Lambda~,
    \quad \phi_*^2 = \Lambda^{-1}~.
\end{equation}
Expanding around this saddle,
\begin{equation}
    \phi= \phi_*+ \delta\phi~,\quad \rho = \rho_*+ \delta\rho~,
\end{equation}
the terms linear in the fluctuations cancel by the equations of motion, and we obtain
\begin{equation}
   - S_{\rm E}^{(\mathrm{2D})} = \frac{\pi}{G_N\Lambda}\chi(S^2)
    +\frac{1}{4G_N}\int_{S^2} \d^2 x \sqrt{\tilde{g}}\,
    \left(\frac{4}{\sqrt{\Lambda}}\delta \phi\!\left(-\widetilde{\nabla}^2  - \frac{2}{r^2}\right)
    \delta \rho + \frac{4}{r^2}\delta \phi^2 + \mathcal{O}(\delta^3) \right)~,
\end{equation}
where the leading term follows from Gauss--Bonnet,
$\int\sqrt{\tilde g}\,\widetilde{R}=4\pi\chi(S^2)$. Note that no $(\delta\rho)^2$ term is
generated and that
$-\widetilde{\nabla}^2-2/r^2$ annihilates the $l=1$ spherical harmonics --- the conformal Killing
modes of the sphere --- which require separate treatment at one loop. Using $\chi(S^2)=2$ and
$\Lambda=3/\ell^2$, the saddle-point value reproduces the total horizon entropy of the Nariai
geometry,
\begin{equation}
   - S_{\rm E,*}^{(\mathrm{2D})} = \frac{\pi}{G_N\Lambda}\,\chi(S^2)
    = \frac{2\pi\ell^2}{3G_N} = 2S_{\mathrm{Nariai}}~,
\end{equation}
one contribution $S_{\mathrm{Nariai}}$ (\ref{eq: Nariai limits SdS}) from each of the two horizons, in
agreement with the extremal entropy found above. (Equivalently, the on-shell Euclidean action of
the $S^2\times S^2$ instanton is $-S^{(\rm 4D)}_{\rm E}= \frac{1}{16\pi G_N}(2\Lambda)\mathrm{vol}(S^2\times S^2) = \frac{1}{16\pi G_N}(2\Lambda) (\frac{4\pi\ell^2}{3})^2$ wwith $\Lambda = \tfrac{3}{\ell^2}$.)

\subsection{Reissner--Nordstr\"om de--Sitter black hole}
The Reissner--Nordström--de Sitter black hole is a solution of the four-dimensional Einstein--Maxwell action
\begin{equation}
    S^{(\mathrm{4D})} = \frac{1}{16\pi G_N}\int \d^4x \sqrt{-g} 
    \left(R- 2\Lambda - F_{\mu\nu}F^{\mu\nu}\right)~,
    \qquad \Lambda>0~.
\end{equation}
The corresponding electrically charged black hole solution is described by the line element
\begin{equation}
    ds^2 = -f(r)\, \d t^2 + \frac{\d r^2}{f(r)} + r^2 \d\Omega_2^2~,
\end{equation}
together with the gauge potential
\begin{equation}
    A = \frac{Q}{r} \d t~,
\end{equation}
where $Q$ denotes the electric charge of the black hole. The metric function takes the form
\begin{equation}
    f(r) = 1- \frac{2G_NM}{r} + \frac{Q^2}{r^2} - \frac{r^2}{\ell^2}
    = -\frac{1}{r^2\ell^2}(r-r_4)(r-r_-)(r-r_+)(r-r_c)~.
\end{equation}
If magnetic charge were included one would obtain the more general dyonic solution.\footnote{This corresponds to $A\rightarrow \frac{Q}{r}\d t + P \cos\theta \d\varphi$ and $Q^2 \rightarrow Q^2 + P^2$ in $f(r)$.} $f(r)$ has four roots, three of them correspond to the various horizons of the spacetime. One of them,
\begin{equation}
r_4 = - r_c-r_--r_+ <0~,
\end{equation}
is negative and therefore unphysical. The remaining three positive roots $0<r_- \leq r_+ \leq r_c$ correspond to the inner (Cauchy) horizon $r_-$, the outer black hole horizon $r_+$ and the cosmological horizon $r_c$. Expressing the parameters of the solution in terms of the horizon radii gives
\begin{align}\label{eq: RN data}
    M G_N &= \frac{1}{2\ell^2}(r_++r_-)(\ell^2-r_+^2 - r_-^2)~,\cr
    Q^2 &= \frac{r_+ r_-}{\ell^2}\big(\ell^2 - r_+^2 -r_-^2 -r_- r_+\big)~,
\end{align}
while the radii obey
\begin{equation}\label{eq: relation l to radii}
     \ell^2 = r_c^2 +r_-^2 + r_+^2 +r_-r_++r_- r_c+ r_c r_+ ~.
\end{equation}

\begin{figure}
\begin{center}
\begin{tikzpicture}[scale=.8]
\begin{axis}[
    xlabel={$M$},
    ylabel={$Q$},
    xmin=0, xmax=5.3,
    ymin=0, ymax=5.9,
    axis lines=left,
    width=8cm, height=8cm,
    xtick={0,1,2,3,4,5},
    ytick={0,1,2,3,4,5},
    xticklabels={},
    yticklabels={},
    tick align=outside,
    axis line style={-latex},
    tick style={thin},
    xlabel style={font=\small},
    ylabel style={font=\small},
    title style={font=\small},
    legend style={
        at={(0.05,0.97)}, anchor=north west,
        draw=none, fill=none,
        font=\small,
        row sep=-2pt,
    },
    legend cell align=left,
]

\addplot[fill=gray!35, draw=none, smooth, forget plot] coordinates {
  (0.0000, 0.0000)
  (0.4992, 0.4994)
  (0.9933, 0.9950)
  (1.4775, 1.4830)
  (1.9467, 1.9596)
  (2.3958, 2.4206)
  (2.8200, 2.8618)
  (3.2142, 3.2786)
  (3.5733, 3.6661)
  (3.8925, 4.0186)
  (4.1667, 4.3301)
  (4.3908, 4.5934)
  (4.5600, 4.8000)
  (4.6692, 4.9396)
  (4.7133, 4.9990)
  (4.7140, 5.0000)  
  (4.6875, 4.9608)
  (4.5867, 4.8000)
  (4.4058, 4.4777)
  (4.1400, 3.9230)
  (3.9376, 3.4183)
  (3.7842, 2.9664)
  (3.6155, 2.3581)
  (3.5254, 1.9502)
  (3.4313, 1.3966)
  (3.3828, 0.9938)
  (3.3333, 0.0000)
} -- cycle;

\addplot[dashed, thick, smooth, black] coordinates {
  (0.0000, 0.0000)
  (0.4992, 0.4994)
  (0.9933, 0.9950)
  (1.4775, 1.4830)
  (1.9467, 1.9596)
  (2.3958, 2.4206)
  (2.8200, 2.8618)
  (3.2142, 3.2786)
  (3.5733, 3.6661)
  (3.8925, 4.0186)
  (4.1667, 4.3301)
  (4.3908, 4.5934)
  (4.5600, 4.8000)
  (4.6692, 4.9396)
  (4.7140, 5.0000)
};
\addlegendentry{cold ($r_- = r_+$)}

\addplot[solid, thick, smooth, black] coordinates {
  (4.7140, 5.0000)
  (4.6875, 4.9608)
  (4.5867, 4.8000)
  (4.4058, 4.4777)
  (4.1400, 3.9230)
  (3.9376, 3.4183)
  (3.7842, 2.9664)
  (3.6155, 2.3581)
  (3.5254, 1.9502)
  (3.4313, 1.3966)
  (3.3828, 0.9938)
  (3.3333, 0.0000)
};
\addlegendentry{Nariai ($r_+ = r_c$)}

\addlegendimage{area legend, fill=gray!35, draw=gray!60}
\addlegendentry{black hole}

\node[star, star points=5, star point ratio=2.25, fill=black, draw=black,
      inner sep=1.8pt]
    at (axis cs: 4.7140, 5.0000) {};

\end{axis}
\end{tikzpicture}
\end{center}
\caption{The Sharkfin diagram of the Reissner--Nordstr\"om--de Sitter black hole.}
\label{fig:shark fin}
\end{figure}
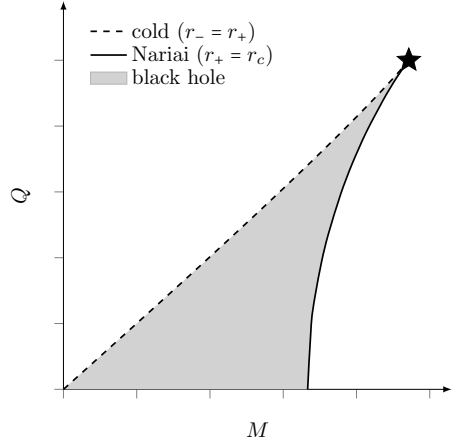

The different ways in which the horizons can coincide lead to several interesting extremal limits of the solution (see figure \ref{fig:shark fin})
\begin{itemize}
    \item \textbf{Cold black hole}: $r_-=r_+$. In this case the inner and outer black hole horizons coincide.  The black hole temperature vanishes while the cosmological horizon still has a non-zero temperature. The near-horizon geometry is $\mathrm{AdS}_2 \times S^2$.
    \item \textbf{Nariai black hole}: $r_+= r_c$. Here the black hole horizon coincides with the cosmological horizon. In the near-horizon limit the geometry approaches $\mathrm{dS}_2\times S^2$, analogous to the Nariai geometry encountered for the Schwarzschild--de Sitter black hole. We have 
    \begin{equation}
        M_{\mathrm{Nar}} G_N = \frac{r_c}{\ell^2}(r_c+r_-)^2~,\quad Q^2_{\mathrm{Nar}}  = \frac{r_c^2 r_-}{\ell^2}(2r_c+r_-)~.
    \end{equation}
    In the limit where $Q_{\mathrm{Nar}}=0$ $r_- = 0$ and from (\ref{eq: relation l to radii}) we obtain $r_c = \frac{\ell}{\sqrt{3}}$. Hence for $Q=0$ the fin ends at the SdS extremal mass (\ref{eq: Nariai limits SdS}).
    \item \textbf{Ultracold black hole}: $r_-=r_+=r_c$. In this maximally degenerate limit all three horizons coincide. The near-horizon geometry is  $\mathbb{M}_2 \times S^2$. From (\ref{eq: RN data}) we obtain
    \begin{equation}
        r_{\mathrm{uc}} = \frac{\ell}{\sqrt{6}}~,\quad Q^2_{\mathrm{uc}}  = \frac{\ell^2}{12}~,\quad M_{\mathrm{uc}} G_N = \frac{2\ell}{3\sqrt{6}}~.
    \end{equation}
    \item \textbf{Lukewarm}: The black hole and cosmological horizon have equal temperature but do not coincide. It corresponds to $Q^2 = G_N^2 M^2$ where
    \begin{equation}
        f(r)= \left(1-\frac{G_N M}{r}\right)^2-\frac{r^2}{\ell^2}~.
    \end{equation}
\end{itemize}

\subsection{Kerr--de Sitter black hole}
The Kerr--de Sitter black hole describes a rotating black hole in a spacetime with positive
cosmological constant. Its line element is given by
\begin{equation}
ds^2 =
- \frac{\Delta_r}{\rho^2}
\left(
\d t-\frac{a}{\Theta}\sin^2\theta\, \d\phi
\right)^2
+\frac{\rho^2}{\Delta_r}\d r^2
+\frac{\rho^2}{\Delta_\theta}\d\theta^2
+\frac{\Delta_\theta}{\rho^2}\sin^2\theta
\left(
a\,\d t-\frac{r^2+a^2}{\Theta}\d\phi
\right)^2 ,
\end{equation}
where
\begin{subequations}
\begin{align}\label{eq:Deltar1}
\Delta_r &= (r^2+a^2)\!\left(1-\frac{r^2}{\ell^2}\right)-2G_NMr ,
\qquad
\Theta = 1+\frac{a^2}{\ell^2},\\
\Delta_\theta &= 1+\frac{a^2}{\ell^2}\cos^2\theta ,
\qquad
\rho^2 = r^2+a^2\cos^2\theta .
\end{align}
\end{subequations}
The spacetime possesses the Killing vectors $\partial_t$ and $\partial_\phi$. However, the null
generator of the horizon is given by the linear combination
\begin{equation}
\chi = \partial_t + \Omega\,\partial_\phi .
\end{equation}
The angular velocity $\Omega$ is determined by requiring the norm $\chi^2$ to vanish at the
horizon,
\begin{equation}
\chi^2 = g_{tt} + 2\Omega g_{t\phi} + \Omega^2 g_{\phi\phi}
= -\frac{\Delta_r}{\rho^2}
\left(1-a\Omega \frac{\sin^2\theta}{\Theta}\right)^2
+\frac{\Delta_\theta}{\rho^2}\sin^2\theta
\left(a-\Omega\frac{r^2+a^2}{\Theta}\right)^2 .
\end{equation}
This condition is satisfied precisely when
\begin{equation}
\Delta_r =0,
\qquad
\Omega = \frac{a\Theta}{r^2+a^2}.
\end{equation}
Evaluated at a horizon $r=r_{\rm hor}$, this quantity corresponds to the angular velocity of the
horizon; we write $\Omega_+\equiv\Omega(r_+)$ and $\Omega_c\equiv\Omega(r_c)$ for the black-hole
and cosmological horizons introduced below. Rotating black holes exhibit the phenomenon of
superradiance, and in the presence of a cosmological horizon the superradiant condition becomes
\begin{equation}
m\Omega_c < \omega < m\Omega_+ ~,
\end{equation}
a window which closes as the two horizons approach each other.

The equation $\Delta_r=0$ generically has four roots which can be written as
\begin{equation}\label{eq:Deltar2}
\Delta_r =
-\frac{1}{\ell^2}(r-r_c)(r-r_+)(r-r_-)(r-r_4),
\qquad
r_c\ge r_+\ge r_->0 ,
\end{equation}
The three positive roots correspond respectively to the inner (Cauchy) horizon $r_-$, the black
hole horizon $r_+$ and the cosmological horizon $r_c$ (for $a\neq0$ the inner horizon radius is
automatically positive). The remaining root $r_4 =-(r_c+r_++r_-)$ is negative and therefore
unphysical. The parameters of the solution satisfy
\begin{equation}
(r_c+r_+)(r_c+r_-)(r_++r_-)=2MG_N\ell^2 ,
\end{equation}
and combining \eqref{eq:Deltar1} with \eqref{eq:Deltar2} gives
\begin{equation}
a^2\ell^2 =
r_c r_- r_+(r_c+r_-+r_+)\quad \mathrm{and}\quad a^2=
\ell^2-(r_c^2+r_-^2+r_+^2+r_c r_-+r_- r_++r_c r_+)~.
\end{equation}
Consistency of the two expressions for $a^2$ imposes one relation among $(r_-,r_+,r_c)$ at fixed
$\ell$: only two of the three radii are independent, matching the two-parameter $(M,a)$ family of
solutions. As in the Reissner--Nordstr\"om--de Sitter case, several interesting limits arise when
horizons coincide or, in the lukewarm case, when their temperatures agree \cite{Anninos:2010gh}:
\begin{itemize}
\item \textbf{Cold limit}: $r_+\rightarrow r_-$, where the inner and outer black hole horizons
coincide and the black hole becomes extremal; the near-horizon geometry is an $S^1$ fibration
over $\mathrm{AdS}_2$.
\item \textbf{Rotating Nariai limit}: $r_+\rightarrow r_c$, where the black hole and cosmological
horizons coincide; the near-horizon geometry is an $S^1$ fibration over $\mathrm{dS}_2$.
\item \textbf{Ultracold limit}: $r_-=r_+=r_c$, where all three horizons coincide and the
two-dimensional factor degenerates to flat space, as for the ultracold
Reissner--Nordstr\"om--de Sitter black hole.
\item \textbf{Lukewarm case}: the black hole and cosmological horizons have the same temperature
without coinciding.
\end{itemize}

\section{de Sitter vs Black hole}\label{sec: dS vs BH}
The central question of this section is: Can the static patch of de Sitter, the region accessible to an observer, be described using the same physics as black holes? In particular, what are the similarities and differences of the de Sitter cosmological horizon and the black hole horizon?

In this section we combine what we have seen in sections \ref{sec: geometry}, \ref{sec:QFT},  \ref{sec:dSentropy} and section \ref{sec:dSBH} to compare and contrast the de Sitter static patch and the black hole horizon.

\subsection{Quasinormal modes in the static patch}\label{subsection:Fields in static patch}
\begin{figure}[ht]
\centering
    \begin{tikzpicture}[scale=1.2]
        \begin{scope}[shift = {(5,0)}]

\coordinate (A) at (0.02, 0.02);
    \coordinate (B) at (0.02,2.98);
    \coordinate (C) at (1.493, 1.493);

\coordinate (D) at (2.98, 0.02);
    \coordinate (E) at (2.98,2.98);
    \coordinate (F) at (1.493, 1.493);    

\shade[right color=gray!10, left color=gray!25] (A) -- (B) -- (C) -- cycle;

\draw[very thick,blue](0,0) -- (0,3.02);   
\draw[very thick](0,3) -- (3.02,3.);       
\draw[very thick](3,0) -- (3,3);   
\draw[very thick](3.02,0) -- (-.02,0);   
\draw[very thick](3,3) -- (0,0);   
\draw[very thick](0,3) -- (3,0);

\node[scale=.8] at (1.5,3.2) {$\mathcal{I}^+$}; 

\node[scale=.8] at (1.5,-0.2) {$\mathcal{I}^-$}; 

  \node[scale=.7, rotate = 90, blue] at (-.2,1.5) {$r=0$};
 \node[scale=.7, rotate = 45] at (.9,.7) {past horizon}; 
\draw[thick,decorate,decoration={snake,amplitude=0.7mm,segment length=4mm},->,red]
    (.7,1.5) -- (1.4,2.2);
\draw[thick,decorate,decoration={snake,amplitude=0.7mm,segment length=4mm},->,red]
    (.5,1.8) -- (1.2,2.5);    
        \end{scope}
    \end{tikzpicture}
\caption{An observer at $r=0$ is surrounded by an event horizon. Quasinormal modes are regular at $r=0$ and purely outgoing at the future horizon.}
\label{fig:QNM}
\end{figure}
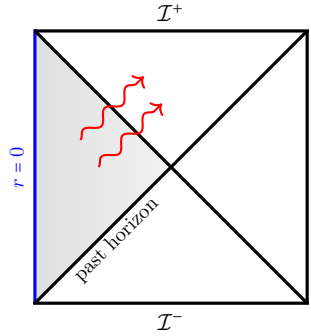

We start by uncovering a beautiful connection between the field content of de Sitter and the Harish-Chandra group characters \cite{Anninos:2020hfj, Sun:2020sgn}. For this we start by looking at the wave equation of a conformally coupled scalar in dS$_4$ with $m^2 \ell^2 = 2$ in the static patch (\ref{eq: static patch}) of dS$_4$. From (\ref{eq:msquared F}) we infer the weight $\Delta=2$ and hence a conformally coupled scalar in dS$_4$ transforms in the complementary series irreducible representation of $\mathrm{SO}(1,4)$. The wave equation is given by
\begin{equation}\label{eq:wave equation}
    (-\nabla^2+m^2)\Phi=0~,\quad -\nabla^2 \equiv - \frac{1}{\sqrt{-g}}\partial_\mu \sqrt{-g}g^{\mu\nu}\partial_\nu~.
\end{equation}
For the static patch metric (\ref{eq: static patch}) we solve (\ref{eq:wave equation}) by separation of variables,
\begin{equation}\label{eq:wavefunction QNM}
    \Phi(t,r,\Omega) = \varphi(r)\, \e^{-i\omega t} Y_{lm}(\Omega)~,
\end{equation}
where $Y_{lm}(\Omega)$ are spherical harmonics on a unit $S^2$, satisfying
\begin{equation}
    -\nabla_{S^2}^2 Y_{lm} = l(l+1) Y_{lm}~, \qquad m=-l,\ldots,l~,\quad l \in \mathbb{Z}_{\geq 0}~.
\end{equation}
The radial equation becomes
\begin{equation}
    -\frac{1}{r^2}\frac{d}{dr}\left(\frac{r^2}{\ell^2}(1-r^2)\varphi'(r)\right)
    + m^2 \varphi(r) + \frac{l(l+1)}{r^2 \ell^2}\varphi(r)
    = \frac{\omega^2}{\ell^2(1-r^2)} \varphi(r)~.
\end{equation}
The wave equation then admits the solutions $\varphi(r) = A \varphi_{\mathrm{n.n.}}(r)+ B\varphi_{\mathrm{n}}(r)$ with \cite{Anninos:2011af, Loganayagam:2023pfb}
\begin{subequations}
    \begin{align}
        \varphi_{\mathrm{n}}(r) &= \left(1-{r^2}\right)^{-\frac{i\omega }{2}}r^{l} \, _2F_1\left(\frac{l-i \omega}{2}+\frac{1}{2},\frac{l-i \omega}{2}+1; \frac{3}{2}+l;r^2\right)~,\\
        \varphi_{\mathrm{n.n.}}(r) &=\left(1-r^2\right)^{-\frac{i\omega }{2}}r^{-1-l}\, _2F_1\left(-\frac{l+i \omega}{2},\frac{1}{2}-\frac{l+i \omega}{2}; \frac{1}{2}-l;{r^2}\right)~.
    \end{align}
\end{subequations}
Near $r=0$ where the static patch observer sits (see figure \ref{fig:QNM}) the solutions behave as
\begin{equation}
    \varphi_{\mathrm{n}}(r) \approx r^l~,\quad \varphi_{\mathrm{n.n.}}(r) \approx r^{-1-l}~.
\end{equation}
We therefore, in line with the notation of \cite{Anninos:2011af} denote the solution regular at $r\rightarrow 0$ normalizable, whereas $\varphi_{\mathrm{n.n.}}$ is not normalizable for $r=0$.
We expand the normalizable solution $\varphi_{\mathrm{n}}$ close to the horizon $r=1$ using the identity (\ref{eq:compendium hypergeometric}) for the hypergeometric function 
\begin{align}\label{eq: F expanded near 1-z}
    \varphi_{\mathrm{n}}(r)=&\left(1-{r^2}\right)^{-\frac{i\omega }{2}}\!\!r^{l}\frac{2^{l+i\omega}}{\sqrt{\pi}} \frac{\Gamma(\frac{3}{2}+l) \Gamma(i \omega)}{\Gamma(1+l+i\omega)}  \,_2F_1 \left(\frac{1}{2}+ \frac{l-i\omega}{2}, 1+\frac{l-i\omega}{2}, 1- i\omega; 1-r^2\right)\cr
    & + \left(1-r^2\right)^{\frac{i\omega }{2}}\!\!r^{l}\frac{2^{l-i\omega}}{\sqrt{\pi}} \frac{\Gamma(\frac{3}{2}+l) \Gamma(-i \omega)}{\Gamma(1+l-i \omega)} \, _2F_1 \left(\frac{1}{2}+ \frac{l+i\omega}{2}, 1+\frac{l+i\omega}{2}, 1+ i\omega; 1-r^2\right).
\end{align}
The two pieces now behave as $(1-r^2)^{\pm i \omega/2}$ close to the horizon. If we impose that the wavefunction is purely outgoing at the future horizon one of the two terms in (\ref{eq: F expanded near 1-z}) needs to vanish. As an example, if we want a  purely outgoing solution at the future cosmological horizon we need to impose the vanishing of the solutions going as $(1-r^2)^{i\omega/2}$, which is equivalent to 
\begin{equation}
     \omega_{k} = -i(1+l+k)~,\quad k=0,1,2,\ldots~.
\end{equation}
This discretuum of poles is known as the quasinormal modes. The corresponding wavefunction is
\begin{equation}
    \Phi(t,r,\Omega) = {r}^l \left(1-r^2\right)^{-\frac{i\omega_{k}}{2}} \, _2 F_1 \left(\frac{1-k}{2}, -\frac{k}{2}; \frac{3}{2}+l; r^2\right) \e^{-i\omega_{k} t}Y_{lm}(\Omega)~.
\end{equation}
Finally, we can count degeneracies of quasinormal modes. The frequency $\omega_{k} = -i$ arises only from $(l,k)=(0,0)$, giving degeneracy $1$. For $\omega_{k}=-2i$, the pairs $(l,k)=(1,0)$ and $(0,1)$ contribute, giving degeneracy $(2+1)+1=4$. For $\omega_{k}=-3i$, the possibilities are
\[
(l,k)\in \{(2,0),\,(1,1),\,(0,2)\}~,
\]
with degeneracies $5$, $3$, and $1$, respectively, summing to $9$. We can keep counting degeneracies, but what we notice is that the numbers $1,4,9,\ldots$ match the coefficients in the expansion of the Harish-Chandra character for a conformally coupled $m^2\ell^2 =2$ scalar in dS$_4$ (\ref{eq: chi20}),
\begin{equation}
    \chi_{2,0}(\mathfrak{t}) = \frac{\e^{-2\mathfrak{t}} + \e^{-\mathfrak{t}}}{(1 - \e^{-\mathfrak{t}})^3}
    = \e^{-\mathfrak{t}} + 4 \e^{-2\mathfrak{t}} + 9 \e^{-3\mathfrak{t}} + \cdots~.
\end{equation}
We can repeat this discussion for arbitrary fields and match the degeneracies of their quasinormal modes to their respective bulk Harish-Chandra characters \cite{Sun:2020sgn}. 
If instead of a conformally coupled scalar we would have considered a scalar with mass $m^2 \ell^2$ we would have instead found 2 strands of quasinormal modes\footnote{In the dS$_2$ case $d=1$ and $l\in \{0,1\}$. Let us also assume that $m^2 \ell^2 = l'(l'+1)$ then $h_{\pm} = \frac{1}{4} \pm \frac{1}{2}\sqrt{\frac{1}{4} - l'(l'+1)} = \frac{1}{4}\pm \frac{i}{2}\nu_{l'}$ for $\nu_{l'}\equiv \sqrt{l'(l'+1)-\frac{1}{4}} \in \mathbb{R}$ if $l'\geq 1$. This in particular implies
\begin{equation}\label{eq: dS2 QNM}
    \omega_k^\pm = - i \Big(\frac{1}{2}\pm i \nu_{l'} +l+2k\Big)~,\quad l\in \{0,1\}\quad \Rightarrow \omega_n^\pm = -\Big(n+\frac{1}{2}\Big)i \pm \nu_{l'}~,\quad n \in \mathbb{N}_0~.
\end{equation} 
For each $n$, $\omega_n^\pm$ has degeneracy exactly one which matches the expansion $\frac{q^\Delta}{1-q}= q^\Delta + q^{\Delta+1} + q^{\Delta+2}+ \ldots$ of the 2D Harish-Chandra character. Moreover after multiplying by $\beta/2$ (\ref{eq: dS2 QNM}) agrees with (\ref{eq: Nariai QNM}).}
\begin{equation}
    \omega_{k}^\pm = -i \left(2h_\pm + l+2k \right)~,\quad h_{\pm} = \frac{d}{4} \pm \frac{1}{2}\sqrt{\frac{d^2}{4}-m^2\ell^2}~,\quad k=0,1,2,\ldots~.
\end{equation}
Quasinormal modes for half-integer fields and how they compare to their corresponding Harish-Chandra characters are still largely unexplored. 
Quasinormal modes have negative imaginary part such that 
\begin{equation}
    \lim_{t\rightarrow \infty} \e^{-i\omega_{k}^\pm t} =0~.
\end{equation}
For the principal series (\ref{eq:higher D rep}), $h_{\pm} = \frac{d}{4}\pm \frac{i\nu}{2}$, $\nu \in \mathbb{R}$ and hence $\Re(\omega_k^\pm) = \pm \nu$. In the large $\nu\rightarrow \infty$ limit we find
\begin{equation}
    \lim_{\nu\rightarrow \infty} \frac{\Im(\omega_{k}^{\pm})}{\Re(\omega_{k}^{\pm})} \rightarrow 0~.
\end{equation}
More explicitly, we observe that the frequency dependence $\e^{-i\omega_{k}^\pm t}$ of the wavefunction (\ref{eq:wavefunction QNM}) for quasinormal modes takes the form
\begin{equation}
    \e^{-i\omega_{k}^\pm t} = \e^{\mp i \nu t}\times \e^{-\Gamma t}~,\quad \Gamma\equiv \frac{d}{2}+l+2k~.
\end{equation}
The first factor oscillates with period $\tfrac{2\pi}{\nu}$. The second factor is exponentially decaying. We thus observe that for small $\nu$ we have a few oscillations before the decay, while for large $\nu$ the wavefunction oscillates rapidly before it decays. In de Sitter, large $l$ quasinormal modes are dissipative and fall into the horizon. A heavy particle contributes a real part to $\omega_{k}^\pm$.

This is different, for example, from BTZ, where the quasinormal modes are given by \cite{Birmingham:2001pj} $\omega_{k,L}= k-4\pi i T_L(n+h_L),$ and $\omega_{k,R}= -k-4\pi i T_R(n+h_R)$.
Increasing the conformal weight increases the imaginary part of the quasinormal modes and therefore the damping rate. General methods for quasinormal mode calculations can be found e.g. in \cite{Horowitz:1999jd, Motl:2003cd}.

\subsection{Gao-Wald theorem}

\begin{figure}[ht]
    \centering
\begin{tikzpicture}[scale=.9,
  sing/.style={very thick, decorate, decoration={snake, amplitude=.5mm, segment length=2.6mm}},
  bdy/.style={very thick}, hor/.style={thick},
  op/.style={circle, fill=red!60!red, inner sep=1.1pt},
  op2/.style={circle, fill=blue!60!blue, inner sep=1.1pt},
  vop/.style={circle, fill=red!70!black, inner sep=1.1pt}]
\begin{scope}
\fill[gray!25] (0,0) -- (1.5,1.5) -- (0,3) -- cycle;
\draw[bdy] (0,0) -- (3,0);
  \draw[bdy] (0,3) -- (3,3);
  \draw[bdy] (0,0) -- (3,0);
  \draw[very thick, blue] (0,0) -- (0,3);
  \draw[bdy] (3,0) -- (3,3);
  \draw[hor] (0,0) -- (3,3);
  \draw[hor] (0,3) -- (3,0);
  \node at (1.5,3.25) {$\mathcal{I}^+$};
  \node at (1.5,-.25) {$\mathcal{I}^-$};
\end{scope}
\begin{scope}[shift={(5,0)}]
\draw[very thick, blue] (0,0) -- (0,4);
\draw[bdy] (3,0) -- (3,4);
\draw[bdy] (0,4) -- (3,4);
\draw[bdy] (0,0) -- (3,0);
\draw[hor] (0,0) -- (3,3);
\draw[thick, densely dashed] (3,0) -- (0,3);
\node at (1.5,4.25) {$\mathcal{I}^+$};
\node at (1.5,-.25) {$\mathcal{I}^-$};
\draw[red, thick] (3,.5)--(0,3.5);
\end{scope}
\end{tikzpicture}
  \caption{Penrose diagram of global de Sitter, with the static patch of the observer shaded in grey and the worldline in blue. A light ray emitted from the pole after $\mathcal{I}^+$ cannot reach the antipodal worldline before $\mathcal{I}^+$. Right: The Gao--Wald effect for de Sitter space. Including backreaction,
the Penrose diagram stretches vertically --- it becomes taller than wide
--- and previously disconnected regions come into causal contact: a light
ray (red) emitted after $\mathcal{I}^-$ crosses the entire space and
reaches the observer's worldline before reaching $\mathcal{I}^+$. }
    \label{fig:GW de Sitter}
\end{figure}
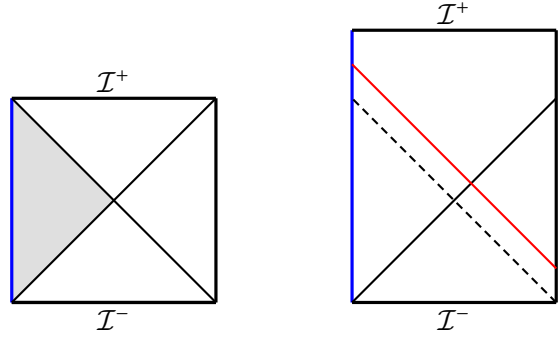

The discussion so far neglects backreaction. 
The Gao–Wald theorem \cite{Gao:2000ga} states that a null geodesically complete, globally hyperbolic spacetime with compact Cauchy surface $\Sigma$, satisfying the Einstein equations with matter obeying the null energy condition and the null generic condition, cannot exhibit a particle horizon.
In particular, at sufficiently late times the past light cone of a spacetime point in a slightly perturbed de Sitter spacetime intersects all of $\Sigma$ (see figure \ref{fig:GW de Sitter}). The requirement of a slightly perturbed de Sitter spacetime is to ensure that the null generic condition is also satisfied.
The behaviour
of the AdS black hole is precisely opposite: infalling positive-energy matter produces a time delay and reduces the causal overlap of signals sent from the two exterior regions; the Penrose diagram stretches horizontally (see figure \ref{fig:GW BH}). Moreover, for the eternal black hole
the boundary correlators obey the KMS condition, and
out-of-time-order correlators exhibit maximal chaos $\lambda_L =
\frac{2\pi}{\beta}$ \cite{Shenker:2013pqa, Maldacena:2015waa}. For de Sitter the status of both statements is far
less clear \cite{Anninos:2018svg,Kolchmeyer:2024fly,Chen:2026boh,Cui:2026bcd,Harlow:2026pwe,Milekhin:2026tbi}.

\begin{figure}[ht]
    \centering
\begin{tikzpicture}[scale=.9,
  sing/.style={very thick, decorate, decoration={snake, amplitude=.5mm, segment length=2.6mm}},
  bdy/.style={very thick}, hor/.style={thick},
  op/.style={circle, fill=red!60!red, inner sep=1.1pt},
  op2/.style={circle, fill=blue!60!blue, inner sep=1.1pt},
  vop/.style={circle, fill=red!70!black, inner sep=1.1pt}]
  \begin{scope}
  \draw[sing] (0,3) -- (3,3);
  \draw[sing] (0,0) -- (3,0);
  \draw[bdy] (0,0) -- (0,3);
  \draw[bdy] (3,0) -- (3,3);
  \draw[hor] (0,0) -- (3,3);
  \draw[hor] (0,3) -- (3,0);
  \draw[red,thick] (0,.5) -- (2.5,3);
  \draw[red,thick] (3,1) -- (1,3);
\node[op] at (1.75,2.27) {};
  \node[op2] at (0,1.5) {};
  \node[op2] at (3,1.5) {};
  \node[scale=.65, left]  at (-.05,1.5) {$\mathcal{O}_L(0)$};
  \node[scale=.65, right] at (3.05,1.5) {$\mathcal{O}_R(0)$};
\end{scope}
\begin{scope}[shift={(5,0)}]
  \draw[sing] (0,0)   -- (4,0);
  \draw[sing] (0.5,3) -- (4.5,3);
  \draw[bdy]  (0,0)   -- (0.5,3);
  \draw[bdy]  (4,0)   -- (4.5,3);
  \draw[thick, double, double distance=1.1pt] (0,0) -- (4.5,3);
  \draw[hor] (0.5,3) -- (1.8,1.2);
  \draw[hor] (4,0)   -- (2.7,1.8);
  \node[scale=.65, rotate=-54] at (1.37,2.1) {$\mathrm{future~horizon}$};
   \node[scale=.65, rotate=-54] at (3.45,1.06) {$\mathrm{past~horizon}$};
  \node[op2] at (0.2,1.2) {};
  \node[op2] at (4.3,1.8) {};
  \node[scale=.7, left]  at (0.15,1.2) {$\mathcal{O}_L$};
  \node[scale=.7, right] at (4.35,1.8) {$\mathcal{O}_R$};
\draw[red, densely dashed, thick] (.2,1.) -- (2.95,3);
\draw[red, densely dashed, thick] (4.28,1.5) -- (3.2,3.02);
\draw[densely dotted, thick, blue!60!black]
    (0.2,1.2) .. controls (1.30,1.28) and (2.0,1.56) .. (2.52,1.82)
              .. controls (3.04,2.08) and (3.52,2.06) .. (4.3,1.8);
\end{scope}
\end{tikzpicture}
\caption{Left: Penrose diagram of the eternal black hole. Null signals (red) sent from the two exterior regions meet behind the horizon (red dot). Right: Including
backreaction of an early perturbation, which sources a shockwave (double line), the Penrose diagram stretches horizontally and previously
connected regions fall out of causal contact: the null signals (red,
dashed) emitted by the two observers reach the singularity without ever
meeting. The two-sided correlator between $\mathcal{O}_L$ and
$\mathcal{O}_R$ --- an out-of-time-order correlator (OTOC)--- is computed in the geodesic approximation by the length of the
blue geodesic, which grows as the diagram stretches and hence the correlator decays.}
    \label{fig:GW BH}
\end{figure}
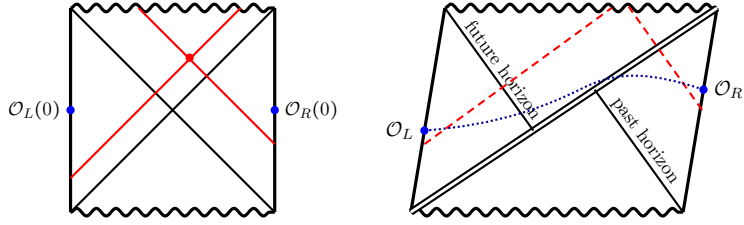

\subsection{State preparation}

\begin{figure}[ht]
\centering
\begin{tikzpicture}[scale=.9]
\begin{scope}[shift={(-4.9,-1.5)}]
  \fill[gray!25] (0,0) -- (1.5,1.5) -- (0,3) -- cycle;
  \draw[very thick] (0,0) -- (3,0);              
  \draw[very thick] (0,3) -- (3,3);              
  \draw[very thick, red] (0,0) -- (0,3);        
  \draw[very thick] (3,0) -- (3,3);              
  \draw[thick] (0,0) -- (3,3);
  \draw[thick] (3,0) -- (0,3);
  \draw[thick, blue] (0,1.5) -- (3,1.5);
  \fill (1.5,1.5) circle (1.3pt);
  \node[scale=.8] at (1.5,3.3)  {$\mathcal{I}^+$};
  \node[scale=.8] at (1.5,-.3)  {$\mathcal{I}^-$};
  \node[scale=.8] at (1.5,-1.15) {dS$_{d+1}$};
\end{scope}
\begin{scope}
  \draw[very thick] (0,0) circle (1.1);
  \draw[thick,red]  (-1.1,0) arc (180:360:1.1 and 0.35);
  \draw[thick, dashed,red] (1.1,0) arc (0:180:1.1 and 0.35);
  \node[scale=.8] at (0,-2.65) {$S^{d+1}$};
\end{scope}
\begin{scope}[shift={(3.6,0)}]
  \draw[very thick]
    (-1.1,0) arc (180:360:1.1) arc (0:-180:1.1 and 0.35);
  \draw[thick,blue] (0,0) ellipse (1.1 and 0.35);
  \node[scale=.8, right] at (1.2,0.25) {$\Sigma\cong S^{d}$};
  \node[scale=.8] at (0,-2.65) {$\Psi_{\rm HH}[\Sigma]$};
\end{scope}
\begin{scope}[shift={(7.9,0)}]
  \draw[very thick] (-1.5,0) arc (180:360:1.5);
  \draw[very thick] (-1.5,0) -- (-1.5,1.5);
  \draw[very thick] (1.5,0)  -- (1.5,1.5);
  \draw[very thick] (-1.5,1.5) -- (1.5,1.5);
  \draw[thick] (0,0) -- (1.5,1.5);
  \draw[thick] (0,0) -- (-1.5,1.5);
  \draw[thick, blue] (-1.5,0) -- (1.5,0);
  \fill (0,0) circle (1.3pt);
  \node[scale=.8] at (0,1.75) {$\mathcal{I}^+$};
  \node[scale=.6, blue, below] at (-0.75,-0.04) {$\Sigma_L$};
  \node[scale=.6, blue, below] at (0.75,-0.04)  {$\Sigma_R$};
  \node[scale=.75] at (0,-0.7) {$\e^{-S_E}$};
  \node[scale=.75] at (0,1.05) {$\e^{\mathrm{i}S}$};
  \node[scale=.8] at (0,-2.15) {$|\Psi_{\rm HH}\rangle$};
\end{scope}
\end{tikzpicture}
\caption{From left to right: dS Penrose diagram. The static patch shaded in gray, the observer's worldline in red. $S^{d+1}$ viewed as the Euclidean continuation of the static patch. The worldline is mapped to a great circle on the sphere. The sphere is also the Euclidean continuation of global de Sitter. Cutting the Euclidean continued global de Sitter sphere -- $S^{d+1}$ -- the boundary is $S^d$, which corresponds to the blue line on the global Penrose diagram. This prepares the HH state. Analytically continuing to Lorentzian we obtain the Hartle-Hawking wavefunction $\Psi_{\mathrm{HH}}$.}
\label{fig: state preparation dS}
\end{figure}
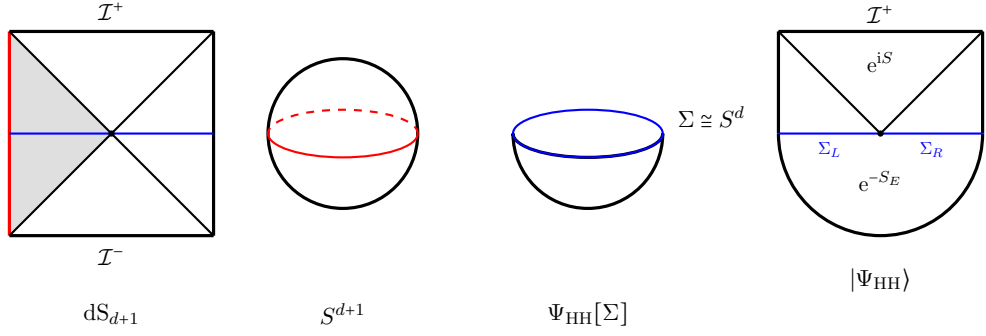

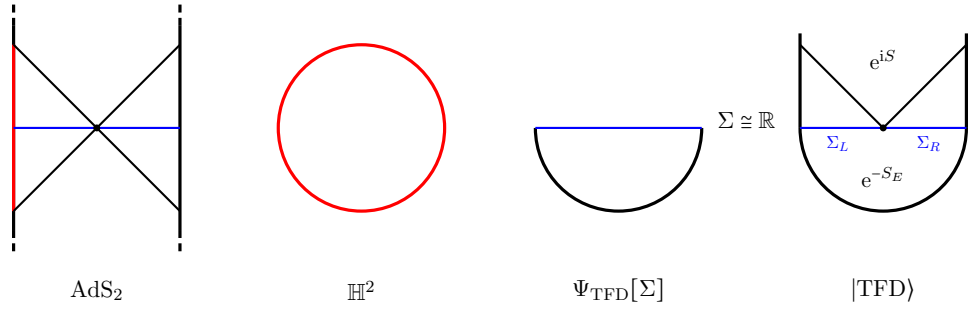
\begin{figure}[ht]
\centering
\begin{tikzpicture}[scale=1]
\begin{scope}
  \draw[very thick] (-1.1,-1.35) -- (-1.1,1.35);
  \draw[very thick] (1.1,-1.35)  -- (1.1,1.35);
   \draw[very thick,red] (-1.1,-1.1)  -- (-1.1,1.1);
  \draw[very thick, densely dashed] (-1.1,1.35) -- (-1.1,1.7);
  \draw[very thick, densely dashed] (1.1,1.35)  -- (1.1,1.7);
  \draw[very thick, densely dashed] (-1.1,-1.35) -- (-1.1,-1.7);
  \draw[very thick, densely dashed] (1.1,-1.35)  -- (1.1,-1.7);
  \draw[thick] (-1.1,-1.1) -- (1.1,1.1);
  \draw[thick] (-1.1,1.1) -- (1.1,-1.1);
  \draw[thick, blue] (-1.1,0) -- (1.1,0);
  \fill (0,0) circle (1.3pt);
  \node[scale=.8] at (0,-2.15) {AdS$_2$};
\end{scope}
\begin{scope}[shift={(3.5,0)}]
  \draw[very thick, red] (0,0) circle (1.1);
  \node[scale=.8] at (0,-2.15) {$\mathbb{H}^2$};
\end{scope}
\begin{scope}[shift={(6.9,0)}]
  \draw[very thick] (-1.1,0) arc (180:360:1.1);
  \draw[thick, blue] (-1.1,0) -- (1.1,0);
  \node[scale=.8, right] at (1.2,0.1) {$\Sigma\cong \mathbb{R}$};
  \node[scale=.8] at (0,-2.15) {$\Psi_{\rm TFD}[\Sigma]$};
\end{scope}
\begin{scope}[shift={(10.4,-1.1)}]
  \draw[very thick] (-1.1,1.1) arc (180:360:1.1);
  \draw[very thick] (-1.1,1.1) -- (-1.1,2.35);
  \draw[very thick] (1.1,1.1)  -- (1.1,2.35);
  \draw[thick] (0,1.1) -- (1.1,2.2);
  \draw[thick] (0,1.1) -- (-1.1,2.2);
  \draw[thick, blue] (-1.1,1.1) -- (1.1,1.1);
  \fill (0,1.1) circle (1.3pt);
  \node[scale=.6, blue, below] at (-0.6,1.07) {$\Sigma_L$};
  \node[scale=.6, blue, below] at (0.6,1.07)  {$\Sigma_R$};
  \node[scale=.75] at (0,0.45) {$\e^{-S_E}$};
  \node[scale=.75] at (0,2.0)  {$\e^{\mathrm{i}S}$};
  \node[scale=.8] at (0,-1.05) {$|\mathrm{TFD}\rangle$};
\end{scope}
\end{tikzpicture}
\caption{From left to right: Penrose diagram of global AdS$_2$ with two Rindler wedges. Continuing the Rindler wedge to Euclidean signature we obtain the Poincare disk. Cutting the disk along the equator prepares the $|\mathrm{TFD}\rangle$.}
\label{fig: state preparation AdS}
\end{figure}

In figures \ref{fig: state preparation dS} and \ref{fig: state preparation AdS} we show graphically the comparison between the preparation of the HH state in dS and the TFD state, which is dual to the eternal black hole. In Lorentzian signature (far left) the Penrose diagram of AdS is a strip with two Rindler wedges. The Penrose diagram of Lorentzian dS is a square (see section \ref{sec: geometry}) with the left and right static patch. Timelike boundaries in AdS are analogous to antipodal worldlines in de Sitter space. Continuing to Euclidean signature we obtain a sphere for de Sitter, and the Poincare disk for AdS$_2$. Cutting the sphere at the equator we obtain a hemisphere with boundary $\Sigma \cong S^{d}$, whereas for AdS$_2$ cutting the disk we obtain $\Sigma \cong \mathbb{R}$. The topology of these cutting surfaces distinguishes the state obtained after Lorentzian continuation. For the half-disk the analytic continuation yields the $|\mathrm{TFD}\rangle$. For the hemisphere with $S^d$ boundary Lorentzian continuation yields the Hartle-Hawking wavefunction $|\Psi_{\mathrm{HH}}\rangle$.

\subsection{Thermodynamics}
As already mentioned in section~\ref{sec:dSentropy}, for an AdS black hole the Euclidean path integral at inverse
temperature $\beta$ computes
\begin{equation}\label{eq: thermo Law 2}
    S_{\mathrm{BH}} -\beta E= \log \mathcal{Z} ~.
\end{equation}
Euclidean de Sitter space is a sphere: both the spatial slices and the whole space, so there is no asymptotic region in which to anchor an ADM
energy. Without boundaries there is no independently specified
$\beta$ either. The relation \eqref{eq: thermo Law 2} therefore reduces to
\begin{equation}
    S_{\mathrm{dS}} = \log \mathcal{Z}~.
\end{equation}
Here $\mathcal{Z}$ is the Euclidean gravitational path integral. The leading semiclassical saddle for
de Sitter is a path integral over metrics on the sphere, discussed in
section~\ref{sec:dSentropy}. In particular it exhibits the conformal mode
problem: the kinetic term of the conformal factor comes with the wrong
sign, so the Euclidean action is unbounded from below and the Gaussian
integral over this mode is unsuppressed. The problem is independent of
the sign of $\Lambda$; only its resolution differs. In AdS the conformal
mode is Wick rotated uniformly and the resulting phase is absorbed into
the normalization of the measure, whereas on the sphere a finite number
of modes must be rotated back, leaving behind an overall phase of the
gravitational path integral \cite{Polchinski:1988ua}.
\subsection{Von Neumann algebra}
Finally, as already discussed in section~\ref{sec:dSentropy}, the algebra of gravitationally dressed observables for an AdS black hole is of type II$_\infty$, while the analogous observer algebra in de Sitter was shown to be of type II$_1$ \cite{Chandrasekaran:2022cip}. A recent refinement that incorporates the anti-scrambling behaviour of an observer conjectures that the algebra is type I \cite{Cui:2026bcd}. 
\newline\newline
We summmarize some of the comparison discussed in this section in table \ref{tab:BH_vs_dS}.

\begin{table}[ht]
    \centering
    \small
    \setlength{\arrayrulewidth}{1.2pt}
    \renewcommand{\arraystretch}{1.6}
    \begin{tabular}{|c!{\vrule width .8pt}c!{\vrule width .8pt}c|}
        \hline
        Feature & AdS black hole & dS horizon \\
        \hhline{|=|=|=|}


        Euc. partition function
        & $\log \mathcal Z=S_{\rm BH}-\beta E$
        & $\log \mathcal Z=S_{\rm dS}$ \\
        \hhline{|=|=|=|}

        KMS condition
        & $\checkmark$
        & $\checkmark$ (QFT), subtle with dynamical observer \\
        \hhline{|=|=|=|}

        OTOCs
        & scrambling, $\lambda_L=\dfrac{2\pi}{\beta}$
        & anti-scrambling,~$\lambda_L$ =$\frac{2\pi}{\beta}$~or $\lambda_L$ = $\frac{4\pi}{\beta}$ \\
        \hhline{|=|=|=|}

        Temperature
        & $\beta$ fixed by boundary ensemble
        & $\beta_{\rm dS}=2\pi\ell$;
          dynamical observer integrates over $\beta$ \\
        \hhline{|=|=|=|}

        Observer algebra
        & type II$_\infty$
        & type II$_1$ (ignoring anti-scrambling)  \\
        \hline
    \end{tabular}
    \caption{Comparison between the AdS black-hole horizon and the
    de Sitter cosmological horizon.}
    \label{tab:BH_vs_dS}
\end{table}

\section{de Sitter in 2D}\label{sec:2D}
In this section we discuss progress on two-dimensional de Sitter space. Many of the conceptual questions which we still need sharp answers for (unitarity, Hilbert space, dS entropy, the conformal mode problem, number of microscopic degrees of freedom,...) are already there for the lower dimensional models. 
There are also direct links between two-dimensional and four-dimensional de Sitter space: The discrete series unitary representation exists in $\mathrm{SO}(1,2)$ and $\mathrm{SO}(1,4)$ (see section \ref{sec:QFT}), the Nariai limit of the Schwarzschild de Sitter black hole is dS$_2\times S^2$ (see section \ref{sec:dSBH}), the edge contribution of four-dimensional de Sitter characters lives in a co-dimension two spacetime (see section \ref{sec:dSentropy}).

\subsection{Liouville theory}

The partition function of the two-dimensional Einstein–Hilbert action coupled to a two-dimensional CFT on a surface $\Sigma_g$ of genus $g$ is 
\begin{equation}
   \mathcal{Z}^{(g)}_{\mathrm{grav}}[\Lambda,c_m] =\int\frac{[\mathcal{D}g]}{\mathrm{vol}(\mathcal{G}_{\mathrm{diff}})}\,\e^{\frac{\vartheta}{4\pi}\int_{\Sigma_g} \d^2 x\sqrt{g}R - \Lambda \int_{\Sigma_g} \d^2 x\sqrt{g}}\mathcal{Z}^{(g)}_{\mathrm{matter}}[g,c_m]~,
\end{equation}
where $\vartheta$ is a coupling and $\Lambda>0$ the cosmological constant. We choose the latter to be bigger than zero to suppress large area fluctuations. The term $\tfrac{1}{4\pi}\int_{\Sigma_g} \d^2 x\sqrt{g}R$ is topological in two dimensions and yields the Euler character $\chi_g = 2-2g$ of the surface $\Sigma_g$. In the absence of matter, the gravity theory has no local degrees of freedom. The matter theory, which for convenience we choose to be a two-dimensional conformal field theory adds fluctuating degrees of freedom.  At least in a small neighbourhood we can always gauge the 2D diffeomorphism group and fix the metric to Weyl gauge
\begin{equation}
    ds^2 = \e^{2\varphi(x)}\tilde{g}_{ij}\d x^i \d x^j~,
\end{equation}
where $\varphi$ is the Weyl factor.
We treat $g=0$ and $g=1$ separately from $g\geq 2$. On the sphere $g=0$ and the matter CFT partition function is fully determined by the conformal anomaly
\begin{equation}
    \mathcal{Z}_{\mathrm{CFT}}^{(0)}[g_{ij}] = \mathcal{N}\e^{-S_{\mathrm{anomaly}}[\varphi]}~,\quad S_{\mathrm{anomaly}}[\varphi] = -\frac{c_m}{48\pi} \int \d^2 x\sqrt{\tilde{g}}\left(2\tilde{g}^{ij}\partial_i\varphi\partial_j\varphi +2 \widetilde{R}\varphi\right)~.
\end{equation}
The effective action governing the two-dimensional gravity theory coupled to a conformal field theory is thus
\begin{equation}
    S_{\mathrm{eff}}[\varphi] = - \frac{c_m}{48\pi}\int \d^2 x\sqrt{\tilde{g}}\left(2\tilde{g}^{ij}\partial_i\varphi\partial_j\varphi +2 \widetilde{R}\varphi\right)  + \Lambda \int \d^2 x\sqrt{\tilde{g}}\e^{2\varphi}
\end{equation}
In the absence of a dimensionful Newton constant the matter central charge $|c_m|\rightarrow \infty$ governs the semiclassical limit. We further distinguish the two limits
\begin{itemize}
    \item case a) $c_m\rightarrow\infty$ leads to a timelike Liouville gravity \cite{Polchinski:1989fn, Bautista:2019jau, Anninos:2021ene, Muhlmann:2022duj}
    \item case b) $c_m\rightarrow -\infty$ leads to a spacelike Liouville  gravity \cite{Polyakov:1981rd, David:1988hj, Distler:1988jt,Knizhnik:1988ak, Seiberg:1990eb}
\end{itemize}
Explicitly, after we impose $c_m+ c_L-26=0$, we obtain on a sphere with radius $r$
\begin{equation}
    \mathcal{Z}_{\mathrm{grav}}^{(0)}[\Lambda] = \e^{2\vartheta} \times \frac{\mathcal{A}}{\mathrm{vol}_{\mathrm{PSL}(2,\mathbb{C})}}\times \left({r}{\Lambda_{\mathrm{uv}}^{1/2}}\right)^{(c_m-26)/3}\times \int [\mathcal{D}\varphi]\e^{-S_{\mathrm{L}}[\varphi]}
\end{equation}
where $\Lambda_{\mathrm{uv}}$ is UV cutoff scaling as inverse length squared and $\mathcal{A}$ is a constant. The central charge $c_{\mathrm{gh}}=-26$ is the central charge of the $\mathfrak{b}\mathfrak{c}$-ghost system introduced when we fix conformal gauge.  Furthermore we distinguish
\begin{subequations}
\begin{align}\label{eq:liouville_tl}
   \mathrm{case}~a)\quad \quad  S_{\mathrm{tL}}[\chi] &= \frac{1}{4\pi}\int \d^2 x\sqrt{\tilde{g}}\left(-\tilde{g}^{ij}\partial_i \chi\partial_j\chi - \widehat{Q}\widetilde{R}\chi + 4\pi \Lambda \e^{2\hat{b}\chi}\right)~,\\ \label{eq:liouville_sp}
    \mathrm{case}~b)\quad \quad S_{\mathrm{sL}}[\varphi] &= \frac{1}{4\pi}\int \d^2 x\sqrt{\tilde{g}}\left(\tilde{g}^{ij}\partial_i \varphi\partial_j\varphi + {Q}\widetilde{R}\varphi + 4\pi \Lambda \e^{2{b}\varphi}\right)~.
\end{align}
\end{subequations}
 Although both actions are in Euclidean signature the timelike Liouville action has a negative sign kinetic term, whereas in the spacelike case it is positive. 
Both spacelike and timelike Liouville theory are two-dimensional conformal field theories.
\paragraph{Spacelike Liouville CFT.}
Spacelike Liouville theory is a unitary CFT and some of its properties are
\begin{itemize}
    \item $Q= b^{-1}+b$, with $b\in (0,1]$ and ${c} = 1+6{Q}^2$
    \item The CFT spectrum of Liouville theory consists of scalar primaries $V_p= \e^{2\alpha\varphi}$, $\alpha = \frac{Q}{2} -p$ with conformal dimension $h_p =\tilde{h}_p =\alpha(Q-\alpha) = \frac{Q^2}{4}-p^2$ and $p\in i \mathbb{R}$. We choose $p\in i \mathbb{R}$ such that the conformal dimensions are bounded from below by $h_p \geq \tfrac{c-1}{24}$. Moreover because it is a unitary CFT we generally take $p \in i\mathbb{R}_{\geq 0}$.
    \item On top of the CFT spectrum with $p \in i \mathbb{R}$, the Virasoro algebra admits degenerate fields which in Liouville variables are given by $p_{\langle r,s\rangle} = \tfrac{r}{2b}+\frac{sb}{2}$, $r,s \in \mathbb{Z}_{\geq 1}$. The degenerate vertex operators are given by $V_{p_{\langle r,s\rangle}}$. Degenerate fields play a crucial role in the Liouville bootstrap approach to the structure constants.
    \item Spacelike Liouville is a unitary, non compact (it has a continuous spectrum) solution to the CFT bootstrap equations, whose structure constants are known and denoted as the DOZZ coefficients \cite{Dorn:1994xn, Zamolodchikov:1995aa, Teschner:2001rv} (see appendix \ref{app:compendium} for details on $\Gamma_b$)
    \begin{align}\label{eq:spDOZZ}
        \langle V_{p_1}(1)V_{p_2}(0)V_{p_3}(\infty) \rangle &= C^{(b)}_{\mathrm{DOZZ}}(p_1,p_2,p_3)\cr
        &= \frac{\Gamma_b(2Q)}{\sqrt{2}\Gamma_b(Q)^3} \frac{\Gamma_b(\frac{Q}{2}\pm p_1 \pm p_2 \pm p_3)}{\prod_{j=1}^3\Gamma_b(Q\pm 2p_j)}~,
    \end{align}~
    where the $\pm$ signs indicate a product over all combinations. For example the numerator is a product of eight Barnes Gamma functions. 
    \item The DOZZ formula has poles and zeros. Since $\Gamma_b(x)$ has poles for $x= -m b - n b^{-1}$, $m,n \in \mathbb{Z}_{\geq 0}$ the zeros of $C_{\mathrm{DOZZ}}$ are located at
    \begin{equation}
        \mathrm{simple~zeros:}\quad p_j = \pm\frac{rb^{-1} +s b}{2}~,\quad  r,s \in \mathbb{Z}_{\geq 1}~.
    \end{equation}
    Poles on the other hand are located at
    \begin{equation}
         \pm p_1 \pm p_2 \pm p_3 = - (m+\frac{1}{2}) b - (n+\frac{1}{2})b^{-1}~,\quad m,n \in \mathbb{Z}_{\geq 0}~.
    \end{equation}
    \item We can extract the two-point function of spacelike Liouville theory from the three-point function by sending one of the vertex operators in (\ref{eq:spDOZZ}) to the identity
    \begin{equation}\label{eq:spDOZZ 2pt}
        \lim_{p_3 \rightarrow -\tfrac{Q}{2}} C^{(b)}_{\mathrm{DOZZ}}(p_1,p_2,p_3) = \frac{1}{\rho_b(p_1)} \left(\delta(p_1- p_2) + \delta(p_1+p_2)\right)~,
    \end{equation}
    where $\rho_b(p)$ is the modular crossing kernel for the torus vacuum character, which is asymptotic to Cardy's formula for the universal density of high-energy states in a unitary compact 2d CFT.
        \item Viewed as a theory of gravity coupled to matter, consistency requires $c_m+ c - 26=0$ which implies
    \begin{equation}
        Q= \sqrt{\frac{25-c_m}{6}}~,\quad b= \frac{\sqrt{25-c_m} -\sqrt{1-c_m}}{2\sqrt{6}}~.
    \end{equation}
\end{itemize}
\paragraph{Timelike Liouville CFT.}
Timelike Liouville theory on the other hand is a non-unitary two-dimensional CFT and some of its properties are
\begin{itemize}
\item $\widehat{Q}= \hat{b}^{-1}-\hat{b}$ with $\hat{b}\in (0,1]$ and $\hat{c}=1-6\widehat{Q}^2$
    \item The spectrum of timelike Liouville theory consists of scalar primaries $\widehat{V}_{\hat{p}} = \e^{2\hat{\alpha}\chi}$, $\hat{\alpha} = -\frac{\widehat{Q}}{2} -i  \hat{p}$ with conformal dimension $\hat{h}_{\hat{p}} = - \frac{\widehat{Q}^2}{4} - \hat{p}^2$. For $\hat p \in i\mathbb{R}$ the spectrum is bounded from below by $\hat{h}_{\hat p} \geq \tfrac{\hat{c}-1}{24}$.
     \item Timelike Liouville CFT is a non-unitary, non compact solution to the CFT bootstrap equations, whose structure constants are given by \cite{Zamolodchikov:2005fy, Kostov:2005av, Schomerus:2003vv, Ribault:2015sxa}
    \begin{equation}\label{eq: tDOZZ}
        \langle \widehat{V}_{\hat{p}_1}(1)\widehat{V}_{\hat{p}_2}(0)\widehat{V}_{\hat{p}_3}(\infty) \rangle = C^{(\hat{b})}_{\mathrm{tDOZZ}}(\hat{p}_1,\hat{p}_2,\hat{p}_3) = \frac{1}{C^{(\hat b)}_{\mathrm{DOZZ}}(i\hat p_1,i\hat p_2,i\hat p_3)}~.
    \end{equation}~
    \item We observe that
    \begin{equation}
       \lim_{\hat p_3 \rightarrow \frac{i\widehat Q}{2}} \langle \widehat V_{\hat p_1} \widehat V_{\hat p_2}\widehat V_{\hat p_3} \rangle  \neq \delta(\hat p_1 \pm \hat p_2) ~.
    \end{equation}
    In other words the ‘diagonal’ structure of the two-point functions is not recovered and the primary with $\hat h_{\hat p_3} =0$ is not the identity operator in the theory.
    \item In spacelike Liouville theory bootstrap quantities are invariant under $p \leftrightarrow -p$. It is then custom to choose $p \in i\mathbb{R}_{\geq 0}$ for the physical spectrum. However, this comes largely from the fact that it is a unitary CFT. Timelike Liouville theory on the other hand is a non-unitary CFT, and a priori there is no reason to restrict $\hat{p}$.
    \item Poles of the structure constants follow from the poles of $\Gamma_{\hat{b}}$ and lie on the lattice
    \begin{equation}\label{eq: poles tDOZZ}
      \mathrm{simple~poles}:\quad   \hat{p}_j = \frac{\pm i}{2}(r\hat b^{-1} + s\hat b)~.
    \end{equation}
   \item The physical momenta in timelike Liouville theory are $\hat p\in i \mathbb{R}$. As we have seen above (\ref{eq: poles tDOZZ}), the poles of the structure constant also lie on the imaginary axis. It has hence been conjectured \cite{Ribault:2015sxa} that the physical spectrum gets shifted $\hat p\in i \mathbb{R} - \epsilon$.
\end{itemize}

\begin{table}[ht]
    \centering
    \small
         \setlength{\arrayrulewidth}{1.2pt}
    \renewcommand{\arraystretch}{1.5}
    \begin{tabular}{|c!{\vrule width .8pt}c!{\vrule width .8pt}c|}
        \hline
        Features & spacelike Liouville & timelike Liouville  \\
        \hhline{|=|=|=|}
        Vertex operators 
        & $V_p=\e^{2\alpha\varphi}, \alpha = \frac{Q}{2}-p ~$ 
        & $\widehat{V}_{\hat{p}}=\e^{2\hat{\alpha}\chi}, \hat{\alpha} = -\frac{\widehat{Q}}{2}-i\hat{p} ~$ \\
        \hhline{|=|=|=|}
        Normalizable 
        & $p\in i \mathbb{R}_{\geq 0}$
        & $\hat{p}\in i\mathbb{R} -\epsilon$ \\
        \hhline{|=|=|=|}
                 \multirow{2}{*}{Conformal dimension}
        & $h_p=\tilde{h}_p = \alpha(Q-\alpha)$
        & $\hat{h}_{\hat{p}}=\hat{\tilde{h}}_{\hat{p}} = \hat{\alpha}(\widehat{Q}+\hat{\alpha})$ \\
        \hhline{|~|~|~|}
        & $= \frac{c-1}{24}-p^2$
        & $=\frac{\hat{c}-1}{24}- \hat{p}^2$ \\
         \hhline{|=|=|=|}
        Reflection symmetry
        & $p\leftrightarrow -p$ 
        & $\mathrm{open}$ \\
       \hhline{|=|=|=|}
       Degenerate fields
        & $p_{\langle r,s\rangle} = \frac{r}{2b} + \frac{sb}{2},~ r,s \in \mathbb{Z}_{\geq 1}$ 
        & $\hat{p}_{\langle r,s\rangle} = i(\frac{r}{2\hat{b}} - \frac{s\hat{b}}{2}),~ r,s \in \mathbb{Z}_{\geq 1}$  \\
      \hhline{|=|=|=|}
        Area operator
        & $V_{p_{\langle 1,-1\rangle}} = \e^{2b\varphi}$ 
        & $\widehat{V}_{\hat{p}_{\langle 1,-1 \rangle}} = \e^{2\hat{b}\chi} $ \\
       \hline
    \end{tabular}
    \caption{Spacelike vs timelike Liouville theory.}
    \label{tab:Liouville}
\end{table}
We summarize various properties in table \ref{tab:Liouville}.
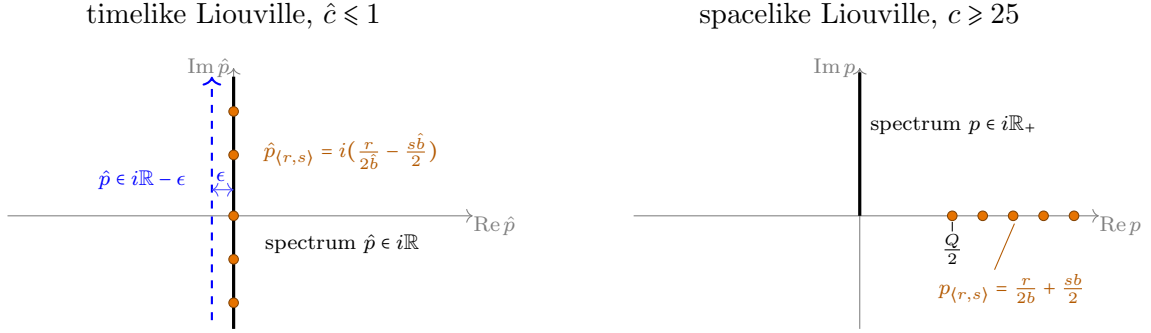
\begin{figure}[ht]
    \centering
\begin{tikzpicture}[scale=1.15,
    dot/.style={orange!90!black, draw=orange!50!black, line width=0.3pt},
    every node/.style={inner sep=1.5pt}]

  \begin{scope}
    \node at (0,2.3) {timelike Liouville, $\hat c \leq 1$};
    \draw[->,gray] (-2.6,0) -- (2.75,0) node[below right=-2pt] {\scriptsize $\mathrm{Re}\,\hat p$};
    \draw[->,gray] (0,-1.3) -- (0,1.7) node[left] {\scriptsize $\mathrm{Im}\,\hat p$};
    \draw[very thick] (0,-1.3) -- (0,1.6);
    \node[below] at (1.25,-0.16) {\scriptsize spectrum $\hat p\in i\mathbb{R}$};
    \draw[blue, dashed, thick, ->] (-.25,-1.2) -- (-.25,1.6);
    \node[above, blue] at (-1.05,0.26) {\scriptsize $\hat p\in i\mathbb{R}-\epsilon$};
     \node[above, blue] at (-.15,0.35) {\scriptsize $\epsilon$};
    \draw[<->, blue!70] (-.25,0.3) -- (0,0.3);
    \foreach \px in {-1,-.5,0,.7,1.2}
      \fill[dot] (0,\px) circle (0.055);
    \node[below, orange!70!black] at (1.35,1)
      {\scriptsize $\hat p_{\langle r,s\rangle}=i(\frac{r}{2\hat b}-\frac{s\hat b}{2})$};
  \end{scope}

  \begin{scope}[xshift=7.2cm]
    \node at (0,2.3) {spacelike Liouville, $c \geq 25$};
    \draw[->,gray] (-2.6,0) -- (2.75,0) node[below right=-2pt] {\scriptsize $\mathrm{Re}\,p$};
    \draw[->,gray] (0,-1.3) -- (0,1.7) node[left] {\scriptsize $\mathrm{Im}\,p$};
    \draw[very thick] (0,0) -- (0,1.65);
    \node[right] at (0.07,1.05) {\scriptsize spectrum $p\in i\mathbb{R}_+$};
    \foreach \px in {1.064,1.414,1.764,2.114,2.464}
      \fill[dot] (\px,0) circle (0.055);
    \draw (1.064,-0.07) -- (1.064,-0.18);
    \node[below] at (1.064,-0.16) {\scriptsize $\tfrac{Q}{2}$};
    \draw[orange!70!black] (1.779,-0.09) -- (1.55,-0.62);
    \node[below, orange!70!black] at (1.75,-0.62)
      {\scriptsize $p_{\langle r,s\rangle}=\frac{r}{2b}+\frac{sb}{2}$};
  \end{scope}
\end{tikzpicture}

\caption{Timelike (left) vs spacelike (right) regime of the momentum $\hat p$ and $p$ respectively. }
\label{fig:contour}
\end{figure}

Above we have seen various properties of spacelike and timelike Liouville CFT. In the spacelike regime we can reproduce the CFT data at least in a semiclassical regime. By contrast in the timelike regime the path integral predictions do not necessarily match the $\hat b\rightarrow 0$ limit of the CFT data. We will show some of these subtleties in the next subsection. In this spirit it is not clear whether timelike Liouville CFT and timelike Liouville gravity are the same theory. 
\subsection{Timelike Liouville gravity}
Timelike Liouville theory provides a model of two-dimensional de Sitter quantum gravity with the following properties: 
\begin{enumerate}
\item It exhibits a finite sphere partition function \cite{Anninos:2021ene}.
\item It contains propagating degrees of freedom.
\item It incorporates the conformal mode problem \cite{Polchinski:1988ua,Anninos:2021ene}.
\item It admits a semiclassical limit with small but nonvanishing quantum fluctuations \cite{Anninos:2021ene,Muhlmann:2022duj}.
\item It admits a ``higher spin" extension \cite{upcomingToda}.
\item It allows for a supersymmetric extension \cite{Anninos:2023exn}.
\item Its Wheeler--DeWitt equation admits solutions associated to cosmological singularities \cite{Martinec:2003ka,CarneirodaCunha:2003mxy,Martinec:2014uva,Anninos:2024iwf, Anninos:2025fer}.
\end{enumerate}
In the following we elaborate on each of these points, and refer for more details to the corresponding references. We start with 1.$-$4. and consider a two-dimensional CFT matter theory with central charge $c_m$ coupled to gravity. In Weyl gauge the metric takes the form 
\begin{equation}\label{eq:Weyl gauge}
    ds^2= \e^{2\hat b \chi} d\tilde{s}^2~. 
\end{equation}
We see that this gauge leaves a final redundancy unfixed. The rescaling $\tilde{g}_{\mu\nu}\rightarrow \e^{2\sigma} \tilde{g}_{\mu\nu}$ can be compensated by $\chi \rightarrow \chi -\sigma/\hat b$. Consequently the theory should be invariant under Weyl rescaling and we obtain the additional constraint
\begin{equation}\label{eq:anomaly cancellation}
    c_m+ c_{\mathrm{tL}} + c_{\mathrm{gh}}=0~,
\end{equation}
where $c_{\mathrm{gh}}=-26$ is the central charge of a $\mathfrak{b}\mathfrak{c}$-ghost system, that appears because we fixed conformal gauge (\ref{eq:Weyl gauge}). Taking the matter theory to be a 2D CFT is a convenient choice. In the limit where $c_m \rightarrow \infty$ the theory contains a large number of propagating degrees of freedom ($2. \checkmark$) and the sphere partition function takes the form 
\begin{equation}
    \mathcal{Z}_{\mathrm{grav}}^{(0)}[\Lambda] = \e^{2\vartheta} \times \frac{\mathcal{A}}{\mathrm{vol}_{\mathrm{PSL}(2,\mathbb{C})}}\times (r\Lambda^{\frac{1}{2}}_{\mathrm{uv}})^{(c_m-26)/3}\times \int [\mathcal{D}\chi]\e^{-S_{\mathrm{tL}}[\chi]}~,
\end{equation}
where the timelike Liouville action is given in (\ref{eq:liouville_tl}). We can further rescale $\chi \rightarrow \frac{1}{\hat b}\chi$, $\Lambda \rightarrow \frac{1}{\hat b^2}\Lambda$. The timelike action then takes the form
\begin{equation}
    S_{\mathrm{tL}}[\chi]=\frac{1}{4\pi \hat b^2}\int_{S^2} \d^2 x\sqrt{\tilde{g}}\left(-\partial_\mu \chi \partial^\mu \chi -\widetilde{R}(1-\hat b^2)\chi + 4\pi \Lambda \e^{2\chi}\right)~.
\end{equation}
From this it is clear that $\hat b\rightarrow 0$ ($\leftrightarrow c_{\mathrm{tL}} \rightarrow -\infty$) is the semiclassical parameter of the theory (first part of $4. \checkmark$). 
The timelike Liouville theory arises because the $c_m\rightarrow \infty$ limit in (\ref{eq:anomaly cancellation}) requires that the Liouville central charge goes to negative infinity which is possible only on the timelike branch. 
On a sphere with radius $r$ the Ricci scalar is $\widetilde R = 2/r^2$ leading to the timelike equations of motion 
\begin{equation}
    2\widetilde{\nabla}^2 \chi -\frac{2}{r^2}(1-\hat b^2) + 8 \pi \Lambda \e^{2 \chi} =0 ~,
\end{equation}
whose solution is a real valued constant $\chi_*= \frac{1}{2}\log (\frac{1-\hat b^2}{4\pi \Lambda r^2})$\footnote{By contrast, the spacelike Liouville theory does not have a real saddle on a sphere topology.} and the on-shell action evaluates to
\begin{equation}
    \int [\mathcal{D}\chi]\e^{-S_{\mathrm{tL}}[\chi]} \approx \e^{-S_{\mathrm{on-shell}}} = \left(\frac{1-\hat b^2}{4\pi \Lambda r^2 \e}\right)^{\frac{1}{\hat b^2}-1} \approx (4\pi \Lambda r^2 \e)^{-\frac{1}{\hat b^2}}~.
\end{equation}
Expanding around the sphere saddle, $\chi = \chi_* + \delta\chi$, the potential generates the full tower of fluctuation vertices,
\begin{align}
    S_{\mathrm{tL}}[\chi_*+\delta\chi] &= S_{\mathrm{tL}}[\chi_*]
    + \frac{1}{4\pi \hat b^2}\int_{S^2}\d^2x\sqrt{\tilde g}\,
    \Big[-\delta\chi\big(-\widetilde\nabla^2 - \tfrac{2-2\hat b^2}{r^2}\big)\delta\chi
    + \frac{1-\hat b^2}{r^2}\Big(\e^{2\delta\chi}-1-2\delta\chi-2\delta\chi^2\Big)\Big] \cr
    &= S_{\mathrm{tL}}[\chi_*]
    + \frac{1}{4\pi \hat b^2}\int_{S^2}\d^2x\sqrt{\tilde g}\,
    \Big[-\delta\chi\big(-\widetilde\nabla^2 - \tfrac{2 - 2\hat b^2}{r^2}\big)\delta\chi
    + \frac{1-\hat b^2}{r^2}\Big(\frac{4}{3}\delta\chi^3 + \frac{2}{3}\delta\chi^4
    + \dots\Big)\Big]\,,
\end{align}
with the $n$-th vertex carrying coefficient $2^n/n!$. In terms of the
canonically normalized fluctuation $\delta\chi = \hat b\,\delta \chi_c$ the cubic,
quartic, \dots\ vertices scale as $\hat b,\ \hat b^2,\ \dots,\ \hat b^{\,n-2}$:
the interactions are suppressed by powers of the semiclassical parameter, so the
quantum fluctuations are small but non-vanishing (second part of $4.\checkmark$),
and it is these vertices that generate the higher-loop corrections to
$\mathcal{Z}^{(0)}_{\mathrm{grav}}$ computed in \cite{Anninos:2021ene, Muhlmann:2022duj}.
The quadratic fluctuation is the 2D version of the conformal mode problem ($3. \checkmark$). Expanding $\delta \chi$ in spherical harmonics we can easily see that all the modes with $l\geq 2$ are Gaussian unsuppressed, the $l=1$ modes are zero modes (lifted at order $\mathcal{O}(\hat b^2)$ which is a feature of the semiclassical path integral), which correspond precisely to the $\mathrm{PSL}(2,\mathbb{C})$ redundancy we divide by in $\mathcal{Z}_{\mathrm{grav}}^{(0)}$, while the $l=0$ mode is Gaussian suppressed. The one-loop problem we already discussed in example 1 in subsection (\ref{subsec:Technical details}), in particular we see that the quadratic fluctuation $\delta \chi$ transforms in the discrete series irreducible representation of $\mathrm{SO}(1,2)$ \cite{Anninos:2021ene}. Higher loops are tractable. Putting everything together we obtain
\begin{equation}\label{eq: Ztl PI}
    Z_{\mathrm{tL}}[\Lambda] \simeq (4\pi \Lambda\e )^{-\frac{1}{\hat b^2}+1} \Lambda_{\mathrm{uv}}^{\frac{7}{6} -\hat b^2} r^{\frac{\hat c}{3}}\left(\frac{1}{\hat b} + \mathcal{O}(\hat b)\right)~,
\end{equation}
where $\Lambda_{\mathrm{uv}}$ is the UV cutoff we introduce in the heat-kernel analysis of the one-loop determinant. Moreover combining the saddle and the one-loop contribution we obtain the correct conformal anomaly of a CFT with central charge $\hat c$ on a sphere. 
This gives our last checkmark $1\checkmark$. Finally the WdW equation for timelike Liouville theory can be found in \cite{Anninos:2024iwf}.
\paragraph{Comparison to the CFT.}
It is interesting to compare this to the structure constants of timelike Liouville theory. For example we can evaluate the three-point function (\ref{eq: tDOZZ}) for  $\hat{p}_{\langle 1,-1 \rangle}$, which implies $\hat \alpha = \hat b$ (see table \ref{tab:Liouville}). This leads to 
\begin{equation}\label{eq: Ztl CFT}
   C_{\mathrm{tDOZZ}}^{(\hat b)}(\hat p_{\langle 1,-1\rangle}, \hat p_{\langle 1,-1\rangle}, \hat p_{\langle 1,-1\rangle}) =\langle \e^{2\hat b \chi}(0)\e^{2\hat b \chi}(1)\e^{2\hat b \chi}(\infty)\rangle = 0~.
\end{equation}
Since the correlator $\langle \e^{2\hat b \chi}(0)\e^{2\hat b \chi}(1)\e^{2\hat b \chi}(\infty)\rangle  \propto \frac{\d^3}{\d\Lambda^3}Z_{\mathrm{tL}}[\Lambda]$, the two results (\ref{eq: Ztl CFT}) and (\ref{eq: Ztl PI}) are potentially incompatible. 

We will propose a solution: In general we cannot analytically continue structure constants from spacelike to timelike Liouville theory, however in the special instance considered here where $p_i = p_{\langle 1,-1\rangle} $ the Barnes Gamma functions $\Gamma_b$ in the DOZZ formula become ordinary $\Gamma$-functions, whose $b\rightarrow i \hat b$ continuation is well defined. 

To compare path integral and CFT we also note that  (\ref{eq:spDOZZ}) exhibits a normalization convenient for CFT purposes, for example it is invariant under $p\leftrightarrow - p$ and the two-point function (\ref{eq:spDOZZ 2pt}) normalization is the Plancherel density $\rho_b(p)$. When comparing with the path integral we need to restore the normalization \cite{Collier:2019weq}
\begin{equation}
    C_b(p_1,p_2,p_3) = \left(\frac{(\pi\Lambda\gamma(b^2)b^{2-2b^2})^{\frac{Q}{2b}}}{2^{\frac{3}{4}}\pi}\frac{\Gamma_b(2Q)}{\Gamma_b(Q)}\right)^{-1} {}{\prod_{j=1}^3(S^{(b)}(p_j)\rho_b(p_j))^{\frac{1}{2}}} C^{(b)}_{\text{DOZZ}}(p_1,p_2,p_3)\, ,
\end{equation}
where 
\begin{equation}\label{eq: Sbp}
    S^{(b)}(p) \equiv -\left(\pi \Lambda\gamma(b^2)\right)^{\frac{2p}{b}}\frac{\Gamma(1-2b^{-1} p)\Gamma(1-2bp)}{\Gamma(1+2b^{-1} p)\Gamma(1+2bp)}~,
\end{equation}
is the so called reflection coefficient and 
\begin{equation}
    \rho_b(p) = -4\sqrt{2}\sin(2\pi b p)\sin(2\pi b^{-1}p) ~. 
\end{equation}
We can now declare that
\begin{equation}\label{eq: analytic continuatio}
   C_b(p_{\langle 1,-1\rangle}, p_{\langle 1,-1\rangle}, p_{\langle 1,-1\rangle})\big|_{b \rightarrow i \hat b} = \frac{\d^3}{\d\Lambda^3}Z_{\mathrm{tL}}[\Lambda]~.  
\end{equation}
If this looks suspicious we can instead consider the spacelike two-point function in the path integral normalization. It is given by
\begin{equation}
    \lim_{p_3 \rightarrow -\frac{Q}{2}} C_{b}(p_1,p_2,p_3) = 2\pi \Big[\delta(p_1+p_2) + S^{(b)}(p_1)\delta(p_1-p_2)\Big]~.
\end{equation}
The right hand side contains nothing that cannot be analytically continued, and indeed in \cite{Harlow:2011ny} it was conjectured that we can analytically continue this formula to timelike. Instead of (\ref{eq: analytic continuatio}) we can then compare $ C_b(p_{\langle 1,-1\rangle},p_{\langle 1,-1\rangle}, -\frac{Q}{2})\big|_{b\rightarrow i \hat b}$ and $\frac{\d^2}{\d\Lambda^2}Z_{\mathrm{tL}}[\Lambda]$. 

\begin{figure}[ht]
\begin{tikzpicture}
    \begin{scope}
        \draw[thick] (0,0) -- (7,0);
        \draw[thick, white] (0,0) --(0,1);
        \draw[thick, white] (3.5,0) -- (3.5,1);
        \filldraw[fill=violet!15, draw=white, thick] (0,0) rectangle (3.5,.95);
        \filldraw[fill=violet!10, draw=white, thick] (0,-1) rectangle (3.5,0);
          \filldraw[fill=green!15, draw=white, thick] (3.5,0) rectangle (7,.95);
         \filldraw[fill=green!10, draw=white, thick] (3.5,-1) rectangle (7,0);
          \draw[thick] (0,-1) -- (0,1.5);
          \draw[thick] (-1,-1) -- (7,-1);
        \draw[thick] (-1,0)--(3.5,0);
        \draw[thick] (-1,1) -- (7,1);
        \draw[thick] (-1,.95) -- (7,.95);
        \draw[thick] (3.5, 1.5) -- (3.5,-1);
        \draw[thick] (3.5,0)--(7,0);
        
        \draw[thick] (8.5,0) -- (14,0);
        \filldraw[fill=violet!20, draw=white, thick] (12.5,-.09) rectangle (14.,.09);
        \filldraw[fill=green!20, draw=white, thick] (8.5,-.09) rectangle (10.8,.09);
        \draw[thick] (10.8,-.2) --(10.8,.2);
        \draw[thick] (12.5,-.2) --(12.5,.2);
         \draw[thick,->] (10.5,-1)--(10.5,1.8);
         \draw[] (13.5,1.4) -- (13.5,1.7);
         \draw[] (13.5,1.4) -- (13.8,1.4);
         \draw[->] (8.4,0)--(14.1,0);
 \node[scale=.7] at (10.9,-.35) {$c=\frac{3}{2}$}; 
  \node[scale=.7] at (12.6,-.35) {$c=\frac{27}{2}$}; 
 \node[scale=.8] at (13.65,1.55) {$c$}; 
 \node[scale=.9] at (-.55,.5) {$\mathrm{CFT}$}; 
  \node[scale=1] at (1.75,.5) {\cite{Rashkov:1996np, Poghossian:1996agj, Fukuda:2002bv,Belavin:2007gz}}; 
  \node[scale=1] at (1.75,-.5) {\cite{Polyakov:1981re, Distler:1989nt}}; 
   \node[scale=1] at (5.25,-.5) {\cite{Anninos:2023exn}};
  \node[scale=1] at (5.25,.5) {\cite{Muhlmann:2025ngz, Rangamani:2025wfa}}; 
  \node[scale=.9] at (-.65,-.5) {$\mathrm{Gravity}$}; 
 \node[scale=.9] at (-.55,1.25) {$\mathcal{N}=1$}; 
 \node[scale=.9] at (1.75,1.25) {$\mathrm{spacelike~Liouville}$};
  \node[scale=.9] at (5.25,1.25) {$\mathrm{timelike~Liouville}$};
    \end{scope}
\end{tikzpicture}
\caption{We show the main references for $\mathcal{N}=1$ Liouville theory. We distinguish spacelike and timelike Liouville, as well as the CFT and the gravity aspect of the theory. From the CFT perspective we restrict to bulk correlators. }
\label{fig:N=1}
\end{figure}
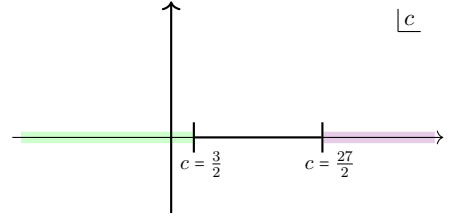

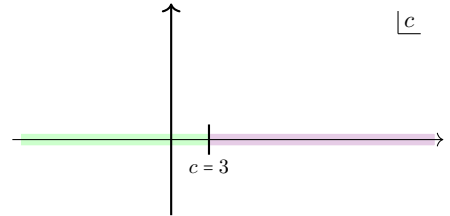
\begin{figure}[ht]
\begin{tikzpicture}
    \begin{scope}
        \draw[thick] (0,0) -- (7,0);
        \draw[thick, white] (0,0) --(0,1);
        \draw[thick, white] (3.5,0) -- (3.5,1);
        \filldraw[fill=violet!15, draw=white, thick] (0,0) rectangle (3.5,.95);
        \filldraw[fill=violet!10, draw=white, thick] (0,-1) rectangle (3.5,0);
          \filldraw[fill=green!15, draw=white, thick] (3.5,0) rectangle (7,.95);
         \filldraw[fill=green!10, draw=white, thick] (3.5,-1) rectangle (7,0);
          \draw[thick] (0,-1) -- (0,1.5);
          \draw[thick] (-1,-1) -- (7,-1);
        \draw[thick] (-1,0)--(3.5,0);
        \draw[thick] (-1,1) -- (7,1);
        \draw[thick] (-1,.95) -- (7,.95);
        \draw[thick] (3.5, 1.5) -- (3.5,-1);
        \draw[thick] (3.5,0)--(7,0);
        
        \draw[thick] (8.5,0) -- (14,0);
        \filldraw[fill=violet!20, draw=white, thick] (11.,-.09) rectangle (14.,.09);
        \filldraw[fill=green!20, draw=white, thick] (8.5,-.09) rectangle (11.,.09);
        \draw[thick] (11.,-.2) --(11.,.2);
         \draw[thick,->] (10.5,-1)--(10.5,1.8);
         \draw[] (13.5,1.4) -- (13.5,1.7);
         \draw[] (13.5,1.4) -- (13.8,1.4);
         \draw[->] (8.4,0)--(14.1,0);
 \node[scale=.7] at (11,-.35) {$c=3$}; 
 \node[scale=.8] at (13.65,1.55) {$c$}; 
 \node[scale=.9] at (-.55,.5) {$\mathrm{CFT}$}; 
  \node[scale=1] at (1.75,.5) {\cite{Hosomichi:2004ph, Hori:2001ax, Muhlmann:2026pdc}}; 
  \node[scale=1] at (1.75,-.5) {\cite{Antoniadis:1990mx}}; 
   \node[scale=1] at (5.25,-.5) {\cite{Anninos:2023exn}};
  \node[scale=1] at (5.25,.5) {$\mathrm{?}$}; 
  \node[scale=.9] at (-.65,-.5) {$\mathrm{Gravity}$}; 
 \node[scale=.9] at (-.55,1.25) {$\mathcal{N}=2$}; 
 \node[scale=.9] at (1.75,1.25) {$\mathrm{spacelike~Liouville}$};
  \node[scale=.9] at (5.25,1.25) {$\mathrm{timelike~Liouville}$};
    \end{scope}
\end{tikzpicture}
\caption{We show the main references for $\mathcal{N}=2$ Liouville theory. We distinguish spacelike and timelike Liouville, as well as the CFT and the gravity aspect of the theory. From the CFT perspective we restrict to bulk correlators. }
\label{fig:N=2}
\end{figure}

Liouville theory admits two natural extensions:
\begin{enumerate}
    \item \emph{Supersymmetry:} We summarize the main references and show the central charge plane in figures \ref{fig:N=1} and \ref{fig:N=2}
    \begin{itemize}
        \item  The $\mathcal{N}=1$\footnote{Throughout this work, $\mathcal{N}=k$ denotes $\mathcal{N}=(k,k)$ supersymmetry.} supersymmetric versions of both spacelike and
        timelike Liouville theory have been studied in the literature (see figures \ref{fig:N=0}, \ref{fig:N=1} and \ref{fig:N=2} for some of the main references).
        In contrast to the bosonic theory, with its single independent structure
        constant, the $\mathcal{N}=1$ theory possesses two independent structure
        constants in the NS sector and two in the R sector. From a path integral
        perspective, $\mathcal{N}=1$ Liouville theory consists of a real scalar
        $\varphi$, a Majorana fermion $\psi$, and an auxiliary field $F$. In the
        timelike theory, the two propagating fields describe particles
        transforming in discrete series irreducible representations of the
        universal cover of the two-dimensional de Sitter isometry group \cite{Anninos:2023exn}.
        \item By contrast, $\mathcal{N}=2$ Liouville theory is under far less
        theoretical control: \cite{Hosomichi:2004ph, Hori:2001ax, Muhlmann:2026pdc} has only scratched the
        surface.
    \end{itemize}
    \item \emph{Higher-spin structure:} Extending $\mathfrak{sl}(2)$ to
    $\mathfrak{sl}(n)$ yields Toda theory \cite{Zamolodchikov:1985wn, Fateev:2007ab}.  
\end{enumerate}
\paragraph{Timelike Toda.} Toda theory generalizes Liouville theory from $\mathfrak{sl}(2)$ to $\mathfrak{sl}(n)$, and coupling it to matter and ghosts provides a two-dimensional
analogue of a higher spin theory \cite{upcomingToda}:
\begin{equation}\label{eq:Todagrav}
     \mathcal{Z}_{\mathrm{h.s.}}[S^2] = \int \frac{[\mathcal{D}\boldsymbol{\chi}]}{\mathrm{vol}_{\mathrm{PSL}(n,\mathbb{C})}}
    \e^{-S^{(n)}_{\mathrm{tToda}}[\boldsymbol{\chi}]}\,
    \mathcal{Z}_{\mathrm{matter}}[\widetilde{\mathcal{B}}]\,
    \mathcal{Z}_{\mathrm{ghost}}[\widetilde{\mathcal{B}}]~,
\end{equation}
where
\begin{equation}\label{eq:tToda}
    S^{(n)}_{\mathrm{tToda}}[\boldsymbol{\chi}]= \frac{1}{4\pi}\int_{S^2} \d^2 x \sqrt{\tilde{g}}\left(-\frac{1}{2}\tilde{g}^{\mu\nu}(\partial_\mu \boldsymbol{\chi},\partial_\nu\boldsymbol{\chi}) -(\widehat{\boldsymbol{Q}},\boldsymbol{\chi})\widetilde{R}+ 4\pi \Lambda \sum_{k=1}^{n-1} \e^{\hat b(e_k,\boldsymbol{\chi})}\right)~.
\end{equation}
In (\ref{eq:Todagrav})
$\widetilde{\mathcal{B}}$ denotes the gauge-fixed higher spin background, and the action
(\ref{eq:tToda}) refers to the principal embedding of $\mathfrak{sl}(2)$ into
$\mathfrak{sl}(n)$, under which the adjoint representation decomposes into fields with conformal dimensions
$\Delta=2,\ldots,n$, which are exactly the values of the two-dimensional discrete series irreducible representation. The Toda field $\boldsymbol{\chi}=\sum_{i=1}^{n-1}\chi_i e_i$ takes
values in the root space of $\mathfrak{sl}(n)$, the $e_k$ are the simple roots with
$(e_j,e_k)$ the Cartan matrix, and $\widehat{\boldsymbol{Q}}=\widehat Q \,\boldsymbol{\rho}$ with
$\widehat Q=\hat b^{-1}-\hat b$ and $\boldsymbol{\rho}=\sum_{k=1}^{n-1}\frac{k(n-k)}{2}\,e_k$ the
Weyl vector, $(\boldsymbol{\rho},\boldsymbol{\rho})=n(n^2-1)/12$. For $n=2$,
(\ref{eq:tToda}) is the timelike Liouville action (\ref{eq:liouville_tl}); This theory is further explored in \cite{upcomingToda}.

\begin{figure}[ht]
\begin{tikzpicture}
    \begin{scope}
        \draw[thick] (0,0) -- (7,0);
        \draw[thick, white] (0,0) --(0,1);
        \draw[thick, white] (3.5,0) -- (3.5,1);
        \filldraw[fill=violet!15, draw=white, thick] (0,0) rectangle (3.5,.95);
        \filldraw[fill=violet!10, draw=white, thick] (0,-1) rectangle (3.5,0);
          \filldraw[fill=green!15, draw=white, thick] (3.5,0) rectangle (7,.95);
         \filldraw[fill=green!10, draw=white, thick] (3.5,-1) rectangle (7,0);
          \draw[thick] (0,-1) -- (0,1.5);
          \draw[thick] (-1,-1) -- (7,-1);
        \draw[thick] (-1,0)--(3.5,0);
        \draw[thick] (-1,1) -- (7,1);
        \draw[thick] (-1,.95) -- (7,.95);
        \draw[thick] (3.5, 1.5) -- (3.5,-1);
        \draw[thick] (3.5,0)--(7,0);
        
        \draw[thick] (8.5,0) -- (14,0);
        \filldraw[fill=violet!20, draw=white, thick] (12.5,-.09) rectangle (14.,.09);
        \filldraw[fill=green!20, draw=white, thick] (8.5,-.09) rectangle (10.8,.09);
        \draw[thick] (10.8,-.2) --(10.8,.2);
        \draw[thick] (12.5,-.2) --(12.5,.2);
         \draw[thick,->] (10.5,-1)--(10.5,1.8);
         \draw[] (13.5,1.4) -- (13.5,1.7);
         \draw[] (13.5,1.4) -- (13.8,1.4);
         \draw[->] (8.4,0)--(14.1,0);
 \node[scale=.7] at (10.9,-.35) {$c=1$}; 
  \node[scale=.7] at (12.6,-.35) {$c=25$}; 
 \node[scale=.8] at (13.65,1.55) {$c$}; 
 \node[scale=.9] at (-.55,.5) {$\mathrm{CFT}$}; 
  \node[scale=1] at (1.75,.5) {\cite{Dorn:1994xn, Zamolodchikov:1995aa, Teschner:2001rv}}; 
  \node[scale=1] at (1.75,-.5) {\cite{Polyakov:1981rd,Knizhnik:1988ak, David:1988hj, Distler:1988jt, Seiberg:1990eb}}; 
   \node[scale=1] at (5.25,-.5) {\cite{Bautista:2019jau, Anninos:2021ene, Muhlmann:2022duj}};
  \node[scale=1] at (5.25,.5) {\cite{Zamolodchikov:2005fy, Kostov:2005av,  Ribault:2015sxa}}; 
  \node[scale=.9] at (-.65,-.5) {$\mathrm{Gravity}$}; 
 \node[scale=.9] at (-.55,1.25) {$\mathcal{N}=0$}; 
 \node[scale=.9] at (1.75,1.25) {$\mathrm{spacelike~Liouville}$};
  \node[scale=.9] at (5.25,1.25) {$\mathrm{timelike~Liouville}$};
    \end{scope}
\end{tikzpicture}
\caption{We show the main references for bosonic Liouville theory. We distinguish spacelike and timelike Liouville, as well as the CFT and the gravity aspect of the theory. From the CFT perspective we restrict to bulk correlators. }
\label{fig:N=0}
\end{figure}
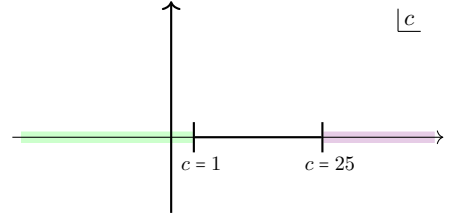

\noindent

\subsection{Dilaton gravity models}
Apart from timelike Liouville theory, dilaton gravity models have also made an important appearance in low-dimensional de Sitter quantum gravity. However, Dilaton gravity models have no local degrees of freedom — all the physics sits in zero modes, boundaries and topology — whereas timelike Liouville coupled to a $c_m \rightarrow \infty$ CFT propagates.
Dilaton gravity models take the form
\begin{multline}
S_\Sigma[g,\Phi]= -\frac{1}{2} \int_\Sigma \d^2 x \, \sqrt{g}\, \big(\Phi R+W(\Phi)\big)-\int_{\partial \Sigma} \d x \, \sqrt{h}\, \Phi (K-1)\\
-\frac{S_0}{2\pi}\left(\frac{1}{2}\int_{\Sigma} \d^2 x \sqrt{g}\,R + \int_{\partial \Sigma} \d x\, \sqrt{h} K \right)~. \label{eq:dilaton gravity action}
\end{multline}
In the above $\Phi$ is the dilaton, $R$ the Ricci scalar and $K$ the extrinsic curvature. For general potential $W$ variation of $\Phi$ gives 
\begin{subequations}
    \begin{align}
        R &= - W'(\Phi)~,\cr
        \nabla_\mu \nabla_\nu \Phi -g_{\mu\nu} \nabla^2 \Phi + \frac{1}{2}g_{\mu\nu} W(\Phi)&=0~.
    \end{align}
\end{subequations}
Each of these models is solved by \cite{Witten:2020ert, Witten:2020wvy, Grumiller:2002nm}
\begin{eqnarray}
    ds^2 = f(r)\d \tau^2 + \frac{\d r^2}{f(r)}~,\quad \Phi = r~,\quad f(r) = \int_{r_h}^r \d\Phi W(\Phi)~.
\end{eqnarray}
Some potentials $W(\Phi)$ studied in the literature are 
\begin{enumerate}
    \item $W(\Phi) = 2\Phi$: AdS$_2$ JT-gravity \cite{Jackiw:1984je, Teitelboim:1983ux} (see also \cite{Almheiri:2014cka, Maldacena:2016upp, Saad:2019lba})
    \item $W(\Phi) = -\frac{1}{\pi}\sinh(2\pi \Phi)$: Sinh-dilaton gravity \cite{Mertens:2020hbs, Collier:2023cyw}
         \item $W(\Phi) \propto \frac{1}{\sqrt{\Phi}} -\Lambda \sqrt{\Phi}$:  SdS reduction \cite{Svesko:2022txo, Maldacena:2019cbz, Cotler:2026jdz}
    \item $W(\Phi) = -2\Phi$: dS$_2$ JT-gravity \cite{Maldacena:2019cbz, Cotler:2019nbi, Cotler:2023eza, Cotler:2024xzz}
   \item $W(\Phi) = -\frac{1}{\pi}\sin(2\pi \Phi)$: Sine-dilaton gravity \cite{Blommaert:2025qrw, Blommaert:2025eps, Verlinde:2024znh, Collier:2025pbm}
   \item $W(\Phi) = 2|\Phi|$: Centaur geometries \cite{Anninos:2017hhn, Anninos:2018svg, Anninos:2022hqo}
\end{enumerate}
While 1. and 2. relate to two-dimensional gravity with $\Lambda <0$, the potentials 3.-6. relate to two-dimensional de Sitter gravity, 6. interpolates between an AdS$_2$ region at
large positive dilaton glued at $\Phi=0$ to a dS$_2$ interior \cite{Anninos:2017hhn, Anninos:2020cwo} — a laboratory for asking de Sitter questions with nearly-AdS$_2$ boundary conditions.  3. is the reduction of Schwarzschild de Sitter discussed in section \ref{sec:dSBH} where schematically $\Phi \propto \phi^2$. 

In contrast to timelike Liouville theory coupled to matter, dilaton models (barring 3. and 6.) in general do not admit a finite sphere partition function \cite{Maldacena:2019cbz, Mahajan:2021nsd}. Each of them has six zero modes on the two-sphere (the $l=1$ conformal Killing modes, as in the timelike Liouville analysis above); the PSL$(2,\mathbb{C})$ quotient absorbs only three of the six, and the remaining three render the sphere partition function divergent. A more careful analysis can for example be found in \cite{Nanda:2023wne}.

Sine-dilaton gravity theory (5.) is also the bulk description of double-scaled SYK (DSSYK) \cite{Cotler:2016fpe, Berkooz:2018qkz, Berkooz:2018jqr}, which is another proposal for microscopic de Sitter holography \cite{Blommaert:2025eps, Verlinde:2024znh}. 
To give an example of the flavour of the equation of motion we consider Sine-dilaton gravity:
\begin{equation}
    S = \int_{\Sigma_{g,n}} \d^2 x\sqrt{g} \left(\Phi R - \frac{1}{\pi}\sin(2\pi \Phi)\right)~.
\end{equation}
The equations of motion of sine-dilaton theory are
\begin{subequations}
    \begin{align}
        R -2\cos(2\pi \Phi)=0~,\\
        \nabla_\mu\nabla_\nu \Phi -g_{\mu\nu}\nabla^2 \Phi - \frac{1}{2\pi}g_{\mu\nu}\sin(2\pi \Phi)=0~.
    \end{align}
\end{subequations}
$\zeta^\mu =\epsilon^{\mu\nu}\partial_\nu \Phi$ is a Killing vector of these equations. Since a generic surface $\Sigma_{g,n}$ does not have any Killing vectors $\zeta^\mu$ needs to vanish and hence the solutions are constant values of $\Phi$ given by
\begin{equation}
    \Phi_*= \frac{m}{2}~,\quad R_*= 2(-1)^m~,\quad m\in \mathbb{Z}~.
\end{equation}
The theory hence admits AdS$_2$ and dS$_2$ vacua, however the dS$_2$ vacua come with a $(-,-)$ signature.

\section{de Sitter in 3D}\label{sec:3D}

The last few years have seen significant progress on dS$_3$: the Complex Liouville string, DSSYK, and $T\overline{T}$ constructions have equipped three-dimensional de Sitter space with explicit models and mechanisms. In this section we won't be able to discuss all of these developments with extensive detail but instead explain the set-up of the complex Liouville string and provide a reference list for DSSYK and $T\overline{T}$ constructions in the end.

\subsection{The wavefunction of three-dimensional de Sitter gravity}
In Lorentzian signature, pure three-dimensional Einstein gravity with positive
cosmological constant\footnote{Massive particles can be added to this setup as Wilson lines in the Chern--Simons formulation discussed below, 
see e.g. \cite{Castro:2023dxp, Castro:2023bvo}} is defined by the Einstein--Hilbert action
\begin{equation}
    S = \frac{1}{16\pi G_N} \int \mathrm{d}^3x \, \sqrt{-g}\,(R - 2\Lambda)\,,
    \qquad \Lambda = \frac{1}{\ell^2} > 0\,.
\end{equation}
At the level of the action, this theory can be recast as an
$\mathrm{SL}(2,\mathbb{C})$ Chern--Simons theory with gauge fields $A$ and
$\bar{A}$ \cite{Witten:1988hc}. The equivalence comes with well-known caveats: gravity
requires an invertible metric while Chern--Simons theory does not, the global
structure of the two theories differs, and large diffeomorphisms and large
gauge transformations are treated differently in the two descriptions.

Our goal is to canonically quantize this theory on spatial slices of topology
$\Sigma_{g,n}$, a genus-$g$ surface with $n$ punctures satisfying $2g-2+n>0$,
so that $\Sigma_{g,n}$ admits a hyperbolic metric; the punctures correspond to
worldlines of massive particles. Quantization begins with a choice of
polarization on the phase space $(g_{ij}, K_{ij})$ --- the analogue of choosing
between the position and momentum representations in quantum mechanics. Taking
the metric as the configuration variable, states become wavefunctionals
$\Psi[g_{ij}]$. Imposing the Wheeler--DeWitt and momentum constraints,\footnote{For a nice discussion of the WdW equations on a spatial slice which is the torus see the recent work \cite{Godet:2024ich, Godet:2025bju}.}
\begin{equation}\label{eq:constraints}
    H \Psi[g_{ij}] = 0\,, \quad H_i \Psi[g_{ij}] = 0
    \quad \Rightarrow \quad
    \Psi\Big[\frac{g_{ij}}{\mathrm{Diff} \times \mathrm{Weyl}}\Big]
    \in \mathcal{H}_{g,n}\,,
\end{equation}
the wavefunction descends to the moduli space $\mathcal{M}_{g,n}$ of
$\Sigma_{g,n}$; we denote it $\Psi_{g,n}$. The Hilbert space
$\mathcal{H}_{g,n}$ is the space of functions --- more precisely, sections of
a line bundle; see \cite{Collier:2023fwi} --- on $\mathcal{M}_{g,n}$. Whether $\Psi_{g,n}$
lives on the moduli space
$\mathcal{M}_{g,n} = \mathcal{T}_{g,n}/\mathrm{MCG}(\Sigma_{g,n})$ or on
Teichm\"uller space $\mathcal{T}_{g,n}$ depends on whether the large
diffeomorphisms, which form the mapping class group, are gauged.

In metric variables, the constraints \eqref{eq:constraints} are second-order
functional differential equations; in the Chern--Simons formulation they
become first order. They can moreover be recast as Virasoro Ward identities
\cite{Verlinde:1989ua}, from which it follows that $\Psi_{g,n}$ transforms as a correlation
function of a two-dimensional CFT with
\begin{subequations}
\begin{align}
    c &= 13 + \frac{3 i \ell}{2 G_N}\,, \\ \label{eq: external momenta 3D}
    \Delta_i &= h_i + \tilde{h}_i = 1 \pm \sqrt{1 - m_i^2 \ell^2}
    \in 1 + i \mathbb{R}\,, \qquad s_i = h_i - \tilde{h}_i\,,
\end{align}
\end{subequations}
for particles of mass $m_i \ell > 1$. These are the weights of
$\mathrm{SL}(2,\mathbb{C})$ principal series representations, as appropriate
for massive particles in dS$_3$. The inner product on $\mathcal{H}_{g,n}$ is
\begin{equation}\label{eq: inner product 3D}
    \langle \Psi_{g,n} | \Psi'_{g,n} \rangle
    = g_s^{2g-2} \int_{\mathcal{M}_{g,n}} \Psi^*_{g,n} \Psi'_{g,n}\,,
\end{equation}
and it is well defined thanks to
\begin{equation}
    c + c^* = 26\,, \qquad \Delta_i + \Delta_i^* = 2\,, \qquad
    s_i - s_i^* = 0\,.
\end{equation}
Finally $g_s$ is the effective string coupling. 

\begin{figure}[ht]
\centering
\begin{tikzpicture}
[
    scale=1.1,
    boundary/.style={thick},
    bulk/.style={thick, dashed},
    curve/.style={thick},
    dcurve/.style={thick, dashed},
    label/.style={scale=0.7}
]
      \draw[opacity=0.8, very thick, black] (0,0) arc [start angle=180, end angle=360, x radius=2cm, y radius=0.5cm];

  \draw[very thick, opacity=0.8, black] (4,0) arc [start angle=0, end angle=180, x radius=2cm, y radius=0.5cm];

\coordinate (E) at (2,-3);
\coordinate (F) at (0,0);
\coordinate (G) at (4,0);
\draw[curve, bend left=15] (F) to (E);
\draw[curve, bend right=15] (G) to (E);

\coordinate (E) at (2,-3);
\coordinate (H) at (.5,0);
\coordinate (I) at (2.3,.3);
\coordinate (J) at (3.3,-.25);
\draw[curve, bend left=15, red] (H) to (E);
\draw[curve, bend right=0, red] (I) to (E);
\draw[curve, bend right=10, red] (J) to (E);

  \fill[gray!20, opacity=0.5]
  (0,0) 
  arc [start angle=180, end angle=360, x radius=2cm, y radius=0.5cm] 
  -- (4,0)                                                        
  arc [start angle=0, end angle=180, x radius=2cm, y radius=0.5cm] 
  -- cycle; 

  \node[red,scale=.8] at (.5,0) {$\mathsf{x}$};
\node[red, scale=.8] at (2.3,.3) {$\mathsf{x}$};
\node[red, scale=.8] at (3.3,-.25) {$\mathsf{x}$};

\coordinate (A) at (1,-.1);
\coordinate (B) at (2,-.1);
\coordinate (C) at (1.15,-.18);
\coordinate (D) at (1.85,-.18);
\draw[curve, bend right=45] (A) to (B);
\draw[curve, bend left=45] (C) to (D);

\coordinate (A) at (2.5,0);
\coordinate (B) at (3.5,0);
\coordinate (C) at (2.65,-.08);
\coordinate (D) at (3.35,-.08);
\draw[curve, bend right=45] (A) to (B);
\draw[curve, bend left=45] (C) to (D);
\node[
    star,
    star points=5,
    rotate=40,
    star point ratio=2.5, 
    draw=black, 
    fill=yellow, 
    minimum size=0.5cm,
    inner sep=0pt
] at (2,-3) {};
\end{tikzpicture}
\caption{The wavefunction of 3D de Sitter quantum gravity is given by the Liouville correlator, with a Milne type singularity at $t=0$. The surface at $\mathcal{I}^+$ is a genus $g$ Riemann surface with $n$ punctures. In the example $g=2$ and $n=3$. The punctures extend to Wilson lines of non-dynamical particles in the past which fall into the singularity at $t=0$.}
\label{fig: Milne CLS}
\end{figure}
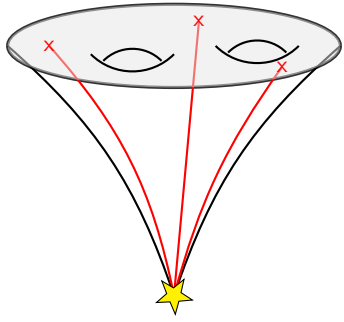

A classical solution of the three-dimensional Einstein--Hilbert action with
positive cosmological constant and spatial slices of topology $\Sigma_{g,n}$
is, in units $\ell=1$,
\begin{equation}\label{eq:MilnedS}
    \mathrm{d}s^2 = -\mathrm{d}t^2 + \sinh^2(t)\, \mathrm{d}\Sigma_{g,n}^2\,,
    \qquad R_{\Sigma_{g,n}} = -2\,,
\end{equation}
where $\mathrm{d}\Sigma_{g,n}^2$ is the hyperbolic metric on $\Sigma_{g,n}$ and
$R_{\Sigma_{g,n}}$ its scalar curvature. Near $t=0$ the metric
degenerates as $\mathrm{d}s^2 \approx -\mathrm{d}t^2 + t^2\,
\mathrm{d}\Sigma_{g,n}^2$: the spatial surface shrinks linearly, and the
geometry ends in a Milne-type singularity (fig.~\ref{fig: Milne CLS}). In
Chern--Simons language this locus is harmless: at $t=0$ we may impose the
topological (gapped) boundary condition $A = \overline{A}$. With this boundary
condition, the wavefunction at $\mathcal{I}^+$ is uniquely fixed to be a
Liouville correlator \cite{Collier:2025lux},
\begin{equation}\label{eq:PsiLiouville}
    \Psi_{g,n}^{(b)}(\boldsymbol{p})
    = \big\langle V_{p_1} \cdots V_{p_n} \big\rangle_{\Sigma_g}^{(b)}
    \in \mathcal{H}_{g,n}^{(b)}(\boldsymbol{p})\,,
    \qquad \boldsymbol{p} = (p_1,\ldots,p_n)\,,
\end{equation}
with external weights
\begin{equation}
    \Delta_i = 1 + \frac{c-13}{12} - 2 p_i^2\,, \qquad p_i^2 \in i\mathbb{R}\,.
\end{equation}
The superscript refers to the standard Liouville parametrization of the
central charge, $c = 1 + 6(b+b^{-1})^2$, to which we return momentarily. The
wavefunction \eqref{eq:PsiLiouville} enjoys several desirable properties:
\begin{enumerate}
    \item It factorizes under degenerations of $\mathcal{I}^+$, i.e.\ at the
    boundaries of $\mathcal{M}_{g,n}$ where a cycle of $\Sigma_{g,n}$ pinches.
    \item It is crossing symmetric --- equivalently, invariant under the
    mapping class group --- so that no sum over different bulk fillings at fixed $\mathcal{I}^+$ is necessary.
    \item It is normalizable with respect to the inner product \eqref{eq: inner product 3D}.
\end{enumerate}
The gauge-invariant observables of the theory are integrated cosmological
correlators \cite{Maldacena:2002vr}. Taking the wavefunction \eqref{eq:PsiLiouville}, they read
\begin{equation}\label{eq:correlators}
    \mathsf{A}_n^{(b)}(\boldsymbol{p}) \equiv \sum_{g=0}^\infty g_s^{2g-2}
    \int_{\mathrm{metrics~on~}\mathcal{I}^+}
    \frac{[\mathcal{D}g]}{\mathrm{Diff}\times \mathrm{Weyl}}\,
    \big| \Psi_{g,n}^{(b)}(\boldsymbol{p}) \big|^2\,~.
\end{equation}
We will explain the reason for summing over topologies of $\mathcal{I}^+$ when we discuss the relation between the CLS and matrix models. Note also that the massive particles
backreact --- they deform $\Sigma_{g,n}$ and hence the wavefunction itself ---
in contrast with a QFT on a rigid de Sitter background.
As a final comment, we observe that the norm (\ref{eq:correlators}) is a limiting case of the York norm (\ref{eq: York norm}). 

\subsection{The complex Liouville string}
The form of the cosmological correlators \eqref{eq:correlators} is
reminiscent of two-dimensional string amplitudes. Indeed, consider a
two-dimensional string theory with worldsheet CFT
\begin{equation}\label{eq:worldsheetCLS}
\begin{array}{c}
\text{$c=13+i\lambda$} \\ \text{Liouville CFT}
\end{array}
\ \oplus\
\begin{array}{c}
\text{$c_-\equiv c^*=13-i\lambda$} \\ \text{Liouville CFT}
\end{array} \ \oplus\  \text{$\mathfrak{b}\mathfrak{c}$-ghosts},
\quad \lambda \in \mathbb{R}_+~,
\end{equation}
whose string amplitudes take the form
\begin{equation}\label{eq:AgnCLS}
    \mathsf{A}_{g,n}^{(b)}(\boldsymbol{p}) = \int_{\mathcal{M}_{g,n}}
    \bigg|\Big\langle \prod_{j=1}^n V_{p_j}\Big\rangle\bigg|^2
    \times \mathrm{ghosts}~.
\end{equation}
In view of \eqref{eq:PsiLiouville}, this is nothing other than the
fixed-$(g,n)$ cosmological correlator:
$\mathsf{A}_n^{(b)} = \sum_g g_s^{2g-2}\, \mathsf{A}_{g,n}^{(b)}$. As reviewed
in section \ref{sec:2D}, the Liouville central charge is parametrized as
$c = 1 + 6(b+b^{-1})^2 = 13 + 6(b^2+b^{-2})$; a central charge of the form
$c = 13 + i\lambda$ thus requires $b \in \e^{\frac{i\pi}{4}}\mathbb{R}$.
Criticality of \eqref{eq:worldsheetCLS} --- cancellation of the conformal
anomaly against the ghosts --- requires $c + c_- = 26$, which fixes
$c_- = c^*$, while physical vertex operators must satisfy the mass-shell
condition $h + h_- = 1$. The momenta of the two Liouville factors are
identified as $p_- = p^*$, so that
$h_p = \tfrac{1}{2} + i\nu$, $\nu \in \mathbb{R}$, or equivalently
$p \in \e^{-\frac{i\pi}{4}}\mathbb{R}$; consequently
$\Delta = h + \tilde h \in 1 + i\mathbb{R}$, in agreement with \eqref{eq: external momenta 3D}. Comparing
$c = 13 + i\lambda$ with $c = 13 + \frac{3i\ell}{2G_N}$ identifies
\begin{equation}
    \lambda = \frac{3\ell}{2G_N}\,.
\end{equation}
Although the moduli-space integral \eqref{eq:AgnCLS} is absolutely convergent
\cite{Collier:2024kwt}, evaluating it directly remains challenging. Fortunately, the complex
Liouville string admits a dual description as a Hermitian two-matrix model
\cite{Collier:2024lys}, to which we now turn.

\subsection{Dual matrix integral}
The dual matrix model is a Hermitian two-matrix integral,
\begin{equation}\label{eq: ZMM}
    \mathcal{Z}_{\mathrm{MM}} = \int_{\mathbb{R}^{2N^2}} \mathrm{d}M_1\,
    \mathrm{d}M_2\, \exp\Big[-N\,\mathrm{tr}\big(V_1(M_1)+V_2(M_2)-M_1M_2\big)\Big]\,,
\end{equation}
with $M_{1,2}$ Hermitian $N\times N$ matrices. The specific potentials
$V_{1,2}$ will not be needed: in the double-scaling limit the model is
characterized entirely by its spectral curve, introduced below. The natural
observables are the resolvents
\begin{equation}\label{eq:resolvent}
    R^{(i)}(x) = \mathrm{tr}\,\frac{1}{x-M_i}\,, \qquad i=1,2\,,
\end{equation}
whose connected correlators admit the genus expansion
\begin{equation}
    \big\langle R(x_1)\cdots R(x_n)\big\rangle_{\mathrm{c}}
    = \sum_{g=0}^\infty \e^{-S_0(2g-2+n)}\, R_{g,n}(x_1,\ldots,x_n)\,,
\end{equation}
with $\e^{-S_0}$ playing the role of the string coupling $g_s$.
They are related as following \cite{Collier:2025lux}
\begin{equation}\label{eq: def gs}
    g_s^{-2} \sim \e^{2S_0}\frac{\sin(\pi b^2)^2 \sin(\pi b^{-2})^2}{(b^{-2}-b^2)^2}~,
\end{equation}
where the tilde indicates an overall $b$ independent constant. 
At finite $N$
the resolvent \eqref{eq:resolvent} has poles at the eigenvalues; at large $N$
these merge into branch cuts along the support of the eigenvalue
distribution, and the resolvent becomes multivalued on the complex plane. The
spectral curve is the Riemann surface on which it is single-valued. For the
complex Liouville string \cite{Collier:2024lys},
\begin{equation}
    \mathsf{x}(z) = -2\cos\big(\pi b^{-1}\sqrt{z}\big)\,, \qquad
    \mathsf{y}(z) = 2\cos\big(\pi b\sqrt{z}\big)\,,
\end{equation}
which has infinitely many branch points,
\begin{equation}\label{eq: branch points}
    \mathrm{d}\mathsf{x}(z_m^*) = 0 \quad\Leftrightarrow\quad
    z_m^* = (mb)^2\,, \quad m\in\mathbb{Z}_{\geq 1}\,.
\end{equation}
From the discontinuity of the resolvent across the cuts one obtains the
leading eigenvalue density
\begin{align}\label{eq: eigenvalues 1}
    \rho_0^{(1)}(x_1) &= |g_s|^{-1} \frac{(b^{-2}-b^2)\sin(-ib^2 \arccosh(\frac{x_1}{2}))}{2\sin(\pi b^{-2})}~,
\end{align}
and equivalently for $\rho_0^{(2)}$.

The higher $R_{g,n}$ are computed by topological recursion \cite{Eynard:2004mh, Chekhov:2006vd,Eynard:2007kz}. The initial data
are
\begin{equation}
    \omega_{0,1}(z) = -\mathsf{y}(z)\,\mathrm{d}\mathsf{x}(z)\,, \qquad
    \omega_{0,2}(z_1,z_2) = \frac{\mathrm{d}z_1\,\mathrm{d}z_2}{(z_1-z_2)^2}\,,
\end{equation}
and for $2g-2+n>0$ the recursion produces multi-differentials related to the
resolvents by
\begin{equation}
    \omega_{g,n}(z_1,\ldots,z_n)
    = R_{g,n}\big(\mathsf{x}(z_1),\ldots,\mathsf{x}(z_n)\big)\,
    \mathrm{d}\mathsf{x}(z_1)\cdots\mathrm{d}\mathsf{x}(z_n)\,.
\end{equation}
The distinctive feature of the complex Liouville string is that the recursion
runs over the infinitely many branch points $z_m^*$ (\ref{eq: branch points}). String
amplitudes are extracted from the $\omega_{g,n}$ via
\begin{equation}
    \mathsf{A}_{g,n}^{(b)}(\boldsymbol{p})
    = \sum_{m_1,\ldots,m_n=1}^\infty
    \mathrm{Res}_{z_1=z^*_{m_1}}\cdots\mathrm{Res}_{z_n=z^*_{m_n}}
    \prod_{j=1}^n \frac{\cos(2\pi p_j\sqrt{z_j})}{p_j}\,
    \omega_{g,n}(z_1,\ldots,z_n)\,.
\end{equation}
Carrying out the recursion yields the string amplitudes --- and with them the
cosmological correlators --- in closed form for all $(g,n)$:
\begin{align}
    \mathsf{A}_{g,n}^{(b)}(\boldsymbol{p})
    &= \sum_{\Gamma\in\mathcal{G}_{g,n}^\infty}
    \frac{1}{|\mathrm{Aut}(\Gamma)|}\, {\int}'
    \prod_{e\in\mathcal{E}_\Gamma} (-2p_e\,\mathrm{d}p_e)
    \prod_{\nu\in\mathcal{V}_\Gamma}
    \left(\frac{b(-1)^{m_\nu}}{\sqrt{2}\sin(\pi m_\nu b^2)}\right)^{2g_\nu-2+n_\nu}
    \nonumber\\
    &\quad\times \prod_{j\in I_\nu} \sqrt{2}\sin(2\pi m_\nu b\, p_j)\,
    \mathsf{V}^{(b)}_{g_\nu,n_\nu}(i\boldsymbol{p}_\nu)\,.
\end{align}
Here the sum runs over stable graphs $\Gamma$ of total genus $g$ with $n$
external legs: each vertex $\nu$ carries a genus $g_\nu$, a branch-point
label $m_\nu$, and $n_\nu$ momenta $\mathsf{p}_\nu$; $I_\nu$ is the set of momenta associated with the vertex $\nu$. $\mathcal{E}_\Gamma$ and $\mathcal{V}_\Gamma$
denote the edges and vertices of $\Gamma$; the primed integral is over the
internal momenta $p_e$, along the contour specified in \cite{Collier:2024lys}; and
$\mathsf{V}^{(b)}_{g,n}$ are the quantum volumes of the Virasoro minimal
string \cite{Collier:2023cyw}.

A further payoff of the matrix model is that it reproduces the
three-dimensional de Sitter entropy. The Euclidean gravity computation of
section~\ref{sec:dSentropy} gave \cite{Anninos:2020hfj, Anninos:2021ihe, Castro:2011xb, Guadagnini:1994ahx,Carlip:1992wg}
\begin{equation}\label{eq:SdS3GH}
    S_{\mathrm{dS}_3} = \log\mathcal{Z}_{\mathrm{grav}}
    = \frac{\pi\ell}{2G_N} - 3\log\frac{\pi\ell}{2G_N} + 5\log(2\pi)
    + \sum_{n\geq 1} c_n \Big(\frac{G_N}{\ell}\Big)^n \pm \frac{5\pi i}{2}~,
\end{equation}
whose leading term is the Gibbons--Hawking entropy $A_{\mathrm{hor}}/4G_N$ and whose
imaginary part is the phase of the gravitational path integral (see
section~\ref{sec:dSentropy}). On the matrix side, the densities $\rho_0^{(i)}$ oscillate,
and we count eigenvalues only up to their first zeros,
$\rho_0^{(i)}(x_{i,*}) = 0$. This defines an effective number of eigenvalues 
\begin{equation}
    N_{\mathrm{eff}} = \int_2^{x_{1,*}}\mathrm{d}x\,
    \rho_0^{(1)}(x) = g_s^{-1} ~,
\end{equation}
where we used (\ref{eq: eigenvalues 1}) and $x_{1,*} = 2\cos(\pi/b^2)$. Finally $g_s$ is given in (\ref{eq: def gs}).
As argued in \cite{Collier:2025lux} in pure dS$_3$ gravity $g_s$ or equivalently $\e^{S_0}$ and the central charge (or equivalently $b$) are not independent parameters. Indeed \cite{Collier:2025lux} argued that
\begin{equation}
    \mathcal{Z}_{\mathrm{grav}}^{S^3}\sim g_{s}^{-2} ~.
\end{equation}
The last unknown parameter is $\e^{S_0}$. In dS$_3$ however there is only one dimensionless parameter $c=13+ 6(b^2+b^{-2})$ which is related to $G_N/\ell_{\mathrm{dS}}$ in the semiclassical limit $c \approx 6 b^{-2}= 3i\ell/(2G_N)$. Indeed, using cutting and gluing rules (Heegaard splitting) one finds $\e^{S_0} \sim \pi/b$, where the tilde again means $b$ independent terms. Combining this $S_0$ with $g_s$ we obtain 
\begin{equation}
    N_{\mathrm{eff}} = \frac{\pi}{b} \frac{\sin(\pi b^2)\sin(\pi b^{-2})}{(b^{-2}-b^2)}~.
\end{equation}
Taking into account that we have one $N_{\mathrm{eff}}$ for each matrix, we can define an entropy
\begin{align}
    S_{\mathrm{matrix}} &\equiv 2\log N_{\mathrm{eff}} = \frac{\pi \ell}{2G_N} - 3\log \frac{\pi \ell}{2G_N} +5 \log 2\pi + 2\log \frac{\pi}{4} \pm \frac{5\pi i }{2} + \mathcal{O}(G_N^2/\ell^2) ~.
\end{align}
where we used $b^{-2} \sim \frac{i\ell}{4G_N}$. Comparing $S_{\mathrm{matrix}}$ and the sphere partition function (\ref{eq: SdS3 GH}) we see that the two match remarkably well. In three dimensions odd powers of $G_N/\ell$ can be removed through local counterterms, whereas even powers cannot. Understanding the microscopic building blocks of $S_{\mathrm{matrix}}$ would amount to understand the finite $N$ matrix model (\ref{eq: ZMM}) which would go beyond the knowledge of the spectral curve.

\begin{figure}[ht]
    \centering
\begin{tikzpicture}
\def\XR{0.25}; 
\def\YR{0.5};  

\begin{scope}[shift={(-6,0)}]

\node[scale=.9] at (-1,1) {$\sum_{m_1=1}^\infty$}; 
\node[scale=.9] at (0,-1) {$\sqrt{2}\sin(2\pi m_1 b p_1)$}; 
\node[scale=.9] at (0,3) {$\sqrt{2}\sin(2\pi m_1 b p_2)$}; 
\node[scale=.9] at (3,-1) {$\sqrt{2}\sin(2\pi m_1 b p_3)$}; 
\node[scale=.9] at (3,3) {$\sqrt{2}\sin(2\pi m_1 b p_4)$};
\node[scale=1.5] at (1.5,1) {$m_1$};
\draw[->,thick] (1.5,4)-- (1.5,1.7);
\node[scale=.9] at (1.5,4.5) {$\left(\frac{b(-1)^{m_1}}{\sqrt{2}\sin(\pi m_1 b^2)}\right)^2\mathsf{V}_{0,4}^{(b)}(ip_1,ip_2,ip_3,ip_4)$};

    \draw[fill=blue,draw=blue,opacity=.4] (0,2.5) to[out=0, in=180] (1.5,2) to[out=0, in =180] (3,2.5) to[out=200,in=160,looseness=.95] (3,1.5) to[out=180, in = 180] (3,.5) to[out=200, in = 160,looseness=.95] (3,-.5) to[out=180, in = 0] (1.5,0) to[out=180, in = 0] (0,-.5) to[out=20, in = -20,looseness=.95] (0,.5) to[out=0, in = 0] (0,1.5) to[out=20, in =-20,looseness=.95] (0,2.5);

\draw[very thick] (0,0) ellipse (0.25 and .5);
\draw[very thick] (0,2) ellipse (0.25 and .5);

\draw[very thick] (3,0) ellipse (0.25 and .5);

\draw[very thick] (3,2) ellipse (0.25 and .5);
\draw[thick] (0,.5) to[out=0, in = 0] (0,1.5);
\draw[thick] (3,.5) to[out=180, in = 180] (3,1.5);
\draw[thick] (0,2.5) to[out=0, in = 180] (1.5,2) to[out=0, in = 180] (3,2.5);
\draw[thick] (0,-.5) to[out=0, in = 180] (1.5,0) to[out=0, in=180] (3,-.5);
\end{scope}
\begin{scope}[shift={(-5,0)}]

\node[scale=.9] at (4.5,1) {$+\sum_{m_1,m_2=1}^\infty$}; 
\node[scale=.9] at (6,-1) {$\sqrt{2}\sin(2\pi m_1 b p_1)$}; 
\node[scale=.9] at (6,3) {$\sqrt{2}\sin(2\pi m_1 b p_2)$}; 
\node[scale=.9] at (10.5,-1) {$\sqrt{2}\sin(2\pi m_2 b p_3)$}; 
\node[scale=.9] at (10.5,3) {$\sqrt{2}\sin(2\pi m_2 b p_4)$};
\draw[->,thick] (8.25,3.8)-- (8.25,1.2);
\node[scale=.9] at (8.25,4) {${\int}^{'} (-2q\d q) 2\sin(2\pi m_1 bq)\sin(2\pi m_2 bq)$};
\draw[->,thick] (7.5,-2)-- (7.5,1);
\draw[->,thick] (9,-2)-- (9,1);
\node[scale=.9] at (6.5,-2.2) {$\frac{b(-1)^{m_1}}{\sqrt{2}\sin(\pi m_1 b^2)}$};
\node[scale=.9] at (10.7,-2.2) {$\frac{b(-1)^{m_2}}{\sqrt{2}\sin(\pi m_2 b^2)}$};

\draw[fill=blue, draw=blue, opacity=.4] (6,2.5) to[out=0, in = 180] (8.25,1) to [out=180, in =0] (6,-.5) to [out=20, in = -20, looseness=.95] (6,0.5) to[out= 0, in = 0] (6,1.5) to [out=20, in = -20, looseness=.95] (6,2.5);

\draw[fill=vert, draw=vert, opacity=.4] (10.5,2.5) to[out=180, in = 0] (8.25,1) to[out=0, in = 180] (10.5,-.5) to[out=160, in = 200, looseness=.95] (10.5,0.5) to[out=180, in = 180] (10.5,1.5) to[out=160, in = 200, looseness=.95] (10.5,2.5);

\draw[very thick] (6,0) ellipse (0.25 and .5);
\draw[very thick] (6,2) ellipse (0.25 and .5);

\draw[very thick] (10.5,0) ellipse (0.25 and .5);

\draw[very thick] (10.5,2) ellipse (0.25 and .5);
\draw [domain=270:440,thick] plot ({10.5+\XR*cos(\x)}, {\YR*sin(\x)}); 

\draw[thick] (6,.5) to[out=0, in =0] (6,1.5); 
\draw[thick] (10.5,.5) to[out=180, in =180] (10.5,1.5); 
\draw[thick] (6,2.5) to[out=0, in = 180] (8.25,1) to[out=0, in=180] (10.5,2.5);
\draw[thick] (6,-.5) to[out=0, in = 180] (8.25,1) to[out=0, in = 180] (10.5,-.5);

\node[scale=1.5] (A) at (6.9,1) {$m_1$};
\node[scale=1.5] (B) at (9.6,1) {$m_2$};
\fill (8.25,1.) ellipse (0.1 and 0.1);

\node[scale=.9] at (12,1) {$+2~\mathrm{perm}$}; 

\end{scope}
\end{tikzpicture}
\caption{Four point amplitude $\mathsf{A}_{0,4}^{(b)}(p_1,p_2,p_3,p_4)$ of the complex Liouville string.}
\end{figure}

\subsection{DSSYK \& dS$_3$}
In \cite{HVerlindetalk, Susskind:2021esx,Susskind:2022bia,Lin:2022nss}
a connection between three-dimensional dS space and double scaled SYK (DSSYK) \cite{Cotler:2016fpe, Berkooz:2018qkz, Berkooz:2018jqr}\footnote{DSSYK also admits a two-dimensional cosmological description \cite{Blommaert:2025rgw,Okuyama:2025hsd,Heller:2025ddj}.} was proposed. It has since then been understood that a more precise connection is that (two copies of) DSSYK describes physics from the perspective of an observer in the 3d static patch \cite{Narovlansky:2023lfz,Verlinde:2024znh,Narovlansky:2025tpb,Tietto:2025oxn,Blommaert:2025eps,Marini:2026zjk,Verlinde:2024zrh,Blommaert:2026ofx}.  The detailed holographic map in the dS$_3$/DSSYK duality has however yet to be completely established.

\subsection{$T\overline{T}$, timelike boundaries $\&$ dS$_3$}
Although we do not study them in detail in these notes, attempts to understand three-dimensional de Sitter by incorporating a finite size timelike boundary have been developed in \cite{Batra:2024kjl,Coleman:2021nor}. The mechanics follows a $T\overline{T}$-type construction. In these constructions the role of timelike boundary is meant to be analogous to the boundary of AdS.

\section{de Sitter in 4D}\label{sec:4D}

The ultimate target in the effort to understand our own Universe is quantum gravity in four-dimensional de Sitter space. Its importance notwithstanding, controlled avenues toward it are few. String constructions such as KKLT
\cite{Kachru:2003aw} are promising but to date do not admit a complete non-perturbative mathematical description. Inspired by holography in AdS, one may hope to gain control by introducing a worldtube --- a timelike boundary at finite distance which is an actively researched topic
\cite{Anderson:2006lqb, Witten:2018lgb,An:2021fcq,Anninos:2024wpy, Liu:2025xij}.
Moreover, in contrast with the two- and three-dimensional models of the
previous sections gravity in four
spacetime dimensions is perturbatively non-renormalizable \cite{tHooft:1974toh,Goroff:1985th}: the
Einstein--Hilbert action is merely the leading term in a derivative expansion
containing infinitely many higher-curvature operators, each with an a priori
independent coupling. As in lower dimensions, Euclidean
gravity offers a possible avenue, inspired by its successes with black hole
thermodynamics \cite{Gibbons:1976ue}. Parallel to these efforts stands dS/CFT
\cite{Strominger:2001pn,Witten:2001kn,Maldacena:2002vr}. Its basic observation is that the
isometries of dS$_4$ act on $\mathcal{I}^+$ as the conformal group of
three-dimensional Euclidean space, so that late-time correlators transform as
those of a Euclidean CFT, with non standard unitarity properties. From this starting point various
proposals have emerged \cite{Maldacena:2002vr,Strominger:2001pn, McFadden:2009fg, Anninos:2011ui, Hertog:2011ky, Collier:2025lux}.
\paragraph{Proposal I.} There exists a Euclidean CFT which computes the correlators \cite{Strominger:2001pn}
\begin{equation}
    \langle 0| \mathcal{O}(x_1)\mathcal{O}(x_2)\ldots \mathcal{O}(x_n)|0\rangle~.
\end{equation}
Time emerges from the CFT. The drawback is that these correlators are not diffeomorphism invariant and the CFT does not obey the usual rules. Moreover it is unclear how to make sense of the static patch.

\paragraph{Proposal II.} The de Sitter Hartle-Hawking wavefunction at late times is computed by a CFT \cite{Maldacena:2002vr}
\begin{equation}
    Z_{\mathrm{CFT}}[\mathrm{sources}] = \Psi_{\mathrm{HH}}[\mathrm{late~time~structure}]~.
\end{equation}
The drawback of this proposal is that it equips us with a wavefunction, but not with a norm.\\ \\
A rather exotic, but very concrete candidate four-dimensional theory of quantum gravity is given by what is called a higher spin theory \cite{Vasiliev:1990en}. It has an infinite set of symmetries, making it likely a UV finite theory. It builds on proposal II \cite{Anninos:2011ui}, but importantly adds to it by equipping it with an inner product \cite{Anninos:2017eib}. We now review some features of this theory. 
\subsection{Higher spin theories}
The perturbative field content of higher spin theories \cite{Vasiliev:1990en} (see for a review \cite{Giombi:2016ejx}) consists of a scalar field $\rho$ together with totally symmetric fields of spin $s\geq 1$:
\begin{equation}\label{eq: higher spin fields}
    \{\rho,A_{\mu_1},g_{\mu_1\mu_2},b_{\mu_1\mu_2\mu_3}, b_{\mu_1\mu_2\mu_3\mu_4},\ldots\}~.
\end{equation}
The spin $s\geq 1$ fields are gauge fields invariant under the transformations
\begin{equation}
    b_{\mu_1 \ldots \mu_s} \rightarrow b_{\mu_1 \ldots \mu_s} + \nabla_{(\mu_1}\xi_{\mu_2 \ldots \mu_s)}~,\quad\quad {\xi}^\nu{}_{\nu\mu_4 \ldots \mu_{s}}=0~,
\end{equation}
where the traceless condition on the gauge parameter $\xi_{\mu_2 \ldots \mu_s}$ appears at $s\ge 3$. For example, for a spin two field
\begin{equation}
    g_{\mu_1 \mu_2} \rightarrow g_{\mu_1 \mu_2}+ \nabla_{\mu_1}\xi_{\mu_2} + \nabla_{\mu_2}\xi_{\mu_1}~,
\end{equation}
where $\xi_{\mu_1}$ is a vector field. Vector fields $\xi_{\mu}$ which satisfy $\nabla_{\mu_1}\xi_{\mu_2} + \nabla_{\mu_2}\xi_{\mu_1}=0$ are Killing vectors.
Higher spin theories were first studied for spacetimes with $\Lambda<0$; indeed, there was hope to prove AdS/CFT from this perspective. In AdS, higher spin theories fall into four categories: minimal type A or type B, non-minimal type A or type B. Minimal higher spin theories only contain the even spin $s$ fields, while type A and type B differ by the parity transformation of the scalar:
\begin{subequations}
\begin{align}
\mathrm{Type ~A}: \quad P\rho(t,\vec{x}) P^{-1} &= + \rho(t,-\vec{x})~,\\
\mathrm{Type ~B}: \quad P\tilde{\rho}(t,\vec{x}) P^{-1} &= - \tilde{\rho}(t,-\vec{x})~.
\end{align}
\end{subequations}
We summarize the four categories in Table~\ref{tab:HST}.
\begin{table}[h!]
\centering
\renewcommand{\arraystretch}{1.4}
\begin{tabular}{c|c|c}
 & \textbf{Type A} & \textbf{Type B} \\ \hline
\textbf{Minimal} &
$\{\rho, g_{\mu_1\mu_2},b_{\mu_1\mu_2\mu_3\mu_4}, \ldots\}$ &
$\{\tilde{\rho}, g_{\mu_1\mu_2}, b_{\mu_1\mu_2\mu_3\mu_4}, \ldots\}$ \\ \hline
\textbf{Non-minimal} &
$\{\rho, A_{\mu_1},g_{\mu_1\mu_2},b_{\mu_1\mu_2\mu_3},  \ldots\}$ &
$\{\tilde{\rho}, A_{\mu_1},g_{\mu_1\mu_2},b_{\mu_1\mu_2\mu_3}, \ldots\}$ \\
\end{tabular}
\caption{The four categories of higher spin theories and their perturbative field content.}
\label{tab:HST}
\end{table}
Higher spin theories are conjectured to be dual to vector models, and in the AdS case \cite{Klebanov:2002ja, Sezgin:2002rt} the dualities are as follows:
\begin{align*}
        \mathrm{type~A~minimal}~ &\Leftrightarrow~ O(N)~\mathrm{vector~model}:~ S=\!\int\!\d^3x\, \phi^I (-\nabla^2)\phi^I~,~\phi^I \in \mathbb{R}\\
        \mathrm{type~A~non~minimal}~ &\Leftrightarrow~U(N)~\mathrm{vector~model}:~S=\!\int\!\d^3x\, \bar{\phi}^I (-\nabla^2)\phi^I~,~ \phi^I \in \mathbb{C}\\
        \mathrm{type~B~minimal}~ &\Leftrightarrow~O(N)~\mathrm{vector~model}:~S= \!\int\!\! \d^3 x\, \overline{\psi}^I \slashed{\partial}\psi^I~,~\psi^I~\mathrm{Majorana}\\
        \mathrm{type~B~non~minimal}~ &\Leftrightarrow~ U(N)~\mathrm{vector~model}:~S= \!\int\!\! \d^3 x\, \overline{\psi}^I \slashed{\partial}\psi^I~,~\psi^I~\mathrm{Dirac}
\end{align*}
In the above we also included the Gaussian parts of the dual vector models, with $I=1,\ldots, N$ and $N\approx \ell^2/\ell_{\mathrm{Planck}}^2$, making the large $N$ limit a semiclassical limit.
The bulk higher spin fields are dual to composite operators in the boundary theory. More precisely, the duality involves only the singlet sector of the vector models: the bulk fields are dual to $O(N)$ (respectively $U(N)$) invariant bilinears. This creates an interacting theory in the bulk with non-trivial Witten diagrams. As an example we can read off $\Delta_{\phi}=1/2$, such that the bulk-boundary correspondence is
\begin{equation}
\rho=\phi^I\phi^I ~,\quad b_{i_1 i_2}= \phi^I\partial_{i_1}\partial_{i_2}\phi^I ~,\ldots~.
\end{equation}
Since for the $U(N)$ model $\phi^I \in \mathbb{C}$, we can also construct conserved currents of odd spin. At spin one we have
\begin{equation}
J_i = i(\phi^I\partial_i \bar{\phi}^I - \bar{\phi}^I \partial_i \phi^I)~,\quad \partial_i J_i=0~,
\end{equation}
and so on.
The $\psi^I$, on the other hand, have conformal dimension $\Delta_\psi=1$ and we build the composite operator
\begin{equation}
\tilde{\rho}= \overline{\psi}^I \psi^I~.
\end{equation}
From the conformal dimensions $\Delta_\phi=1/2$ and $\Delta_{\psi}=1$ we immediately infer that $\Delta_\rho=1$ and $\Delta_{\tilde{\rho}}=2$, such that $\tilde{\rho}$ transforms like the shadow operator of $\rho$.
\paragraph{Higher spin theories in dS.}
In \cite{Anninos:2011ui} (see also \cite{Neiman:2022qfq}) the authors argued that there is no obstruction to realizing higher spin dualities with $\Lambda>0$. In four spacetime dimensions, going from AdS to dS amounts to $\Lambda \rightarrow - \Lambda$. In terms of the rank of the dual CFT, it switches the sign of $N$:
\begin{equation}
    O(N) \overset{N\rightarrow -N}{\longrightarrow} \mathrm{Sp}(N)~,\quad \quad U(N) \overset{N\rightarrow -N}{\longrightarrow} U(N)~.
\end{equation}
It also switches the spin-statistics assignment, e.g.~type A minimal higher spin theory is dual to an $\mathrm{Sp}(N)$ vector model with Gaussian action
\begin{equation}
    S= \int \d^3 x\, \phi^I (-\nabla^2)\phi^J\Omega_{IJ}~,\quad I,J=1,\ldots ,N~,
\end{equation}
where the $\phi^I$ are real but anticommuting. On the other hand, type B minimal higher spin theory is dual to an $\mathrm{Sp}(N)$ vector model of commuting Majorana fermions.
In de Sitter, the higher spin spectrum
\begin{equation}\label{eq: higher spin fields dS}
    \{\rho,A_{\mu_1},g_{\mu_1\mu_2},b_{\mu_1\mu_2\mu_3}, b_{\mu_1\mu_2\mu_3\mu_4},\ldots\}
\end{equation}
consists of a conformally coupled scalar $\rho$ in 4D together with spin $s\geq 1$ fields, which are the highest depth partially massless fields $D_{s-1,s}^\pm$ (see section \ref{sec:QFT}). The superscript $\pm$ denotes the helicity components of the fields.

\paragraph{Wavefunction perspective of higher spin.}
Proposal II relates the bulk Hartle--Hawking wavefunction to the boundary CFT partition function. The boundary is future infinity $\mathcal{I}^+$, where the de Sitter isometries act as conformal transformations.
As an example, for the type A minimal higher spin theory the Hartle--Hawking wavefunction takes the following form \cite{Anninos:2017eib}:
\begin{equation}
\Psi_{\mathrm{HH}}[\mathcal{B}] = Z_{\mathrm{CFT}}^{\mathrm{Sp}(N)}[\mathcal{B}] = \int [\mathcal{D}\phi^I]\,\e^{-S_{\mathrm{Sp}(N)}[\phi^I, \mathcal{B}]}~,
\end{equation}
where $\mathcal{B}$ collectively denotes the boundary profiles of the bulk fields.
Note that in the above we are not integrating over the boundary profiles. If we were to integrate over them, the sources $\mathcal{B}$, which couple to the (conserved) currents, would become dynamical conformal higher spin gauge fields.
The action for the $\mathrm{Sp}(N)$ model is given by\footnote{The skew symmetric matrix $\Omega_{IJ}$ is given by 
\begin{equation}
    \Omega_{IJ} = \begin{pmatrix}
    0 & \mathbb{I}_{\frac{N}{2}}\\
    -\mathbb{I}_{\frac{N}{2}} & 0
    \end{pmatrix}~,
\end{equation}
where $N$ is even. }
\begin{equation}
S_{\mathrm{Sp}(N)} = \int \d^3 x\left(\phi^I (-\nabla_c^2)\phi^J \Omega_{IJ} + \rho\, \phi^I \Omega_{IJ}\phi^J + g_{ij} T^{ij} + \ldots \right)~,
\end{equation}
where $-\nabla_c^2 \equiv -\nabla^2 + \tfrac{R}{8}$ is the conformal Laplacian in three dimensions, and $\rho$ and $g_{ij}$ denote the boundary profiles of the bulk scalar and metric. We are using Latin indices to indicate the projection to the 3D boundary of the bulk fields $b_{\mu_1\ldots \mu_s}$.
The most general expression is
\begin{equation}
\int [\mathcal{D}\phi^I]\, \e^{-\int \d^3x\, \d^3 y\, \phi^I(x) \left[-\nabla_c^2\,\delta^3(x-y) +\mathcal{B}(x,y)\right] \phi^J(y)\,\Omega_{IJ}}= {\det}^{N/2}\!\left(-\nabla_c^2 + \mathcal{B}\right)~.
\end{equation}
The above is a wavefunction of de Sitter space. However, a wavefunction is only an element of a Hilbert space; without an inner product we cannot perform any calculations. In particular we need to understand the measure $[\mathcal{D}\mathcal{B}]$.
To do so we take a small detour and consider the expectation value 
\begin{equation}
    \langle \e^{N a^{i_1\ldots i_s}\tilde{b}_{i_1 \ldots i_s}}\rangle \equiv \langle \Psi_{\mathrm{HH}}|\e^{N a^{i_1\ldots i_s}\tilde{b}_{i_1 \ldots i_s}}|\Psi_{\mathrm{HH}}\rangle = \int [\mathcal{D}\mathcal{B}]\,|\Psi_{\mathrm{HH}}[\mathcal{B}]|^2\, \e^{N\tr\left[\left(\mathcal{D}^{-1}\mathcal{A}\mathcal{D}^{-1}-\mathcal{D}^{-1}\right)\mathcal{B}\right]}~,
\end{equation}
where $\mathcal{D}\equiv -\nabla_c^2$ and the $-N\tr(\mathcal{D}^{-1}\mathcal{B})$ term in the exponent normal-orders the insertion. We also defined
\begin{equation}
\mathcal{A} \equiv a^{i_1\ldots i_s}\mathcal{D}_{i_1\ldots i_s}
\end{equation}
where indices are raised and lowered using the two-point function of $\Psi_{\mathrm{HH}}$ \cite{Anninos:2017eib}. 
Although we do not know the measure, we can evaluate the above in a large $N$ saddle point approximation.
Variation with respect to $\mathcal{B}$ gives the saddle point equation
\begin{equation}
    (1+\mathcal{D}^{-1}\mathcal{B})^{-1} = 1-\mathcal{D}^{-1}\mathcal{A}~,
\end{equation}
which leads to
\begin{equation}
     \langle \e^{N a^I\tilde{b}_I}\rangle\approx \det(1-\mathcal{D}^{-1}\mathcal{A})^{-N}\, \e^{-N \tr(\mathcal{D}^{-1}\mathcal{A})}~,
\end{equation}
where the second factor is generated on-shell by the normal-ordering term.
But this is nothing but the generating function of a bosonic $O(2N)$ vector model
\begin{equation}
  \det(1-\mathcal{D}^{-1}\mathcal{A})^{-N}\, \e^{-N \tr(\mathcal{D}^{-1}\mathcal{A})} = \frac{1}{Z_0} \int [\mathcal{D}\mathcal{Q}]\,\e^{-\int \d^3x\, \d^3y\, \mathcal{Q}^I(x)\left(\mathcal{D}-\mathcal{A}\right)\!(x,y)\, \mathcal{Q}^I(y)}\,\e^{-N \tr(\mathcal{D}^{-1}\mathcal{A})}~,
\end{equation}
where $[\mathcal{D}\mathcal{Q}]$ is now a flat measure over $2N$ real commuting fields $\mathcal{Q}^I$, $I=1,\ldots,2N$, and $Z_0$ denotes the same Gaussian integral at $\mathcal{A}=0$.
This suggests the identification \cite{Anninos:2017eib, De:2026stn}
\begin{equation}
\mathcal{B}(x,y) = \mathcal{Q}^I(x) \mathcal{Q}^I(y)~,\quad I = 1,2,\ldots ,2N~.
\end{equation}
The $\mathcal{Q}^I(x)$ transform like vectors under $O(2N)$, while the higher spin symmetries act on the spatial argument,
\begin{equation}
 \mathcal{Q}^I(x) \rightarrow \int \d^3 y\, \mathcal{U}(x,y)\, \mathcal{Q}^I(y)~.
\end{equation}
The norm of the wavefunction can now be expressed as
\begin{equation}
\langle \Psi_{\mathrm{HH}}|\Psi_{\mathrm{HH}}\rangle = \int [\mathcal{D}\mathcal{Q}^I]\, |\Psi_{\mathrm{HH}}[\mathcal{Q}^I]|^2~.
\end{equation}
The higher spin wavefunction in terms of $\mathcal{Q}$ is given by
\begin{equation}
\Psi_{\mathrm{HH}}[\mathcal{Q}^I] = \e^{-\frac{1}{2}\int \d^3 x\, \mathcal{Q}^I (-\nabla_c^2)\, \mathcal{Q}^I}~.
\end{equation}
Say we are at $\mathcal{I}^+$ with $k$ spatial points. The bilocal $\mathcal{B}(x,y) = \mathcal{Q}^I(x)\mathcal{Q}^I(y)$ is then a $k\times k$ matrix of rank at most $\min(2N,k)$: independent of how large $k$ is, the $O(2N)$ and higher spin invariant data depends only on $\min(2N,k)$. Since in the continuum limit $k\rightarrow \infty$, if we gauge the higher spin symmetries \cite{Higuchi:1991tm} the invariant content of the theory scales with $N \approx \ell^2/\ell^2_{\mathrm{Planck}}$ which scales like the Gibbons-Hawking entropy (see section \ref{sec:dSentropy}).

\begin{figure}[ht]
\centering
\begin{tikzpicture}
\begin{scope}
[
    scale=1,
    rotate=180,
    shift = {(-2,-5.2)},
    boundary/.style={thick},
    bulk/.style={thick, dashed},
    curve/.style={thick},
    dcurve/.style={thick, dashed},
    label/.style={scale=0.7}
]
      \draw[dashed, thick, opacity=0.6] (1.8,1) arc [start angle=0, end angle=180, x radius=.8cm, y radius=0.2cm];
  \draw[opacity=0.8,  thick, opacity=0.6] (0.2,1) arc [start angle=180, end angle=360, x radius=.8cm, y radius=0.2cm];
  \draw[opacity=0.8, thick, blue] (0.2,1) arc [start angle=180, end angle=360, x radius=.8cm, y radius=.8cm];  
\coordinate (A) at (1.8,1);
\coordinate (B) at (0.2,1);
\coordinate (C) at (-1.5,2.5);
\coordinate (D) at (3.5,2.5);
\coordinate (E) at (3.5,-.5);
\draw[curve, bend right=25] (B) to (C);
\draw[curve, bend left=25] (A) to (D);
      \draw[very thick] (3.5,2.5) arc [start angle=0, end angle=180, x radius=2.5cm, y radius=0.3cm];
  \draw[opacity=0.9, very thick] (-1.5,2.5) arc [start angle=180, end angle=360, x radius=2.5cm, y radius=0.3cm];
 \fill[gray!30, opacity=0.8]
  (-1.5,2.5)
  arc[start angle=180, end angle=360, x radius=2.5cm, y radius=0.3cm]
  arc[start angle=0, end angle=180, x radius=2.5cm, y radius=0.3cm]
  -- cycle; 
\end{scope}
\begin{scope}
[
    scale=1,
    boundary/.style={thick},
    bulk/.style={thick, dashed},
    curve/.style={thick},
    dcurve/.style={thick, dashed},
    label/.style={scale=0.7}
]
      \draw[dashed,  thick, opacity=0.6] (1.8,1) arc [start angle=0, end angle=180, x radius=.8cm, y radius=0.2cm];
  \draw[opacity=0.6, thick] (0.2,1) arc [start angle=180, end angle=360, x radius=.8cm, y radius=0.2cm];
  \draw[opacity=0.8, thick, blue] (0.2,1) arc [start angle=180, end angle=360, x radius=.8cm, y radius=.8cm];  
\coordinate (A) at (1.8,1);
\coordinate (B) at (0.2,1);
\coordinate (C) at (-1.5,2.5);
\coordinate (D) at (3.5,2.5);
\coordinate (E) at (3.5,-.5);
\draw[curve, bend right=25] (B) to (C);
\draw[curve, bend left=25] (A) to (D);
      \draw[dashed, very thick, opacity=0.9] (3.5,2.5) arc [start angle=0, end angle=180, x radius=2.5cm, y radius=0.3cm];
  \draw[opacity=0.9, very thick] (-1.5,2.5) arc [start angle=180, end angle=360, x radius=2.5cm, y radius=0.3cm];
 \fill[gray!30, opacity=0.8]
  (-1.5,2.5)
  arc[start angle=180, end angle=360, x radius=2.5cm, y radius=0.3cm]
  arc[start angle=0, end angle=180, x radius=2.5cm, y radius=0.3cm]
  -- cycle; 
\node[label] at (1,2.6) {$\mathcal{I}^+$}; 
\end{scope}
\begin{scope}
[
    scale=1.,
    rotate=180,
    shift = {(-9.8,-3.7)},
    boundary/.style={thick},
    bulk/.style={thick, dashed},
    curve/.style={thick},
    dcurve/.style={thick, dashed},
    label/.style={scale=0.7}
]
      \draw[ very thick] (1.8,1) arc [start angle=0, end angle=180, x radius=.8cm, y radius=0.2cm];
  \draw[opacity=0.8,  very thick] (0.2,1) arc [start angle=180, end angle=360, x radius=.8cm, y radius=0.2cm];
  \draw[opacity=0.8, thick, blue] (0.2,1) arc [start angle=180, end angle=360, x radius=.8cm, y radius=.8cm];  
 \fill[gray!30, opacity=0.8]
  (1.8,1)
  arc[start angle=0, end angle=180, x radius=.8cm, y radius=0.2cm]
  arc[start angle=180, end angle=360, x radius=.8cm, y radius=0.2cm]
  -- cycle; 
  
\end{scope}
\begin{scope}
[
    scale=1.,
    shift = {(7.8,1.5)},
    boundary/.style={thick},
    bulk/.style={thick, dashed},
    curve/.style={thick},
    dcurve/.style={thick, dashed},
    label/.style={scale=0.7}
]
      \draw[dashed, very thick] (1.8,1) arc [start angle=0, end angle=180, x radius=.8cm, y radius=0.2cm];
  \draw[very thick] (0.2,1) arc [start angle=180, end angle=360, x radius=.8cm, y radius=0.2cm];
  \draw[opacity=0.8, thick, blue] (0.2,1) arc [start angle=180, end angle=360, x radius=.8cm, y radius=.8cm];  
 \fill[gray!30, opacity=0.8]
  (1.8,1)
  arc[start angle=0, end angle=180, x radius=.8cm, y radius=0.2cm]
  arc[start angle=180, end angle=360, x radius=.8cm, y radius=0.2cm]
  -- cycle; 
  
\end{scope}
\begin{scope}
[
    scale=1.,
    shift = {(4.4,1.6)},
    boundary/.style={thick},
    bulk/.style={thick, dashed},
    curve/.style={thick},
    dcurve/.style={thick, dashed},
    label/.style={scale=0.7}
]
  \draw[opacity=0.8, thick, blue] (0.2,1) arc [start angle=180, end angle=360, x radius=.8cm, y radius=.8cm];  
        \draw[dashed, thick, opacity=0.6] (1.8,1) arc [start angle=0, end angle=180, x radius=.8cm, y radius=0.2cm];
  \draw[opacity=0.8,  thick, opacity=0.6] (0.2,1) arc [start angle=180, end angle=360, x radius=.8cm, y radius=0.2cm];
\end{scope}
\begin{scope}
[
    scale=1.,
    rotate=180,
    shift = {(-6.4,-3.6)},
    boundary/.style={thick},
    bulk/.style={thick, dashed},
    curve/.style={thick},
    dcurve/.style={thick, dashed},
    label/.style={scale=0.7}
]
  \draw[opacity=0.8, thick, blue] (0.2,1) arc [start angle=180, end angle=360, x radius=.8cm, y radius=.8cm];  
\end{scope}
\draw[<->,thick] (3.75,2.6) -- node[above, scale=.9] {$?$} (4.45,2.6);
\draw[<->,thick] (6.35,2.6) -- node[above, scale=.9] {eq. (\ref{eq: fluing formula})} (7.85,2.6);
\end{tikzpicture}
\caption{Norm of the Hartle-Hawking wavefunction versus the gluing formula (\ref{eq: fluing formula}).}
\label{fig:gluing HH}
\end{figure}
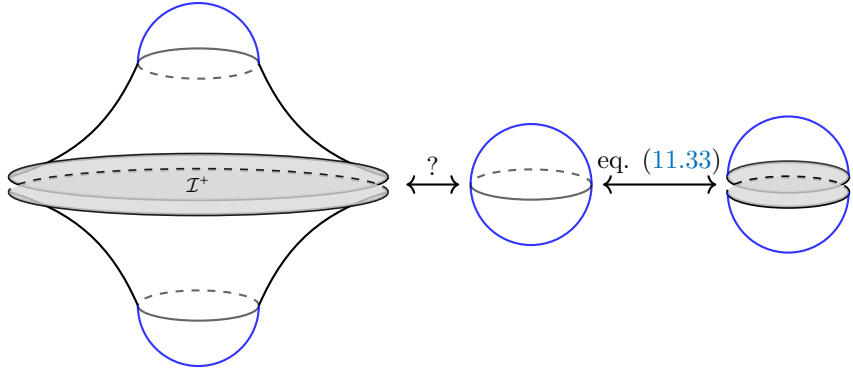

\paragraph{Euclidean perspective of higher spin.}
We are prompted to sharpen the question about the sphere partition function of higher spin theories and the static patch \cite{Neiman:2017zdr, Neiman:2018ufb}. The main obstruction to that is the absence of a Lagrangian formulation of higher spin theories. Instead we will pursue a different route \cite{Giombi:2013fka}. In section \ref{sec:QFT} we obtained the Harish-Chandra characters for the higher spin fields. As a reminder, $\rho$ is a conformally coupled scalar in 4D i.e. $m^2\ell^2 =2$, which means it transforms in the complementary series of $\mathrm{SO}(1,4)$. The other fields with spin $s\geq 1$ are highest depth partially massless fields (these are the truly massless fields). Below we summarize their Harish-Chandra characters: 
\begin{subequations}\label{eq:chis}
\begin{align}\label{eq:chi10}
\chi_{1, 0}(\mathfrak{t}) &= \frac{\e^{-\mathfrak{t}} + \e^{-2\mathfrak{t}}}{(1-\e^{-\mathfrak{t}})^3}~,\\ \label{eq:chissp1}
\chi_{1+s, s}(\mathfrak{t}) &= \chi_{s,\mathrm{bulk}}(\mathfrak{t}) - \chi_{s,\mathrm{edge}}(\mathfrak{t}) ~,\\ \label{chibulks}
     \chi_{s,\mathrm{bulk}}(\mathfrak{t})&= 2\frac{(2s+1)\e^{-(s+1)\mathfrak{t}} -(2s-1)\e^{-(s+2)\mathfrak{t}}}{(1-\e^{-\mathfrak{t}})^3}~, \\ \label{eq:chiedge}
    \chi_{s,\mathrm{edge}} (\mathfrak{t})&=\frac{1}{3} \frac{s(s+1)(2s+1) \e^{-s \mathfrak{t}} -s(s-1)(2s-1) \e^{-(s+1)\mathfrak{t}}}{(1-\e^{-\mathfrak{t}})}~.
 \end{align}   
\end{subequations}
At one-loop order the sphere partition function is the sum over all these characters. It takes the following form \cite{Anninos:2020hfj, Anninos:2026hia}
\begin{align}\label{oneloophs}
     \log \mathcal{Z}_{\mathrm{h.s.}}^{(1)} &=\int_0^\infty\! \frac{\d \mathfrak{t}}{2\mathfrak{t}} \frac{1+\e^{-\mathfrak{t}}}{1-\e^{-\mathfrak{t}}}  \sum_{s\in 2\mathbb{N}}\chi_{1+s,s}(\mathfrak{t}) 
     = \int_0^\infty \frac{\d \mathfrak{t}}{2\mathfrak{t}} \frac{1+\e^{-\mathfrak{t}}}{1-\e^{-\mathfrak{t}}} \bigg(\frac{-\e^{-\mathfrak{t}}}{(1-\e^{-\mathfrak{t}})^2} +2 \frac{\e^{-\frac{3\mathfrak{t}}{2}}+\e^{-\frac{\mathfrak{t}}{2}}}{(1-\e^{-\mathfrak{t}})^2}\bigg)~\cr
     &=\frac{\zeta(3)}{8\pi^2} +2 \left(\frac{3\zeta(3)}{16\pi^2} -\frac{1}{8}\log 2\right)~,
\end{align}
where in going to the second line we only kept the UV finite part. The higher spin sphere partition function thus takes the form
\begin{equation}
    \mathcal Z_{\mathrm{h.s.}}[S^4] \approx \e^{-S_{\mathrm{EdS}}^{(N)}}\mathcal{Z}_{\mathrm{h.s.}}^{(1)}~,
\end{equation}
where EdS denotes the Euclidean higher spin de Sitter action. As mentioned above it is unknown. To access it we use two tricks. First, as mentioned above AdS and dS higher spin theories are related by $N\rightarrow -N$. Assuming that the higher spin AdS/CFT duality holds the Euclidean AdS on-shell action needs to take a particular form, given by \cite{Klebanov:2002ja, Sun:2020ame}
\begin{equation}
    -S_{\mathrm{EAdS}}^{(N)}= (1-N)\left(\frac{1}{8}\log 2 -\frac{3\zeta(3)}{16\pi^2}\right)~.
\end{equation}
Second the Euclidean dS and renormalized Euclidean AdS on-shell action are related at least at the level of the Einstein Hilbert action by $S_\mathrm{EdS}^{(N)} = 2 S^{(-N)}_{\mathrm{EAdS}}$. Extrapolating that this holds true also for higher spin actions we conjecture that
\begin{equation}
    -S_{\mathrm{EdS}}^{(N)} = - 2S_{\mathrm{EAdS}}^{(-N)}>0~.
\end{equation}
Combining the saddle and one-loop contribution leads to 
\begin{equation}
    \mathcal Z_{\mathrm{h.s.}}[S^4] \approx \frac{(-i)^{\mathcal{P}}}{\mathrm{vol}_N(G_{\mathrm{h.s.}})} \mathrm{exp}\Bigg({2N\underbrace{\left(\frac{1}{8}\log 2-\frac{3\zeta(3)}{16\pi^2}\right)}_{>0} + \frac{\zeta(3)}{8\pi^2}}\Bigg)~.
\end{equation}
In this expression $\mathcal{P}$ is the generalization of Polchinski's phase for all spin $s$ fields, whereas $G_{\mathrm{h.s.}}$ is the higher spin group generated by the spin $s$ Killing tensors. 
\paragraph{Gluing formula.}
The sum of all the four-sphere one-loop contributions gives two pieces: The one-loop partition function of a three-dimensional conformal higher spin
gauge theory, and the one-loop partition function of a conformally coupled scalar on $S^3$.
In \cite{Anninos:2026hia} this is interpreted as a gluing formula for the four sphere partition function, explicitly
\begin{equation}
   \mathcal Z_{\mathrm{glue}} = \frac{(-i)^{\mathcal{P}}}{\mathrm{vol}_N(G_{\mathrm{h.s.}})} \int [\mathcal{D}\mathcal B]\big|Z_{\mathrm{Sp}(N)}[\mathcal B]\big|^2  ~,
\end{equation}
where $Z_{\mathrm{Sp}(N)}[B]$ is the partition function of an $\mathrm{Sp}(N)$
vector model built from anticommuting scalars. $G_{\mathrm{h.s.}}$ is both the 4D Vasiliev algebra, as well as the 3D Eastwood algebra \cite{Eastwood:2002su}, a priori it depends on the coupling $N$ indicated by the subscript.\footnote{As an analogy $\mathrm{SO}(1,4)$ is both the dS$_4$ isometry group as well as the 3D Euclidean conformal group.} We have
\begin{align}\label{eq: fluing formula}
\mathcal Z_{\mathrm{glue}} &= \frac{(-i)^{\mathcal{P}}}{\mathrm{vol}_N(G_{\mathrm{h.s.}})} \int [\mathcal{D}\mathcal B]\big|Z_{\mathrm{Sp}(N)}[\mathcal B]\big|^2 =\frac{(-i)^{\mathcal{P}}}{\mathrm{vol}_N(G_{\mathrm{h.s.}})} \int [\mathcal{D}\mathcal B]{\det}^{N} \big( -\nabla_c^2 +\mathcal B\big)  \cr
 &\approx \frac{(-i)^{\mathcal{P}}}{\mathrm{vol}_N(G_{\mathrm{h.s.}})} \times {\det}^N(-\nabla_c^2) \times \left(\mathrm{higher~spin~2pt~function} ~\cite{Giombi:2013yva}\right) \cr
 &= \frac{(-i)^{\mathcal{P}}}{\mathrm{vol}_N(G_{\mathrm{h.s.}})} \mathrm{exp}\Bigg({2N\left(\frac{1}{8}\log 2-\frac{3\zeta(3)}{16\pi^2}\right) + \frac{\zeta(3)}{8\pi^2}}\Bigg) = \mathcal Z_{\mathrm{h.s.}}[S^4]~,
\end{align}
which confirms our conjecture to one-loop order.
The boundary theory is a free $\mathrm{Sp}(N)$ vector model. This is the same vector model that appeared in the dS/CFT correspondence \cite{Anninos:2011ui}, however their origin is different: In dS/CFT the vector model lives at $\mathcal{I}^+$ where the dS isometries become conformal transformations. In the Euclidean sphere partition function calculation the $\mathrm{Sp}(N)$ model lives on a finite $S^3$ hypersurface (see figure \ref{fig:gluing HH}). 

The gluing formula becomes even more intriguing if we add half-integer higher spin fields \cite{Letsios:2023qzq, Anninos:2025mje}. On the $S^3$ hypersurface we now find a $\mathcal{N}=2$ superconformal field theory and combining saddle and one-loop contributions we obtain 
\begin{equation}\label{eq: ZSUSY HS}
    \mathcal Z^{(N)}_{\mathcal{N}=2~\mathrm{h.s.}}[S^4] = 2^N\times \frac{N^{-\frac{1}{16}}}{\mathrm{vol}(G_{\mathrm{s.hs.}})}~,
\end{equation}
where $G_{\mathrm{s.hs.}}$ is the supersymmetric analog of the higher spin group and we extracted the $N$ dependence from its generators, hence the volume in (\ref{eq: ZSUSY HS}) is independent of $N$.
In particular we can consider ratios where the volume of the $G_{\mathrm{s.hs.}}$ drops out. As an example we have
\begin{equation}
    r_{N,M} \equiv \frac{\mathcal Z^{(N)}_{\mathcal{N}=2~\mathrm{h.s.}}[S^4]}{\mathcal Z^{(M)}_{\mathcal{N}=2~\mathrm{h.s.}}[S^4]}\approx 2^{N-M}\left(\frac{M}{N}\right)^{\frac{1}{16}} ~,
\end{equation}
which for special values of $N,M$ is an integer.

\section*{Acknowledgments}
It is a great pleasure to thank Dionysios Anninos, Tarek Anous, Andreas Blommaert, Jonah Kudler-Flam, Victor Ivo, David Kolchmeyer, Zimo Sun, as well as ChatGPT and Claude for useful discussions. 
I am  particularly indebted to Dionysios Anninos for many illuminating discussions.
I would also like to thank the students and organizers of the 31st W.E.~Heraeus
``Saalburg'' Summer School for providing a stimulating environment.
B.M. gratefully acknowledges funding provided by the Leinweber foundation at the Institute for Advanced Study and the National Science Foundation with grant number PHY-2514611.  B.M. also acknowledges fruitful discussions during the workshop ``Observers, wormholes and complex saddles in cosmology", organized at the Bernoulli Center for Fundamental Studies (EPFL, Lausanne) from 18--22 May 2026.

\appendix

\section{Compendium of useful formulas}\label{app:compendium}

\subsection{Special functions}
\paragraph{Hypergeometric function.} The connection formula relating $z$ and $1-z$ (valid for $c-a-b\notin\mathbb{Z}$),
\begin{align}\label{eq:compendium hypergeometric}
    {}_2F_1(a,b;c;z) &= \frac{\Gamma(c)\Gamma(c-a-b)}{\Gamma(c-a)\Gamma(c-b)}\,{}_2F_1(a,b;a+b-c+1;1-z)\cr
    &\quad+ (1-z)^{c-a-b}\,\frac{\Gamma(c)\Gamma(a+b-c)}{\Gamma(a)\Gamma(b)}\,{}_2F_1(c-a,c-b;c-a-b+1;1-z)~,
\end{align}
and Gauss' summation at $z=1$ (for $\mathrm{Re}(c-a-b)>0$),
\begin{equation}
    {}_2F_1(a,b;c;1) = \frac{\Gamma(c)\Gamma(c-a-b)}{\Gamma(c-a)\Gamma(c-b)}~.
\end{equation}
The associated Legendre function on the cut $-1<x<1$ is
\begin{equation}\label{eq:Legendre to hypergeometric}
    P_\nu^\mu(x) = \frac{1}{\Gamma(1-\mu)}\left(\frac{1+x}{1-x}\right)^{\frac{\mu}{2}}\,{}_2F_1\!\left(-\nu,\,\nu+1;\,1-\mu;\,\frac{1-x}{2}\right)~.
\end{equation}
\paragraph{Gamma functions.} We use the shorthand $\gamma(x)\equiv \Gamma(x)/\Gamma(1-x)$ and Euler's reflection formula
\begin{equation}
    \Gamma(z)\,\Gamma\!\left(1-z\right) = \frac{\pi}{\sin(\pi z)}~,\quad z \notin \mathbb{Z}~.
\end{equation}
\paragraph{Barnes double Gamma function.} $\Gamma_b(x)$ is defined for $\mathrm{Re}\,x>0$ by ($Q=b+b^{-1}$)
\begin{equation}\label{eq:compendium Gammab def}
    \log \Gamma_b(x) = \int_0^\infty \frac{\d t}{t}\left[\frac{\e^{-xt} - \e^{-\frac{Qt}{2}}}{\big(1-\e^{-bt}\big)\big(1-\e^{-t/b}\big)} - \frac{1}{2}\Big(\frac{Q}{2}-x\Big)^{\!2}\e^{-t} - \frac{1}{t}\Big(\frac{Q}{2}-x\Big)\right]~,
\end{equation}
and extends to a meromorphic function of $x$ with \emph{no zeros} and simple poles at
\begin{equation}
    x = -mb - n b^{-1}~,\qquad m,n\in\mathbb{Z}_{\geq 0}~.
\end{equation}
It is invariant under $b\rightarrow b^{-1}$, $\Gamma_b = \Gamma_{b^{-1}}$, and satisfies the two shift equations
\begin{equation}\label{eq:compendium Gammab shifts}
    \Gamma_b(x+b) = \frac{\sqrt{2\pi}\, b^{bx-\frac{1}{2}}}{\Gamma(bx)}\,\Gamma_b(x)~,\qquad
    \Gamma_b(x+b^{-1}) = \frac{\sqrt{2\pi}\, b^{-\frac{x}{b}+\frac{1}{2}}}{\Gamma(x/b)}\,\Gamma_b(x)~,
\end{equation}
which determine it recursively from the normalization and special values
\begin{equation}
    \Gamma_b\Big(\frac{Q}{2}\Big) = 1~,\qquad
    \Gamma_b(b) = \frac{b^{1/2}}{\sqrt{2\pi}}\,\Gamma_b(Q)~,\qquad
    \Gamma_b(x)\sim \frac{\Gamma_b(Q)}{2\pi x}~,\quad x \rightarrow 0~.
\end{equation}
The Upsilon function of the DOZZ literature is related to $\Gamma_b$ as following
\begin{equation}
    \Upsilon_b(x) \equiv \frac{1}{\Gamma_b(x)\,\Gamma_b(Q-x)}~,\qquad
    \Upsilon_b(x+b) = \gamma(bx)\, b^{1-2bx}\,\Upsilon_b(x)~,
\end{equation}
which obeys $\Upsilon_b(x)=\Upsilon_b(Q-x)$, $\Upsilon_b(Q/2)=1$, and has zeros (no poles) on both lattices $x=-mb-nb^{-1}$ and $x=Q+mb+nb^{-1}$ --- the origin of the pole/zero pattern of the DOZZ formula quoted in section \ref{sec:2D}.
\paragraph{Hankel functions.} For $z\rightarrow \infty$,
\begin{equation}
    H_\nu^{(1,2)}(z) \sim \sqrt{\frac{2}{\pi z}}\,\e^{\pm i \left(z - \frac{\nu\pi}{2}-\frac{\pi}{4}\right)}~,
\end{equation}
and at half-integer order they truncate to elementary functions,
\begin{equation}
    H^{(1)}_{\frac{1}{2}}(z) = -i\sqrt{\frac{2}{\pi z}}\,\e^{iz}~,\qquad
    H^{(1)}_{\frac{3}{2}}(z) = -\sqrt{\frac{2}{\pi z}}\,\e^{iz}\left(1+\frac{i}{z}\right)~,
\end{equation}
with $H^{(2)}_\nu(z) = \big(H^{(1)}_\nu(z^*)\big)^*$.
\paragraph{Heat-kernel regularization.} For $\epsilon>0$,
\begin{equation}\label{eq:compendium log trick}
    -\frac{1}{2}\log \frac{x\,\epsilon^2\,\e^{2\gamma_E}}{4} = \int_0^\infty \frac{\d t}{2t}\,\e^{-\frac{\epsilon^2}{4t}}\,\e^{-xt}~.
\end{equation}
\paragraph{Hurwitz zeta function.} $\zeta(s,a)=\sum_{n\geq0}(n+a)^{-s}$, with
\begin{equation}
    \zeta'(0,a) = \log\Gamma(a) - \tfrac{1}{2}\log(2\pi)~,\qquad \zeta(-n,a) = -\frac{B_{n+1}(a)}{n+1}~,
\end{equation}
and the special values
\begin{equation}
    \zeta'(0) = -\tfrac{1}{2}\log(2\pi)~,\qquad \zeta'(-1)= \tfrac{1}{12}-\log A~,\qquad \zeta'(-2) = -\frac{\zeta(3)}{4\pi^2}~,
\end{equation}
where $A$ is Glaisher's constant.
\paragraph{Harmonic numbers and the Beta function.} $H_x = \gamma_E + \psi(x+1)$, in particular
\begin{equation}
    H_{\frac12} = 2-2\log 2~,\quad H_1 = 1~,\quad H_{\frac32}=\tfrac{8}{3}-2\log2~,\quad H_2 = \tfrac{3}{2}~,\quad H_4 = \tfrac{25}{12}~,
\end{equation}
and $B(x,y) = \Gamma(x)\Gamma(y)/\Gamma(x+y)$.
\paragraph{Fourier sums on the circle.} ($\theta\sim\theta+2\pi$, $\overline{\Delta}\equiv 1-\Delta$):
\begin{equation}
    \sum_{n\in\mathbb{Z}} \e^{in\theta} = 2\pi\,\delta(\theta)~,\qquad
    \sum_{n\in \mathbb{Z}} \frac{\Gamma(\overline{\Delta}+n)}{\Gamma(\Delta+n)}\, \e^{in\theta}
    = \frac{\Gamma(2\overline{\Delta})\cos(\pi \overline{\Delta})}{2^{2\overline{\Delta}-1}}
    \left(\frac{1}{\sin^2\frac{\theta}{2}}\right)^{\overline{\Delta}}~.
\end{equation}

\subsection{Spheres, harmonics and group volumes}
\paragraph{Volumes and dimensions.}
\begin{equation}
    \mathrm{vol}(S^n) = \frac{2\pi^{\frac{n+1}{2}}}{\Gamma\!\left(\frac{n+1}{2}\right)}~,\qquad
    \mathrm{vol}\big(\mathrm{SO}(n)\big) = \prod_{k=2}^{n}\mathrm{vol}(S^{k-1})~,\qquad
    \dim \mathrm{SO}(n) = \frac{n(n-1)}{2}~,
\end{equation}
in the normalization used throughout these notes; explicitly
\begin{equation}
    \mathrm{vol}(\mathrm{SO}(3)) = 8\pi^2~,\qquad
    \mathrm{vol}(\mathrm{SO}(4)) = (2\pi)^4~,\qquad
    \mathrm{vol}(\mathrm{SO}(5)) = \frac{128\pi^6}{3}~.
\end{equation}
\paragraph{Scalar harmonics on $S^{d+1}$} (unit radius):
\begin{equation}\label{eq:degeneracy Sdp1}
    -\nabla^2 Y_{l} = l(l+d)\,Y_{l}~,\qquad
    D_l^{(d+2)} = \frac{(2l+d)\,\Gamma(l+d)}{\Gamma(d+1)\,\Gamma(l+1)}~,\qquad l \geq 0~,
\end{equation}
e.g.\ $D^{(3)}_l = 2l+1$ on $S^2$, $D^{(4)}_l = (l+1)^2$ on $S^3$, and $D^{(5)}_l = \tfrac{1}{6}(l+1)(l+2)(2l+3)$ on $S^4$.
\paragraph{Transverse traceless spin-$s$ harmonics on $S^{d+1}$.}
\begin{equation}\label{eq:eigenvalues spin s Laplacian}
    -\nabla^2_{(s)} f_{l,(s)} = \big[l(l+d)-s\big]\,f_{l,(s)}~,\qquad l\geq s~,\qquad
    \nabla\cdot f_{l,(s)}=0~,\quad \mathrm{tr}\, f_{l,(s)}=0~.
\end{equation}
The degeneracy is
\begin{align}\label{eq:eigenvalues spin s Laplacian}
    D_{l,s}^{(d+2)}&= g_s \frac{(l-s+1)(l+s+d-1)(2l+d)(l+d-2)!}{d!(l+1)!}~,\cr
    g_s&= \frac{(2s+d-2)(s+d-3)!}{(d-2)!s!}
\end{align}
For $s=1$, $l=1$ these are the Killing vectors: their number is $\dim\mathrm{SO}(d+2)$, and they are zero modes of $-\nabla^2_{(1)}-d$.

\subsection{Representation data and characters}
\paragraph{Casimir and masses.}
\begin{equation}
    \mathcal{C}_2 = \Delta(\Delta-d) + s(s+d-2)~,
\end{equation}
\begin{equation}
    m^2\ell^2 = \Delta(d-\Delta)~~(s=0)~,\qquad
    m^2 \ell^2= (\Delta+s-2)(d+s-2-\Delta)~~(s\geq1)~,
\end{equation}
with the partially massless values and the Higuchi bound ($t$ the depth)
\begin{equation}
    m^2_{s,t} \ell^2= (s-t-1)(d+s+t-3)~,\quad t = 0,1,\ldots,s-1~,\qquad m^2_{\mathrm{H.b.}} = m^2_{s,0}~.
\end{equation}
In dS$_2$: $\Delta = \tfrac{1}{2}\big(1\pm\sqrt{1-4m^2\ell^2}\big)$, and the discrete series has $m^2\ell^2 = -t(t+1)$, $\Delta = 1+t$, $t \in \mathbb{N}_0$.

\subsection{Coordinates, embeddings and the invariant distance}
All slicings solve $-(X^0)^2 + (X^1)^2+\ldots+(X^{d+1})^2 = \ell^2$.
\begin{center}
\renewcommand{\arraystretch}{1.7}
\begin{tabular}{l l l}
\hline
patch & embedding & metric $ds^2/\ell^2$ \\
\hline
global & $X^0 = \sinh\tau~,\; X^i = \cosh\tau\, y^i$ & $-\d\tau^2 + \cosh^2\!\tau\, \d\Omega_d^2$\\
static & $X^0 = \cos\rho\sinh t~,\; X^{d+1}=\cos\rho\cosh t~,$ & $-\cos^2\!\rho\,\d t^2 + \d\rho^2 + \sin^2\!\rho\,\d\Omega_{d-1}^2$\\
& $X^i = \sin\rho\, n^i$ & \\
planar & $X^0 = \frac{1-\eta^2+\vec{x}^{\,2}}{-2\eta}~,\; X^{d+1} = \frac{1+\eta^2-\vec{x}^{\,2}}{-2\eta}~,$ & $\frac{-\d\eta^2 + \d\vec{x}^{\,2}}{\eta^2}~,\quad \eta<0$\\
& $X^i = \frac{x^i}{-\eta}$ & \\
\hline
\end{tabular}
\end{center}
(embedding coordinates in units of $\ell$; $y^iy_i = n^in_i = 1$).
\paragraph{Curvature data.} For dS$_{d+1}$ / $S^{d+1}$ of radius $\ell$:
\begin{equation}
    \Lambda = \frac{d(d-1)}{2\ell^2}~,\quad
    R = \frac{d(d+1)}{\ell^2}~,\quad
    R_{\mu\nu} = \frac{d}{\ell^2}g_{\mu\nu}~,\quad
    A_{\mathrm{hor}} = \ell^{d-1}\mathrm{vol}(S^{d-1})~.
\end{equation}

\subsection{Flipping formulas}
For a character $\chi(q) = \sum_k c_k q^k$ with finitely many negative powers,
\begin{equation}
    [\chi]_+ \equiv \chi - c_0 - \sum_{k<0} c_k\big(q^{k}+q^{-k}\big)
\end{equation}
(drop the constant, reflect negative powers with a sign flip), while for a full integrand $F(q)=\sum_k c_k q^k$,
\begin{equation}
    \{F\}_+ \equiv \sum_{k<0}c_k\,q^{-k} + \sum_{k\geq0}c_k\,q^k
\end{equation}
(keep the constant, reflect negative powers keeping signs). They are related by
\begin{equation}
    \{F\}_+ = \frac{1+q}{1-q}\Big([\chi_{\mathrm{bulk}}]_+ - [\chi_{\mathrm{edge}}]_+ + c_0\Big)~,\qquad
    F = \frac{1+q}{1-q}\big(\chi_{\mathrm{bulk}}-\chi_{\mathrm{edge}}\big)~,
\end{equation}
with $c_0$ the constant term of $F$; for a depth-$(s-1)$ massless field $c_0 = -2N^{\mathrm{KT}}_{1}$ with $N^{\mathrm{KT}}_1 = \dim\mathrm{SO}(d+2)$ the number of Killing vectors. The negative coefficients of $F$ count the phase of the gravitational path integral, $\mathcal{Z}\supset \e^{\pm\frac{i\pi}{2}P_s}$ with $P_s = \sum_{\alpha<0}|d_\alpha|$.

\bibliographystyle{JHEP}
\bibliography{bib}
\end{document}